\documentclass[twocolumn, nofootinbib]{revtex4-2}
\usepackage{hyperref}
\newcounter{daggerfootnote}
\newcommand*{\daggerfootnote}[1]{%
    \setcounter{daggerfootnote}{\value{footnote}}%
    \renewcommand*{\thefootnote}{\fnsymbol{footnote}}%
    \footnote[2]{#1}%
    \setcounter{footnote}{\value{daggerfootnote}}%
    \renewcommand*{\thefootnote}{\arabic{footnote}}%
    }

\makeatletter
\def\maketitle{
\@author@finish
\title@column\titleblock@produce
\suppressfloats[t]
}
\makeatother

\usepackage{amsmath,amsopn}
\usepackage{amssymb}
\usepackage{graphicx}
\usepackage{dcolumn}
\usepackage{bm}
\usepackage[usenames, dvipsnames]{color}

\newcommand{\QATOP}[2]{\genfrac{}{}{0pt}{}{#1}{#2}}
\newcommand{\func}[1]{\operatorname{#1}}
\input{tcilatex}

\begin{document}

\title{Wannier-Orbital theory and ARPES for the quasi-1D conductor LiMo$_{6}$%
O$_{17}$.\\
Part I: Six-band $t_{2g}$ Hamiltonian.}
\author{L. Dudy}
\affiliation{Randall Laboratory, University of Michigan, Ann Arbor, MI 48109, USA}
\affiliation{Physikalisches Institut und R\"ontgen Center for Complex Material Systems,
Universit\"at W\"urzburg, D-97074 W\"urzburg, Germany\\
}
\affiliation{Synchrotron SOLEIL, L'Orme des Merisiers, 91190 Saint-Aubin, France}
\author{J.W. Allen}
\affiliation{Randall Laboratory, University of Michigan, Ann Arbor, MI 48109, USA}
\author{J.D. Denlinger}
\affiliation{Advanced Light Source, Lawrence Berkeley National Laboratory, Berkeley, CA
94270, USA}
\author{J. He\daggerfootnote{deceased in 2021.}}
\affiliation{Department of Physics and Astronomy, Clemson University, Clemson, SC 29534,
USA}
\author{M. Greenblatt}
\affiliation{Department of Chemistry \& Chemical Biology, Rutgers University, 123 Bevier
Rd. Piscataway, NJ 08854, USA}
\author{M.W. Haverkort}
\affiliation{Max-Planck-Institut f\"ur Festk\"orperforschung, Heisenbergstrasse 1,
D-70569 Stuttgart, Germany}
\affiliation{Max-Planck-Institut f\"ur Chemische Physik fester Stoffe, N\"othnitzer Str.
40, D-01187 Dresden, Germany\\
}
\affiliation{Institut f\"ur Theoretische Physik, Universit\"at Heidelberg, Philosophenweg
16, D-69120 Heidelberg, Germany}
\author{Y. Nohara}
\affiliation{Max-Planck-Institut f\"ur Festk\"orperforschung, Heisenbergstrasse 1,
D-70569 Stuttgart, Germany}
\author{O.K. Andersen}
\affiliation{Max-Planck-Institut f\"ur Festk\"orperforschung, Heisenbergstrasse 1,
D-70569 Stuttgart, Germany}
\email{oka@fkf.mpg.de}

\begin{abstract}
In this and the two following papers, we present the results of a combined
study by density-functional (LDA) band theory and angle-resolved
photoemission spectroscopy (ARPES) of lithium purple bronze, Li$_{1x}$Mo$%
_{6} $O$_{17}$. This material is particularly notable for its unusually
robust quasi-one-dimensional (quasi-1D) behavior. The band structure, in a
large energy window around the Fermi energy, is basically 2D and formed by
three Mo $t_{2g}$-like extended Wannier orbitals (WOs), each one giving rise
to a 1D band running at a 120$^{\circ }$ angle to the two others. A
structural "dimerization" from $\mathbf{c}/2$ to $\mathbf{c}$ gaps the $xz$
and $yz$ bands while leaving the $xy$ bands metallic in the gap but
resonantly coupled to the gap edges and, hence, to the two other directions.
The resulting complex shape of the quasi-1D Fermi surface (FS), verified by
our ARPES, thus depends strongly on the Fermi energy position in the gap,
implying a great sensitivity to Li stoichiometry of properties dependent on
the FS, such as FS nesting or superconductivity. The theory is verified in
detail by the recognition and application of an ARPES selection rule that
enables, for the first time, the separation in ARPES spectra of the two
barely split $xy$ bands and the observation of their complex split FS. The
strong resonances prevent either a two-band tight-binding (TB) model or a
related real-space ladder picture from giving a valid description of the
low-energy electronic structure. Down to a temperature of 6$\,$K we find no
evidence for a theoretically expected downward renormalization of
perpendicular single particle hopping due to LL fluctuations in the quasi-1D
chains. This paper I introduces the material, motivates our study,
summarizes the Nth-order muffin-tin orbital (NMTO) method that we use,
analyzes the crystal structure and the basic electronic structure, and
presents our NMTO calculation of the $t_{2g}$ low-energy WOs and the
resulting tight-binding (TB) Hamiltonian for the six lowest energy bands,
only the four lowest being occupied. Thus this paper sets the theoretical
framework and nomenclature for the following two papers.
\end{abstract}

\date{\today }
\pacs{Valid PACS appear here}
\maketitle

\mbox{} \clearpage
\section{Introduction}

The present paper (I) and its two companion papers (II and III) are devoted
to a detailed study of the band structure of the lithium purple bronze
(LiPB) LiMo$_{6}$O$_{17}$\footnote{{We do not use the conventional name, Li$%
_{0.9}$Mo$_{6}$O}$_{17},$ because the highly accurate ARPES bands to be
described here are filled corresponding to the stoichiometry Li$_{1.02}$Mo$%
_{6}$O$_{17}$ (see Paper III)}, combining angle-resolved photoemission
spectroscopy (ARPES) and Wannier function band theory using the Nth-order
muffin-tin orbital method (NMTO). Since its discovery \cite{Greenblatt1984}
and structure determination \cite{Onoda1987} LiPB has been heavily studied
as a quasi-1D material\footnote{%
Ref.\onlinecite{Dudy2013} summarizes and references prior work dating back
to Ref. \onlinecite{Greenblatt1984}.} \cite{Merino2012,Merino2015, Cho2015,
Lera2015,Wu2016,Platt2016, Chudzinski2017, Lu2019}. Thus, it is notable as
an unusually good and interesting example of the non-Fermi liquid (non-FL)
properties exhibited by one-dimensional (1D) interacting electron systems,
such as in the exactly solvable Tomonaga-Luttinger (TL) model \cite%
{Tomonaga1950,Luttinger1963} or in the more generalized notion of the
Luttinger Liquid (LL) \cite{Giamarchi2004}. A highly non-intuitive example
of such non-FL properties is that the energy ($E$) dependence of the
momentum ($\kappa $) integrated single-particle spectral function, which
would give simply the one-electron density of states in a non-interacting
system, goes to zero upon approaching the Fermi-energy ($E_{F}$) as a power
law $\left( E_{F}-E\right) ^{\alpha }$, with $\alpha $ interpreted as the
anomalous exponent of the TL model.\footnote{%
The power law is valid for $T$=0. For nonzero $T$, the exact dependence
evolves to be quantitatively more complicated but qualitatively similar.}

\begin{figure}[hbt]
\includegraphics[width=\linewidth]{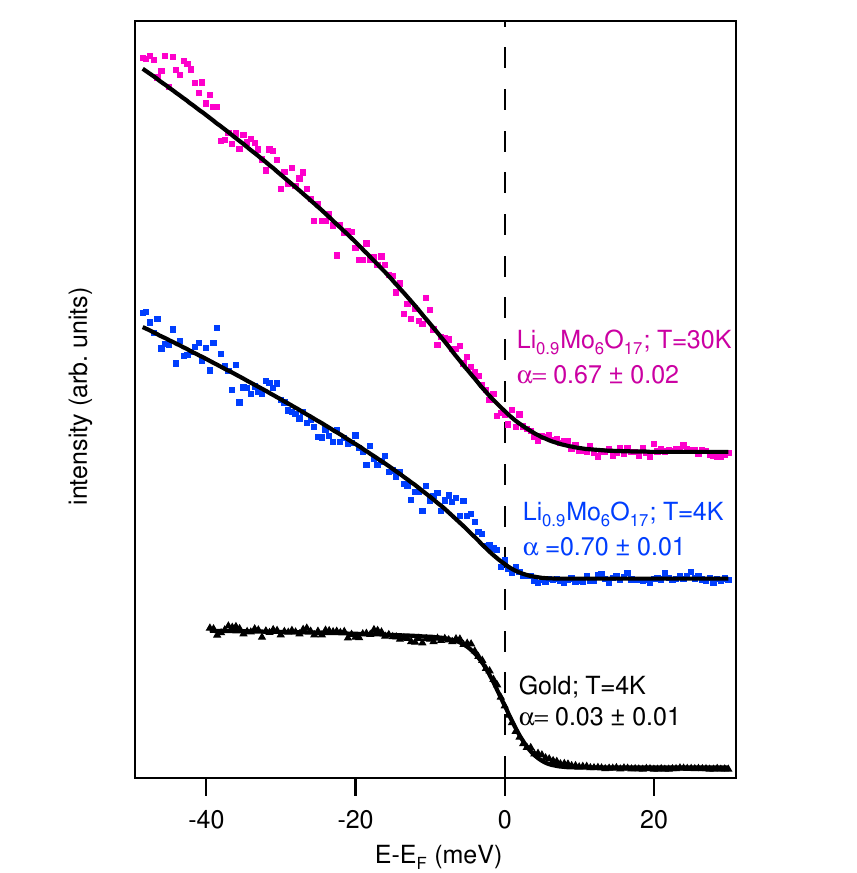}
\caption{Angle integrated photoemission spectra of lithium purple bronze for 
$T$=4 and 30~K taken with a resolution of 5 meV with photon energy $h\protect%
\nu $=8.4~ eV \protect\cite{Dudy2013}. For reference, a gold spectrum with
the same settings is also shown. All the spectra are generally well fitted
by the TL-model lineshape, showing for LiMo$_{6}$O$_{17}$ a value of $%
\protect\alpha \approx 0.7$. In Ref. \onlinecite{Dudy2013}, it was deemed
ambiguous whether a very noisy feature around 5 meV in the fit residuals is
intrinsic or arises from some systematic experimental error. The gold
spectrum fits well with $\protect\alpha $ essentially zero, corresponding
mathematically to a Fermi edge. }
\label{TLexp_old}
\end{figure}

FIG. \ref{TLexp_old} reproduces angle-integrated data from a previous
photoemission study \cite{Dudy2013}, showing this unusual property for LiPB
for the spectral function below $E_{F}$, as probed by ARPES. The spectra for 
$\kappa $-integration along the quasi-1D direction, for temperatures $T$=4~K
and 30~K and resolution 5~meV, are well described by a power law with $%
\alpha $=0.7 over at least 40 meV, compared to the Fermi edge of a $T$=4~K
gold reference spectrum. Also, scanning tunneling spectroscopy \cite%
{Hager2005}, which probes the spectral function on both sides of $E_{F}$,
shows the power law "V-shape" down to 4~K \footnote{%
In Ref. \onlinecite{Dudy2013}, it was deemed ambiguous whether a very noisy
feature around 5 meV in the fit residuals is intrinsic or arises from some
systematic experimental error.}. It is then equally non-intuitive in the TL
model that, nonetheless, the underlying band-structure Fermi momentum $k_{F}$
and thus, the Fermi surface (FS) remains well-defined \cite{Orgad2001_2}.
Paper III presents a detailed determination of the FS for LiPB.

The band structure and, in particular, the magnitude(s) of the transverse
hoppings ($t_{\perp }$) between its 1D chains, and the resulting FS, are
especially interesting and important for LiPB. The general theoretical
expectation \cite{Giamarchi2004} is that, for $T$ decreasing below a scale
set by $t_{\perp }$, LL behavior is unstable against dimensional crossover
from 1D to some sort of 3D Fermi-liquid (FL) behavior, typically by a phase
transition to some 3D ordered state like a charge or spin density wave (CDW
or SDW). Band calculations to date suggest values of $t_{\perp }$=20~meV
(232 K) and yet the data of FIG.~\ref{TLexp_old} indicate that its non-FL 1D
properties likely last until the material goes superconducting (SC) at $%
T_{SC}$=1.9~K. Indeed other properties of LiPB, albeit novel and
interesting, exhibit no clear evidence for dimensional crossover above $%
T_{SC}$ \cite{Dudy2013}. However, theory \cite{Giamarchi2004} also suggests
that LL fluctuations on the chains can strongly suppress the single-particle
hopping and consequently the crossover $T$. For example, in the case of one
chain per primitive cell and hopping only to the nearest chain, the
suppression of $t_{\perp }$ is by the factor $(t_{\perp }/t)^{\alpha
/(1-\alpha )}$, where $t\gg t_{\perp }$\ is the hopping along the chains.%
\textbf{\ }For a typical band-theory value of $\left\vert t\right\vert $=0.8
eV \cite{Chudzinski2012}\textbf{\ }and the value of $\alpha $=0.7 cited
above, one obtains $t_{\mathrm{eff}\,\perp }$ = 4$\,\mu $eV or 0.04$\,$K,
even smaller than $T_{SC}$.\footnote{%
Some measurements \cite{Wang2006} have yielded a smaller $\alpha $=0.6, for
which the effective $t_{\perp }$ is larger, 80$\,\mu $eV or 0.9$\,$K, still
smaller than $T_{SC}.$}. Such a small value might thus account for the
exceptional stability of 1D physics in this material, and should be manifest
in the low-$T$ single-particle electronic structure. However, up to now, the
transverse hopping and resulting FS have never been measured experimentally
or characterized theoretically as fully as is needed and possible.

There is additional motivation for our study. As described in detail further
below, LiPB is complex in having two approximately 1D bands associated with
there being two equivalent chains (and two formula units) per primitive
cell, each half filled for stoichiometric LiMo$_{6}$O$_{17}$. Thus most LiPB
theories to date \cite{Chudzinski2012,Merino2012,Merino2015, Cho2015,
Lera2015, Platt2016, Chudzinski2017, Lu2019} have modeled the quasi-1D
electrons as a lattice of pairs of chains regarded as ladders, with simple
tight-binding (TB) $t_{\perp }$ and $t_{\perp }^{\prime }$ parameters for
nearest neighbor intra- and inter-ladder hopping, respectively. So it is of
great interest to check the validity of the ladder picture, which involves
the relative magnitudes of the perpendicular hoppings within and between
primitive cells. These hoppings determine the perpendicular dispersion and
splitting of the two bands forming the FS. Of particular interest is the
normal state FS giving rise to SC. The FS also gives the clearest
experimental access to the details of the transverse hoppings.

Another motivation is to demonstrate the use of the NMTO method for creating
chemically meaningful Wannier functions --in the present case Wannier
orbitals (WOs) centered on Mo1, the only octahedrally fully coordinated
molybdenum (Sect. \ref{SectDims})-- and their TB Hamiltonian, and to
establish them as important tools for predicting and interpreting the ARPES
data. As summarized in more detail below, like the study in Ref. %
\onlinecite{Nuss2014}, our theory uses the local density-functional
approximation (DFT-LDA) to derive a set of localized Wannier functions,
which, however, in our case, is complete in the sense that it contains all
three Mo1$\,$4$d\,t_{2g}$-like orbitals per formula unit, and thereby spans
the occupied as well as the lowest empty bands. The two quasi-1D metallic
bands are $xy$-like and situated in a 0.4 eV gap between valence bands
formed by the $xz$ and $yz$ WOs, bonding between the ladder rungs, and
conduction bands formed by the same WOs, but antibonding between rungs.
After integrating out the $xz$ and $yz$ degrees of freedom (in Paper III),
our theory leads to the conclusions, that the effective transverse couplings
between the two quasi-1D bands cannot be described by a simple TB model, and
also that they have very long range, making ladders ill-defined. In this
respect all previous TB ladder models are very unrealistic.

The theory also leads to a selection rule (in Paper II) that enables the two
barely split quasi-1D bands to be separated in the ARPES spectra near $E_{F}$
for the first time. The split and warped FS obtained thereby in ARPES at 6~K
is in excellent agreement with the predicted FS, giving a detailed
confirmation of the theory (Paper III). This means that the predicted LL
renormalization of the perpendicular hoppings with decreased temperature
does not occur, and so cannot be the origin of the robustness of the LiPB 1D
behavior. We can also infer that the LDA FS is the normal state FS relevant
for theories \cite{Cho2015,Lera2015,Platt2016,Sepper2013,Lebed2013} of the
SC. We note that the occurrence of SC is sample dependent \cite{Xu2009},
and, in this context, that the details of the theoretical FS shape are
extremely sensitive to the position of $E_{F}$, which is controlled by the
Li-content (or the content of oxygen vacancies). For our samples, the $E_{F}$
position indicates that they are very nearly stoichiometric, which is the
circumstance found in theory to give the most 1D FS. Although we have not
explicitly verified SC for our samples, these findings are consistent with
the hypothesis that SC has a 1D origin\footnote{%
The SC upper critical field is much larger than the Pauli-limiting value\cite%
{Wakeham2011}, suggesting unconventional pairing arising from an essentially
1D normal state} and that the absence of SC in some samples may be linked to
sample stoichiometry through the sensitivity of the FS. Finally, although $T$
dependence was not a particular focus of the experiments, we find the \emph{%
same} FS at 30K, implying that a mysterious resistivity upturn below $%
T_{M}\approx $ 25K is not likely to be associated with a gross change in
electronic structure \cite{Dudy2013}.

Before proceeding we emphasize that our purpose is not merely to present the numerical results of yet another DFT-LDA calculation for LiPB.  Rather, DFT-LDA is a tool to implement the overall program described in the three papers.  We use it to obtain chemically meaningful NMTO WOs, which, in turn, we use to gain new insight into how the numerical results come about.  The central theory result is a portable six-band analytic TB Hamiltonian to describe the low-energy band structure.  The agreement of its eigenvalues with the ARPES band structure is already generally good using LDA parameter values and can be made excellent by some additional adjustments, showing that its functional forms are faithful to the physics.  Further, along with its underlying WOs, it can be used to understand the complex ARPES intensity variations in unprecedented detail, in particular the selection rule already mentioned.  Ultimately the combination of new theory insight and ARPES experiment yields new knowledge of the details of the Fermi surface and the magnitude and range of the perpendicular hopping for the two metallic bands.

In the remainder of this introductory section, we give a more detailed
overview of the theory relative to previous work, and describe the division
of content between the three papers.

The basic band structure in the vicinity of $E_{F}$ has been known for many
years from pioneering TB calculations based on the semi-empirical extended-H%
\"{u}ckel method \cite{Whangbo1988}. There are two approximately 1D-bands
dispersing across $E_{F}$, associated with there being two equivalent chains
of Mo atoms having a zigzag arrangement (zig-Mo1-zag-Mo4-zig), and two
formula units per primitive cell. The two bands have Mo 4$d_{xy}$ character
and for stoichiometric LiMo$_{6}$O$_{17}$ they are half-filled. There are
also two filled bands not far below $E_{F}$.

Quantitatively correct band structures require charge-self-consistent DFT
calculations, not a small task for a transition-metal oxide with 48 atoms
per cell, so it took nearly twenty years for the first self-consistent DFT
(LDA) band structure to appear \cite{Satpathy2006} and another six for the
second \cite{Jarlborg2012}. Both calculations were performed with the linear
muffin-tin orbital method (LMTO) in the atomic-spheres approximation. Such
LDA-LMTO band structures provided guidance for the TB band-structure
parameters used in early many-body models \cite{Merino2012,Chudzinski2012}.
Higher-resolution low-temperature ARPES data and more accurate NMTO
calculations show agreement even on the details \cite{13CORPESAllenHaverkort}
of the filled bands.

An alternative TB model \cite{Nuss2014} has been derived by first using the
highly accurate full-potential linear augmented-plane-wave (LAPW) method to
perform a charge self-consistent DFT (LDA) calculation of the band structure
over a wide energy range, and then projecting from it a set of four
so-called maximally localized Wannier functions, which describe the two
quasi-1D bands and the two valence bands. The Wannier functions of this
model are therefore not atomic, but essentially the bonding linear
combination of those on Mo1 and Mo4, and the integral for hopping between
these $xy$-bond orbitals is only about half the one for hopping between the
atomic orbitals considered in the TB models previously used \cite%
{Merino2012,Chudzinski2012}. The study of Nuss and Aichhorn \cite{Nuss2014}
also provides a simplified two-band TB Hamiltonian by folding the two
occupied $xz$ and $yz$ bands down into the two $xy$ bands, thereby becoming $%
\widetilde{xy}$ bands, and fitting their hybridization such as to modify the 
$t_{\perp }$ parameters. The result is said to be in good agreement with
those discussed in Ref. \onlinecite{Chudzinski2012}.

For the theory of the present paper and its two companion papers, early
results of which were given in Ref. \cite{13CORPESAllenHaverkort}, we need
and provide an improved 3D visualization of the crystal structure, with an
associated wording (Sect. \ref{crystal_structure} of the present paper):
ribbons containing Mo1, Mo2, Mo4, and Mo5 for zigzag chains and bi-ribbons
for ladders, and an overview of the electronic structure (Sect. \ref%
{SectElStruc} below). In the theory, we perform an LDA Wannier-function
calculation with the new full-potential version \cite{Nohara2016} of the
NMTO method. We obtain the set of all three (per formula unit) Mo1$%
\,4d\,t_{2g}$ WOs, not only the $xy$ orbitals, but also the $xz$ and $yz$
orbitals. Also the latter form 1D bands, but with primitive translations $(%
\mathbf{c\pm b})/2$ until the dimerization to $\mathbf{c\pm b}$ gaps them
around $E_{F}$. Indeed (Sect. \ref{SectDims} below), the structural reason
why LiPB is 1D while (most) other Mo bronzes are 2D \cite{Pouget1991,
Whangbo1991, Denlinger1999,Breuer1995} is exactly this $c/2$ to $c$
dimerization of the ribbons (zigzag chains) into bi-ribbons (the two zigzag
chains are not related by translation). Note that this dimerization of the $%
xz$ and $yz$ bands causing them to gap at $2k_{F}$ is distinct from the $b/2$
to $b$ dimerization causing the $xy$ bands to gap at $4k_{F}$ --and which we
neglect, as did Nuss and Aichhorn (Sect.~\ref{Sectbims} below).
Hybridization between the resulting valence and conduction bands and the
metallic $xy$ bands\footnote{%
We call the $xy$ band the metallic band and, like for semiconductors, call
the gapped $xz$ and $yz$ bands valence and conduction bands.} induces
striking $k_{\perp }$=$k_{c}$-dependent features (FIG.~\ref{ThreeBands}
below, FIG.s$~$II$~$\ref{CEC} and \ref{ARPES_Bandstructure_TB}~(c2), and
FIG.~s~III~\ref{Theo_CEC} and \ref{Fig:NewFSExtract}\footnote{%
I, II, and III refer to sections, figures, and equations in Paper I, II, and
III, respectively.\label{Paper I}}). These features depend strongly on the
energy position in the gap. Therefore the resulting FS warping and splitting
also has features that depend strongly on the value of $E_{F}$, as set by
the effective Li stoichiometry. Furthermore, this $E_{F}$ dependence of the
perpendicular dispersion cannot be captured with a Wannier basis which in
addition the metallic $xy$ orbitals contain only the \emph{occupied} $xz$
and $yz$ orbitals \cite{Nuss2014}. We, therefore, include WOs which account
not only for the valence but also for the conduction bands, leading to a
very accurate and yet portable (i.e. analytical) $t_{2g}$ six-band TB
Hamiltonian. Subsequent analytical L\"{o}wdin downfolding to a two-band
Hamiltonian, which has resonance- rather than TB form, enables a new and
detailed understanding of all the various microscopic contributions to the
perpendicular dispersion, and their relation to the crystal structure and to
the FS.

The details of the theoretical method including all six Mo1 $t_{2g}$ WOs and
their TB Hamiltonian are presented in Sect.s~\ref{SectElCalc}, \ref{SectLowE}%
, and \ref{SectH} of this Paper I. The theory of the ARPES intensity
variations and its application to LiPB are presented in Sect.~\ref%
{SectIntensity} of the following Paper II, and the details of the downfolded
two-band Hamiltonian and the resulting FS are presented in Paper III. The
theory is validated in detail by new higher resolution ARPES experiments for
two different samples, down to temperatures of 6~K and 30~K. The data and
the analysis results are presented at appropriate places in the course of
the presentation of the theory in the three papers. As found previously \cite%
{13CORPESAllenHaverkort, Nuss2014} there is very good general agreement with
LDA dispersions up to 150~meV below $E_{F}$. Refinement of the LDA-derived
parameters of the six-band Hamiltonian yields an accurate and detailed
description of the ARPES low-energy band structure (Sect.~\ref{SectAgreement}
in Paper II), including the striking features of the $xy$-like bands and the
associated distinctive FS features (FIG.~\ref{Fig:NewFSExtract} in Paper
III). As mentioned already, the direct observation of these features, not
identified in our previous ARPES studies, is enabled by the recognition and
application of a selection rule (Sect. \ref{Sectzoneselect} in Paper II)
according to which the $c$-axis dimerization gaps the energy bands, but
--for a range of photon energies-- has negligible effects on the ARPES
intensities.

\section{NMTO Method}

\label{SectElCalc}

The electronic-structure calculations were performed for the stoichiometric
crystal with the structure determined for LiMo$_{6}$O$_{17}$~\cite{Onoda1987}%
. Doping --which is small due to the opposing effects of Li intercalation
and O deficiencies-- was treated in the rigid-band approximation.

For the DFT-LDA \cite{Barth1972} calculations, including the generation of
Wannier functions and their TB Hamiltonian,  for the Kohn-Sham \cite{Kohn1965}  one-electron energies, $E_j^{\mathbf{k}}$, and eigenvectors, $u_{Rlm,j}^{\mathbf{k}}$, we used the recently developed
self consistent full-potential version \cite{Nohara2016} of the $N$th-order
--also called 3rd-generation$-$ muffin-tin orbital (NMTO) method \cite%
{Andersen2000,Tank2000}, a descendant \cite{00Odile} of the classical linear
muffin-tin orbital (LMTO) method \cite{Andersen1975}\cite{Andersen1984}.
Since NMTOs were hitherto generated for overlapping MT potentials imported
from self-consistent LMTO-ASA or linear augmented plane wave (LAPW)
calculations \cite%
{Pavarini2005,Zurek2005,Yamasaki2006,Lechermann2006,Boeri2007,Liu2008,Kent2008,Tanusri2009,Zurek2010,Andersen2011,Haverkort2012}
rather than self consistently in full-potential calculations, and since NMTO
Wannier orbitals (WOs) are generated in a very different way than maximally
localized Wannier functions \cite{Marzari2012}, making them useful for
many-body calculations also for $d$- and $f$-electron atoms at low-symmetry
positions\footnote{%
For materials with $d$- or $f$-electron atoms exclusively at high-symmetry
positions, maximally localized and NMTO Wannier functions (WFs) give similar
results when settings are similar \cite{Lechermann2006}. However, maximally
localized WFs are usually not centered at low-symmetry sites, and if forced
to, they generally do not transform according to the irreducible
representations of the point group. As a consequence, crystal fields depend
strongly on the settings. The software found on www.quanty.org interfaces
several methods for generating WFs and allows users to compare results.\label%
{CompMaxLoc copy(1)}}, here follows a concise description of our method as
applied to LiPB. More complete and pedagogical accounts of the formalism may
be found e.g. in Ref.s \cite{Tank2000,00Odile} and \cite{12Juelich}.

As illustrated in Chart (\ref{SCF}) we first generate the full potential, $%
V\left( \mathbf{r}\right) ,$ by charge self-consistent LDA calculation using
a relatively large basis set, $\chi _{Rlm}^{\mathbf{k}}\left( \mathbf{r}%
\right) ,$ consisting of the Bloch sums of the two Li~2$s$ NMTOs per
primitive cell, of all 60 Mo~4$d$ NMTOs, of all 136 O~2$s$ and 2$p$ NMTOs,
plus 138 1$s$ NMTOs on the interstitial sites (\textrm{E}) with MT radii
exceeding 1~Bohr radius. The resulting number of 336 NMTOs/cell is smaller
than the number of LMTOs \cite{Satpathy2006,Jarlborg2012} -- and an order of
magnitude smaller than the number of LAPWs \cite{Nuss2014} needed for LiPB.

\begin{figure}[h]
\begin{center}
Calculating $V\left( \mathbf{r}\right) $ and $v_{R}\left( \left\vert \mathbf{%
r-R}\right\vert \right) $ self-consistently
\par
using the LDA with the basis of 336 NMTOs/cell:%
\begin{equation}
\begin{tabular}{ccc}
$H_{Rlm,R^{\prime }l^{\prime }m^{\prime }}^{\mathbf{k}}\,\&\,O_{Rlm,R^{%
\prime }l^{\prime }m^{\prime }}^{\mathbf{k}}$ & $\rightarrow $ & $E_{j}^{%
\mathbf{k}}\,\&\,u_{Rlm,j}^{\mathbf{k}}$ \\ 
$\uparrow $ & $Rlm$ & $\downarrow $ \\ 
$E_{0,.,N}\,\&\,\chi _{Rlm}^{\mathbf{k}}\left( \mathbf{r}\right) $ & $\in $
& $\rho \left( \mathbf{r}\right) $ \\ 
$\uparrow $ & $336$ & $\downarrow $ \\ 
$v_{R}\left( \left\vert \mathbf{r-R}\right\vert \right) $ & $\leftarrow $ & $%
V\left( \mathbf{r}\right) $ \\ 
$\downarrow $ &  & $\downarrow $ \\ 
(\ref{6WO}) &  & (\ref{6WO})%
\end{tabular}
\label{SCF}
\end{equation}%
\end{center}
\end{figure}

After each iteration towards self-consistency, $V\left( \mathbf{r}\right) $
is least-squares fitted to an \emph{overlapping} MT potential (OMTP) \cite%
{Zwierzycki2009}, which is a constant, the MT zero, plus a superposition of
spherically symmetric potential wells, $\sum_{R}v_{R}\left( \left\vert 
\mathbf{r-R}\right\vert \right) ,$ centered at the atoms and larger
interstitials. The ranges of the potential wells, the MT radii $s_{R},$ were
chosen to overlap by 25\%. Specifically: $s_{\mathrm{Li}}$=2.87, $s_{\mathrm{%
Mo}}$=2.34-2.55, $s_{\mathrm{O}}$=1.72-1.89, and $s_{\mathrm{E}}$=1.03-2.48
Bohr radii. The overlaps considerably improve the fit to the full potential
and reduce the MT discontinuities of the potential and, hence, the
curvatures of the basis functions\footnote{%
The LMTOs of Methfessel and Schilfgaarde \cite{00Methfessel} are defined for
a conventional MT potential, but are modified in the interstitial near the
MTs to avoid large discontinuities of the orbital curvatures. Also, the
LMTOs of Wills et al. \cite{00Wills,Wills2010,Lejaeghere2016} are defined
for MTs without overlap, but are not modified. As a consequence, multiple-$%
\kappa $ sets are needed.}$.$ The OMTP is used to generate the NMTO basis
set for the next iteration towards charge self consistency and --this being
reached-- to generate the massively downfolded basis set consisting of the 6
Bloch sums of the Mo1~4$d\left( t_{2g}\right) $ NMTOs which --after
symmetrical orthonormalization and Fourier transformation (FT) (\ref{FT})
back to real space [see Chart (\ref{6WO})]-- becomes the set of WOs
describing the 6 bands around the Fermi level. The full potential, $V\left( 
\mathbf{r}\right) ,$ enables us to accurately include in the 6-band TB
Hamiltonian crystal-field terms, such as the one between the $xy$- and the $%
xz$- or the $yz$ WOs which decisively influence the resonance peak in the
metallic $xy$-like band [see Sect.~\ref{Sectxy-xz} in Paper III]. 
\begin{figure}[h]
\begin{center}
Constructing the 6 WOs and their TB Hamiltonian:%
\begin{equation}
\begin{tabular}{ccc}
$v_{R}\left( \left\vert \mathbf{r-R}\right\vert \right) $ & From (\ref{SCF})
& $V\left( \mathbf{r}\right) $ \\ 
$\downarrow $ & and Eq.$\,$(\ref{Dwf}) & $\downarrow $ \\ 
$E_{0,.,N}\,\&\,\chi _{Rm}^{\mathbf{k}}\left( \mathbf{r}\right) $ & $%
\longrightarrow $ & $H_{Rm,R^{\prime }m^{\prime }}^{\mathbf{k}%
}\,\&\,O_{Rm,R^{\prime }m^{\prime }}^{\mathbf{k}}$ \\ 
& $Rm\in 6$ & $\downarrow $ \\ 
$w_{Rm}\left( \mathbf{r-R}\right) $ & $\leftarrow \,$FT & $\chi ^{\mathbf{k}%
}\left( \mathbf{r}\right) \left( O^{\mathbf{k}}\right) ^{-\frac{1}{2}}$ \\ 
$\tilde{H}_{Rm,R^{\prime }m^{\prime }}^{TB}$ & $\leftarrow \,$FT & $\left(
O^{\mathbf{k}}\right) ^{-\frac{1}{2}}H^{\mathbf{k}}\left( O^{\mathbf{k}%
}\right) ^{-\frac{1}{2}}$%
\end{tabular}
\label{6WO}
\end{equation}%
\end{center}
\end{figure}

We now describe the construction (\ref{NMTOs}) of the NMTOs which is more
complex than that of e.g. LAPWs, but achieves order(s)-of-magnitude
reduction in the size of the basis-set. Admittedly, some understanding of
solid-state chemistry is required to use NMTOs efficiently to generate WO
sets, but they can provide insights not usually obtained by use of
plane-wave sets and projection of maximally localized Wannier functions \cite%
{Marzari2012}.

\begin{figure}[h]
\begin{center}
Constructing the NMTO set:%
\begin{equation}
\begin{tabular}{l}
Hard-sphere sites \& radii:~$\mathbf{R,~}a_{R}$ \\ 
OMTP wells \& radii:~$v_{R}\left( r\right) ,$ $s_{R}\approx 1.5a_{R}$ \\ 
Energy $(E)$ mesh:~$E_{0,..,\,N}$ \\ 
Radial wave functions:~$\varphi _{Rl}\left( E,r\right) $ \\ 
Phase shifts:~$\eta _{Rl}\left( E\right) $ \\ 
P$\text{artial\ waves}$:~$\left[ \varphi _{Rl}\left( E,r\right) -\varphi
_{Rl}^{o}\left( E,r\right) \right] Y_{lm}\left( \mathbf{\hat{r}}\right) $ \\ 
Screened spherical waves:~$\psi _{Rlm}\left( E,\mathbf{r}\right) $ \\ 
Screened structure (or slope) matrix:~$S_{R^{\prime }l^{\prime }m^{\prime
},Rlm}\left( E\right) $ \\ 
Kinked partial waves:~$\phi _{Rlm}\left( E,\mathbf{r}\right) ,\;$Eq.$\,$(\ref%
{KPW}) \\ 
Kink matrix:~$K_{R^{\prime }l^{\prime }m^{\prime },Rlm}\left( E\right) ,$ Eq.%
$\,$(\ref{K}) \\ 
Downfolding from $K^{336}\left( E\right) $ to $K^{6}\left( E\right) ,$ Eq.$%
\, $(\ref{Dwf}) \\ 
Green matrix:$~G\left( E\right) =K\left( E\right) ^{-1}$ \\ 
Lagrange matrix:~$L_{Rlm,R^{\prime }l^{\prime }m^{\prime }}\left( E_{\nu
}\right) $ \\ 
NMTOs:~$\chi _{Rlm}\left( \mathbf{r}\right) ,\;$Eq.$\,$(\ref{NMTO}) \\ 
Overlap matrix:~$\left\langle \chi _{Rlm}^{\mathbf{k}}\mid \chi _{R^{\prime
}l^{\prime }m^{\prime }}^{\mathbf{k}}\right\rangle \equiv O^{\mathbf{k}}$ \\ 
Hamiltonian matrix:~$\left\langle \chi _{Rlm}^{\mathbf{k}}\left\vert -\Delta
+V\left( \mathbf{r}\right) \right\vert \chi _{R^{\prime }l^{\prime
}m^{\prime }}^{\mathbf{k}}\right\rangle \equiv H^{\mathbf{k}}$%
\end{tabular}
\label{NMTOs}
\end{equation}%
\end{center}
\end{figure}

For each MT well, $v_{R}\left( r\right) ,$ and energy, $E,$ on a $(N+1)$%
\emph{\ point mesh}, the radial Schr\"{o}dinger equations\footnote{%
Actually, the scalar-relativistic Dirac equations.} for $l$=$0,..,l_{R\,\max
}$ are integrated outwards from the origin to the MT radius, $s_{R},$ thus
yielding the radial functions, $\varphi _{Rl}\left( E,r\right) ,$ and their
phase-shifts, $\eta _{Rl}\left( E\right) ,$ which due to the centrifugal
term vanish for all $l\geq l_{\max }\left( R\right) $. Continuing the
integration smoothly inwards --this time over the MT zero-- yields the
phase-shifted free waves, $\varphi _{Rl}^{o}\left( E,r\right) ,$ which we
truncate at and inside the so-called \emph{hard }sphere with radius, $%
a_{R}\approx 0.65s_{R}.$ The differences, $\varphi _{Rl}\left( E,r\right)
-\varphi _{Rl}^{o}\left( E,r\right) ,$ often referred to as \emph{tongues,}
tend smoothly to zero when going outside the MT sphere, and jump
discontinuously to $\varphi _{Rl}\left( E,r\right) $ when going inside the
hard sphere. After multiplication by the appropriate cubic harmonic, $%
Y_{lm}\left( \mathbf{\hat{r}}\right) ,$ these \emph{discontinuous and tongued%
} \emph{partial waves} will be used together with the screened spherical
waves (SSWs), $\psi _{Rlm}\left( E,\mathbf{r}\right) ,$ to be defined below,
to form a set of \emph{kinked partial waves} (KPWs)\footnote{%
KPWs are also called exact, energy-dependent MTOs (EMTOs)\cite%
{94EMTO,Vitos2007}.}, $\phi _{Rlm}\left( E,\mathbf{r}\right) ,$ analogous to
Slater's augmented plane waves (APWs), and --eventually-- of smooth and
energy-\emph{in}dependent NMTOs [see Eq.s (\ref{KPW}) and \ref{NMTO}].
Partial waves with the same $Rlm$ as one of the NMTOs in the basis set are
called \emph{active }($A$) and the remaining partial waves with non-zero
phase shifts \emph{passive.} Since $l_{\max }\left( R\right) \,$=$%
\,4,\,3,\,3,$ and $2$ for $R=$ Mo, O, Li, and E, the vast majority of
partial waves are passive.

In order to combine the many partial waves to the set of KPWs, we first form
the set of \emph{tail-} or \emph{envelope} functions, $\psi _{Rlm}\left( E,%
\mathbf{r}\right) ,$ also called \emph{screened} \emph{spherical waves}
(SSWs): They are wave-equation solutions that satisfy the boundary
conditions that any cubic-harmonic projection around any site, $\hat{P}%
_{R^{\prime }l^{\prime }m^{\prime }}\left( r_{R^{\prime }}\right) \psi
_{Rlm}\left( E,\mathbf{r}\right) ,$ has a node at the hard-sphere radius if $%
R^{\prime }l^{\prime }m^{\prime }$ is active and differs from $Rlm,$ and has
the proper phase shift, $\eta _{R^{\prime }l^{\prime }}\left( E\right) ,$ if 
$R^{\prime }l^{\prime }m^{\prime }$ is passive. This node condition is what
makes the SSW localized --and the more, the larger the basis set, i.e. the
number of active channels. The input to a screening calculation (see Sect.
3.3 in Ref. \cite{12Juelich} or II.B in Ref. \cite{Nohara2016}) is the
energy, the hard-sphere structure, and the passive phase shifts. The output
is the \emph{screened structure-} or \emph{slope matrix} whose element, $%
S_{R^{\prime }l^{\prime }m^{\prime },Rlm}\left( E\right) ,$ gives the slope
of $\psi _{Rlm}\left( E,\mathbf{r}\right) $ at the hard sphere in the active 
$R^{\prime }l^{\prime }m^{\prime }$ channel. The set of screened spherical
waves is then augmented by the partial waves to become the basis set of KPWs
(see e.g. FIG.s 4-6 in Ref. \cite{12Juelich}):%
\begin{equation}
\phi _{Rlm}\left( E,\mathbf{r}\right) =\psi _{Rlm}\left( E,\mathbf{r}\right)
+\left[ \varphi _{Rl}\left( E,r\right) -\varphi _{Rl}^{o}\left( E,r\right) %
\right] Y_{lm}\left( \mathbf{\hat{r}}\right) .  \label{KPW}
\end{equation}%
The KPW, $\phi _{Rlm}\left( E,\mathbf{r}\right) ,$ has a head formed by the
active partial wave with the \emph{same} $Rlm,$ plus passive waves, and a
tail which inside the other MT spheres is formed solely by passive partial
waves. Hence, all active projections of $\phi _{Rlm}\left( E,\mathbf{r}%
\right) $, except its own, vanish. Such a KPW is localized, everywhere
continuous, and everywhere a solution of Schr\"{o}dinger's equation for the
MT potential --\emph{except} at all hard spheres where it has \emph{kinks }%
in the active channels. The kink, $K_{R^{\prime }l^{\prime }m^{\prime
},Rlm}\left( E\right) ,$ at the hard sphere in channel $R^{\prime }l^{\prime
}m^{\prime }$ is 
\begin{equation}
S_{R^{\prime }l^{\prime }m^{\prime },Rlm}\left( E\right) -\delta _{R^{\prime
}l^{\prime }m^{\prime },Rlm}\left. \frac{\partial \ln \varphi
_{Rl}^{o}\left( E,r\right) }{\partial \ln r}\right\vert _{a_{R}}.  \label{K}
\end{equation}%
This kink matrix also equals the MT Hamiltonian minus the energy, i.e. the
kinetic energy, in the KPW representation: $K\left( E\right) =$ $%
\left\langle \phi \left( E\right) \left\vert -\Delta
+\sum_{R}v_{R}-E\right\vert \phi \left( E\right) \right\rangle $. Any linear
combination of KPWs with the property that the kinks from all heads and
tails cancel, is smooth and therefore, by construction, a solution with
energy $E$ of Schr\"{o}dinger's equation for the OMTP --except for the
tongues sticking into neighboring MT spheres and thereby causing errors of
merely 2nd and higher order in the potential overlap. This kink-cancelation
condition gives rise to the screened Korringa, Kohn, Rostoker (KKR) \emph{%
secular} equations of band theory: $K\left( E\right) u=0$.

\emph{Downfolding} of a large to a small set of KPWs corresponds to changing
the phase shifts in the channels to be downfolded (denoted $I$ for
"integrated out") from those of hard spheres, $\eta _{A}\left( E\right) ,$
to the proper phase shifts, $\eta _{Rl}\left( E\right) ,$ and is performed
on the kink matrix (\ref{K}). For example is the kink matrix for the set of
6 KPWs in terms of the blocks of the kink matrix for the 336 set:%
\begin{equation}
K_{AA}^{6}\left( E\right) =K_{AA}^{336}\left( E\right) -K_{AI}^{336}\left(
E\right) K_{II}^{336}\left( E\right) ^{-1}K_{IA}^{336}\left( E\right) .
\label{Dwf}
\end{equation}%
Note, that this downfolding, which is done \emph{prior} to N-ization (\ref%
{NMTO}) [see Chart (\ref{NMTOs})], makes the resulting NMTO set far better
localized and far more accurate than the set obtained by standard L\"{o}wdin
downfolding of a basis of energy-\emph{in}dependent orbitals, e.g. LMTOs 
\cite{Lambrecht1986}, Slater type obitals \cite{Zurek2007}, or NMTOs,
followed by linearization of the energy dependence of the denominators (e.g.
Eq.~(\ref{H2}) in Paper III).

For a \emph{Hamiltonian} formulation of the band-structure problem, we need
a basis set of energy-\emph{in}dependent \emph{smooth} functions analogous
to the well-known \emph{linear} APWs (LAPWs) and MTOs (LMTOs) \cite%
{Andersen1975}. This set \cite{Andersen2000,Tank2000}\cite{00Odile} is
arrived at by $N$th-order polynomial interpolation (Lagrange) in the Hilbert
space of KPWs with energies at a chosen mesh of $N+1$ energies, $%
E_{0},..,E_{N}:$%
\begin{equation}
\chi _{R^{\prime }l^{\prime }m^{\prime }}\left( \mathbf{r}\right) =\sum_{\nu
=0}^{N}\sum_{Rlm}^{\mathrm{active}}\phi _{Rlm}\left( E_{\nu },\mathbf{r}%
\right) \,L_{Rlm,R^{\prime }l^{\prime }m^{\prime }}\left( E_{\nu }\right) .
\label{NMTO}
\end{equation}%
Here, $\chi _{R^{\prime }l^{\prime }m^{\prime }}\left( \mathbf{r}\right) $
is a member of the active set of NMTOs and $L\left( E_{\nu }\right) $ is the
matrix of Lagrange coefficients, which is given by the kink matrix (\ref{K})
evaluated at the points of the energy mesh. For an NMTO, the kinks at the
hard spheres are reduced to discontinuities of the (2N+1)st derivatives and
for a quadratic (N=2) MTO (QMTO), as used for LiPB, this means that the 4
lowest radial derivatives are continuous, i.e. the QMTO is "supersmooth".
Also the MT-Hamiltonian- and overlap matrices in the NMTO representation, $%
H= $ $\left\langle \chi \left\vert -\Delta +\sum_{R}v_{R}\right\vert \chi
\right\rangle $ and $O=$ $\left\langle \chi \mid \chi \right\rangle ,$ are
given by the kink matrix and its first energy derivative evaluated at the
energy mesh --or more conveniently-- as divided differences of its inverse,
the Green matrix $G\left( E\right) \equiv K\left( E\right) ^{-1}$ [see Eq.s
(91), (94), and (95) in Ref. \cite{00Odile}].

The NMTO set may be arbitrarily small and, nevertheless, span the exact
solutions at the N+1 chosen energies of Schr\"{o}dinger's equation for the
MT potential to 1st order in the potential overlap. Specifically, a set with 
$n$ NMTOs (per cell) yields $n$ eigenfunctions and eigenvalues (energy
bands), $E,$ whose errors are proportional to respectively $\left(
E-E_{0}\right) ..\left( E-E_{N}\right) $ and $\left( E-E_{0}\right)
^{2}..\left( E-E_{N}\right) ^{2}.$ The choice of NMTO set, i.e. which
orbitals to place on which atoms, merely determines the prefactors of these
errors and the range of the orbitals. But only with chemically sound
choices, will the delocalization of the KPWs, caused by the N-ization in Eq.$%
\,$(\ref{NMTO}), be negligible.

In order to explain this, we now consider the simple example of
NaCl-structured NiO: Placing the three $p$-orbitals on every O, the five $d$%
-orbitals on every Ni, and letting the energy mesh span the 10 eV region of
the $pd$-bands, generates a basis set of eight atomic-like NMTOs yielding
the eight $pd$-bands and wave functions (see FIG.s$\,$2 and 4 in Ref.$\,$%
\cite{Haverkort2012} and FIG.$\,7$ top in Ref. \cite{12Juelich}). Placing
merely the three $p$-orbitals on every O and letting the mesh span the 5 eV
region of the O $p$-bands, generates a basis set, consisting of O $p$-like
NMTOs with bonding $d$-like tails on the Ni neighbors, which yields accurate 
$p$-bands and wave functions (FIG.$\,7$ bottom in Ref. \cite{12Juelich}).
Placing, instead, the five $d$-orbitals on every Ni and letting the mesh
span the 4 eV region of the Ni $d$-bands, generates a basis set of Ni $d$%
-like NMTOs which have antibonding $p$-like tails on the O neighbors and
yields accurate $d$-bands and wave functions (FIG.$\,7$ center in Ref. \cite%
{12Juelich}). With the five $d$-orbitals on Ni, \emph{but} a mesh spanning
the three O $p$-bands, we get three $d\left( t_{2g}\right) $-like Ni NMTOs, $%
xy,$ $xz,$ and $yz,$ with large $pd\pi$-bonding $p$-tails on the four O
neighbors in the plane of the $t_{2g}$-orbital, plus the two $d\left(
e_{g}\right) $-like Ni NMTOs with huge $pd\sigma$-bonding tails -- on the
two apical oxygens for $3z^{2}-1,$ and on the four oxygens in the $xy$ plane
for $x^{2}-y^{2}$. These fairly delocalized Ni $d$-NMTOs clearly exhibit the
Ni-O bonding, but they form a schizophrenic basis set which yields the three
O $p$-bands connected across the $pd$-gap to two of the five Ni $d$-bands by
steep "ghost" bands.

This example indicates how the NMTO method can be used to explore covalent
interactions in complex materials. Other examples may be found in Refs. \cite%
{Pavarini2005,Zurek2005,Yamasaki2006,Lechermann2006,Boeri2007,Liu2008,Kent2008,Tanusri2009,Zurek2010,Andersen2011,Haverkort2012}%
. Note that the fewer the bands to be picked out of a manifold, i.e. the
more diluted the basis set, the more extended are its orbitals because the
set is required to solve Schr\"{o}dingers equation in all space. The
increased extent leads to an (exponentially) increased energy dependence of
the KPWs and that requires using NMTOs with a finer energy mesh. As a
consequence, the smaller the set, the more complicated its orbitals.

Generalized Wannier functions are finally obtained by orthonormalization of
the corresponding NMTO set [see Chart (\ref{6WO})]. \emph{Symmetrical
orthonormalization} yields the set of Wannier functions, which we refer to
as \emph{Wannier orbitals }(WOs) because they are atom-centered with
specific orbital characters. The localization of these WOs hinges on the
fact that each KPW in the set vanishes (with a kink) inside the hard sphere
of any other KPW in the set. This condition essentially maximizes the
on-site and minimizes the off-site Coulomb integrals and has the same spirit
as the condition of minimizing the spread, $\left\langle \chi \left\vert
\left\vert \mathbf{r-\Re }\right\vert ^{2}\right\vert \chi \right\rangle ,$
used to define the maximally localized Wannier functions \cite{Marzari2012}.

For LiPB, we used quadratic N(=2)MTOs and for the large-basis-set
calculation chose the three energies $E_{\nu }$=$\pm 1$ and 0 Ry with
respect to the MT zero, i.e. $-22,$ $-8$ and $6$ eV with respect to the
center of the gap, which is approximately the Fermi level (see FIG.~\ref%
{FIG2}). For the six-orbital calculation, we took $E_{\nu }$=$-0.8,$ $-0.4,$
and $0.2$ eV with respect to the center of the gap (see FIG.~\ref%
{ARPES_Bandstructure_LDA} in Paper II).

For the low-energy electronic structure of LiMo$_{6}$O$_{17}$ we need to
pick from the sixty Mo$~$4$d$ bands above the O$~$2$p$ -- Mo$~$4$d$ gap (see
FIG.~\ref{FIG2}) a conveniently small and yet separable set of bands around
the Fermi level. In this case, where no visible gap separates such bands
from the rest of an upwards-extending continuum, the NMTO method is uniquely
suited for picking a subset of bands for which the Wannier set is
intelligible and as localized as possible. This \emph{direct generation} of
WOs (through trial and error by inspecting the resulting bands like we
discussed above for NiO) differs from the procedures for projecting\emph{\ }%
localized Wannier functions from the Bloch functions of the computed band
structure by judiciously choosing their phases \cite{Satpathy1988,Ku2002} or
by minimizing the spread \cite{Marzari2012}\cite{Nuss2014}. We shall return
to it in Sect.~\ref{SectLowE} after the crystal structure and the basic
electronic structure of LiPB has been discussed.

Since the resulting set of six NMTOs may have a fairly long range, all LDA
calculations were performed in the representation of Bloch sums,%
\begin{equation}
\chi_{Rm}\left( \mathbf{k,r}\right) \equiv\sum\nolimits_{\mathbf{T}%
}\chi_{Rm}\left( \mathbf{r-T}\right) e^{2\pi i\mathbf{k\cdot T}},
\label{Bloch}
\end{equation}
of orbitals translated by the appropriate lattice vector, $\mathbf{T.}$
Specifically, the screening of the structure matrix was done $\mathbf{k}$-by-%
$\mathbf{k.}$ In order to obtain printable WOs --obtained by symmetrical
orthonormalization of the NMTOs-- and a portable Hamiltonian whose $%
H_{R^{\prime}m^{\prime},\,R+T\,m}$ element is the integral for hopping
between the WOs centered at respectively $\mathbf{R}^{\prime}$ and $\mathbf{%
R+T,}$ we need to Fourier transform back to real space: 
\begin{equation}
H_{R^{\prime}m^{\prime},\,R+T\,m}=\left\vert \mathbf{a\cdot b\times c}%
\right\vert \int_{\mathrm{BZ}}d^{3}k\,e^{-2\pi i\mathbf{k\cdot T}%
}H_{R^{\prime}m^{\prime},\,R\,m}\left( \mathbf{k}\right) .  \label{FT}
\end{equation}
Here, the integral with its prefactor is the average over the BZ, as is
appropriate when the localized orbital is normalized to unity. Moreover, $%
H_{Rm^{\prime},\,Rm}$ is the energy of the orbital when $m$=$m^{\prime},$
and the crystal-field term when $m\neq m^{\prime}$. The Hamiltonian (\ref{FT}%
), truncated after $\left\vert \mathbf{R-R}^{\prime}-\mathbf{T}\right\vert $
exceeds some distance, the lattice constant $a$ for LiPB, we shall refer to
as the \emph{tight-binding} (TB) Hamiltonian (Sect$\,.\,$\ref{SectH}). This
truncation makes its energy-band eigenvalues more smooth and wavy than those
of $H_{R^{\prime}m^{\prime},\,R\,m}\left( \mathbf{k}\right) .$

To the MT Hamiltonian, we finally add the second-order correction for the
tongue-overlap and the full-potential perturbation \cite{00Odile,Nohara2018}%
. Products of NMTOs --as needed for evaluation of matrix elements and the
charge density-- are evaluated as products of partial waves limited to their
MT spheres plus products of screened spherical waves \cite{00Odile,12Juelich}%
. The latter are smooth functions and are interpolated across the
interstitial from their first three radial derivatives at the hard sphere 
\cite{Nohara2016}. In order to make it trivial to solve Poisson's equation,
this interpolation uses spherical waves which --in order to make the
matching at the hard spheres explicitly-- are screened to have all phase
shifts with $l\leq 4$ equal to those of hard spheres.

The band structure obtained from our full-potential LDA calculation with the
large NMTO basis set agrees well with LDA and GGA control calculations
performed with the LAPW method. We did not include the spin-orbit coupling
in the NMTO calculation, but did so with the LAPW method and show the result
in Paper II FIG.~\ref{LAPWwoSOC} together with the LDA TB bands.

\begin{figure*}[tbh]
\includegraphics[width=\linewidth]{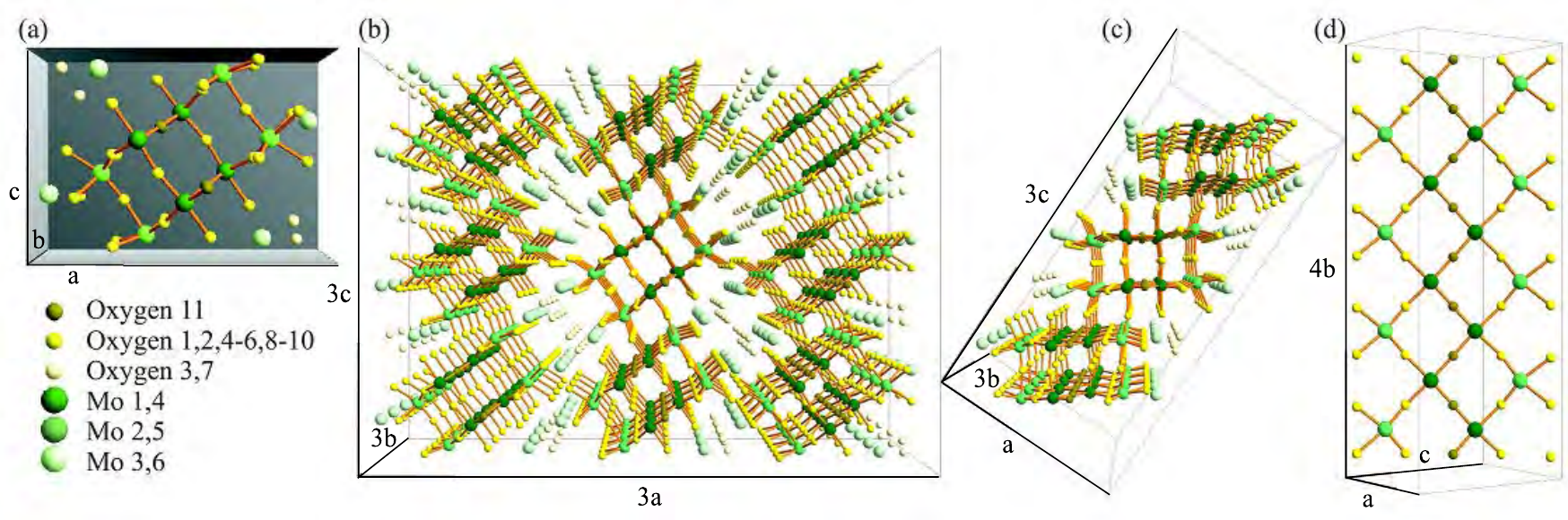}
\caption{Crystal structure of LiMo$_{6}$O$_{17}$. Li (not shown) is
intercalated in the hollows near light-green Mo and light-yellow O atoms. 
\textbf{(a)} primitive cell spanned by the translation vectors $\mathbf{a},$ 
$\mathbf{b},$ and $\mathbf{c.}$ Whereas $\mathbf{b}$ is orthogonal to both $%
\mathbf{c}$ and $\mathbf{a,}$ the latter has a one-per-cent component along $%
\mathbf{c.}$ The relative lengths of the translation vectors are: $%
a/b\approx 2.31$ and $c/b\approx \protect\sqrt{3}.$ The primitive cell
contains two strings: (Mo3) - Mo2 - Mo1 - Mo4 - Mo5 - (Mo6) and the inverted
one: (MO6) - MO5 - MO4 - MO1 - MO2 - (MO3). When we need to distinguish
between two equivalent sites (related by inversion), we use upper-case
letters for the one in the upper string. Together, the two strings form a
bi-string. \textbf{(b)} a 3$\times $3$\times $3 supercell showing $bc$-slabs
translated by $\pm \mathbf{a}$. The slabs are separated by Mo3 and Mo6
(light green). \textbf{(c)} A single slab, rotated such that six-fold
coordinated Mo (green and dark green) have their bonds to O in the vertical, 
$z,$ and two horizontal, $x$ and $y,$ directions Eq.~(\protect\ref{xyz}).
Oriented this way, see also Chart (\protect\ref{ac}), the slab forms a
staircase running up the $c$-direction, i.e. with the steps translated by $%
\mathbf{c.}$ A single step is a bi-ribbon formed by translating a bi-string
infinitely many times by $\mathbf{b}$. The midpoint between Mo1 and a
nearest MO1 in the same bi-ribbon is a center of inversion. The planes
perpendicular to $\mathbf{b}$ containing Mo1 and Mo5, as well as those
containing Mo4 and Mo2, are mirror planes. The sequence along the vertical,
almost straight lines along $\mathbf{z}$ is: (MO6) - Mo5 - MO2 - Mo1 - MO4 -
(Mo3) and Li intercalates between the Mo$_{3}$O$_{4}$ tetrahedron and the MO$%
_{3}$O$_{4}$ tetrahedron right above it \protect\cite{Onoda1987}. \textbf{(d)%
} 4 primitive cells along $b$ of a single ribbon. Along its center runs the
dark-green $_{\text{Mo1}}\diagup ^{\text{Mo4}}$ zigzag chain. Parallel
herewith and shifted by $\mathbf{z}$ [see (c) and Chart (\protect\ref{xy})]
is the partner ribbon with its $_{\text{MO4}}\diagup ^{\text{MO1}}$ zigzag
chain [see also FIG. 1 in Ref.~\protect\cite{Satpathy2006}].}
\label{FIG1}
\end{figure*}

\section{Crystal structure\label{crystal_structure}}

The crystal structure of LiPB was determined at room temperature and
described by Onoda et al \cite{Onoda1987}. As shown in FIG. \ref{FIG1},
there are two LiMo$_{6}$O$_{17}$ units in the primitive cell spanned by the
translations $\mathbf{a},$ $\mathbf{b},$ and $\mathbf{c}$ shown in (a).
Whereas $\mathbf{b}$ is orthogonal to both $\mathbf{c}$ and $\mathbf{a,}$
the latter has a tiny component along $\mathbf{c.}$ The relative lengths of
the primitive translation vectors are: $a/b\approx 2.311$ and $c/b\approx
1.720,$ with $b=5.523\,\mathring{A}.$ We note that in much of the
literature, especially experimental papers, an alternate axis labeling \cite%
{Greenblatt1984} is used\footnote{%
It is, therefore, essential to check any particular article for these
definitions.} with the definitions of $a$ and $c$ interchanged. Here we
follow Onoda et al. \cite{Onoda1987}. Since $\mathbf{a},$ $\mathbf{b},$ and $%
\mathbf{c}$ are nearly orthogonal, so are the primitive translations, $%
\mathbf{a}^{\ast },$\textbf{\ }$\mathbf{b}^{\ast },$\textbf{\ }and $\mathbf{c%
}^{\ast },$ of the reciprocal lattice. They are defined by:%
\begin{equation}
\left( 
\begin{array}{ccc}
\mathbf{a\cdot a}^{\ast } & \mathbf{a\cdot b}^{\ast } & \mathbf{a\cdot c}%
^{\ast } \\ 
\mathbf{b\cdot a}^{\ast } & \mathbf{b\cdot b}^{\ast } & \mathbf{b\cdot c}%
^{\ast } \\ 
\mathbf{c\cdot a}^{\ast } & \mathbf{c\cdot b}^{\ast } & \mathbf{c\cdot c}%
^{\ast }%
\end{array}%
\right) \equiv \left( 
\begin{array}{ccc}
1 & 0 & 0 \\ 
0 & 1 & 0 \\ 
0 & 0 & 1%
\end{array}%
\right) ,  \label{recip}
\end{equation}%
where we use the \emph{crystallographic} definition of the scale of
reciprocal space without the factor $2\pi $ on the right-hand side used in
the \emph{solid-state} definition. The former is traditionally used in
diffraction and the latter in spectroscopy. In this paper, we use the
crystallographic definition unless otherwise stated. The top of FIG. \ref%
{FIG2} shows half the Brillouin-zone (BZ) with origin at $\mathrm{\Gamma }$
and spanned by $\pm \frac{\mathbf{a}^{\ast }}{2}$ (\textrm{B}), $\mathbf{\pm 
}\frac{\mathbf{b}^{\ast }}{2}$ (\textrm{Y}), and $\pm \frac{\mathbf{c}^{\ast
}}{2}$ (\textrm{Z}). The Bloch vector,%
\begin{equation}
\mathbf{k}=\,k_{a}\mathbf{a}^{\ast }+k_{b}\mathbf{b}^{\ast }+k_{c}\mathbf{c}%
^{\ast },  \label{k}
\end{equation}%
is specified by its dimensionless $\left( k_{a},k_{b},k_{c}\right) $%
-components which, according to Eq.s (\ref{recip}) and (\ref{k}), are the
projections of $\mathbf{k}$\textbf{\ }onto respectively $\mathbf{a}$, $%
\mathbf{b}$, and $\mathbf{c,}$ or equivalently: they are the projections
onto the respective \emph{directions} in units of $a^{-1}$, $b^{-1}$, and $%
c^{-1}$. Occasionally, we shall use the solid-state definition where the $k$
components are the same, but $\mathbf{a}^{\ast }$, $\mathbf{b}^{\ast }$, $%
\mathbf{c}^{\ast }$, and $\mathbf{k}$ are $2\pi $ larger, e.g. the Fermi
vector for stoichiometric LiPB has length $(\pi /2)b^{-1}=0.2844$ \AA $^{-1}$
instead of $(1/4)b^{-1}$.

The most relevant symmetry points have $k_{a}$=$0,$ and are: $\left(
k_{b},k_{c}\right) $=$\mathrm{Z}\left( 0,\frac{1}{2}\right) ,$ $\mathrm{Y}%
\left( \frac{1}{2},0\right) ,$ $\mathrm{\Gamma }\left( 0,0\right) ,$ $%
\mathrm{C}\left( \frac{1}{2},\frac{1}{2}\right) ,$ $\mathrm{W}\left( \frac{1%
}{2},\frac{1}{4}\right) ,$ $\mathrm{\Lambda }\left( 0,\frac{1}{4}\right) ,$
plus their equivalents. Higher BZs are shifted by reciprocal lattice
vectors, $\mathbf{G,}$ which means that $k_{a},$ $k_{b},$ and $k_{c}$ are
shifted by integers, which we name respectively $L,$ $M,$ and $N$ and shall
use in Sect.~\ref{SectH}, and in Sect.s \ref{Sectzoneselect}, and \ref%
{Secthv} in Paper II.

A simplifying --hitherto overlooked-- \emph{approximate} view of the
complicated structure in FIG. \ref{FIG1} is that all Mo atoms are on a
lattice spanned by the primitive translations:%
\begin{equation}
\frac{\mathbf{c+a}}{6}\mp \frac{\mathbf{b}}{2}\equiv \QATOP{\mathbf{x}}{%
\mathbf{y}}\quad \mathrm{and\quad }\frac{\mathbf{c}}{2}\mathrm{-}\frac{%
\mathbf{c+a}}{6}\equiv \mathbf{z}\,.  \label{xyz}
\end{equation}%
These are orthogonal to within a few degrees and their lengths, 3.82$\,$\AA %
, are equal to within 0.3\%. This means that all 12 Mo atoms approximately
form a simple cubic lattice, 12 times finer than the proper lattice. In FIG.~%
\ref{FIG1}~(c), the structure is turned to have $\mathbf{z}$ in the vertical
direction, and $\mathbf{x}$ and $\mathbf{y}$ in the horizontal plane. This
view is useful for understanding the structure, the computed Wannier
orbitals and the measured ARPES, but should not be overstretched. Using for
instance the inverse to the transformation (\ref{xyz}),%
\begin{equation}
\mathbf{a}=2\left( \mathbf{x+y-z}\right) ,\;\mathbf{b=y-x,\;c}=\mathbf{x+y}+2%
\mathbf{z,}  \label{abc}
\end{equation}%
\emph{and assuming }the $xyz$ system to be orthonormal leads to: $a=2\sqrt{3}%
,$ $b=\sqrt{2},$ and $c=\sqrt{6}$ times 3.82$\,$\AA , which are wrong by
respectively $+$3.7, $-$2.2, and $-$2.1 per cent.

As specified in FIG.$~$\ref{FIG1}~(a), of the twelve Mo sites in the
primitive cell, six are inequivalent. Four of these (dark-green Mo1 and Mo4,
and green Mo2 and Mo5) are six-fold coordinated with oxygen (dark yellow and
yellow) in the $\pm x$, $\pm y$, and $\pm z$ directions and form a network
of \emph{corner-sharing} MoO$_{6}$ \emph{octahedra.} We shall call them
octahedral molybdenums. The remaining two types of Mo (light-green Mo3 and
Mo6) are four-fold coordinated with oxygen (yellow and light yellow). The
latter, \emph{tetrahedrally }coordinated Mo atoms (light green, set in
parentheses in the following) form double layers, which separate the network
of corner-sharing MoO$_{6}$ octahedra into \emph{slabs.} The crystals cleave
between slabs.

Such a slab has the form of a \emph{staircase} with steps of \emph{bi-ribbons%
} stacked with period $\mathbf{c}$ as seen in~FIG.$~$\ref{FIG1}~(c).
Schematically, this is: 
\begin{equation}
\begin{array}{cc}
&  \\ 
&  \\ 
\mathbf{c} & \nearrow \\ 
\mathbf{a} & \searrow \\ 
\mathbf{z} & \uparrow \\ 
&  \\ 
& 
\end{array}%
\,\fbox{$%
\begin{array}{ccccccccc}
\mathbf{2} & 1 & \mathbf{4} & 5 &  &  & \mathbf{2} & 1 & \mathbf{4} \\ 
\mathbf{1} & 2 &  &  & \mathbf{5} & 4 & \mathbf{1} & 2 &  \\ 
\mathbf{4} & 5 &  &  & \mathbf{2} & 1 & \mathbf{4} & 5 &  \\ 
&  & \mathbf{5} & 4 & \mathbf{1} & 2 &  &  & \mathbf{5} \\ 
&  & \mathbf{2} & 1 & \mathbf{4} & 5 &  &  & \mathbf{2} \\ 
\mathbf{5} & 4 & \mathbf{1} & 2 &  &  & \mathbf{5} & 4 & \mathbf{1} \\ 
\mathbf{2} & 1 & \mathbf{4} & 5 &  &  & \mathbf{2} & 1 & \mathbf{4}%
\end{array}%
$},  \label{ac}
\end{equation}%
where the octahedral molybdenums lying in the same $ac$-plane are either
normal- or bold-faced. The distance between such $ac$-planes is $\frac{%
\mathbf{b}}{2}.$ A \emph{single ribbon} is \emph{four} octahedral
molybdenums \emph{wide} and, as seen here:

\begin{equation}
{%
\begin{array}{cc}
&  \\ 
\mathbf{b} & \uparrow \\ 
\mathbf{c+a} & \longrightarrow \\ 
\mathbf{y} & \nearrow \\ 
\mathbf{x} & \searrow \\ 
& 
\end{array}%
}\,{%
\begin{array}{cc}
\mathrm{Mo:} & \mathrm{MO:} \\ 
\fbox{$%
\begin{array}{cccccc}
& 2 &  & 4 &  &  \\ 
&  & 1 &  & 5 &  \\ 
& 2 &  & 4 &  &  \\ 
&  & 1 &  & 5 &  \\ 
& 2 &  & 4 &  &  \\ 
&  & 1 &  & 5 & 
\end{array}%
$} & \,\fbox{$%
\begin{array}{cccccc}
& 5 &  & 1 &  &  \\ 
&  & 4 &  & 2 &  \\ 
& 5 &  & 1 &  &  \\ 
&  & 4 &  & 2 &  \\ 
& 5 &  & 1 &  &  \\ 
&  & 4 &  & 2 & 
\end{array}%
$}%
\end{array}%
},  \label{xy}
\end{equation}%
and in FIG.s$~$\ref{FIG1}~(c) and (d), extends indefinitely in the $b$%
-direction and lies in the horizontal $xy$-plane containing the vectors $%
\mathbf{b=y-x}$ and $\mathbf{c+a=}$ $3(\mathbf{x+y})$. The lower half of a
bi-ribbon, seen in the left-hand panel of Chart (\ref{xy}), consist of (Mo3)
- Mo2 - Mo1 - Mo4 - Mo5 - (Mo6) \emph{strings }separated by $\mathbf{b}$ and
can be taken either as a zigzag line changing translation between $\mathbf{y}
$ and $\mathbf{x,}$ and thus running along $\mathbf{c+a}$, or as a nearly
straight line running along $\mathbf{x,}$ or as one running along $\mathbf{y}
$ [see (d) and (\ref{xy})]. In the following, we shall refer to these as
respectively $\left( \mathbf{c+a}\right) $-zigzag, $\mathbf{x}$-, and $%
\mathbf{y}$ strings.

The \emph{upper} ribbon is shown to the right in Chart (\ref{xy}). Its Mo
sequence, (MO6) - MO5 - MO4 - MO1 - MO2 - (MO3), is \emph{inverted} such
that e.g. MO4 is on top of Mo1. When we \emph{need} to distinguish between
two equivalent sites related by inversion in their midpoint --a center of
inversion for the entire crystal-- we use upper-case letters for the one in
the upper ribbon.

Note that the $\left( \mathbf{c+a}\right) $-zigzag string is different from
--and perpendicular to-- the $\diagdown _{\text{Mo1}}\diagup ^{\text{Mo4}%
}\diagdown $ zigzag \emph{chain} along $\mathbf{b},$ the backbone of the
electronic 1D $xy$-band shown in FIG.$\,$1 of Ref. \cite{Satpathy2006}
together with its partner $\diagdown _{\text{MO4}}\diagup ^{\text{MO1}%
}\diagdown $ in the upper ribbon.

\subsection{$c$-dimerization\label{SectDims}}

The vectors from Mo1 to its two nearest MO1 neighbors inside and outside the
bi-ribbon are respectively $\frac{\mathbf{c\pm b}}{2}-\mathbf{d}$ and $%
-\left( \frac{\mathbf{c\pm b}}{2}+\mathbf{d}\right) $ where%
\begin{equation}
\mathbf{d}=0.012\mathbf{a}+0.033\mathbf{c}  \label{dim}
\end{equation}%
is the \emph{displacement dimerization}.\emph{\ }Hence, the distances
measured along $\mathbf{c}$ from a ribbon to its neighbors inside and
outside the bi-ribbon are respectively 6.6\% smaller and 6.6\% larger than
the average distance $\frac{c}{2}.$

Due to the stacking (\ref{ac}) into a staircase of bi-ribbons, Mo4 differs
from Mo1, and Mo5 differs from Mo2, in having \emph{no }neighbor belonging
to the next bi-ribbon, i.e., they have only \emph{one} octahedral Mo
neighbor along $z.$ As seen in Charts (\ref{ac}) and (\ref{xy}), Mo1 has
six, Mo4 five, Mo2 four, and Mo5 three nearest Mo neighbors which are
octahedrally coordinated with oxygen.

In the next section, we shall explain --and later demonstrate by computation
and experiment-- that the six lowest energy bands are described by the six 
\emph{planar} $t_{2g}$ Wannier orbitals (WOs), $w_{m}\left( \mathbf{r}%
\right) $ and $W_{m}\left( \mathbf{r}\right) $ $\left( m\mathrm{=}%
xy,xz,yz\right) ,$ centered\footnote{%
We use a notation according to which a function, e.g., $\phi _{R}\left( 
\mathbf{r}\right) ,$ $w\left( \mathbf{r}\right) \equiv $ $w_{\mathrm{Mo1}%
}\left( \mathbf{r}\right) ,$ or $W\left( \mathbf{r}\right) \equiv W_{\mathrm{%
MO1}}\left( \mathbf{r}\right) ,$ of the space vector $\mathbf{r}$ is
centered at $\mathbf{r=R\equiv R}_{R}\mathbf{,}$ whereas a function such as $%
\varphi _{R}\left( r\right) Y_{lm}\left( \mathbf{\hat{r}}\right) $ of $%
r\equiv $ $\left\vert \mathbf{r}\right\vert $ and $\mathbf{\hat{r}}$\textbf{%
\ }$\equiv $ $\mathbf{r/}\left\vert \mathbf{r}\right\vert $ is centered at
the origin. \label{Rcenter}} on respectively Mo1 and on MO1. These sites,
separated by $\frac{\mathbf{c+b}}{2}-\mathbf{d,}$ are \emph{special} in
having a \emph{full} nearest-neighbor shell of octahedral molybdenums and
therefore best preserve the $t_{2g}$ symmetry of the WO and are least
sensitive to the \emph{steps} of the staircase, the second cause for the 
\emph{dimerization}. Such a WO (FIG.$\,$\ref{Wannier}) spreads substantially
onto the four nearest Mo neighbors in the orbital's plane with amplitudes
falling in the same order as the above-mentioned Mo coordination of those
neighbors. As a result of this, and the smallness of the displacement
dimerization (\ref{dim})$,$ the two $t_{2g,m}$ WOs are \emph{approximately}
related by\emph{\ half }a \emph{lattice translation:}%
\begin{equation}
W_{m}\left( \mathbf{r-}\frac{\mathbf{c+b}}{2}\right) ~\approx ~w_{m}\left( 
\mathbf{r}\right) .  \label{undim}
\end{equation}%
However, the \emph{exact} relation is:%
\begin{equation}
W_{m}\left( \mathbf{r-}\frac{\mathbf{c+b}}{2}+\mathbf{d}\right) =w_{m}\left(
-\mathbf{r}\right) ,  \label{inv}
\end{equation}%
and its differences, $W_{m}\left( \mathbf{r-}\frac{\mathbf{c+b}}{2}+\mathbf{d%
}\right) -W_{m}\left( \mathbf{r-}\frac{\mathbf{c+b}}{2}\right) $ and $%
w_{m}\left( -\mathbf{r}\right) -w_{m}\left( \mathbf{r}\right) ,$ to the
approximate relation (\ref{undim}) will be\textbf{\ }referred to as
respectively the \emph{displacement-} and the \emph{inversion dimerization.}

A consequence of the approximate translational equivalence (\ref{undim}),
which may be seen to hold far better for the $xy$- than for the $xz$- and $%
yz $ WOs, is that the low-energy band structure, $E_{j}\left( \mathbf{k}%
\right) $ with $j$=1-6 (e.g. FIG.~\ref{ThreePureBands}), approximately
consists of 3 bands, $E_{m}\left( \mathbf{k}\right) ,$ one with each $m$%
-character, and extending in a \emph{double zone} of the \emph{sparse}
reciprocal sublattice spanned by%
\begin{equation}
\left( \mathbf{a}^{\ast },\,\mathbf{c}^{\ast }+\mathbf{b}^{\ast },\,\mathbf{c%
}^{\ast }-\mathbf{b}^{\ast }\right) .  \label{undimrecip}
\end{equation}%
This is the reciprocal of the \emph{un}-dimerized lattice spanned by%
\begin{equation}
\left( \mathbf{a,}\,\frac{\mathbf{c+b}}{2},\,\frac{\mathbf{c-b}}{2}\right)
=\left( \mathbf{a,\,y+z},\,\mathbf{x+z}\right)  \label{undimprim}
\end{equation}%
with only \emph{one} LiMo$_{6}$O$_{17}$ unit per primitive cell. The two
last translations (\ref{undimprim}), we shall call \emph{pseudo }%
translations. If expression (\ref{undim}) were true, the $E_{m}\left( 
\mathbf{k}\right) $ band would be equivalent to the one translated by (any
odd number times) $-\mathbf{c}^{\ast },$ e.g. with $E_{m}\left( \mathbf{k+c}%
^{\ast }\right) ,$ but the presence of the inversion- and displacement
dimerizations, (\ref{inv}) and (\ref{dim}), cause these two bands to gap
where they cross, i.e. at the boundaries of the small zones. The resulting
band structure is periodic in the proper (small) zone, corresponding to the
proper primitive cell with \emph{two} LiMo$_{6}$O$_{17}$ units, and has six
continuous bands, two for each $m,$ of which the lower is approximately $%
E_{m}\left( \mathbf{k}\right) $ and the higher is approximately $E_{m}\left( 
\mathbf{k+c}^{\ast }\right) $ in the odd-numbered zones; and the other way
around in the even-numbered zones. As we shall see in Paper II, ARPES
approximately sees only the $E_{m}\left( \mathbf{k}\right) $-like band, i.e.
both bands, but separated in zones, and this will allow us to resolve the
splitting and the perpendicular dispersion of the two quasi-1D bands in the
gap.

The \emph{un-dimerized }lattice has \emph{one }LiMo$_{6}$O$_{17}$ unit per
primitive cell and is spanned by the primitive translations (\ref{undimprim}%
) where, on the right-hand side, we have used the approximate relation (\ref%
{abc}). This shows that the un-dimerized lattice is 2D and hexagonal in the
planes perpendicular to $\mathbf{a.}$ This is the structure of the purple
bronzes isoelectronic with LiPB, NaMo$_{6}$O$_{17}$ and KMo$_{6}$O$_{17},$
where CDW fluctuations with wavevector $\mathbf{c}^{\ast }$ have been
observed below 120$\,$K and have been explained as driven by the
simultaneous gapping of the 1D $xz$- and the $yz$ Fermi-surface (FS) sheets
by one and the same nesting vector, $\mathbf{c}^{\ast }$ \cite%
{Whangbo1991,Foury1993}. The lattice reciprocal to the un-dimerized one is
spanned by (\ref{undimrecip}), and its BZ is the double zone of the
dimerized structure, i.e. that of LiPB shown in FIG.$\,$\ref{FIGDoubleZone}.
Hence, we may consider the structure of quasi-1D LiPB as the CDW
dimerization of quasi-2D Na- or KPB, whose electronic structure consists of
the 1D $xy$-, $yz$-, and $xz$-bands dispersing at 120$%
{{}^\circ}%
$ relatively to each other in the plane perpendicular to $\mathbf{a}$. The
reason why not also the $xy$ FS sheets gap away is that relation (\ref{undim}%
) holds much better for the $xy$ WOs than for the $xz$ and $yz$ WOs.

For comparison with Charts (\ref{ac}) and (\ref{xy}), the un-dimerized slab
is:%
\begin{equation}
\begin{array}{cc}
&  \\ 
&  \\ 
\mathbf{c} & \nearrow \\ 
\mathbf{a} & \searrow \\ 
\mathbf{z} & \uparrow \\ 
&  \\ 
& 
\end{array}%
\,\fbox{$%
\begin{array}{ccccccccc}
\mathbf{4} & 1 & \mathbf{2} &  &  & 5 & \mathbf{4} & 1 & \mathbf{2} \\ 
\mathbf{1} & 2 &  &  & \mathbf{5} & 4 & \mathbf{1} & 2 &  \\ 
\mathbf{4} &  &  & 5 & \mathbf{4} & 1 & \mathbf{2} &  &  \\ 
&  & \mathbf{5} & 4 & \mathbf{1} & 2 &  &  & \mathbf{5} \\ 
& 5 & \mathbf{4} & 1 & \mathbf{2} &  &  & 5 & \mathbf{4} \\ 
\mathbf{5} & 4 & \mathbf{1} & 2 &  &  & \mathbf{5} & 4 & \mathbf{1} \\ 
\mathbf{4} & 1 & \mathbf{2} &  &  & 5 & \mathbf{4} & 1 & \mathbf{2}%
\end{array}%
$},  \label{undimac}
\end{equation}%
where the steps of the staircase are smoothed out to a ramp, and the two
ribbons are identical:%
\begin{equation}
{%
\begin{array}{cc}
&  \\ 
\mathbf{b} & \uparrow \\ 
\mathbf{c+a} & \longrightarrow \\ 
\mathbf{y} & \nearrow \\ 
\mathbf{x} & \searrow \\ 
& 
\end{array}%
}{%
\begin{array}{cc}
\mathrm{Mo:} & \mathrm{MO:} \\ 
\,\fbox{$%
\begin{array}{cccccc}
&  & 4 &  & 2 &  \\ 
& 5 &  & 1 &  &  \\ 
&  & 4 &  & 2 &  \\ 
& 5 &  & 1 &  &  \\ 
&  & 4 &  & 2 &  \\ 
& 5 &  & 1 &  & 
\end{array}%
$}\, & \fbox{$%
\begin{array}{cccccc}
&  & 5 &  & 1 &  \\ 
&  &  & 4 &  & 2 \\ 
&  & 5 &  & 1 &  \\ 
&  &  & 4 &  & 2 \\ 
&  & 5 &  & 1 &  \\ 
&  &  & 4 &  & 2%
\end{array}%
$}%
\end{array}%
}  \label{undimxy}
\end{equation}

\subsection{$b$-dimerization\label{Sectbims}}

An unrelated and different dimerization is the one known from the
description of the 1D band structure as two, approximately 4~eV broad, $%
\frac{1}{4}$-filled $xy$ bands [FIG. \ref{FIG2} and Eq.s~(\ref{exy})-(\ref%
{exyu})] running on zigzag chains along $\mathbf{b}$ [24,25] and with the
nearest neighbor Mo1-Mo4 hopping integral $t\sim -1$~eV$.$ In this view, Mo1
and Mo4 are inequivalent because of a dimerization from $\frac{\mathbf{b}}{2}
$ to $\mathbf{b}$. In reciprocal space$\mathbf{,}$ this dimerization is from 
$2\mathbf{b}^{\ast }$ to $\mathbf{b}^{\ast }$ and causes gaps at $k_{b}=\pm 
\frac{1}{2}\approx \pm 2k_{F}$ which separate the broad $xy$ bands into two
lower $\frac{1}{2}$-filled and two higher empty bands. The two latter bands
will \emph{not} be described by our set of six WOs, which are essentially
Mo1-Mo4 bonding orbitals (FIG.~\ref{Wannier}), but would require the
inclusion of also Mo1-Mo4 anti-bonding orbitals, thus leading to a basis set
unpractically large for our purpose of understanding the photoemission from
the occupied bands.

\section{Basic electronic structure\label{SectElStruc}}

Shortly after the structural determination, Whangbo and Canadell \cite%
{Whangbo1988} used the extended H\"{u}ckel method to calculate and explain
the basic electronic structure, but it took almost twenty years before a
charge-self-consistent calculation could be performed. This was done by
Popovi\'{c} and Satpathy \cite{Satpathy2006} who used the LDA-DFT and the
LMTO method. In the following, we shall explain and expand on these works
using the insights gained from the view of the structure given in the
previous section and from the results of the WO calculations to be presented
in Sect.s~\ref{SectLowE} and \ref{SectH}.

\begin{figure}[tbh]
\includegraphics[width=\linewidth]{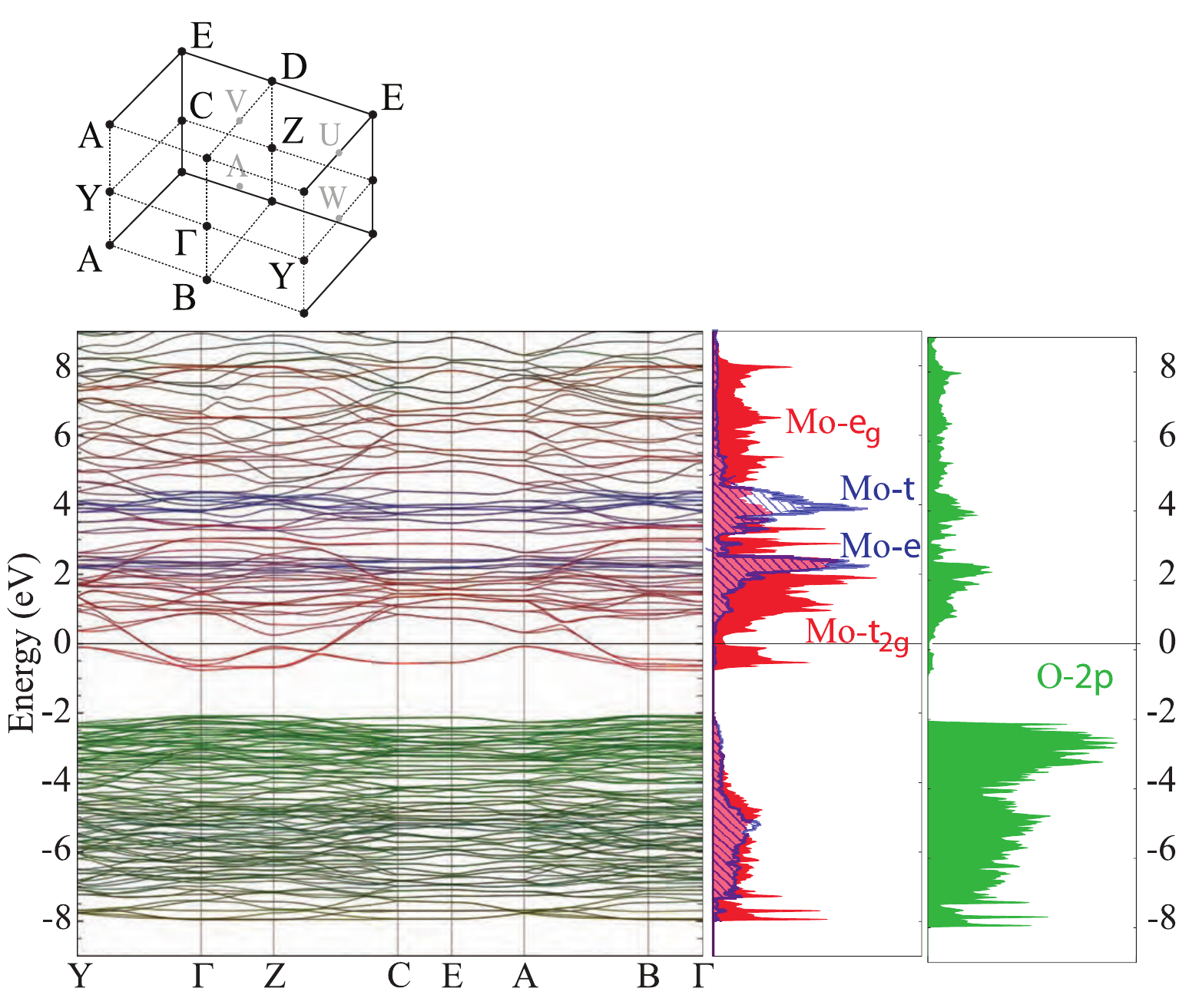}
\caption{The top shows half ($k_{c}\geq 0$) the (1st) Brillouin-zone of LiMo$%
_{6}$O$_{17}$\ with the labeling of the symmetry points for P2$_{1}$/m
obtained from the Bilbao crystal server \protect\cite{Aroyo2014}. The most
relevant symmetry points have $k_{a}$=$0$ and are: $\left(
k_{b},k_{c}\right) $=$\mathrm{\Gamma }\left( 0,0\right) ,$ $\mathrm{Y}\left( 
\frac{1}{2},0\right) ,$ $\mathrm{W}\left( \frac{1}{2},\frac{1}{4}\right) ,$ $%
\mathrm{C}\left( \frac{1}{2},\frac{1}{2}\right) ,$ $\mathrm{Z}\left( 0,\frac{%
1}{2}\right) ,$ $\mathrm{\Lambda }\left( 0,\frac{1}{4}\right) ,$ and their
equivalents. Below is the LDA band-structure calculated with the large basis
set of 336 NMTOs (left) and its partial densities of states (right) over a
wide energy range. The zero of energy is the Fermi level of the
stoichiometric compound. }
\label{FIG2}
\end{figure}

In FIG.$~$\ref{FIG2}, we show the LDA energy bands over a range of $\pm 9$
eV around the Fermi level, together with their density of states projected
onto O (green) and onto tetrahedrally- (blue) and octahedrally- (red)
coordinated Mo. The bands between $-8$ and $-2$ eV have predominantly O $2p$
character and those extending upwards from $B\sim -$0.7 eV predominantly Mo $%
4d$ character and, above +0.8 eV, also Mo $5s$ and $5p$ characters. The
states in the O $2p$ band are bonding linear combinations with Mo $5s,5p,$
and $4d$ orbitals, the more bonding, the lower their energy. The states in
the Mo $4d$ band are anti-bonding linear combinations with O $2s$ and $2p$
orbitals; the more anti-bonding, the higher their energy.

The 2~eV gap between the O $2p$-like and Mo $4d$-like bands is --for the
purpose of counting-- ionic with Li donating one and Mo six electrons to-
and O acquiring two electrons from the Mo$\,4d$-like bands above the gap,
which thereby hold 2$\left( 1+6\times 6-17\times 2\right) $ $=6$ electrons
per 2(LiMo$_{6}$O$_{17}$). Had this charge been spread uniformly over all
molybdenums, this would correspond to a Mo $d^{0.5}$ occupation.

The $4d$ orbitals forming the most anti-bonding and bonding states with O
are the $e_{g}$ orbitals, $3z^{2}-1$ and $x^{2}-y^{2},$ on the \emph{%
octahedrally} coordinated Mo because their lobes point \emph{towards} the
two O neighbors along $z$ and the four O neighbors in the $xy$ plane,
respectively, and thereby form $pd\sigma $ bonds and anti-bonds. Not only
the $e_{g}$ orbitals on the octahedrally coordinated Mo, but all $4d$
orbitals ($t$ and $e)$ on the \emph{tetrahedrally} coordinated Mo form
filled bonding and empty anti-bonding states with their O neighbors, and
thereby contributing to the stability of the crystal. However, as seen from
the projected densities of states in FIG.$\,$\ref{FIG2}, none of them
contribute to the LDA bands within an eV around the Fermi level, which are
the ones of our primary interest. So as long as there are no additional
perturbations or correlations with energies in excess of this,\ which is
assumed in the $\frac{1}{2}$-filled models, the Mo $t$ and $e$ orbitals are
uninteresting for the low-energy electronics, and so are the Li $2s$
orbitals which contribute two bands several eV above the Fermi level, mix a
bit with the oxygen states several eV below $E_{F}$, and donate their two
electrons to them. Due to the Pauli principle, changing the Li content
(doping) will change the position of the Fermi level, which is inside the Mo 
$t_{2g}$ bands. 

Although the MoO$_{4}$ tetrahedra do not contribute any electrons near the
Fermi level, their arrangement in double layers perpendicular to $\mathbf{a,}
$ separating the staircases of corner-sharing octahedra, has an important
impact on the low-energy electronic structure: It suppresses the hopping
between the low-energy orbitals across the double layer to the extent that
we shall neglect it in our TB model for the six lowest bands\footnote{{As
seen in FIG.~\ref{FIG1} (b) and Chart (\ref{ac}), the shortest path for
hopping of low-energy electrons across the double layer of tetrahedrally
coordinated molybdenums is Mo5 - (MO6) - MO5, i.e. from Mo5 in a bottom
ribbon, along $-\mathbf{z}$ to (MO6) in the top ribbon of the neighboring
staircase, and then along $\mathbf{-x}$ or $\mathbf{+y}$ to MO5 in that top
ribbon. This zigzag path thus passes via merely \emph{one} tetrahedrally
coordinated Mo atom and gives rise to the slight $k_{a}$-dispersion of the
two nearly degenerate }$xy$ bands {seen most clearly in FIG.~\ref{FIG2}
along CE and 1 eV above the Fermi level. \label{slabhop}}}.

\subsection{The $t_{2g}$ bands\label{Sectt2g}}

The remaining $4d$ orbitals on octahedrally coordinated Mo are the $t_{2g}$
orbitals, $xy,$ $xz,$ and $yz,$ whose lobes point \emph{between} the four O
neighbors in respectively the $xy,$ $xz,$ and $yz$ planes and therefore form
relatively weak $pd\pi $ bonds and antibonds, e.g. $\mathrm{Mo}\,xy\pm 
\mathrm{O}\,y$ on the $\frac{\mathbf{x}}{2}$ bond. Whereas the bonds are
dominated by oxygen and form bands that are part of the O $2p$ continuum
below about $-4$ eV, the \emph{antibonds} are dominated by Mo and form $%
4\times 2\times 3=24$ bands which extend from $+3.0~$eV down to the bottom
of the Mo $4d$ continuum at $-$0.7 eV. This spread in energy is \emph{%
conventionally} described as due to hopping between \emph{dressed} Mo$%
\,t_{2g}$ orbitals, where the dressing consists of the $pd\pi $ anti-bonding
tails on the four oxygens in the plane of the orbital. Since the dressed
orbitals are planar, the strongest hoppings are between \emph{like} $t_{2g}$
orbitals which are nearest neighbors in the same plane, e.g. as seen in the
right-hand panel of FIG.$~$\ref{FIGyzb&xyamc} between the dressed $xy$
orbital on Mo1$_{\text{\textbf{o}}}$ and those on Mo4$_{\mathbf{x}},$ Mo4$_{%
\mathbf{y}}$, Mo2$_{\mathbf{-x}},$ and Mo2$_{\mathbf{-y}}$ , or as seen in
the left-hand panel between the dressed $yz$ orbital on Mo1$_{\mathbf{o}}$
and those on Mo4$_{\mathbf{y}},$ Mo4$_{\mathbf{z}}$, Mo2$_{\mathbf{-y}},$
and Mo2$_{\mathbf{-z}}$. These hoppings are $dd\pi $-like and of magnitude $%
t=-1~$eV$.$

The main dispersion of the $xy$ band is, therefore, in the direction of the $%
xy $ lobe pointing along $\mathbf{y-x}=\mathbf{b,}$ that of the $yz$ band is
in the direction of the $yz$ lobe pointing along $\mathbf{y+z=}\frac{\mathbf{%
c+b}}{2}\mathbf{\ }$(see left-hand FIG.$~$\ref{FIGyzb&xyamc}), and that of
the $xz$ band is in the direction of the $xz$ lobe pointing along the $%
\mathbf{x+z=}\frac{\mathbf{c-b}}{2};$ see Charts (\ref{ac}) and (\ref{xy})

\begin{figure}[tbh]
\includegraphics[width=\linewidth]{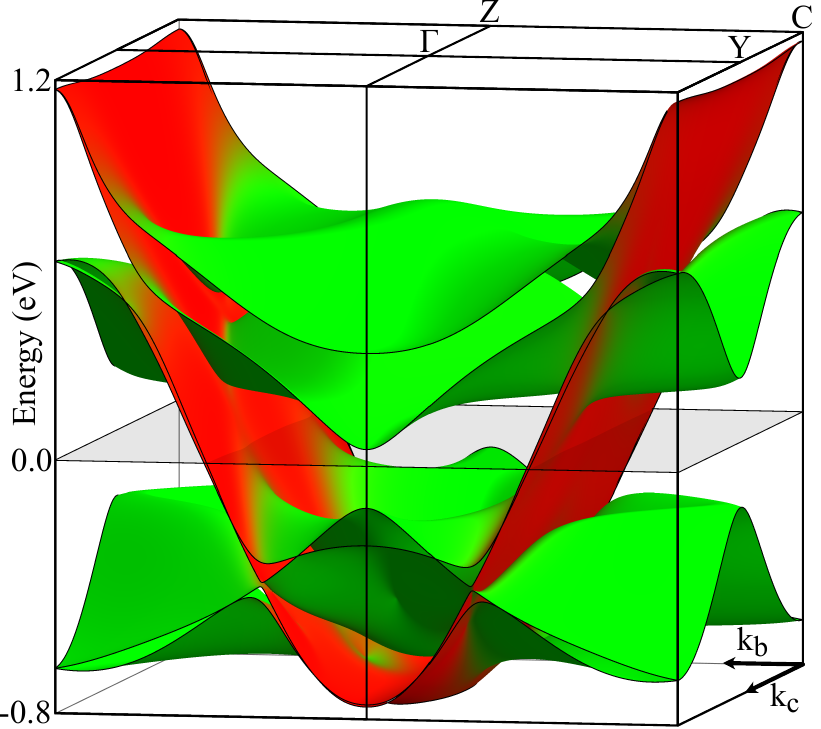}
\caption{LDA $t_{2g}$ energy band structure and Fermi level of LiPB at half
filling. Red and green colors indicate respectively $xy$ and $xz/yz$
characters. The energy region is from 0.8 eV below to 1.2 eV above $E_{F}$
(see FIG.~ \protect\ref{FIG2}) and the $\mathbf{k}$-space region is the BZ
(see FIG.~ \protect\ref{FIGDoubleZone}). The LDA TB parameters listed in
Tables (\protect\ref{taup})-(\protect\ref{lp}) were used, like in FIG.~ 
\protect\ref{ARPES_Bandstructure_LDA}~(a) in Paper II.}
\label{3Dt2gBands}
\end{figure}

\begin{figure}[tbh]
\includegraphics[width=\linewidth]{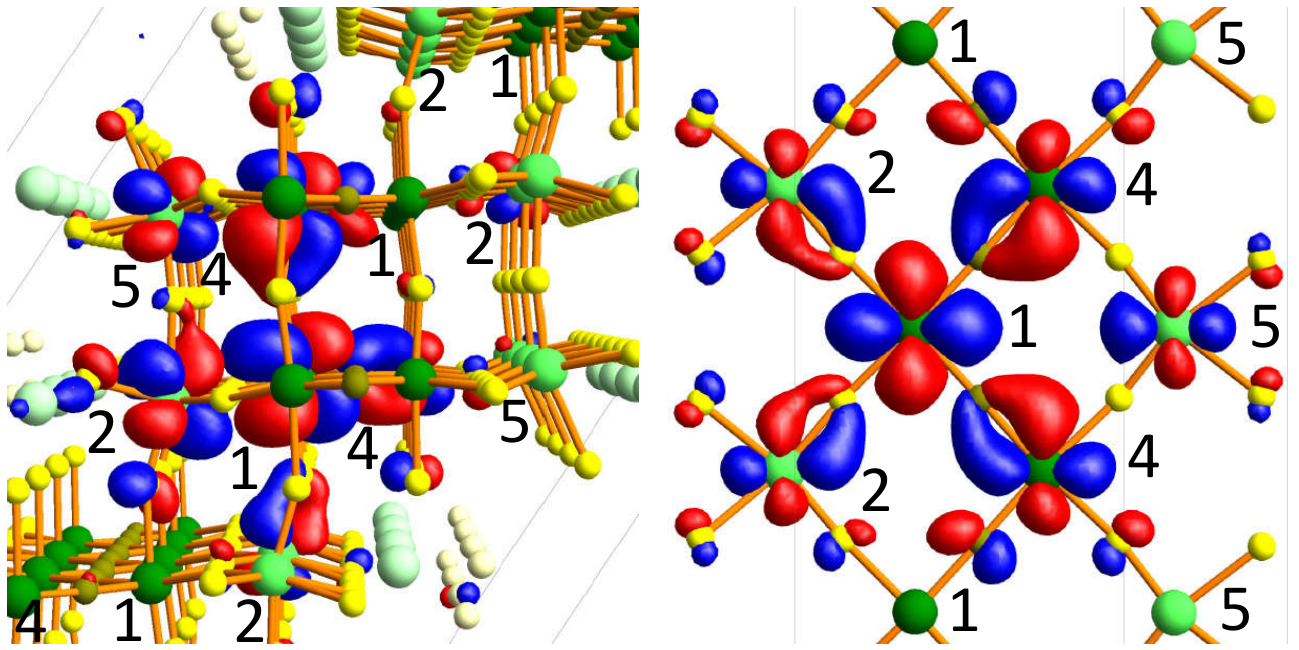}
\caption{\emph{Left:} $yz$ standing-wave state with $k_{c}+k_{b}=$ $\pm 
\frac{1}{2}$ which behaves like $\cos 2\protect\pi \frac{1}{2}\left(
r_{c}+r_{b}\right) .$ View along $\mathbf{b}$ as in FIG.~\protect\ref{FIG1}%
~(c) and Chart (\protect\ref{ac}). \emph{Right: }$xy$ standing-wave state
with $\left\vert k_{b}\right\vert =$ $\frac{1}{4}\sim k_{F}$ which behaves
like $\cos 2\protect\pi \frac{1}{4}r_{b}.$ View along $-\mathbf{z\sim a-c}$
as in (d) and Chart (\protect\ref{xy}). The sign of a lobe is indicated by
its color. See also FIG.~\protect\ref{Wannier} with caption. As explained in
Sect.~VI, these figures are identical with those shown in FIG.~ 9 of the $yz$
and $xy$ Wannier Orbitals (WOs) on site 1. The $xy$ WO has a halo consisting
of $xy$ partial waves on the four nearest Mo-neighbor sites, 4 and 2, plus a
weaker one on the next-nearest Mo-neighbor site 5. The $yz$ WO has a halo
consisting of $yz$ patial waves (of unequal magnitude) on the two nearest-Mo
neighbor sites 4, plus weaker ones on Mo2 and 5.}
\label{FIGyzb&xyamc}
\end{figure}

\subsubsection{The \emph{four} $xy$ bands.}

The right-hand panel of FIG.$~$\ref{FIGyzb&xyamc} shows the $xy$
standing-wave state with $\left\vert k_{b}\right\vert $=$\frac{1}{4}\sim
k_{F}$ which behaves like $\cos 2\pi k_{b}\mathbf{b}^{\ast }\mathbf{\cdot b}%
r_{b}\equiv $ $\cos 2\pi \frac{1}{4}r_{b},$ i.e. is even around the $\mathbf{%
c+a}$ line through Mo1 and Mo5 and has nodes at the Mo1-Mo5 lines translated
by $\pm \mathbf{b}$. Here, the orientation is like in FIG.$~$\ref{FIG1}~(d)
with the Mo numbering given in the left-hand panel of Chart~(\ref{xy}). We
see that the amplitudes of the dressed $xy$ orbitals are largest at Mo1 and
decrease in the order Mo4, Mo2, and Mo5, thus\textbf{\ }following the
decrease of the Mo coordination mentioned in Sect.$~$\ref{SectDims}. The
dressed orbitals on the four nearest neighbors (Mo4 at $x$\ and $y,$\ and
Mo2 at $-x$\ and $-y$) anti-bond to the central orbital, i.e.
nearest-neighbor lobes have different colors. This is the reason why the
overlap from the neighboring dressed $xy$ orbital weakens the O $y$ (or $x$)
amplitude on the common oxygen such that its contour merges with that of the
weaker Mo neighbor. Hence, O $p$ is $pd\pi $ anti-bonding with Mo1$\,xy$ and
bonding with Mo4$\,xy$ and Mo2$\,xy.$ The net result is $pd\pi $
non-bonding, essentially.

The $xy$ band disperses almost exclusively in the $\mathbf{b}$ direction,
and now, we imagine going to the $xy$ state with $\left\vert
k_{b}\right\vert $=$\frac{3}{4}$=$\frac{1}{4}\left[ \mathrm{mod}\frac{1}{2}%
\right] $ and energy $\sim $2.8 eV above $E_{F},$ i.e. to the state in the 
\emph{next} $xy$ band. Here, the signs (colors) of the dressed $xy$ orbitals
on the four nearest neighbors (Mo4 and Mo2) will have changed, whereby the
overlaps on the common oxygens shared with Mo1 will have their amplitudes
enhanced and the O$\,y$ (or $x$) contour will be separated by a node, not
only from the stronger Mo1-$xy$ contour, but also from the weaker Mo4-$xy$
(or Mo2-$xy)$ contour. In the following, we shall refer to the $xy$ band
with the lower/higher energy as the Mo1-Mo4 bonding/anti-bonding band,
although both of these bands are $pd\pi $ non- or anti-bonding; but the one
with the lower energy has fewer $pd\pi $ nodes.

The dressed $xy$ orbitals lie in the plane of a ribbon, and those along the
infinite zigzag chain, $\overset{\mathbf{-x}}{\rightarrow }$~Mo1 $\overset{%
\mathbf{y}}{\rightarrow }$ Mo4 $\overset{\mathbf{-x}}{\rightarrow }$, with 
\emph{pseudo} translation $\frac{\mathbf{y-x}}{2}\mathrm{=}\frac{\mathbf{b}}{%
2}$ form the well-known \cite{Whangbo1988}\cite{Satpathy2006}\cite%
{Chudzinski2012} quasi-1D band with dispersion\footnote{%
We denote energy bands $E\left( \mathbf{k}\right) =\varepsilon \left( 
\mathbf{k}\right) +E_{0}$, and their dispersions $\varepsilon \left( \mathbf{%
k}\right) $. Here, $E_{0}$ is the center of the band.\label{dispersion}}:%
\begin{equation}
\varepsilon _{xy}\left( \mathbf{k}\right) \sim 2t\cos \left( 2\pi \mathbf{k}%
\cdot \frac{\mathbf{b}}{2}\right) =2t\cos \pi k_{b},  \label{exy}
\end{equation}%
where $t\sim -1~$eV. Since $\mathbf{b,}$ and not $\frac{\mathbf{b}}{2},$ is
the proper lattice translation because Mo1 and Mo4 are not equivalent, the
band must be folded from the large BZ bound by the midplanes $\left( k_{b}%
\text{=}\pm \text{1}\right) $ of the reciprocal-lattice vectors $\pm 2%
\mathbf{b}^{\ast },$ into the proper, small BZ bound by the midplanes $%
\left( k_{b}=\pm \frac{1}{2}\right) $ of $\pm \mathbf{b}^{\ast }$ whereby it
becomes $-2t\cos \pi k_{b}.$ An equivalent prescription --more useful than
BZ-folding, as we shall see below for the bonding $yz$ and $xz$ bands,-- is
to say that if $\mathbf{k}$ must be limited to the proper, small BZ, then we
must also consider the band, 
\begin{equation}
\varepsilon _{xy}\left( \mathbf{k+b}^{\ast }\right) \sim 2t\cos \left[ \pi
\left( k_{b}+1\right) \right] =-2t\cos \pi k_{b},  \label{exyu}
\end{equation}%
translated by the proper reciprocal lattice vector, $-\mathbf{b}^{\ast }.$
Finally, the \emph{in}equivalence of --or "dimerization into"-- Mo1 and Mo4,
couples the $xy\left( \mathbf{k}\right) $ and $xy\left( \mathbf{k+b}^{\ast
}\right) $ bands, and where they are degenerate --which is for $\left\vert
k_{b}\right\vert $=$\frac{1}{2},$ i.e. at the boundaries of the proper BZ
(YC)-- they are gapped by $\pm $0.3~eV. Since this gap is relatively large,
the $xy\left( \mathbf{k}\right) $ band is bonding and the $xy\left( \mathbf{%
k+b}^{\ast }\right) $ band anti-bonding between Mo1 and Mo4 for $\mathbf{k}$
inside the proper BZ. The latter, empty $xy$ band, which extends from
approximately 1.7 to 3.4~eV above $E_{F},$ we shall neglect in the bulk of
the present papers, as was already mentioned in Sect. \ref{Sectbims}.

Degenerate and parallel with the Mo1-Mo4 bonding and anti-bonding $xy$ bands
running along the lower ribbon are MO4-MO1 bonding and anti-bonding $XY$
bands running along the upper ribbon [see FIG.$~$\ref{FIG1}~(d), Chart~(\ref%
{ac}), and the right-hand panel of Chart~(\ref{xy})]. Their $\left\vert
k_{b}\right\vert $=$\frac{1}{4}\sim k_{F}$ standing-wave state looks like
the one shown on the right-hand side of FIG.$~$\ref{FIGyzb&xyamc}, but has
MO1 on top of Mo4 and vice versa. Viewed along $\mathbf{b,}$ the appearance
of the $xy$ and $XY$ states is like that of the $xy$ and $XY$ WOs in the
first two columns on the top row of FIG.$~$\ref{Wannier}. From there, we
realize that these flat, parallel states are well separated, each one on its
own ribbon, with no contribution on the oxygens in between. The $dd\delta $%
-like hops between the $xy$ and $XY$ orbitals inside the same bi--ribbon $%
\left( t_{\perp }\equiv t_{1}+u_{1}\sim -14\,\mathrm{meV}\right) $ and
between the $XY$ and $xy$ orbitals in different bi-ribbons $\left( t_{\perp
}^{\prime }\equiv t_{1}-u_{1}\sim -8\,\mathrm{meV}\right) ,$ give the bands
a \emph{perpendicular} ($k_{c}$-)$\ $\emph{dispersion}, which is two orders
of magnitude smaller than the $k_{b}$-dispersion given by Eq.$\,$(\ref{exy}%
). If all ribbons were translationally equivalent, i.e.$\,$if the primitive
translations (neglecting $\mathbf{a)}$ were $\mathbf{x+z}=\frac{\mathbf{c-b}%
}{2}$ and $\mathbf{y+z=}\,\frac{\mathbf{c+b}}{2}$ with reciprocal-lattice
translations $\mathbf{c}^{\ast }\mathbf{-b}^{\ast }$ and $\mathbf{c}^{\ast }+%
\mathbf{b}^{\ast },$ the $dd\delta $ hopping would add 
\begin{equation*}
2t_{1}\left[ \cos \pi \left( k_{c}-k_{b}\right) +\cos \pi \left(
k_{c}+k_{b}\right) \right] =4t_{1}\cos \pi k_{b}\cos \pi k_{c}
\end{equation*}%
to Eq.$\,$(\ref{exy}). But since the primitive translations are really $%
\mathbf{b}$ and $\mathbf{c,}$ we must --if we want to confine $\mathbf{k}$
to the proper BZ-- also add the equivalent term translated by the proper
reciprocal lattice vector\footnote{%
Substituting $\mathbf{c}^{\ast }$ by $-\mathbf{c}^{\ast }$ gives the same
result because their difference, $2\mathbf{c}^{\ast },$ is a translation of
the reciprocal lattice spanned by $\mathbf{c}^{\ast }\mathbf{-b}^{\ast }$
and $\mathbf{c}^{\ast }\mathbf{+b}^{\ast }.$\label{c}}, $-\mathbf{c}^{\ast
}. $ As a result, we get for the two $\frac{1}{2}$-filled $xy$ bands:%
\begin{equation}
\varepsilon _{xy}\left( 
\begin{array}{c}
\mathbf{k} \\ 
\mathbf{k+c}^{\ast }%
\end{array}%
\right) \sim 2t\left[ 1\pm \left( 2t_{1}/t\right) \cos \pi k_{c}\right] \cos
\pi k_{b},  \label{epsxy}
\end{equation}%
where the distortion caused by the\textbf{\ }gap extending upwards from $%
\sim 1.1~$eV above $E_{F}$ has been neglected. As long as $\mathbf{k}$ is
inside the 1st BZ $\left( \left\vert k_{c}\right\vert \leq \frac{1}{2}%
\right) ,$ the $xy\left( \mathbf{k}\right) $ band is bonding and the $%
xy\left( \mathbf{k+c}^{\ast }\right) $ band anti-bonding between ribbons,
i.e. between $xy$ and $XY.$ In the 2nd BZ $\left( \left\vert
k_{c}-1\right\vert \leq \frac{1}{2}\right) ,$ the opposite is true (see FIG.$%
\,$\ref{ThreePureBands}). The translational \emph{in}equivalence of the two
ribbons --i.e. the dimerization into bi-ribbons-- finally splits the
degeneracy of the $xy\left( \mathbf{k}\right) $ and $xy\left( \mathbf{k+c}%
^{\ast }\right) $ bands at the BZ boundaries $\left\vert k_{c}\right\vert $=$%
\frac{1}{2}$ (the ZCED planes) by $\pm 2\sqrt{2}u_{1}\cos \pi k_{b},$ which
for $k_{b}$=$\frac{1}{4}\sim k_{F}$ is a mere $\pm 8~$meV.

\subsubsection{The two $yz$ and the two $xz$ bands.}

In the planes perpendicular to the bi-ribbons [FIG.~\ref{FIG1}~(c) and
Chart~(\ref{ac})] and cutting them along the $\mathbf{y}$-strings [FIG.~\ref%
{FIG1}~(d) and Chart~(\ref{xy})], lie the dressed $yz$ orbitals, and in the
planes cutting along the $\mathbf{x}$-strings, lie the dressed $xz$
orbitals. The left-hand panel of FIG.$\,$\ref{FIGyzb&xyamc} shows that $%
\left\vert k_{c}+k_{b}\right\vert $=$\frac{1}{2}$ standing-wave state of the 
$yz$ band which behaves like $\cos 2\pi \frac{1}{2}\left( r_{c}+r_{b}\right)
,$ i.e. is even around the Mo1-containing planes which are perpendicular to $%
\mathbf{c}^{\ast }+\mathbf{b}^{\ast }$ and has nodes in the MO1-containing
planes. Like for $xy$ state in the right-hand panel, the dressed $yz$
orbitals on the four nearest neighbors in the plane of the orbital (MO4 at $%
\mathbf{z}$, Mo4 at $\mathbf{y,}$ Mo2 at $-\mathbf{y}$, and MO2 at $-\mathbf{%
z}$) are $dd\pi $ anti-bonding with the dressed $yz$ on the central Mo1,
which means $pd\pi $ non-bonding with the oxygen. Here again, the amplitudes
of the dressed $yz$ orbitals decrease like the Mo coordination.

\begin{figure}[tbh]
\includegraphics[width=\linewidth]{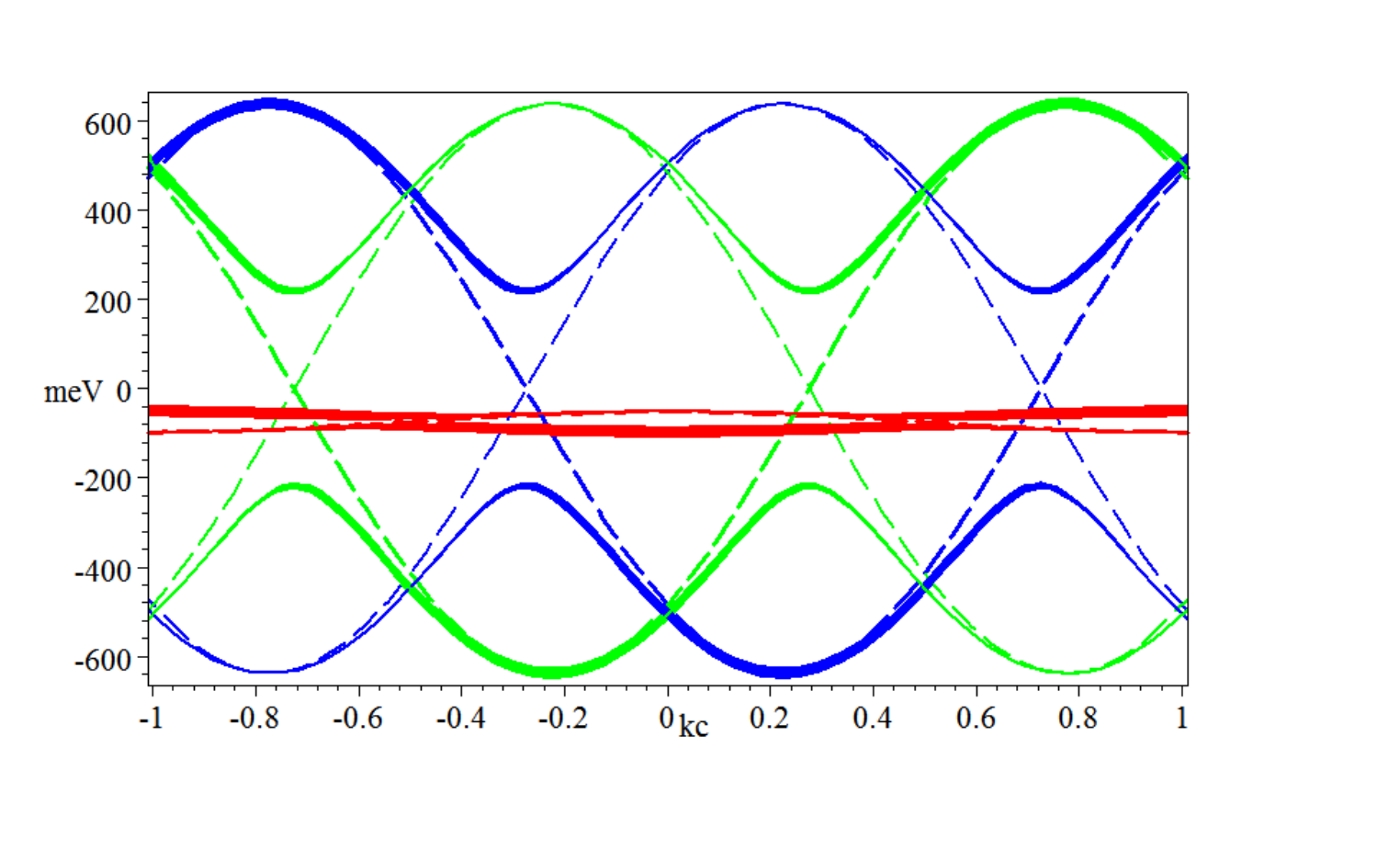}
\caption{The six pure-$m$ $t_{2g}$ bands ($m=yz$ green, $xz$ blue, and $xy$
red) as functions of $k_{c}$ in the double zone $\left( -1<k_{c}\leq
1\right) $ and for $k_{b}$=0.225, i.e. along the brown dot-dashed line
containing P and Q in FIG.~\protect\ref{FIGDoubleZone}. 10\% hole doping
will place the Fermi level inside the red bands. The $m\left( \mathbf{k}%
\right) $ and $m\left( \mathbf{k+c}^{\ast }\right) $ bands, whose
dispersions are described in Sect.~\protect\ref{Sectt2g}, have period $%
\Delta k_{c}$=2 and are shown in respectively fat- and normal dashed lines.
The $\left\{ m\left( \mathbf{k}\right) ,m\left( \mathbf{k+c}^{\ast }\right)
\right\} $-hybridized bands have period 1 and are shown in full lines. Their 
\emph{additional} (vertical) fatness is proportional to the $\left\vert 
\mathbf{k}\right\rangle $-character and has period 2. The ARPES-refined TB
parameters listed in Tables (\protect\ref{taup}), (\protect\ref{tp}), (%
\protect\ref{A&Gp}), and (\protect\ref{EFARPES}) were used and, accordingly,
the zero of energy is the center of the gap, i.e., the common energy of the $%
xz,$ $XZ,$ $yz,$ and $YZ$ WOs. The Fermi level of samples G and H is at
+75\thinspace meV.}
\label{ThreePureBands}
\end{figure}

Whereas in the plane of the $xy$ orbital, Mo4 --like Mo1-- has four nearest
neighbors of molybdenums coordinated octahedrally with oxygen, in the plane
of the $yz$ orbital, Mo4 has only three neighbors, and so does Mo2, while
Mo5 has merely two. As noted in Sect.~\ref{SectDims}, this is due to the
stacking into a staircase of bi-ribbons (\ref{ac}). As a result, the $yz$
orbitals on the Mo1- and MO1-sharing zigzag double chain,%
\begin{equation}
\begin{array}{cc}
\mathbf{c} & \nearrow \\ 
\mathbf{a} & \searrow \\ 
\mathbf{z} & \uparrow%
\end{array}%
\,\fbox{$%
\begin{array}{cccccc}
&  & \mathbf{5} & \rightarrow 4 & \mathbf{\uparrow 1\rightarrow } & 2\uparrow
\\ 
&  & \mathbf{\rightarrow 2} & \uparrow 1\rightarrow & \uparrow \mathbf{4} & 5
\\ 
\mathbf{5} & \rightarrow 4 & \mathbf{\uparrow 1\rightarrow } & 2\uparrow & 
&  \\ 
\mathbf{\rightarrow 2} & \uparrow 1\rightarrow & \mathbf{4\uparrow } & 5 & 
& 
\end{array}%
$},  \label{doublechain}
\end{equation}%
running up the staircase with pseudo translation $\mathbf{z+y}=\frac{\mathbf{%
c+b}}{2},$ form a quasi-1D band dispersing like%
\begin{equation}
\varepsilon _{yz}\left( \mathbf{k}\right) =2A_{1}\cos \left( 2\pi \mathbf{%
k\cdot }\frac{\mathbf{c+b}}{2}\right) =2A_{1}\cos \pi \left(
k_{c}+k_{b}\right) ,  \label{eyz}
\end{equation}%
with an effective hopping integral, $A_{1}\sim -0.3$\thinspace eV and
bandwidth 4$\left\vert A_{1}\right\vert \sim 1.2$ eV. Because the hopping
between ribbons proceeds via Mo4 inside the bi-ribbon, but via Mo2 between
bi-ribbons, and because the former distance is shorter than the latter, the
hopping integrals are different, respectively $A_{1}+G_{1}\sim -0.3-0.1\sim $
$-0.4$ eV and $A_{1}-G_{1}\sim $ $-0.2$ eV. This dimerization into
bi-ribbons causes $\mathbf{c,}$ rather than $\frac{\mathbf{c+b}}{2},$ to be
a primitive lattice translation whereby the $yz\left( \mathbf{k}+\mathbf{c}%
^{\ast }\right) $ band with dispersion 
\begin{equation}
\varepsilon _{yz}\left( \mathbf{k+c}^{\ast }\right) =-2A_{1}\cos \pi \left(
k_{c}+k_{b}\right)  \label{eyzc}
\end{equation}%
is equivalent to the $yz\left( \mathbf{k}\right) $ band (\ref{eyz}). Where
these bands are degenerate, i.e. for $\left\vert k_{c}+k_{b}\right\vert $=$%
\frac{1}{2},$ they gap by $\pm 2\left\vert G_{1}\right\vert \sim \pm 0.2~$eV
whereby they become:%
\begin{equation}
\varepsilon _{yz}=\pm \sqrt{\left[ 2A_{1}\cos \pi \left( k_{c}+k_{b}\right) %
\right] ^{2}+\left[ 2G_{1}\sin \pi \left( k_{c}+k_{b}\right) \right] ^{2}}.
\label{gap}
\end{equation}%
For $\mathbf{k}$ between the $\left\vert k_{c}+k_{b}\right\vert $=$\frac{1}{2%
}$ planes, the $yz\left( \mathbf{k}\right) $ and $yz\left( \mathbf{k+c}%
^{\ast }\right) $ bands are respectively bonding and anti-bonding between
neighboring ribbons. The two $yz$ bands, decorated by the $\mathbf{k}$%
-character (\ref{eyz}), are shown in green in FIG.$\,$\ref{ThreePureBands}.

It should be noted that the gapping takes place for $\left\vert
k_{c}+k_{b}\right\vert $=$\frac{1}{2},$ which is \emph{not} at the boundary
of the conventional BZ, $\left\vert k_{b}\right\vert $=$\frac{1}{2}$ and $%
\left\vert k_{c}\right\vert $=$\frac{1}{2}$ shown in FIG.s$\,$\ref{FIG2} and %
\ref{FIGDoubleZone}, but where the $\varepsilon _{yz}\left( \mathbf{k}%
\right) $ and $\varepsilon _{yz}\left( \mathbf{k+c}^{\ast }\right) $ bands
are degenerate. Nevertheless, the zone centered at $\Gamma \left( 0,0\right) 
$ and bound by the planes $\left\vert k_{b}\right\vert $=$\frac{1}{2}$ and $%
\left\vert k_{c}+k_{b}\right\vert $=$\frac{1}{2}$ (ZYAD) \emph{is} a
primitive cell of the reciprocal lattice, and we call it a \emph{physical
zone,} useful for understanding properties of the $yz$ bands.

With the substitution: $\mathbf{b}\rightarrow \mathbf{-b},$ everything said
about the $yz$ bands holds for the $xz$ bands (shown in blue in FIG.$\,$\ref%
{ThreePureBands}).

As regards choices of zones, we can either take:%
\begin{eqnarray}
\left\vert k_{b}\right\vert &\leq &\frac{1}{2}\;\mathrm{and}\;\left\vert
k_{c}\right\vert \leq \frac{1}{2},  \label{xyzone} \\
\left\vert k_{b}\right\vert &\leq &\frac{1}{2}\;\mathrm{and}\;\left\vert
k_{c}+k_{b}\right\vert \leq \frac{1}{2},  \label{yzzone} \\
\mathrm{or\;}\left\vert k_{b}\right\vert &\leq &\frac{1}{2}\;\mathrm{and}%
\;\left\vert k_{c}-k_{b}\right\vert \leq \frac{1}{2},  \label{xzzone}
\end{eqnarray}%
but \emph{not} $\left\vert k_{c}+k_{b}\right\vert \leq \frac{1}{2}$ and $%
\left\vert k_{c}-k_{b}\right\vert \leq \frac{1}{2}$ whose volume (area) is
only \emph{half} the BZ volume (see FIG.$\,$\ref{FIGDoubleZone}).
Expressions (\ref{xyzone})-(\ref{xzzone}) thus define the \emph{physical}
zones for respectively the $xy,$ the $yz,$ and the $xz$ bands.

\subsubsection{Line-up of the six lowest $t_{2g}$ bands}

The bottoms of the $xz$ and $yz$ bands and those of the degenerate $xy$
bands are \emph{all} at $\Gamma \left( \mathbf{k\mathrm{=}0}\right) $ and at
the energy of that linear combination of the dressed $xz,\,yz,$ or $xy$
orbitals which is the least anti-bonding between all octahedral molybdenums
(FIG. \ref{FIG2}). According to the LDA, these energies are: $B_{xz/yz}\sim
E_{F}-0.6$~eV and $B_{xy}\sim E_{F}-0.7$~eV. Since the $4\left\vert
A_{1}\right\vert $ width of the $xz$ and $yz$ bands is only about one third
the $4\left\vert t\right\vert $-width of the $xy$ bands, the $4\left\vert
G_{1}\right\vert $-gap halfway up in the $xz$ and $yz$ bands extends between
the energies $B_{xz/yz}+2\left\vert A_{1}\pm G_{1}\right\vert \sim \pm 0.2~$%
eV with respect to the Fermi level set by the $\frac{1}{2}$-filled, lower $%
xy $ bands. In the following Paper II (e.g. FIG. \ref%
{ARPES_Bandstructure_LDA}), we shall see that agreement with ARPES requires
a 0.1 eV downward shift of the $xz/yz$ bands with respect to the $xy$ bands,
whereby $B_{xz/yz}\sim $ $E_{F}-0.7$~eV$\,\sim B_{xy}\equiv B.$ This
low-energy $t_{2g}$ band structure is shown in FIG.$\,$\ref{ThreePureBands}
along the line $k_{b}$=0.225 perpendicular to $\mathbf{b,}$ the direction of
quasi 1D conductivity.

In summary, since the 6 lowest bands are $t_{2g}$ like, the 6 electrons
would half fill them in case of weak Coulomb correlations, thus
corresponding to a $t_{2g}^{3}$ configuration. Covalency between the $xz$
and $XZ$ orbitals, as well as between the $yz$ and $YZ$ orbitals, together
with the availability of one $xz$ and one $yz$ electron per string, result
in the covalent bonds which dimerize the ribbons into bi-ribbons and thereby
gap the $xz$ and $yz$ bands into filled bonding and empty anti-bonding
bands. The remaining one $xy$ electron per string finally half fills the
quasi-1D band dispersing strongly along $\mathbf{b.}$

The six $t_{2g}$ bands are illustrated in FIG.$\,$\ref{3Dt2gBands}, from
where it is seen that the gaps in the green $xz$ and $yz$ bands are around
the center of the red, metallic $xy$ bands and, hence, around the grey,
transparent Fermi level.

\begin{figure}[tbh]
\includegraphics[width=\linewidth]{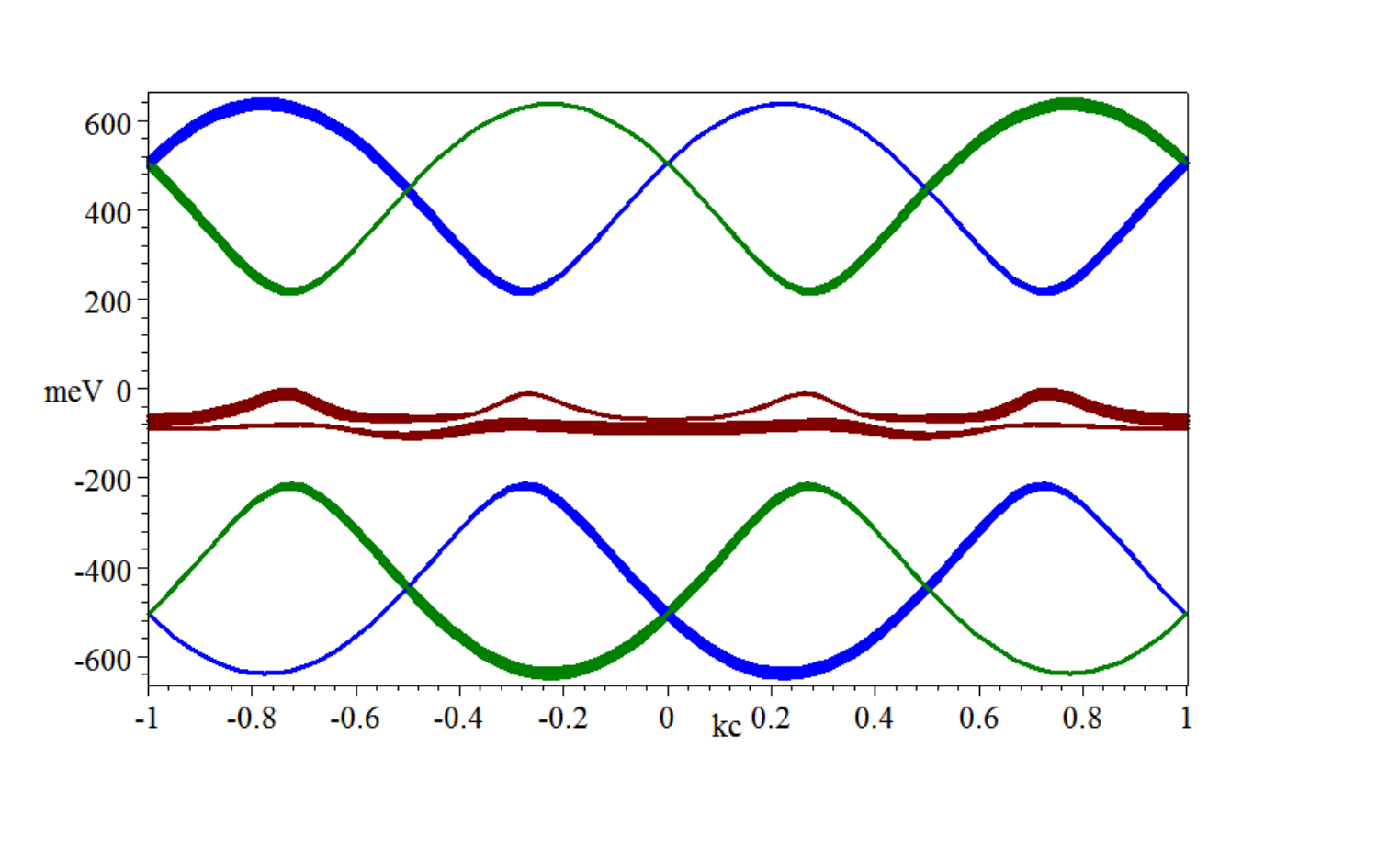}
\caption{As FIG.~\protect\ref{ThreePureBands} ($k_{b}$=0.225), but including
the hybridization of the $\widetilde{xy}$ band (dark red) with the $xz$
(blue) and $yz$ (green) valence and conduction bands via L\"{o}wdin
downfolding as explained in Paper III. This hybridization brings in the
parameters $\protect\alpha _{n}\pm \protect\gamma _{n}$ and $a_{n}\pm g_{n}$
[Eq.s (\protect\ref{ag}) and (\protect\ref{ap})]. Not included in this
figure are the $xy$-hybridizations of the blue $xz$ and the green $yz$
bands, as well as the hybridization between the $xz$ and $yz$ bands. }
\label{ThreeBands}
\end{figure}

\subsubsection{Constant energy contours (CECs)\label{subsubsubCEC}}

\begin{figure}[tbh]
\includegraphics[width=\linewidth]{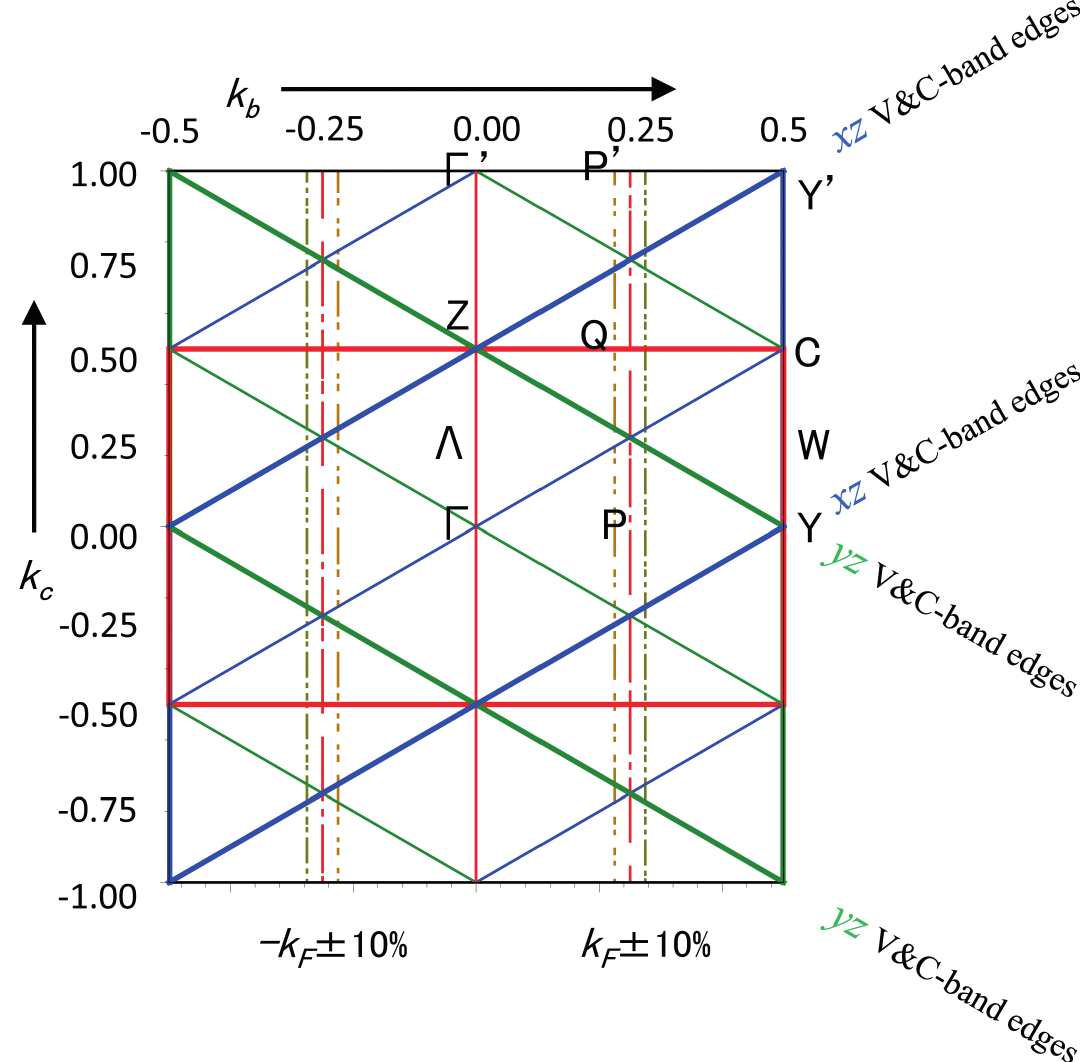}
\caption{Double zone: $-0.5<k_{b}\leq 0.5$ and $-1<k_{c}\leq 1.$ The red,
blue, and green solid lines show the \emph{physical} zones for respectively
the $xy$, $xz$, and $yz$ pure bands; see Eq.s (\protect\ref{xyzone}), (%
\protect\ref{xzzone}), and (\protect\ref{yzzone}). The red zone is the BZ,
and its irreducible part is the one with $0\leq k_{c}\leq 0.5.$ The
reciprocal-lattice points are: $\mathbf{G}=M\mathbf{b}^{\ast }+N\mathbf{c}%
^{\ast },$ i.e. $\left( k_{b},k_{c}\right) =\left( M,N\right) ,$ with $M$
and $N$ integers. Those (\protect\ref{undimrecip}) with $M+N$ \emph{even},
form the lattice reciprocal of the un-dimerized lattice (\protect\ref%
{undimprim}) whose BZ is the double zone. Shifting this even reciprocal
lattice by $\mathbf{c}^{\ast }$ yields the \emph{odd} reciprocal lattice;
see Sect.~\protect\ref{Sectk+c}. Weak lines indicate the positions of
pure-band maxima and minima (see FIG. 6). The red dot-dashed lines indicate
the positions of the left and the right, doubly degenerate Fermi-surface
sheets for stochiometric 2(LiMo$_{6}$O$_{17}$). The brown and olive
dot-dashed lines correspond to 10\%, respectively, hole and electron doping.
Because $c/b=1.720\approx \protect\sqrt{3},$ the $\mathrm{\Gamma C\Gamma }%
^{\prime }$ triangles are almost equilateral, and since this is so in the
present figure, it is to scale. The $k_{b}$ axis, has been turned by $\sim
90^{\circ }$ w.r.t. $\mathbf{b}$ in the real-space FIG.s \protect\ref{FIG1}%
~(d), and \protect\ref{FIGyzb&xyamc} right. }
\label{FIGDoubleZone}
\end{figure}

FIG.$~$\ref{FIGDoubleZone} shows the double zone, $\left\vert
k_{b}\right\vert \leq \frac{1}{2}$ and $\left\vert k_{c}\right\vert \leq 1,$
and --schematically and in weak lines-- constant-energy contours (CECs) for
the $xz\left( \mathbf{k}\right) $ bands in blue, the $yz\left( \mathbf{k}%
\right) $ bands in green, and the (almost) degenerate, $\frac{1}{2}$-filled $%
xy\left( \mathbf{k}\right) $ and $xy\left( \mathbf{k+c}^{\ast }\right) $
bands (\ref{epsxy}) in red. The bottoms of these four bands are along
respectively the blue, green, and red lines passing through $\Gamma \left(
0,0\right) .$ The tops of the $xz\left( \mathbf{k}\right) $ and $yz\left( 
\mathbf{k}\right) $ bands are along the blue and green lines passing through 
$\Gamma ^{\prime }\left( 0,\pm 1\right) .$ The top of the degenerate $xy$
bands (which is a cusp because Eq.~(\ref{epsxy}) neglects the Mo1-Mo4 gap)
is along the red, vertical BZ boundary $\left\vert k_{b}\right\vert $=$\frac{%
1}{2}$.

For the degenerate $xy$ bands, we also show the CECs for three energies
close to the Fermi level corresponding to half-filling (red dot-dash), 10\%
hole- (brown dot-dash), and 10\% electron (olive dot-dash) doping. For the
gapped $xz\left( \mathbf{k}\right) $ and $xz\left( \mathbf{k+c}^{\ast
}\right) $ bands we show the coinciding CECs for the valence- and
conduction-band edges (blue solid lines), and similarly for the $yz$-band
edges (green solid lines). The CECs for the $xz\left( \mathbf{k+c}^{\ast
}\right) $ and $yz\left( \mathbf{k+c}^{\ast }\right) $ bands of course equal
those for respectively the $xz\left( \mathbf{k}\right) $ and $yz\left( 
\mathbf{k}\right) $ bands, but translated along $k_{c}$ by an odd integer.

As seen in FIG.$\,$\ref{ThreeBands} for $k_{b}$=0.225, corresponding to 10\%
hole-doping, the $xz$\ and $yz$\ valence-band edges running along $%
\left\vert k_{c}\mp k_{b}\right\vert $=$\frac{1}{2}$\ ($\mathrm{YZY}^{\prime
}$) and merely 0.2 eV~below the $xy$ bands push \emph{resonance peaks} up at 
$\left\vert k_{c}\right\vert $=0.725 and 0.275 in the upper $xy$\ band. This
gives rise to "notches" pointing towards Z (FIG.$~$\ref{FIGDoubleZone}) in
the inner sheets of $xy$ CECs with energies in the lower half of the gap$.$\
Near $k_{c}$=0 ($\Gamma $Y) and $k_{c}$=$\pm $1 ($\Gamma ^{\prime }\mathrm{Y}%
^{\prime }$), hybridization with the $xz$\ and $yz$\ valence and conduction
bands, which are now 0.5~eV away (FIG.$\,$\ref{ThreeBands}), reduces the $%
dd\delta $-like splitting (\ref{epsxy}) between the $xy$ bands seen FIG.$\,$%
\ref{ThreePureBands} to become almost a \emph{contact} between the two $xy$
bands and CECs.\ Near the BZ boundaries, $\left\vert k_{c}\right\vert $=$%
\frac{1}{2}$ ($\Gamma \mathrm{C}$), mixing of the $xy\left( \mathbf{k}%
\right) $ and $xy\left( \mathbf{k+c}^{\ast }\right) $ bands and
hybridization of the lower with the $xz$\ and $yz$\ conduction bands 0.5~eV
above, pushes down \emph{bulges} in the lower $xy$ band, thus causing the
outer CEC sheets to bulge outwards. We shall search for these features
--predicted here for the first time-- using new ARPES measurements in Paper
III, devoted to the detailed study and explanation of the splitting and
perpendicular dispersion of the quasi-1D metallic bands in the gap.

\section{Low-energy Wannier Orbitals\label{SectLowE}}

In the previous section, our view moved from the energy scale of the Li$%
\,2s, $ Mo$\,5sp\,4d,$ and O$\,2p$\ atomic shells to the decreasing energy
scales of the Li$^{+},$ Mo$^{6+},$ and O$^{-\,-}$ ions, to the covalently
bonded MoO$_{4}$ tetrahedra and MoO$_{6}$ octahedra and, finally, to the
low-energy bands of the MoO$_{6}$ octahedra condensed into strings, ribbons,
and staircases of bi-ribbons by sharing of the $pd\pi $-bonded O corners.
This change of focus from large to small energy scales and, concomitantly,
from small to large spatial scales, we have followed computationally with
the NMTO method in the LDA by using increasingly narrow and fine energy
meshes and increasingly sparse basis sets as was described in Sect.$\,$\ref%
{SectElCalc}.

The \emph{six lowest Mo }$4d$\emph{\ bands,} i.e. those within $\pm 0.75$ eV
around the Fermi level (see FIG.s~\ref{FIG2} and \ref{3Dt2gBands}), we found
to be completely described by the basis set consisting of the three $xy,$ $%
xz,$ and $yz$ NMTOs centered on Mo1 plus the three equivalent ones (\ref{inv}%
) centered on MO1, that is of \emph{one }$t_{2g}$\emph{-set per string,}
which is per LiMo$_{6}$O$_{17}$. Symmetrical orthonormalization yielded the
corresponding set of WOs whose $xy$ and $yz$ orbitals are what was actually
shown in FIG.$\,$\ref{FIGyzb&xyamc}. The centers of the $t_{2g}$ WOs were
chosen at Mo1 and MO1 because those are the only octahedral molybdenums
whose 6 nearest molybdenum neighbors are also octahedrally surrounded by O.

Each WO spreads out to the 4 nearest octahedral molybdenums in the plane of
the orbital and, as explained in the previous sections, this leads to almost
half the WO charge being on Mo1, slightly less on Mo4, considerably less on
Mo2, and much less on Mo5. There is no discrepancy between the $t_{2g}^{3}$
configuration and the Mo $d^{0.5}$ occupancy mentioned at the beginning of
Sect.$\,$\ref{SectElStruc}: The latter is an average over all 6 molybdenums
in a string of which only 4 carry $t_{2g}$ partial waves, which are combined
into \emph{one} set of $t_{2g}$ WOs, each one being effectively spread onto
3 molybdenums. So the occupation is perhaps more like Mo $d^{1}.$

What \emph{localizes} a $t_{2g,m}$ WO in the set of all three $t_{2g}$ WOs
on all Mo1$\,$and MO1 atoms, is the condition that its projections onto all $%
t_{2g}$ partial waves on all Mo1 and MO1 atoms, \emph{except} the $t_{2g,m}$
partial wave on the own site, must \emph{vanish}\footnote{%
Strictly speaking, this holds for the set of KPWs rather than of NMTOs and
of WOs (see Sect. \ref{SectElCalc})}. On the other hand, the WO spreads onto
any \emph{other} site and partial wave in the crystal in such a way that the
WO set spans the solutions of Schr\"{o}dinger's equation at the N+1=3 chosen
energies. For the view (\ref{ac}), this is schematically:%
\begin{equation}
\begin{array}{cc}
&  \\ 
\mathbf{c} & \nearrow \\ 
\mathbf{a} & \searrow \\ 
\mathbf{z} & \uparrow \\ 
& 
\end{array}%
\,\fbox{$%
\begin{array}{ccccccc}
\mathbf{4} & 5 &  &  & \mathbf{2} & \circ & \mathbf{4} \\ 
&  & \mathbf{5} & 4 & \circ & 2 &  \\ 
&  & \mathbf{2} & \bigotimes & \mathbf{4} & 5 &  \\ 
\mathbf{5} & 4 & \circ & 2 &  &  & \mathbf{5} \\ 
\mathbf{2} & \circ & \mathbf{4} & 5 &  &  & \mathbf{2}%
\end{array}%
$},  \label{WOac}
\end{equation}%
with $\bigotimes $ indicating the site (here Mo1) of the WO, and $\circ $
the sites where all $t_{2g}$ characters are required to vanish, i.e. the
sites of the other WOs in the $t_{2g}$ set. For the view (\ref{xy}), the
Mo1-centered WO is:%
\begin{equation}
{%
\begin{array}{cc}
&  \\ 
\mathbf{b} & \uparrow \\ 
\mathbf{c+a} & \longrightarrow \\ 
\mathbf{x} & \searrow \\ 
\mathbf{y} & \nearrow \\ 
& 
\end{array}%
}\,{%
\begin{array}{cc}
\mathrm{Mo:} & \mathrm{MO:} \\ 
\fbox{$%
\begin{array}{cccccc}
&  & \circ &  & 5 &  \\ 
& 2 &  & 4 &  &  \\ 
&  & \bigotimes &  & 5 &  \\ 
& 2 &  & 4 &  &  \\ 
&  & \circ &  & 5 & 
\end{array}%
$} & \fbox{$%
\begin{array}{cccccc}
&  & 4 &  & 2 &  \\ 
& 5 &  & \circ &  &  \\ 
&  & 4 &  & 2 &  \\ 
& 5 &  & \circ &  &  \\ 
&  & 4 &  & 2 & 
\end{array}%
$}.%
\end{array}%
}  \label{WOxy}
\end{equation}

\begin{figure*}[bht]
\includegraphics[width=\linewidth]{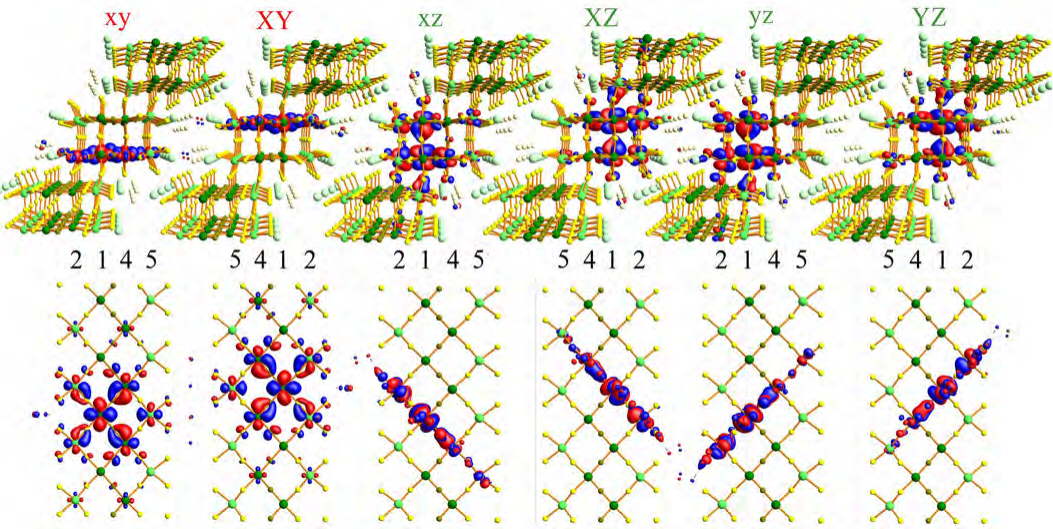}
\caption{The six Mo1- and MO1-centered $t_{2g}$ WOs spanning the six lowest
energy bands of 2(LiMo$_{6}$O$_{17})$ shown in FIG.~\protect\ref{3Dt2gBands}%
. The orientation is as in FIG.~\protect\ref{FIG1}~(c) and (d), with the
numbering of the octahedrally-coordinated molybdenums given in respectively
Charts (\protect\ref{ac}) and (\protect\ref{xy}). Shown are the WO's
constant-density surfaces containing 70\% of the WO's charge with the color
giving the sign of the lobe. With the usual 90\% cut-off, as e.g. used for
the $t_{2g}$ WOs in NiO \protect\cite{Haverkort2012}), the overlaps would
have been obscurely large. This more diffuse character of the LiPB WOs is
needed in order that they accurately describe bands that are not visibly
separated from the higher, more anti-bonding bands [see FIG.~ \protect\ref%
{FIG2}, and the long-range of the hopping integrals in (\protect\ref{taup}),
(\protect\ref{ap}), and (\protect\ref{lp})]. The slices shown in FIG.~%
\protect\ref{FIGyzb&xyamc} of standing-wave $yz$ and $xy$ states equal the
corresponding WO, because for the values of $\left\vert \mathbf{k}%
\right\vert $ chosen, the overlap from the neighbor WOs in (\protect\ref%
{kket}) is invisible.}
\label{Wannier}
\end{figure*}

Our $t_{2g}$ WOs are insensitive to the exact orientation chosen for the $%
xyz $ system --we took the one given by Eq.$\,$(\ref{xyz})-- because they
have all partial waves \emph{other than} $xy,$ $xz,$ and $yz$ on Mo1 and MO1
downfolded\footnote{%
For the $XYZ$ system located on MO1 in the upper ribbon, we merely translate
the $xyz$ system from Mo1 to MO1. These two parallel, local coordinate
systems do \emph{not} follow the space-group symmetry, specifically the
center of inversion between the nearest Mo1-MO1 neighbors. But the $t_{2g}$
projections on the Mo1 and MO1 hard spheres do, because they are even, and
this is all that matters for the WOs.}. The contents of these partial waves
are thus determined uniquely by the requirement that the WO basis set solves
Schr\"{o}dingers equation exactly at the chosen energies for the LDA
potential used to construct the WOs. In this way, the downfolding procedure
ensures that the shape of $t_{2g}$ orbitals is given by the \emph{chemistry}
rather than by the choice of directions. Specifically, the downfolded
content of partial waves with $e_{g}$ character rotates the directions of
the $t_{2g}$ lobes into the proper "chemical" directions \cite{Pavarini2005}%
. Moreover, the downfolded partial-wave contents on the remaining Mo2, Mo4,
and Mo5 atoms in the string ensure that their relative phases are the proper
ones for the energies chosen. Also, a WO on the upper string is correctly
inverted with respect to the one on the lower string [see Eq.$\,$ (\ref{inv}%
)]. Similarly, the downfolded partial waves on all oxygens give the proper O$%
\,2p$ dressing.

The WOs are obtained by symmetrical orthonormalization of the NMTO set and
this causes a delocalization which, however, for our $t_{2g}$ set is small
and invisible in FIG.$\,$\ref{FIGyzb&xyamc}. What we do see, and noted in
the previous section, is that each $t_{2g}$ WO has tails with the same $%
t_{2g}$ character as that of the head on the 4 nearest molybdenums in the
plane of the orbital. These tails are connected to the head via $pd\pi $
tails on the 4 connecting oxygens such that the sign is anti-bonding with
the $t_{2g}$ head and bonding with the $t_{2g}$ tail. In effect, this
results in a $dd\pi $ anti-bond between the oxygen-dressed $t_{2g}$ orbitals
forming the WO head and tail.

Since the $xy$ WO lies \emph{in} the plane of its ribbon, it only spreads
onto a neighboring ribbon via a weak covalent interaction of symmetry $%
dd\delta $ causing no visible tails in the upper rows of FIG.$\,$\ref%
{Wannier}. This is in contrast to the strong inter-ribbon $dd\pi $-spread of
the $xz$ and $yz$ WOs. The consequence for the six-band Hamiltonian to be
presented in the next section is that the $xy$-$XY$ inter-ribbon hopping
integral $t_{1}$ in Eq.$\,$(\ref{epsxy}) and its dimerization $u_{1},$ are
about 30 times smaller than the respective $xz$-$XZ$ and $yz$-$YZ$
inter-ribbon integrals $A_{1}$ and $G_{1}$ in Eq. (\ref{gap}).\ For the same
reason, the selection rule derived in Sect.~\ref{Usingt2gWOs} of Paper II
that ARPES sees the lower band in the 1st and the higher band in the 2nd
zone is better obeyed for the $xy$\ bands than for the (occupied) lower $xz$%
\ and $yz$\ bands (compare FIG.s \ref{FIGxyZoneSelectul} and \ref%
{FIGyzZoneSelect} in Paper II).

With the knowledge that the right-hand panel of FIG.$\,$\ref{FIGyzb&xyamc}
shows the WO, $xy\left( \mathbf{r}\right) ,$ let us now imagine building the
1D Bloch sum $xy\left( k_{b}\mathbf{b}^{\ast },\mathbf{r}\right) $ of WOs (%
\ref{Bloch}) through integer translations by $n\mathbf{b}$, multiplication
with $e^{2\pi ink_{b}},$ and superposition: Around Mo1 and Mo5, only $%
xy\left( \mathbf{r}\right) $ contributes (neglecting the tail outside the
70\% contour), but around Mo2 and Mo4, also $xy\left( \mathbf{r+b}\right)
e^{-2\pi ik_{b}}$ and $xy\left( \mathbf{r-b}\right) e^{2\pi ik_{b}}$
contribute. As a result, at the bottom of the band ($k_{b}$=0) the
amplitudes around Mo1 and Mo4 are nearly equal, and anti-bonding between Mo1
and Mo4, whereas the amplitude around Mo2 is smaller, but also anti-bonding
to Mo1 so that the $p\pi $ character on all 4 oxygens vanishes. At the Fermi
level, $\left\vert k_{b}\right\vert $=$\frac{1}{4},$ whereby the \emph{sum}
of the Bloch waves with positive and negative $k_{b}$ has the same shape as $%
xy\left( \mathbf{r}\right) $ near Mo1 and Mo5, and a node at the neighboring
Mo1 and Mo5 (i.e., those translated by $\pm \mathbf{b).}$ This is the
standing-wave state described in the previous section. The shape of the 
\emph{difference} between the waves with $k_{b}$ positive and negative is
the same, but shifted by $\mathbf{b.}$ At the top of the band, $\left\vert
k_{b}\right\vert $=$\frac{1}{2},$ whereby the Bloch waves change sign upon
translation by $\mathbf{b}$ so that there is a node through Mo2 and Mo4 for
one of the linear combinations, and through Mo1 and Mo5 for the other. If we
finally build the Bloch sums with $\left\vert k_{b}\right\vert $=$\frac{3}{4}%
,$ we find that they are \emph{identical} with those for $\left\vert
k_{b}\right\vert $=$\frac{1}{4},$ because in order to form both the
low-energy Mo1-Mo4 bonding and the high-energy anti-bonding states, we would
need a set containing \emph{two} $xy$ WOs, one centered at Mo1 and the other
at Mo4. In order for a \emph{single} WO to describe the lower, bonding part
of a 4$~$eV wide band, gapped in the middle by merely 0.6 eV, it must in
order to reproduce the strong curvature at the top of the lower band at $%
\left\vert k_{b}\right\vert $=$\frac{1}{2}$ have the ZB here (rather than at
1), as well as long-range in the direction $\left( b\right) $ of the
dispersion. That the latter is not seen in the first panel on the bottom row
of FIG.$\,$\ref{Wannier} is due to our contour cut-off at 70\%. But in the
Hamiltonian [Eq.s$\,$(\ref{Hsub}), (\ref{tautu}), and (\ref{taup})], it
gives rise to $xy$-$xy$ hopping integrals, $\tau _{n},$ which we need to
carry as far as to $n$=12.

For future first-principles studies enabling Mott localization onto Mo1 or
Mo4, WO sets larger than the one of six used in the present papers will be
needed.

In a similar way, we can imagine building the states of the two 1D $yz$
bands (\ref{eyz})-(\ref{gap}) from pseudo Bloch sums of the $yz$ and $YZ$
WOs (FIG.{s}~\ref{FIGyzb&xyamc} and \ref{Wannier}) through pseudo
translations by $n\frac{\mathbf{c+b}}{2}$, multiplication with $e^{\pi
in\left( k_{c}+k_{b}\right) },$ and superposition. These WOs have their
proper positions, i.e. at respectively Mo1 and MO1$^{\ref{Rcenter}}$, and we
use $yz\left( \mathbf{r-}n\frac{\mathbf{c}+\mathbf{b}}{2}\right) $ for $n$
even and $YZ\left( \mathbf{r-}\left( n-1\right) \frac{\mathbf{c}+\mathbf{b}}{%
2}\right) $ for $n$ odd; see Eq.$\,$(\ref{undim}) and also Eq.$\,$(\ref{kket}%
) to which we shall return. These WOs are so localized that each one spills
over only to its neighboring $\mathbf{y}$-string. The integrals for intra
and inter bi-ribbon hops, $A_{1}\pm G_{1},$ whose complicated hopping paths
between elementary, dressed $yz$ orbitals were shown in (\ref{eyz}), are
simply those between nearest-neighbor $yz$ and $YZ$ WOs. All farther-ranged
hopping integrals, $A_{n>1}$ and $G_{n>1}$, are negligible.

The square of a WO, summed over all lattice translations yields the charge
density obtained by filling that band, provided that we neglect its
hybridization with the other bands. Summing this charge density over all six
WOs yields the charge density obtained by filling all six $t_{2g}$ bands,
hybridizations now included. As an example: Squaring the $xy$ WO in FIG.$\,$%
\ref{FIGyzb&xyamc} will remove the colors and enhance the density on Mo1
with respect to that on the two Mo4 atoms, and even more with respect to
that on the two Mo2 atoms, and mostly with respect to that on Mo5.
Translating this charge density by $\pm \mathbf{b}$ and summing, doubles the
charge density on Mo4 and on Mo2 due to overlap. As a result, the charge
density on Mo1 and Mo4 will be nearly equal and larger than that on Mo2,
while the one on Mo5 will be the smallest.

This charge density compares well with the one obtained by Popovic and
Satpathy \cite{Satpathy2006} for the quasi-1D band by filling it in a narrow
range around the Fermi level and shown in the plane of the lower ribbon in
their FIG.$\,$5.\footnote{%
That their \cite{Satpathy2006} density on Mo2 is smaller than the one on Mo5
is presumably due to an erroneous exchange of the labels Mo1 and Mo4.}

Nuss and Aichhorn \cite{Nuss2014} described the \emph{four} lowest bands,
i.e. the two valence bands and the two metallic bands, with a set of
maximally localized Wannier functions obtained numerically by minimizing the
spread $\left\langle \chi \left\vert \left\vert \mathbf{r-\Re }\right\vert
^{2}\right\vert \chi \right\rangle $. Their WFs are bond centered and are
essentially our $yz\left( \mathbf{r}\right) +YZ\left( \mathbf{r}\right) ,$
our $xz\left( \mathbf{r}\right) +XZ\left( \mathbf{r}\right) ,$ and a WF
along each $\diagdown _{\text{Mo1}}\diagup ^{\text{Mo4}}\diagdown _{\text{Mo1%
}}\diagup ^{\text{Mo4}}\diagdown $ chain with $xy$-like, similar-sized
contours on all four sites, smaller contours on the Mo2 and Mo5 sites
closest to the bond, and even smaller contours on the next Mo2 and Mo5
sites. This WF is extended along the chain, but appears from their FIG.$\,$4
to have about the same degree of localization as our disc-shaped WO seen in
the first column of FIG.$\,$\ref{Wannier}.

\section{Six-band $t_{2g}$ tight-binding Hamiltonian\label{SectH}}

Since our TB Hamiltonian is considerably more detailed than those previously
published \cite{Satpathy2006,Jarlborg2012}\cite{Merino2012,Chudzinski2012},
we have been forced to change \emph{notation.} The relation between the
earlier notation and ours is, first of all: $t_{\perp }=t_{1}+u_{1}$ and $%
t_{\perp }^{\prime }=t_{1}-u_{1}$. The integral $t\sim -1~$eV for the
Mo1-Mo4 hopping used in the earlier work --as well as in the previous
sections-- is the coefficient to $\cos \pi k_{b},$ whereas $\tau _{n}$ to be
used in Eq.$~$(\ref{tau}) and in the following is the coefficient to $\cos
2\pi nk_{b}$. The symbol $t$ will from \emph{now} on --unless with explicit
reference to Eq.~(\ref{exy})-- denote the\emph{\ function} of $k_{b}$ and $%
k_{c}$ which is defined in terms of the \emph{perpendicular} 1st an 2nd
nearest hopping integrals $t_{1}$ and $t_{2}$ in Eq.~(\ref{tautu}).

\subsection{Sublattice $\left\{ w,W\right\} $-representation\label{SectwW}}

In the representation of the six Bloch sums (\ref{Bloch}) of the three
Mo1-centered $t_{2g}$ WOs$^{\ref{Rcenter}}$, $w_{m}\left( \mathbf{k,r}%
\right) $= $xy\left( \mathbf{k,r}\right) ,$ $xz\left( \mathbf{k,r}\right) ,$
and $yz\left( \mathbf{k,r}\right) ,$ as well as of the three MO1-centered
WOs times a common phase factor, $W_{m}\left( \mathbf{k,r}\right) e^{\pi
i\left( k_{c}+k_{b}\right) }$= $XY\left( \mathbf{k,r}\right) e^{\pi i\left(
k_{c}+k_{b}\right) },$ $XZ\left( \mathbf{k,r}\right) e^{\pi i\left(
k_{c}+k_{b}\right) },$ and $YZ\left( \mathbf{k,r}\right) e^{\pi i\left(
k_{c}+k_{b}\right) },$ the TB Hamiltonian (\ref{FT}) is:%
\begin{equation}
\fbox{$%
\begin{array}{cccccccc}
H &  & xy & XY & xz & XZ & yz & YZ \\ 
&  &  &  &  &  &  &  \\ 
xy &  & \tau & t-iu & \alpha +i\gamma & a-ig & \bar{\alpha}+i\bar{\gamma} & 
\bar{a}-i\bar{g} \\ 
XY &  & t+iu & \tau & a+ig & \alpha -i\gamma & \bar{a}+i\bar{g} & \bar{\alpha%
}-i\bar{\gamma} \\ 
xz &  & \alpha -i\gamma & a-ig & 0 & A-iG & \lambda -i\mu & l-im \\ 
XZ &  & a+ig & \alpha +i\gamma & A+iG & 0 & l+im & \lambda +i\mu \\ 
yz &  & \bar{\alpha}-i\bar{\gamma} & \bar{a}-i\bar{g} & \lambda +i\mu & l-im
& 0 & \bar{A}-i\bar{G} \\ 
YZ &  & \bar{a}+i\bar{g} & \bar{\alpha}+i\bar{\gamma} & l+im & \lambda -i\mu
& \bar{A}+i\bar{G} & 0%
\end{array}%
$,}  \label{Hsub}
\end{equation}%
using simplified labeling of the rows and columns. The six WOs are
real-valued and shown in FIG. \ref{Wannier}. The common $\mathbf{k}$%
-dependent phase factor, $e^{\pi i\left( k_{c}+k_{b}\right) },$ multiplying
the Bloch sums of the upper-string WOs, has been included in order that
matrix-elements between the two different sublattices take the simple form (%
\ref{Hsub}) where the asymmetry between integrals for hopping in- and
outside a bi-ribbon (electronic dimerization) is given by the imaginary part.

The zero of energy is chosen as the common energy of the $xz,$ $XZ,$ $yz,$
and $YZ$ WOs.

The quantities in (\ref{Hsub}) named by Greek and Latin letters are
real-valued \emph{functions} of the Bloch vector (\ref{k}). Specifically:%
\begin{eqnarray}
\tau \left( k_{b}\right) &=&\tau _{0}+\sum\nolimits_{n=1}^{12}2\tau _{n}\cos
2\pi nk_{b},  \label{tau} \\
t\left( \mathbf{k}\right) &=&\left( 2t_{1}\cos \pi k_{b}+2t_{2}\cos 3\pi
k_{b}\right) 2\cos \pi k_{c},  \label{tautu} \\
u\left( \mathbf{k}\right) &=&\left( 2u_{1}\cos \pi k_{b}+2u_{2}\cos 3\pi
k_{b}\right) 2\sin \pi k_{c},  \notag
\end{eqnarray}%
describe the pure $xy/XY$ bands,%
\begin{eqnarray}
A\left( \mathbf{k}\right) &=&2A_{1}\cos \pi \left( k_{c}-k_{b}\right) ,
\label{AG} \\
G\left( \mathbf{k}\right) &=&2G_{1}\sin \pi \left( k_{c}-k_{b}\right) , 
\notag
\end{eqnarray}%
describe the pure $xz/XZ$ bands$,$ and $\bar{A}$ and $\bar{G}$ describe the
pure $yz/YZ$ bands. An \emph{overbar} is generally used when switching from
an $xz$ to a $yz$ orbital and indicates the mirror operation $%
k_{b}\leftrightarrow -k_{b},$ e.g.: $\bar{a}\left( k_{b},k_{c}\right) \equiv
a\left( -k_{b},k_{c}\right) .$ The hybridizations between the $xy/XY$ and
the $xz/XZ$ bands are given by the Bloch sums:%
\begin{eqnarray}
\alpha \left( \mathbf{k}\right) &=&\alpha _{0}+2\alpha _{1}\cos 2\pi
k_{b}+2\alpha _{2}\cos 2\pi k_{c}  \label{ag} \\
&&+2\alpha _{3}\cos 2\pi \left( k_{c}+k_{b}\right) +2\alpha _{3}^{\prime
}\cos 2\pi \left( k_{c}-k_{b}\right) ,  \notag \\
a\left( \mathbf{k}\right) &=&2a_{1}\cos \pi \left( k_{c}-k_{b}\right)
+2a_{1}^{\prime }\cos \pi \left( k_{c}+k_{b}\right)  \notag \\
&&+2a_{2}\cos \pi \left( k_{c}-3k_{b}\right) +2a_{2}^{\prime }\cos \pi
\left( k_{c}+3k_{b}\right) ,  \notag \\
\gamma \left( \mathbf{k}\right) &=&2\gamma _{1}\sin 2\pi k_{b}+2\gamma
_{2}\sin 2\pi k_{c}  \notag \\
&&+2\gamma _{3}\sin 2\pi \left( k_{c}+k_{b}\right) +2\gamma _{3}^{\prime
}\sin 2\pi \left( k_{c}-k_{b}\right) ,  \notag \\
g\left( \mathbf{k}\right) &=&2g_{1}\sin \pi \left( k_{c}-k_{b}\right)
+2g_{1}^{\prime }\sin \pi \left( k_{c}+k_{b}\right)  \notag \\
&&+2g_{2}\sin \pi \left( k_{c}-3k_{b}\right) +2g_{2}^{\prime }\sin \pi
\left( k_{c}+3k_{b}\right) ,  \notag
\end{eqnarray}%
and the hybridizations between the $xz/XZ$ and $yz/YZ$ bands by:%
\begin{eqnarray}
\lambda \left( \mathbf{k}\right) &=&\lambda _{0}+2\lambda _{1}\cos 2\pi
k_{b}+2\lambda _{2}\cos 2\pi k_{c}+2\lambda _{3}\cos 2\pi 2k_{b},  \notag \\
l\left( \mathbf{k}\right) &=&\left( 2l_{1}\cos \pi k_{b}\right) 2\cos \pi
k_{c},  \notag \\
\mu \left( \mathbf{k}\right) &=&2\mu _{1}\sin 2\pi k_{b}+2\mu _{3}\sin 2\pi
2k_{b}.  \notag \\
m\left( \mathbf{k}\right) &=&\left( 2m_{1}\cos \pi k_{b}\right) 2\sin \pi
k_{c}.  \label{lambdalmum}
\end{eqnarray}%
The dispersion along $\mathbf{a}^{\ast }$ is neglected, and the Bloch sums
are truncated for distances exceeding the lattice constant $a,$ which means
after the 3rd-nearest neighbors. The long-ranged $\tau \left( k_{b}\right) $
is an exception and will be discussed below. Due to the truncation of hops
longer than $a,$ the effective value of $k_{a}$ is not $0,$ but the one for
which $\cos 2\pi k_{a}$=0, i.e. $\frac{1}{4}.$ The truncation also means
that our LDA TB bands are a bit more wavy and smoother than those obtained
from the original LDA NMTO Hamiltonian downfolded in $\mathbf{k}$-space. The 
$A$ and $G$ sums (\ref{AG}) are converged already after 1st-nearest
neighbors.

The Greek-lettered Bloch sums are over hops on the \emph{same} sublattice
whereby their $\mathbf{k}$ dependence is periodic in the reciprocal lattice
spanned by $\mathbf{b}^{\ast }$ and $\mathbf{c}^{\ast },$ e.g. 
\begin{equation}
\alpha \left( k_{b},k_{c}\right) =\alpha \left( k_{b}+M,k_{c}+N\right)
\label{Greek}
\end{equation}%
with $M$ and $N$ any integer. The Latin-lettered Bloch sums are over hops%
\emph{\ between} the Mo1- and MO1-centered sublattices and averaged such
that these Bloch sums are periodic in the double zone spanned by $\mathbf{c}%
^{\ast }\mathbf{+b}^{\ast }$ and $\mathbf{c}^{\ast }-\mathbf{b}^{\ast }$
(see Sect.$\,$\ref{Sectk+c}), but change sign upon odd reciprocal-lattice
translations, e.g. 
\begin{equation}
a\left( k_{b},k_{c}\right) =\left( -\right) ^{M+N}a\left(
k_{b}+M,k_{c}+N\right) .  \label{Latin}
\end{equation}%
Note the difference between $\alpha $ and $a.$

The number of parameters entering Eq.s~(\ref{tau})-(\ref{lambdalmum}) and
whose values are given in Eq.s~(\ref{taup})-(\ref{lp}) below are far more
numerous than those few ($t,$ $t_{\perp },$ $t_{\perp }^{\prime },$ $A_{1},$ 
$G_{1},$ and $B$) used in the simplified description given in Sect.$\,$\ref%
{SectElStruc}; a description which, nevertheless, suffices to understand the
CECs and bands measured by ARPES and shown in respectively FIG.s~\ref{CEC}
and \ref{ARPES_Bandstructure_Cuts} in Paper II. The LDA low-energy TB bands
shown in FIG.$~$II~\ref{ARPES_Bandstructure_LDA}~(a) together with the
occupied bands measured by ARPES (grey circles and black dots) have much
more detail, and the surprisingly good agreement between them proves this
detail to be real. This is emphasized by the nearly perfect agreement seen
in FIG.~\ref{ARPES_Bandstructure_LDA}~(b) and obtained by shifting merely
the on-site energy, $\tau _{0},$ of the degenerate $xy$ and $XY$ WOs upwards
by 100 meV with respect to the energy of the degenerate $xz,$ $XZ,$ $yz,$
and $YZ$ WOs. In Sect.$\,$\ref{SectEk} of Paper II we shall describe the
details of the energy bands while the differences between LDA and ARPES will
be in focus of Sect.$~$\ref{SectAgreement} of Paper II.

Below, we give the values in meV of the on-site energies and hopping
integrals obtained from the first-principles LDA full-potential NMTO
calculation (\ref{FT}) together with the (shifted) values and the [ARPES
refined] values (see Sect.s~\ref{SectShifting}\textbf{\ }and \ref%
{SectRefining} in Paper II), in those cases where they differ:%
\begin{equation}
\fbox{$%
\begin{array}{lll}
\tau _{0}=47\,\left( 147\right) \,\left[ 203\right] &  &  \\ 
\tau _{1}=-422\,\left[ -477\right] & \tau _{5}=-11 & \tau _{9}=-2 \\ 
\tau _{2}=47\,\left[ 87\right] & \tau _{6}=8 & \tau _{10}=1 \\ 
\tau _{3}=-31 & \tau _{7}=-4 & \tau _{11}=-1 \\ 
\tau _{4}=17 & \tau _{8}=3 & \tau _{12}=1%
\end{array}%
$}  \label{taup}
\end{equation}%
\begin{equation}
\fbox{$%
\begin{array}{ll}
t_{1}=-11 & u_{1}=-3 \\ 
t_{2}=-5 & u_{2}=1%
\end{array}%
$}  \label{tp}
\end{equation}%
\begin{equation}
\fbox{$%
\begin{array}{cc}
A_{1}=-319 & G_{1}=-98\,\left[ -109\right]%
\end{array}%
$}  \label{A&Gp}
\end{equation}%
\begin{equation}
\fbox{$%
\begin{array}{cccc}
\alpha _{0}=31 &  &  &  \\ 
\alpha _{1}=20 & a_{1}=-49 & \gamma _{1}=8 & g_{1}=1 \\ 
\alpha _{2}=-5 & a_{1}^{\prime }=-8 & \gamma _{2}=-6 & g_{1}^{\prime }=5 \\ 
\alpha _{3}=10 & a_{2}=-6 & \gamma _{3}=2 & g_{2}=-3 \\ 
\alpha _{3}^{\prime }=-4 & a_{2}^{\prime }=-11 & \gamma _{3}^{\prime }=-4 & 
g_{2}^{\prime }=-11%
\end{array}%
$}  \label{ap}
\end{equation}%
\begin{equation}
\fbox{$%
\begin{array}{ll}
\lambda _{0}=-61 & \mu _{1}=7 \\ 
\lambda _{1}=7 & \mu _{3}=-11 \\ 
\lambda _{2}=22\,\left[ 15\right] & l_{1}=20 \\ 
\lambda _{3}=-11\,\left[ -5\right] & m_{1}=12\,\left[ 6\right]%
\end{array}%
$}.  \label{lp}
\end{equation}%
Subscript $0$ indicates an on-site energy, which is the energy of the WO in
case the two WOs are identical\textbf{,} and an anisotropy energy in case
they are different. Further subscripts indicate 1st, 2nd, and 3rd-nearest
neighbor hops.

As mentioned above, the zero of energy is chosen as the common energy of the 
$xz,XZ,yz,$ and $YZ$ WOs. This is the center of the gap in the approximation
that the hybridizations (\ref{lambdalmum}) between the $xz/XZ$ and $yz/YZ$
bands are neglected. In Sect.$\,$\ref{SectElStruc} and in footnote \ref%
{dispersion} this energy was named $E_{0}\sim B+2\left\vert A_{1}\right\vert
.$ The common energy of the $xy$ and $XY$ WOs, i.e. the center of the
unhybridized $xy$ bands, is $\tau _{0}$ with respect to that of the $%
xz,XZ,yz,$ and $YZ$ WOs.

For our basis containing merely one Mo1- and one MO1 WO, the Fourier series (%
\ref{tau}) for the dominating $k_{b}$ dependence of the two $\frac{1}{2}$%
-filled $xy$ bands converges slowly as explained in Sect.~\ref{SectLowE}.
For many purposes, it suffices to linearize $\tau \left( k_{b}\right) $
around $k_{b}=\frac{1}{4}\approx k_{F}$ or $-\frac{1}{4}:$%
\begin{eqnarray}
\tau \left( k_{b}\right) &\approx &\tau _{0}+2\sum\nolimits_{n=1}\left(
-1\right) ^{n}\tau _{2n}  \label{lin} \\
&&-\left( \left\vert k_{b}\right\vert -\frac{1}{4}\right) \,4\pi
\sum\nolimits_{n=0}\left( -1\right) ^{n}\left( 2n+1\right) \tau _{2n+1}. 
\notag
\end{eqnarray}%
With the values given in (\ref{taup}), the upper line of Eq.$\,$(\ref{lin})
says that --neglecting FS warping and splitting, i.e. the perpendicular (\ref%
{tp}) and hybridization (\ref{ap}) integrals -- the Fermi level at \emph{half%
} filling is 
\begin{equation}
E_{F}\approx \tau \left( \frac{1}{4}\right) =-23~\left( 77\right) ~\left[ 53%
\right] ~\mathrm{meV}  \label{EFARPES}
\end{equation}%
above the center of the gap. According to Eq.~(\ref{lin}), this differs from
the on-site $xy$ energy, $\tau _{0}=$ $47\,\left( 147\right) \,\left[ 203%
\right] ,$ by the alternating sum 2$\left( -\tau _{2}+\tau _{4}-..\right) $.
Hence, the reason why the ARPES-refined value of the Fermi level for
half-filling is 150 meV below $\tau _{0}$ --and thereby closer to the center
of the gap than the LDA value shifted by 100 meV-- is caused by the
refinement of the $\tau _{2}$ value. In Sect.~\ref{SectFS} of Paper III, the
average $k_{Fb}$-value measured by ARPES at $33$ eV is 0.254. The Fermi
level is thus approximately $\tau \left( 0.254\right) $=$75~$meV rather than
53 meV above the center of the gap.

The value of the coefficient to $\left\vert k_{b}\right\vert -\frac{1}{4}$
in the lower line of Eq.$\,$(\ref{lin}), times $b,$ yields the Fermi
velocity at half filling:%
\begin{equation}
v_{F}=4.0~\left( 4.0\right) ~\left[ 4.6\right] ~\mathrm{eV\,\mathring{A}.}
\label{vF}
\end{equation}%
This LDA value is a bit larger than those of Satpathy and Popovic (3.72~eV$~$%
\AA )\cite{Satpathy2006} and of Nuss and Aichhorn (0.93~10$^{5}~$m/s= 3.8~eV$%
~$\AA )\cite{Nuss2014}.\ Our ARPES-refined value, which is consistent with
FIG.$\ $\ref{ARPES_Bandstructure_TB}~(c2) in Paper II, exceeds the LDA value
by 15\%. Reasons for this velocity enhancement will be discussed in Sect.~%
\ref{Section:ExpFermiVelo} of Paper III. The dimensionless coupling constant
used in Ref. \cite{Kopietz1995} has the value%
\begin{equation}
e^{2}/\left( \pi \hbar v_{F}\right) =1.14~\left[ 0.99\right] .  \label{Kurt}
\end{equation}

The splitting-and-warping effects neglected above are considered in detail
in Sect.s~\ref{twoband}, \ref{SectOrigins} and \ref{SectFS}, in Paper III.

Of the matrix elements, $\left\langle xy_{0}\left\vert H\right\vert
xz_{n}\right\rangle =\alpha _{n}\pm \gamma _{n},$ determining the $xz$ and $%
yz$ hybridization of the $xy$ bands, $\alpha _{0}$ is the crystal-field term
and $\alpha _{n}\pm \gamma _{n}$ and $\overline{\alpha }_{n}\pm \overline{%
\gamma }_{n}$ are integrals for hopping between $n$th-nearest neighbors with
the upper sign for forwards- and the lower for backward hopping. Although
these Greek-lettered hops are between WOs on the same sublattice, forwards
and backward hoppings differ because there is no inversion symmetry around
Mo1. These energies, except $\alpha _{0}$ and $\alpha _{1},$ are small but
significant for the detailed $k_{c}$-dispersion of the $xy$ band near the
Fermi level, especially the resonance behavior. The same holds for the
Latin-lettered hopping integrals, $\left\langle xy_{0}\left\vert
H\right\vert XZ_{n}\right\rangle =a_{n}\pm g_{n},$ between WOs on different
sublattices, except for $a_{1}.$ In Paper III, Sect.~\ref{Sectxy-xz} in
particular, we shall see that the crystal-field term $\alpha _{0}$ and the
integral for hopping from $xy$ to an $XZ$ or $YZ$ nearest-neighbor, $a_{1},$
are of major\textbf{\ }importance.

The matrix elements $\left\langle xz_{0}\left\vert H\right\vert
yz_{n}\right\rangle =\lambda _{n}\pm \mu _{n}$ and $\left\langle
xz_{0}\left\vert H\right\vert YZ_{n}\right\rangle =l_{n}\pm m_{n}$ are
larger\ than these, but of minor importance for the $xy$ band near $E_{F}$,
and we shall neglect them in the two-band Hamiltonian derived in Paper III
Sect.~\ref{twoband}. They are decisive for the levels near $Z$ where the
valence (V) and conduction (C) bands come closest. Since the Bloch sums (\ref%
{lambdalmum}) are badly converged, we found it necessary to truncate the sum
and then refine the hopping values as shown in square brackets in Table~\ref%
{lp}. This refinement was enabled by the fact that the six-band Hamiltonian (%
\ref{Hsub}) simplifies at the points of high symmetry such as $\mathrm{Z.}$

\subsection{Reciprocal sublattice $\left\{ \mathbf{k,k+c}^{\ast }\right\} $%
-representation\label{Sectk+c}}

In connection with Eq.s~(\ref{epsxy}) and (\ref{gap}), we noted that the
numerical values of the most important inter-ribbon hoppings have smaller
dimerizations than mean values, i.e. $t_{1}\pm u_{1}\sim -11\mp 3$ meV for
the $xy$ band and $A_{1}\pm G_{1}\sim -0.3\mp 0.1$ eV for the $xz$ and $yz$
bands. If such electronic $c$-axis dimerizations (Sect.$\,$\ref{SectDims})
are neglected, the energy bands are the dashed bands in FIG.~6 and
correspond to all strings being related by primitive translations $\frac{%
\mathbf{c+b}}{2}$ and $\frac{\mathbf{c-b}}{2},\mathbf{\ }$rather than there
being 2 translationally inequivalent strings per primitive cell whose
primitive translations are \textbf{c} and $\mathbf{b}.$ A natural way of
describing the proper electronic structure is therefore in terms of basis
functions which are \emph{pseudo Bloch sums} of WOs with respect to this
--too short-- lattice periodicity. Specifically in LiPB, a pseudo-Bloch sum
is:%
\begin{eqnarray}
&&\left\vert w;\mathbf{k}\right\rangle \equiv  \label{kket} \\
&&\frac{1}{\sqrt{2}}\sum_{\mathbf{T}}e^{2\pi i\mathbf{k\cdot T}}\left[
w\left( \mathbf{r-T}\right) +e^{2\pi i\mathbf{k\cdot }\frac{\mathbf{c+b}}{2}%
}W\left( \mathbf{r-T}\right) \right] ,  \notag
\end{eqnarray}%
where the $\mathbf{T}$-sum is over the proper lattice translations, $w\left( 
\mathbf{r}\right) $ is the WO centered on Mo1, taken as the origin of the
primitive cell, and $W\left( \mathbf{r}\right) $ is the WO on MO1, which is
at $\frac{\mathbf{c+b}}{2}-\mathbf{d}$. Both $\sum_{\mathbf{T}}e^{2\pi i%
\mathbf{k\cdot T}}w\left( \mathbf{r-T}\right) $ and $\sum_{\mathbf{T}%
}e^{2\pi i\mathbf{k\cdot T}}W\left( \mathbf{r-T}\right) $ are proper Bloch
sums (\ref{Bloch}) and, hence, periodic functions of $\mathbf{k}$ in a
single zone. However, the $\mathbf{k}$-dependent phase factor, $e^{\pi
i\left( k_{c}+k_{b}\right) },$ multiplying the second Bloch sum makes this
--and herewith the entire pseudo Bloch sum (\ref{kket})-- a function of $%
\mathbf{k}$ which is merely \emph{periodic in the double} zone, i.e. on the 
\emph{sparse} reciprocal lattice spanned by $\mathbf{c}^{\ast }+\mathbf{b}%
^{\ast }$ and $\mathbf{c}^{\ast }-\mathbf{b}^{\ast }$, that is:%
\begin{eqnarray}
\left\vert w;\mathbf{k}\right\rangle &=&\left\vert w;\mathbf{k}+M^{\prime
}\left( \mathbf{c}^{\ast }+\mathbf{b}^{\ast }\right) +N^{\prime }\left( 
\mathbf{c}^{\ast }-\mathbf{b}^{\ast }\right) \right\rangle  \notag \\
&\equiv &\left\vert w;\mathbf{k}+M\mathbf{b}^{\ast }+N\mathbf{c}^{\ast
}\right\rangle ,  \label{doublezone}
\end{eqnarray}%
with $M^{\prime }$ and $N^{\prime }$ any integers, which means: with $M+N$ 
\emph{even}$.$ Had $W\left( \mathbf{r}\right) $ not been displaced and
inverted as described in Sect.~\ref{SectDims}, the pseudo Bloch sum (\ref%
{kket}) would have been a proper Bloch sum for the \emph{undimerized}
crystal. In FIG.~\ref{ThreePureBands}, the energy bands of the even
pseudo-Bloch sums are the dashed blue, green, and red bands with bottoms
near $k_{c}$=0.

The correct long periodicity in real space --and \emph{single}-zone
periodicity in reciprocal space-- can now be described by including in the
basis set the pseudo Bloch sum with $\mathbf{k}$ translated to the \emph{%
other} sparse sublattice (we may think of reciprocal space as a checkerboard
consisting of 1st and 2nd zones). This second set of Bloch waves$^{\ref{c}}$
is thus $\left\vert w;\mathbf{k+c}^{\ast }\right\rangle ,$ for which $M+N+1$
is even, i.e. $M+N$ is \emph{odd}$.$ Their energy bands are the dashed ones
with bottoms near $k_{c}$=$\pm $1 (and tops near $k_{c}$=0) in FIG.~\ref%
{ThreePureBands}. Finally, in order to diagonalize the Hamiltonian (\ref%
{HRecip}), the even and odd pseudo-Bloch sums with the same value of $%
\mathbf{k}$ are allowed to mix and the bands to gap. In absence of
dimerization, the even and odd pseudo-Bloch functions are identical, apart
from a phase factor, and so are the even and odd energy bands which are
merely separated by $\mathbf{c}^{\ast }$ and cross without gapping (FIG.~\ref%
{ThreePureBands}).

The basis set $\left( \left\vert w;\mathbf{k}\right\rangle ,\,\left\vert w;%
\mathbf{k+c}^{\ast }\right\rangle \right) $ of pseudo-Bloch sums is simply
the unitary transformation:%
\begin{eqnarray}
\left\vert w;\mathbf{k}\right\rangle &=\left[ w\left( \mathbf{k,r}\right)
+e^{\pi i\left( k_{c}+k_{b}\right) }W\left( \mathbf{k,r}\right) \right] /%
\sqrt{2}  \notag \\
\left\vert w;\mathbf{k+c}^{\ast }\right\rangle &=\left[ w\left( \mathbf{k,r}%
\right) -e^{\pi i\left( k_{c}+k_{b}\right) }W\left( \mathbf{k,r}\right) %
\right] /\sqrt{2}  \label{trf}
\end{eqnarray}%
of the set $\left( w\left( \mathbf{k,r}\right) ,\,W\left( \mathbf{k,r}%
\right) \right) $ of proper Bloch sums (\ref{Bloch}) of the two WOs, $%
w\left( \mathbf{r}\right) $ and $W\left( \mathbf{r}\right) .$ We may check
that translation of $\mathbf{k}$ by $\mathbf{c}^{\ast }$ exchanges the
functions on the left-hand side, leaves the proper Bloch functions on the
right-hand side invariant, and --by adding 1 to $k_{c}$-- changes sign for
the second row of the matrix, which correctly exchanges its columns. With
the common phase factor $e^{\pi i\left( k_{c}+k_{b}\right) }$ included in
the definition of $W\left( \mathbf{k,r}\right) $ as done for the Hamiltonian
(\ref{Hsub}) in the $\left\{ w,W\right\} $-representation, the
transformation (\ref{trf}) is simply the bonding-anti-bonding transformation
for each of the three $t_{2g}$ Bloch orbitals, and the inverse
transformation is:%
\begin{eqnarray}
w\left( \mathbf{k,r}\right) &=&\left[ \left\vert w;\mathbf{k}\right\rangle
+\left\vert w;\mathbf{k+c}^{\ast }\right\rangle \right] /\sqrt{2}  \notag \\
e^{\pi i\left( k_{c}+k_{b}\right) }W\left( \mathbf{k,r}\right) &=&\left[
\left\vert w;\mathbf{k}\right\rangle -\left\vert w;\mathbf{k+c}^{\ast
}\right\rangle \right] /\sqrt{2}.  \label{itrf}
\end{eqnarray}

Transformed to this $\left\{ \mathbf{k,k+c}^{\ast }\right\} $%
-representation, the six-band Hamiltonian (\ref{Hsub}) becomes: 
\begin{widetext}
\begin{equation}
\fbox{$%
\begin{array}{cccccccc}
H &  & \left\vert xy;\mathbf{k}\right\rangle & \left\vert xy;\mathbf{k+c}%
^{\ast }\right\rangle & \left\vert xz;\mathbf{k}\right\rangle & \left\vert
xz;\mathbf{k+c}^{\ast }\right\rangle & \left\vert yz;\mathbf{k}\right\rangle
& \left\vert yz;\mathbf{k+c}^{\ast }\right\rangle \\ 
&  &  &  &  &  &  &  \\ 
\left\langle xy;\mathbf{k}\right\vert &  & \tau +t & iu & \alpha +a & 
i\left( \gamma +g\right) & \bar{\alpha}+\bar{a} & i\left( \bar{\gamma}+\bar{g%
}\right) \\ 
\left\langle xy;\mathbf{k+c}^{\ast }\right\vert &  & -iu & \tau -t & i\left(
\gamma -g\right) & \alpha -a & i\left( \bar{\gamma}-\bar{g}\right) & \bar{%
\alpha}-\bar{a} \\ 
\left\langle xz;\mathbf{k}\right\vert &  & \alpha +a & -i\left( \gamma
-g\right) & A & iG & \lambda +l & -i\left( \mu -m\right) \\ 
\left\langle xz;\mathbf{k+c}^{\ast }\right\vert &  & -i\left( \gamma
+g\right) & \alpha -a & -iG & -A & -i\left( \mu +m\right) & \lambda -l \\ 
\left\langle yz;\mathbf{k}\right\vert &  & \bar{\alpha}+\bar{a} & -i\left( 
\bar{\gamma}-\overline{g}\right) & \lambda +l & i\left( \mu +m\right) & \bar{%
A} & i\bar{G} \\ 
\left\langle yz;\mathbf{k+c}^{\ast }\right\vert &  & -i\left( \bar{\gamma}+%
\bar{g}\right) & \bar{\alpha}-\bar{a} & i\left( \mu -m\right) & \lambda -l & 
-i\bar{G} & -\bar{A}%
\end{array}%
$},  \label{HRecip}
\end{equation}%
\end{widetext}
where $\left\vert m;\mathbf{k}\right\rangle \equiv \left\vert w_{m};\mathbf{k%
}\right\rangle .$

The $3\times 3$ blocks $\left\langle \mathbf{k}\left\vert H\right\vert 
\mathbf{k}\right\rangle $ and $\left\langle \mathbf{k+c}^{\ast }\left\vert
H\right\vert \mathbf{k+c}^{\ast }\right\rangle $ are real-valued, symmetric,
and periodic in respectively the even and the odd sublattice. This means
that $\left\langle \mathbf{k+c}^{\ast }\left\vert H\right\vert \mathbf{k+c}%
^{\ast }\right\rangle $ equals $\left\langle \mathbf{k}\left\vert
H\right\vert \mathbf{k}\right\rangle $ with the sign in front of the
Latin-lettered Bloch sum flipped [see Eq.$\,$(\ref{Latin})]. The
off-diagonal blocks $\left\langle \mathbf{k}\left\vert H\right\vert \mathbf{%
k+c}^{\ast }\right\rangle $ are caused by the $c$-axis dimerizations and are
purely imaginary.

The band structure with the $c$-axis dimerizations neglected, consists of
the 3 eigenvalues of the $\left\langle \mathbf{k}\left\vert H\right\vert 
\mathbf{k}\right\rangle $ block in the double zone $\left( \left\vert
k_{c}\right\vert \leq 1\right) .$ The dimerization effects may be included
by translating these 3 \emph{undimerized} bands (dashed in FIG.~\ref%
{ThreePureBands}) by $-1$ along $k_{c},$ whereby the 2nd BZ $\left( \frac{1}{%
2}\leq k_{c}\leq \frac{3}{2}\right) $ falls on top of the 1st $\left( -\frac{%
1}{2}\leq k_{c}\leq \frac{1}{2}\right) ,$ and finally split them by $%
\left\langle \mathbf{k}\left\vert H\right\vert \mathbf{k+c}^{\ast
}\right\rangle $.

In the following, we shall keep the $c$-axis dimerizations, but often
neglect the $mm^{\prime }$ hybridizations.

\subsection{Pure-$m$ bands\label{SectPureBands}}

The $xy$ (red), $xz$ (blue), and $yz$ (green) bands drawn in full lines in
FIG.~\ref{ThreePureBands} have the hybridizations between them neglected.
They are the so-called \emph{pure-}$m$ bands, the eigenvalues of the three 2$%
\times $2 blocks, $H_{m},$ along the diagonal in Eq.~(\ref{Hsub}) or (\ref%
{HRecip}), with elements given as functions of $k_{b}$ and $k_{c}$ in Eq.s~(%
\ref{tau}), (\ref{tautu}), and (\ref{AG}), and numerical values in Eq.s~(\ref%
{tp}) and (\ref{A&Gp}). The pure-$yz$\ band we already met in Eq.~(\ref{gap}%
). Note that $t,$ $A,$ and $\bar{A}$ are negative in the 1st zone and that $%
u,$ $G,$ and $\bar{G}$ are negative in the positive half of the first zone
because nearest-neighbor hopping integrals are negative. After subtraction
from $H_{xy}$ of the diagonal $\tau \left( k_{b}\right) $-term, all three
blocks have the same form (traceless and Hermitian), and so do, therefore,
their upper~$\left( j\text{=}2\right) $ and lower~$\left( j\text{=}1\right) $
eigenvalues:%
\begin{equation}
\pm \sqrt{A^{2}+G^{2}},\,\pm \sqrt{\bar{A}^{2}+\bar{G}^{2}},\,\mathrm{and}%
\,\pm \sqrt{t^{2}+u^{2}}  \label{epsul}
\end{equation}%
for $m$=$xz,$ $yz,$ and $xy,$ respectively. Similarly for the orthonormal
eigenfunctions expressed in terms of the WO Bloch sums used as a basis in
Eq.~(\ref{Hsub}), or of the WO pseudo Bloch sums used in Eq.~(\ref{HRecip}):%
\begin{eqnarray}
&&w_{\QATOP{2}{1}}\left( \mathbf{k,r}\right) =  \notag \\
&&\frac{1}{\sqrt{2}}\left[ w\left( \mathbf{k,r}\right) e^{-i\phi \left( 
\mathbf{k}\right) }\mp W\left( \mathbf{k,r}\right) e^{\pi i\left(
k_{c}+k_{b}\right) }\right]  \label{wW} \\
&=&\frac{1}{2}\left[ 
\begin{array}{l}
\left\vert w;\mathbf{k}\right\rangle \left( e^{-i\phi \left( \mathbf{k}%
\right) }\mp 1\right) + \\ 
\left\vert w;\mathbf{k+c}^{\ast }\right\rangle \left( e^{-i\phi \left( 
\mathbf{k}\right) }\mp 1\right)%
\end{array}%
\right] ,  \label{ww}
\end{eqnarray}%
where%
\begin{equation}
e^{i\phi }\equiv \frac{-A-iG}{\sqrt{A^{2}+G^{2}}},~\frac{-\bar{A}-i\bar{G}}{%
\sqrt{\bar{A}^{2}+\bar{G}^{2}}},\mathrm{and~}\frac{-t-iu}{\sqrt{t^{2}+iu^{2}}%
}.  \label{Phi}
\end{equation}

For quasi-1D structures such as LiPB, the band-structure phase $\phi \left( 
\mathbf{k}\right) $ varies from 0 at the centre of the physical zone (see
Sect.~\ref{Sectzones}) to $\pm \frac{\pi }{2}$ at the $\QATOP{\mathrm{right}%
}{\mathrm{left}}$ zone boundaries (ZB), and to $\pm \pi $ at the centers of
the second zones.

From Eq.~(\ref{ww}) we see that the $\left\vert \mathbf{k}\right\rangle $ 
\emph{characters} of the upper and lower $m$ bands --shown in FIG.~\ref%
{ThreePureBands} as \emph{fatness} added to the respective band dispersions
and computed by perturbing the $\left\langle m;\mathbf{k}\left\vert
H\right\vert m;\mathbf{k}\right\rangle $ element in the matrix (\ref{HRecip}%
) by a small constant energy-- are:%
\begin{equation}
\left\vert \frac{e^{-i\phi \left( \mathbf{k}\right) }\mp 1}{2}\right\vert
^{2}=\frac{1\mp \cos \phi \left( \mathbf{k}\right) }{2}\equiv f_{\QATOP{2}{1}%
}\left( \mathbf{k}\right) ,  \label{fat}
\end{equation}%
and these are the same as the $\left\vert \mathbf{k+c}^{\ast }\right\rangle $
characters of the other bands. In FIG.~\ref{FIGUxy^2}, the solid dark and
light curves give the $\left\vert \mathbf{k}\right\rangle $-character of
respectively the lower and upper $xy$ bands as function of $k_{c}$ in the
double zone and along the same line ($k_{b}$=0.225) as in FIG.~\ref%
{ThreePureBands}. We see the dominant $\left\vert \mathbf{k}\right\rangle $
character switch from the lower band in the 1st zone to the upper band in
the 2nd zone over a range of $k_{c}$ around the zone boundary (ZB), $%
\left\vert k_{c}\right\vert $=$\frac{1}{2}$. On $\phi $-scale (\ref{Phi}),
the switching behavior is independent of $m$ and given by Eq.~(\ref{fat}),
which says that the interval around the ZB, $\left\vert \phi \right\vert $=$%
\frac{\pi }{2},$ where the $\left\vert \mathbf{k}\right\rangle $ character
of \emph{both} bands exceeds e.g. $\frac{1}{7}\approx 14\%$ , is: $\phi
=\left( 0.5\pm 0.253\right) \pi .$ For the two $xy$ bands and $k_{b}$=0.225,
this "overlap interval" obtained from the 3rd Eq.~(\ref{Phi}) with Eq.s~(\ref%
{tautu}) and (\ref{tp}) is: $k_{c}=\left( 0.35|0.65\right) .$ Had there been
no electronic dimerization, i.e. if $u$ or $G$=0, the switching curves (FIG.~%
\ref{FIGUxy^2}) would have been meandering with vertical steps of size 1 at
the zone boundaries where the two $m$ bands would have crossed without
gapping and could have been folded out to a single $m$ band in the double
zone.

\begin{figure}[!bth]
\includegraphics[width=\linewidth]{FIGUxy_2.jpg}
\caption{$\left\vert \mathbf{k}\right\rangle $ character (or relative
intensity), $\left[ 1\mp \cos \protect\phi \left( \mathbf{k}\right) \right]
/2,$ of the upper (light) and lower (dark) pure $xy$ bands as functions of $%
k_{c}$ for $k_{b}$=$0.225$ in the double zone (FIG.~\protect\ref%
{FIGDoubleZone}); from Eq.s~(\protect\ref{ww}), (\protect\ref{fat}), (%
\protect\ref{tautu}), and (\protect\ref{tp}). The dominant (or most
intensive) $\left\vert \mathbf{k}\right\rangle $ character switches from the
lower band in the odd-numbered zones,\textbf{\ }$\left\vert
k_{c}-2n\right\vert <\frac{1}{2},$ to the upper band in the even-numbered
zones, $\left\vert k_{c}-\left( 2n+1\right) \right\vert <\frac{1}{2}$.
Whereas the gapped bands (\protect\ref{epsul}) have single-zone periodicity,
their $\left\vert \mathbf{k}\right\rangle $ characters are periodic in the
double zone. In FIG.s \protect\ref{ThreePureBands} and \protect\ref%
{ThreeBands} the $\left\vert \mathbf{k}\right\rangle $ character is shown as
fatness added to (i.e. decorating) the band.}
\label{FIGUxy^2}
\end{figure}

The switching curves for the two $xy$ bands along lines of constant $k_{b}$
are the same for both signs of $k_{b}$, and while they always cross at the
zone boundary, the steepness of their steps decreases with increasing $%
\left\vert k_{b}\right\vert $ [see Eq.s (\ref{tautu}) and (\ref{AG}), and
the second column of FIG.~II~\ref{FIGxyZoneSelectul} for $\kappa _{a}$=6.4
and $\left\vert k_{b}\right\vert $=0.225, 0.250, and 0.275].

In contrast to the complementarity of the $\left\vert \mathbf{k}%
\right\rangle $ and $\left\vert \mathbf{k+c}^{\ast }\right\rangle $
characters exhibited by Eq.~$\left( \text{\ref{ww}}\right) $, the $w$ and $W$
characters are 50\% for both bands and all $\mathbf{k,}$ as seen from Eq.~(%
\ref{wW}).

Bands with \emph{different} $m$ \emph{do} hybridize with each other: the $%
xy\left( \mathbf{k}\right) $ band with the $xz\left( \mathbf{k}\right) $ and 
$yz\left( \mathbf{k}\right) $ bands due to the $\alpha $ and $a$ hops, and
with the $xz\left( \mathbf{k+c}^{\ast }\right) $ and $yz\left( \mathbf{k+c}%
^{\ast }\right) $ due to the $\gamma $ and $g$ hops. The $xz\left( \mathbf{k}%
\right) $ band hybridizes with the $yz\left( \mathbf{k}\right) $ band due to
the $\lambda $ and $l$ hops, and with the $yz\left( \mathbf{k+c}^{\ast
}\right) $ band due to the $\mu $ and $m$ hops.

When $\left\vert k_{b}\right\vert \sim \frac{1}{4}\approx k_{F},$ the $xy$
bands are situated in the gap between the $xz$ valence and conduction bands
and between the $yz$ valence and conduction bands. The hybridization caused
by the $\left( \alpha ,a\right) $ hops in the $\left\vert \mathbf{k}%
\right\rangle $-conserving part and by the $\left( \gamma ,g\right) $ hops
in the $\left\vert \mathbf{k}\right\rangle $-$\left\vert \mathbf{k+c}^{\ast
}\right\rangle $ mixing part of the Hamiltonian (\ref{HRecip}) makes the
difference between the red pure $xy$ bands in FIG.$~$\ref{ThreePureBands}
and the dark-red hybridized $\widetilde{xy}$ bands in FIG.$~$\ref{ThreeBands}%
. The $\left( \gamma ,g\right) $ hops modify the shape of the $\widetilde{xy}
$ bands, but do not significantly extend the switching region around the BZ
boundary, $\left\vert k_{c}\right\vert $=$\frac{1}{2},$ in which the $%
\left\vert \mathbf{k}\right\rangle $-$\left\vert \mathbf{k+c}^{\ast
}\right\rangle $ mixing occurs.

\subsection{Brillouin- and physical zones \label{Sectzones}}

Bloch functions are characterized by their translational symmetry in
reciprocal space and the choice of the primitive cell (zone) is arbitrary as
long as it contains each $\mathbf{k}$-point once and only once. As we
mentioned after Eq.s$\,$(\ref{gap}) and (\ref{xyzone})-(\ref{xzzone}), it
may be possible and convenient to choose the zone compatible with the
electronic structure, i.e. such that gaps occur at the zone boundaries.

For the familiar free-electron model, where $\left\vert \mathbf{k}%
\right\rangle \sim $ $e^{2\pi i\mathbf{k\cdot r}}$ and $\varepsilon \left( 
\mathbf{k}\right) $ increases isotropically and monotonically with the
distance, $k,$ from $\Gamma $(0,0,0)$,$ one chooses that zone which is
closer to $\Gamma $ than to any other point, $\mathbf{G,}$ of the reciprocal
lattice. This is the Brillouin zone (BZ). Application of a weak
pseudopotential with crystalline symmetry will couple the basis functions $%
\left\vert \mathbf{k}\right\rangle ,$ $\left\vert \mathbf{k-G}%
_{1}\right\rangle ,..,$ and $\left\vert \mathbf{k-G}_{n}\right\rangle ,$
where $\mathbf{G}_{1},$ $..,$ $\mathbf{G}_{n}$ are the reciprocal lattice
points closest to $\Gamma $, and thereby gap bands where they cross, i.e.
where $k=\left\vert \mathbf{k-G}_{n}\right\vert ,$ which is at the
boundaries of the BZ.

For LiPB, the situation is different: Rather than being isotropic, the
low-energy electronic structure consists of three pairs of quasi-1D bands.
For each, we have only two inequivalent pseudo-Bloch sums, $\left\vert w_{m};%
\mathbf{k}\right\rangle $ and $\left\vert w_{m};\mathbf{k+c}^{\ast
}\right\rangle .$ As described in the previous subsection, and already in
the introductory section \ref{Sectt2g}, and as\textbf{\ }seen in FIG.s~\ref%
{ThreePureBands}-\ref{FIGDoubleZone}, the lower (bonding) $m$ band has $%
\left\vert \mathbf{k}\right\rangle $ character inside its zone and $%
\left\vert \mathbf{k+c}^{\ast }\right\rangle $ character outside, while the
upper (anti-bonding) band has $\left\vert \mathbf{k+c}^{\ast }\right\rangle $
character inside and $\left\vert \mathbf{k}\right\rangle $ character
outside. Here, "inside" means around a point, such as $\Gamma \left( k_{b}%
\text{=}0,k_{c}\text{=}0\right) ,$ of the \emph{even} reciprocal sublattice
(confusingly called an odd-numbered BZ), and "outside" means around a point,
such as $\Gamma \left( 0,1\right) $ or $\Gamma \left( 0,-1\right) ,$ of the 
\emph{odd} reciprocal sublattice (called an even-numbered BZ)$.$

The $xz$ (blue) and $yz$ (green) bands are functions with period 2 of
respectively $k_{c}-k_{b}$ and $k_{c}+k_{b}$ whereby the\ $\mathbf{k}$ and $%
\mathbf{k+c}^{\ast }$ bands cross --and gap by $\pm 2G_{1}$-- along
respectively $\left\vert k_{c}-k_{b}\right\vert $=$\frac{1}{2}$ and $%
\left\vert k_{c}+k_{b}\right\vert $=$\frac{1}{2},$ which are then "physical"
zone boundaries of the $xz$ and $yz$ bands [Eq.s (\ref{xzzone}) and (\ref%
{yzzone}) and FIG.~\ref{FIGDoubleZone}].

The $xy$ bands (red) disperse strongly with $k_{b},$ which is normal to the
plane of FIG.~\ref{ThreePureBands}, but since the gap at $\left\vert
k_{b}\right\vert \mathbf{=}\frac{1}{2}$ caused by $b$-axis dimerization is
far above $E_{F},$ we merely consider the two lowest of the four $xy$ bands
(see Sect.~\ref{Sectbims}) and their strong $k_{b}$-dependence is described
merely by $\tau \left( k_{b}\right) $. The small, $\pm 2u,$ gap near $E_{F}$
is due to the $c$-axis dimerization (Sect.~\ref{SectDims}) and occurs at the 
$\left\vert k_{c}\right\vert $=$\frac{1}{2}$ zone boundary. Hence, the
physical zone for the $xy$ bands is the rectangular one, [$\left\vert
k_{b}\right\vert \leq \frac{1}{2}$ and $\left\vert k_{c}\right\vert \leq 
\frac{1}{2},~$Eq.~(\ref{xyzone}$)]$ which is --actually-- the Brillouin zone
(BZ).

While the nearly degenerate, $xy$ bands are half full (metallic), the lower $%
xz$ and $yz$ bands are full and the upper are empty.

For the purpose of \emph{calculating} the electronic structure, i.e. when
diagonalizing the $6\times 6$ Hamiltonian or the L\"{o}wdin downfolded $%
2\times 2$ Hamiltonian in Paper III, we normally use the rectangular BZ.

\section{Summary}

In this paper, we have developed the single-particle framework on the basis
of which we shall discuss and refine the new ARPES measurements of the band
structure and Fermi surface of LiPB to be presented in Papers II and III.

In the Introduction, we gave an overview of the properties and current
theories of this intriguing quasi-1D metal and laid out the plan for --and
gave the main results of-- our three papers.

In Sect.~\ref{SectElCalc}, our DFT method for direct computation of Wannier
functions and their TB Hamiltonian, the full-potential NMTO method, was
explained. Its unique ability to produce physically and chemically
meaningful Wannier \emph{orbitals} (WOs) --multi-center Mo1 $4d\,t_{2g,m}$
orbitals in the present case-- is crucial for our understanding of LiPB
whose crystal structure (FIG.~\ref{FIG1}) consists of MoO$_{6}$ octahedra
connected via corners into slabs perpendicular to $\mathbf{a}^{\ast }.$ Each
slab consists of \emph{ribbons,} 4 molybdenums wide in the $\mathbf{a}+%
\mathbf{c}$ direction and extending indefinitely in the perpendicular
direction, the direction of quasi-1D conductivity Chart~(\ref{ac}). The
well-known zigzag chains, $\diagdown _{\text{Mo1}}\diagup ^{\text{Mo4}%
}\diagdown _{\text{Mo1}}\diagup ^{\text{Mo4}}\diagdown ,$ with primitive
translation $\mathbf{b}$ are the spines of the ribbons. The ribbons, with
every second \emph{displaced} from the position $\frac{\mathbf{c+b}}{2}$ by
a vector $-\mathbf{d}$\textbf{\ }(\ref{dim}) and \emph{inverted} [Eq.~(\ref%
{inv}) and Chart~(\ref{ac})], are stacked on top of each other into \emph{bi}%
-ribbons, whereby the slab forms a \emph{staircase} with steps of bi-ribbons
and primitive translations $\mathbf{c}$. Without this $c$-axis dimerization,
the staircase would have been a smooth ramp [Charts (\ref{undimac})-(\ref%
{undimxy})] with the undimerized crystal lattice spanned by $\left( \mathbf{a%
},\frac{\mathbf{c+b}}{2},\frac{\mathbf{c-b}}{2}\right) $ and its reciprocal
lattice spanned by $\left( \mathbf{a}^{\ast },\mathbf{c}^{\ast }\mathbf{+b}%
^{\ast },\mathbf{c}^{\ast }\mathbf{-b}^{\ast }\right) ,$ as compared with
the primitive translations $\left( \mathbf{a},\mathbf{b},\mathbf{c}\right) $
and $\left( \mathbf{a}^{\ast },\mathbf{b}^{\ast },\mathbf{c}^{\ast }\right) $
of real LiPB. The staircase is terminated by insulating MoO$_{4}$ tetrahedra
and Li intercalates between staircases, which is also where the crystal
cleaves. All 8 octahedral and 4 tetrahedral molybdenums approximately form a
simple cubic lattice [Eq.~(\ref{xyz})].

Anticipating the results of the more technical Sect.s \ref{SectLowE} and \ref%
{SectH} --as well as the band structures and Fermi surface in Papers II and
III-- in Sect.~\ref{SectElStruc} we gave an elementary description of the
electronic structure inside the slab, from the 10~to the 0.1~eV scale around
the Fermi level seen in respectively FIG.s~\ref{FIG2}, \ref{3Dt2gBands}, and %
\ref{ThreePureBands}. The \emph{double} zone, which is the BZ of the
undimerized lattice, was shown in FIG.~\ref{FIGDoubleZone}. The boundary
between the 1st and 2nd \emph{physical} zones, which is where in the absence
of dimerization the $m\left( \mathbf{k}\right) $ and $m\left( \mathbf{k+c}%
^{\ast }\right) $ bands cross, is shown in solid red, blue, and green lines
for $m$=$xy$, $xz,$ and $yz,$ respectively. The minimum and maximum of the
pure-$m\left( \mathbf{k}\right) $ band --as well as of the pure-$m\left( 
\mathbf{k+c}^{\ast }\right) $ band-- are at the weak lines. The brown
dot-dashed line has $k_{b}$=0.9$k_{F},$ and is the one along which the pure-$%
m$ bands in FIG.~\ref{ThreePureBands} were shown.

The DFT-LDA full-potential NMTO calculations in Sect.~\ref{SectLowE} showed
that the six lowest energy bands --half of them occupied-- are described by
a \emph{set of six} $t_{2g}$ WOs per 2LiMo$_{6}$O$_{17}$, namely $%
w_{m}\left( \mathbf{r}\right) $ centered on Mo1 in the lower ribbon and $%
W_{m}\left( \mathbf{r}\right) $ on the equivalent MO1 in the upper ribbon.
These sites, separated by $\frac{\mathbf{c+b}}{2}-\mathbf{d}$, are special
in having a full nearest-neighbor shell of octahedral molybdenums and
therefore best preserve the $t_{2g}$ symmetry of the WO and are least
sensitive to the steps of the staircase. As seen in FIG.s~\ref{FIGyzb&xyamc}
and \ref{Wannier}, the WOs have $t_{2g}$ symmetry around Mo1 (or MO1), and
spill over into neighboring atoms which carry \emph{no} WO [see Charts (\ref%
{WOac}) and (\ref{WOxy})]. This spill-over is necessitated by the
requirement that the six Mo1- and MO1-centered $t_{2g}$ WOs completely span
the wavefunctions of the six lowest bands and causes what we call a halo on
the near atoms.

The $xy$ WOs lie inside their respective ribbon and have strong, long-ranged 
$dd\pi $ intra-ribbon hopping integrals, $\tau _{n},$ along $\mathbf{b,}$
very weak $dd\delta $ inter-ribbon $xy$-$XY$ hopping integrals, $%
t_{n}+u_{n}, $ between partner ribbons, and even weaker, $t_{n}-u_{n},$
between bi-ribbons [Eq.s~(\ref{Hsub})-(\ref{tau}), (\ref{tautu}), and (\ref%
{taup})-(\ref{tp})]. Between slabs, the $xy$ hopping is negligible$^{\ref%
{slabhop}}$.

The equivalent $xz$ and $yz$ WOs stand perpendicular to the ribbons and the $%
dd\pi $ nearest-neighbor $xz$-$XZ$ hopping integral, $A_{1}+G_{1},$ between
partner ribbons is as strong as the $dd\pi $ intra-ribbon hopping integral $%
\tau _{1}$ between $xy$ orbitals, twice as strong as the hopping integral $%
A_{1}-G_{1}$ between bi-ribbons [Eq.s~(\ref{Hsub}), (\ref{AG}) and (\ref%
{A&Gp})] and 30 times stronger than the $dd\delta $ integrals $t_{1}\pm
u_{1} $ for $xy$-$XY$ hopping. The $A_{1}\pm G_{1}$ integrals are for
hopping up or down the staircase with steps $\frac{\mathbf{c-b}}{2}$ for $xz$%
-$XZ$ and $\frac{\mathbf{c+b}}{2}$ for $yz$-$YZ.$ The two $xz$ bands are
gapped by the hopping dimerization, $\pm 2G_{1}\approx \pm 0.2~$eV, and so
are the two $yz$ bands (FIG.s~\ref{ThreePureBands} and \ref{ThreeBands}).

The six-band TB Hamiltonian was given in Eq.~(\ref{Hsub}) in terms of these
and further hopping integrals, Bloch summed as in Eq.~(\ref{Bloch}). The
basis functions were the Bloch-summed WOs, $w_{m}\left( \mathbf{k,r}\right) $
and $e^{\pi i\left( k_{c}+k_{b}\right) }W_{m}\left( \mathbf{k,r}\right) ,$
on respectively the lower and the upper ribbon, and the $\mathbf{k}$%
-dependent phase factor in front of $W_{m}\left( \mathbf{k,r}\right) $ was
included in order to make the electronic dimerizations purely imaginary.
Further insight was gained in Sect.~\ref{Sectk+c} by transforming from this
basis set --which for each $m$ consists of two Bloch sums, one over the Mo1
positions and the other over the MO1 positions,-- to one with two \emph{%
pseudo} Bloch sums (\ref{kket}), $\left\vert m;\mathbf{k}\right\rangle
\equiv \left\vert w_{m},\mathbf{k}\right\rangle $ and $\left\vert m;\mathbf{%
k+c}^{\ast }\right\rangle $, each of which is a Bloch sum over both Mo1 and
MO1 with every second phase factor along $\mathbf{c}$ chosen as $e^{2\pi i%
\mathbf{k\cdot }\left( \mathbf{T+}\frac{\mathbf{c+b}}{2}\right) },$\ i.e. as
if there were no displacement dimerization, $\mathbf{d}$\textbf{=0.} The
transformation is (\ref{trf}). Considered as a function of $\mathbf{k,}$ the
pseudo Bloch sum $\left\vert \mathbf{k}\right\rangle $ is a periodic
function on the sparse, so-called $\emph{even}\mathbf{,}$ reciprocal lattice
spanned by $\left( \mathbf{a}^{\ast },\mathbf{c}^{\ast }\mathbf{+b}^{\ast },%
\mathbf{c}^{\ast }\mathbf{-b}^{\ast }\right) $ whose BZ is the \emph{double }%
zone shown in FIG.~\ref{FIGDoubleZone}. Together with the function $%
\left\vert \mathbf{k+c}^{\ast }\right\rangle ,$ periodic on the \emph{odd }%
reciprocal lattice, they form a complete, orthonormal basis set for the
proper, dimerized crystal. In the absence of dimerization, $\left\vert 
\mathbf{k}\right\rangle $ and $\left\vert \mathbf{k+c}^{\ast }\right\rangle $
are identical apart from a phase factor, but they become linearly \emph{in}%
dependent in the presence of dimerization and will mix near the boundaries
of the appropriate physical zone. The six-band TB Hamiltonian (\ref{HRecip})
in this so-called $\left\{ \mathbf{k},\mathbf{k+c}^{\ast }\right\} $%
-representation was used to visualize the $\left\vert \mathbf{k}%
\right\rangle $ characters of the band structures in FIG.s~\ref%
{ThreePureBands} and \ref{ThreeBands} (and in Paper II FIG. \ref%
{ThreeBandskonly}) as their (additional) \emph{fatness} (\ref{fat}). FIG.~(%
\ref{FIGUxy^2}) showed how the $\left\vert \mathbf{k}\right\rangle $
characters of the lower and upper $m$ bands switch between 0 and 1, and back
again, as the Bloch vector crosses the boundaries of the physical ($m$%
-dependent) zone.

In the following Paper II, we shall find the important result that this
interesting $\left\vert \mathbf{k}\right\rangle $-character variation is
experimentally manifested as an ARPES intensity selection rule. As mentioned
already in the Introduction, when this selection rule is combined with our
new ARPES data, it enables the separation of the two bands that disperse to
define the FS. In Paper III we will also use the selection rule to reveal
the FS features that are peculiar to each of the two bands. Thereby our new
ARPES results both confirm, and are greatly aided by, our new theory. But
first, at the beginning of Paper II, we shall give the complete theory which
includes the distortion of the ARPES intensity variations caused by the $c$%
-axis displacement- and inversion dimerizations. The latter depends on the
photon energy, which we have chosen such that they basically cancel.

\begin{acknowledgments}
We are indebted to Tanusri Saha-Dasgupta, Sashi Satpathy, and Zoran Popovic
for their active participation at the initial stage of this project. JWA
acknowledges past support of this work by the U.S. National Science
Foundation (grant DMR-07-04480).

This research used resources of the Advanced Light Source, which is a DOE
Office of Science User Facility under contract no. DE-AC02-05CH11231. MG
acknowledges support by NSF-DMR-1507252 grant.
\end{acknowledgments}

\mbox{} \clearpage

\title{Wannier-Orbital theory and ARPES for the quasi-1D conductor LiMo$_{6}$%
O$_{17}$. \\
Part II: Intensity variations and the six $t_{2g}$-bands}

\begin{abstract}
This is the second paper of a series of three papers presenting a combined
study by band theory and angle-resolved photoemission spectroscopy (ARPES)
of lithium purple bronze. The $t_{2g}$ Wannier Orbitals (WOs) and resulting
six-band tight-binding (TB) Hamiltonian found in paper~I are here used to
develop a theory of the ARPES intensity variations, including a selection
rule whose validity relies on the smallness of and the cancellation between the displacement- and
inversion-dimerizations of the zig-zag chains (ribbons) in regions of the
final-state wavevector, $\mathbf{\kappa }$. We then present the ARPES
results for the band structure of the four occupied $t_{2g}$-bands (gapped $%
xz$, $yz$, and split metallic $xy$). A detailed comparison to the theory
validates the selection rule. We present the Fermi surface (FS) as seen
directly in the raw ARPES data, both parallel and perpendicular (using
photon-energy dependence) to the sample surface, and show that the selection
rule can enable separation of the barely split and highly
quasi-one-dimensional $xy$-bands. We adjust the energy of the $xy$ WO energy
by 0.1~eV ($\approx \frac{1}{4}$ of the gap) with respect to that of the
gapped $xz$ and $yz$ WOs and, in a second step, fine-tune merely 7 out of the more than 40 TB
parameters to achieve an excellent fit to the ARPES bands lying more than
0.15 eV below the Fermi level. So doing then also gives nearly perfect
agreement closer to the Fermi level. 
\end{abstract}

\date{\today }
\pacs{Valid PACS appear here}
\maketitle

\section{Introduction}

This is the second paper in a series of three presenting a detailed study of
the band structure of the quasi-1D lithium purple bronze (LiPB) combining
LDA-NMTO band theory and angle-resolved photoemission spectroscopy (ARPES).

In Paper I we explained the NMTO method (Sect.~\ref{SectElCalc}) and used it
to derive, for the occupied and lowest unoccupied bands of LiPB (Sect.~\ref%
{SectElStruc})$,$ a chemically meaningful set of Wannier functions (Sect.~%
\ref{SectLowE}) --called Wannier orbitals (WO)-- and their tight-binding
(TB) Hamiltonian in portable, i.e. analytical, form (Sect.~\ref{SectH}).

The monoclinic crystal structure of LiMo$_{6}$O$_{17}$ (Sect.~\ref%
{crystal_structure} of Paper I) consists of MoO$_{6}$ octahedra connected by
corners into slabs perpendicular to the reciprocal-lattice vector $\mathbf{a}%
^{\ast }.$ Each slab consists of \emph{bi-ribbons}, 4 molybdenums wide in
the \textbf{a+}$\mathbf{c}$ direction, $^{\text{Mo2}}\diagdown _{\text{Mo1}%
}\diagup ^{\text{Mo4}}\diagdown _{\text{Mo5}}$ and $^{\text{MO5}}\diagdown _{%
\text{MO4}}\diagup ^{\text{MO1}}\diagdown _{\text{MO2}}$ (using
lower/upper-case letters for the lower/upper \emph{string}), and extending
indefinitely along the direction $\mathbf{b}$ of quasi-1D conductivity [See
FIG.~I~\ref{FIG1} together with Charts~I~$($\ref{ac}) and~(\ref{xy})]%
\footnote{%
I, II, and III refer to sections, figures, equations, and footnotes in Paper
I, II, and III, respectively.\label{Paper I copy(1)}}. The \emph{spines} of
the ribbons are the well-known $_{\text{Mo1}}\diagup ^{\text{Mo4}}\diagdown
_{\text{Mo1}}\diagup ^{\text{Mo4}}$ and $_{\text{MO4}}\diagup ^{\text{MO1}%
}\diagdown _{\text{MO4}}\diagup ^{\text{MO1}}$ \emph{zigzag chains} along $%
\mathbf{b}$. The upper string is related to the lower by translation of Mo1
to MO1 by the vector $\frac{\mathbf{c+b}}{2}-\mathbf{d,}$ followed by \emph{%
inversion} around their midpoint, $\frac{1}{2}\left( \frac{\mathbf{c+b}}{2}-%
\mathbf{d}\right) .$ Had there been \emph{no} displacement- ($\mathbf{d}$=0)
and \emph{no} inversion-, i.e. no $c$-axis dimerization (Sect.~I~\ref%
{SectDims}), all ribbons would have been related by a primitive translation
vector $\frac{\mathbf{c+b}}{2},$ and thus stacked into a ramp [Chart~I~$($%
\ref{undimac})]. \emph{With} displacement- and inversion dimerizations, the
slab forms a \emph{staircase} [Chart~I~$($\ref{ac})] with steps of
bi-ribbons and running up and down along $\pm \mathbf{c}$. A staircase is
terminated by insulating MoO$_{4}$ tetrahedra, and Li intercalates between
staircases which is also where the crystal cleaves.

We found that the six lowest energy bands (FIG.s ~I~\ref{FIG2} and \ref%
{3Dt2gBands}) --half occupied-- are accurately described by the set of six,
real-valued $t_{2g}$ WOs,$^{\text{I}\,\ref{Rcenter}}$ $w_{m}=xy,$ $xz,$ $yz$
centered on Mo1, and $W_{m}=XY,$ $XZ,$ $YZ,$ centered on MO1, and with $x,$ $%
y,$ and $z$ directions as indicated in Charts~I~(\ref{ac}) or (\ref{WOac})
and (\ref{xy}) or (\ref{WOxy}). Such a $t_{2g}$ WO has tails with the same $%
m $ on the nearest Mo neighbors in its plane which contribute to the halo of
the WO (FIG.s~I~\ref{FIGyzb&xyamc} and \ref{Wannier}).

The $xy$ WOs lie well inside their respective ribbon and have strong,
long-ranged $dd\pi $ $xy$-$xy$ or $XY$-$XY$ hopping integrals, $\tau _{n},$
with Bloch sum $\tau \left( k_{b}\right) $ along the ribbon, very weak $%
dd\delta $ $xy$-$XY$ hopping integrals, $t_{n}+u_{n},$ between partner
ribbons, and even weaker hopping integrals, $t_{n}-u_{n},$ between
bi-ribbons [Eq.s~I~(\ref{tau})$,$ (\ref{tautu}), (\ref{taup}), and (\ref{tp}%
)]. Between slabs, the hopping is negligible. The $xz$ and $yz$ WOs, which
are equivalent, stand perpendicular to the ribbons and the $dd\pi $
nearest-neighbor hopping integral, $A_{1}+G_{1},$ between partner ribbons
--up or down the staircase with steps $\frac{\mathbf{c-b}}{2}$ for $xz$-$XZ$
and $\frac{\mathbf{c+b}}{2}$ for $yz$-$YZ-$ is twice as strong as the
hopping integral, $A_{1}-G_{1},$ between bi-ribbons. The two $xz$-bands are
gapped by the hopping dimerization, $\pm 2G_{1}\mathrm{\sim \mp }0.2~$eV,
and so are the two $yz$-bands [Eq.s~I~(\ref{AG}) and (\ref{A&Gp})].

In our notation, greek-lettered Bloch sums, e.g. $\tau ,$ are over hops on
the ribbon, whereby they are real and single-zone periodic in $\mathbf{k}$.
Latin-lettered Bloch sums, such as $t,$ $u,$ $A,$ and $G,$ are over hops%
\emph{\ between} ribbons and are therefore real and double-zone periodic
[Eq.s~I~(\ref{Greek}) and (\ref{Latin})].

\begin{figure}[tbh]
\includegraphics[width=\linewidth]{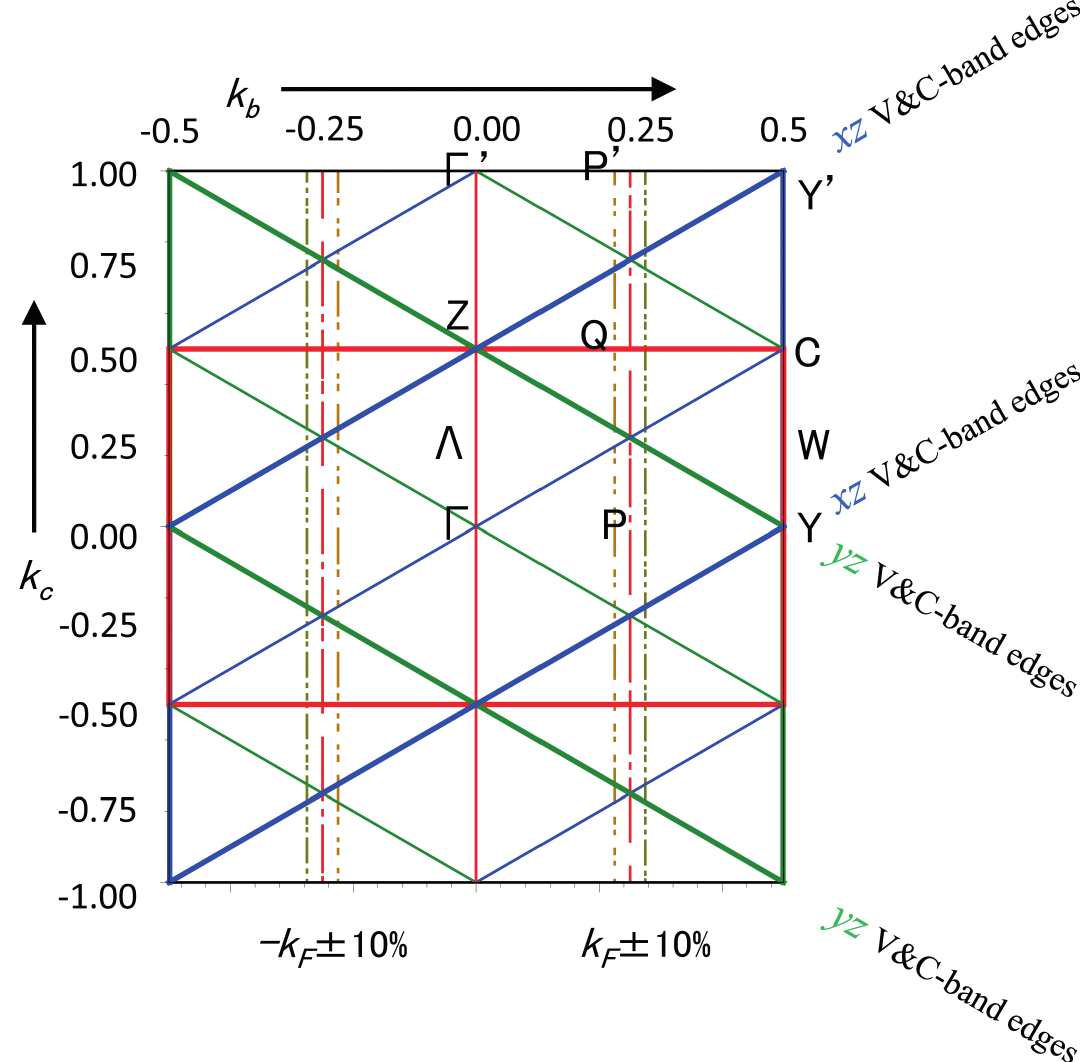}
\caption{Double zone, same as FIG. 8 in Paper I. The $\QATOP{1\mathrm{st}}{2%
\mathrm{nd}}$ physical zones shown in solid lines for the $xy$ (red), $xz$
(blue)$,$ and $yz$ (green) pure bands are respectively: $\left\vert
k_{c}\right\vert \lessgtr \frac{1}{2}$, $\left\vert k_{c}+k_{b}\right\vert
\lessgtr \frac{1}{2},$ and $\left\vert k_{c}-k_{b}\right\vert \lessgtr \frac{%
1}{2}.$ The red zone is the BZ, and its irreducible part is the one with $%
0\leq k_{c}\leq 0.5.$ Weak lines indicate the positions of pure-band maxima
and minima (see FIG. I~6). The red dot-dashed lines indicate the positions
of the left and the right doubly degenerate Fermi-surface sheets for
stochiometric 2(LiMo$_{6}$O$_{17}$). The brown and olive dot-dashed lines
respectively correspond to 10\% hole and electron doping. See Sect.$\,$I$\,$%
\protect\ref{subsubsubCEC} and Sect.$\,$\protect\ref{SectCEC}. The figure is
to scale because the triangles are almost equilateral.}
\label{FIGPhysicalZones2}
\end{figure}

In Eq.~I~(\ref{Hsub}) we gave the \emph{six-band TB Hamiltonian} in the
representation of the Bloch-summed WOs, $w_{m}\left( \mathbf{r,k}\right)
\equiv $ $\sum_{\mathbf{T}}e^{2\pi i\mathbf{k\cdot T}}w_{m}\left( \mathbf{r-T%
}\right) $ and $W_{m}\left( \mathbf{r,k}\right) \equiv $ $\sum_{\mathbf{T}%
}e^{2\pi i\mathbf{k\cdot T}}W_{m}\left( \mathbf{r-T}\right) ,$ with the
latter multiplied by $e^{2\pi i\mathbf{k\cdot }\frac{\mathbf{c+b}}{2}}=$ $%
e^{\pi i\left( k_{c}+k_{b}\right) }$ so that the \emph{hopping dimerizations
are purely imaginary}; numerically, they are about 30\% of the corresponding
hopping integral. Further insight was gained by transforming from this
sublattice $\{w,W\}$-representation (Sect.~I~\ref{SectwW}) to the reciprocal
sublattice $\left\{ \mathbf{k,k+c}^{\ast }\right\} $-representation (Sect.~I~%
\ref{Sectk+c}), i.e. from a basis with two sets of Bloch sums, $w_{m}\left( 
\mathbf{r,k}\right) $ and $W_{m}\left( \mathbf{r,k}\right) ,$ each a
periodic function of $\mathbf{k}$ in the single zone, to a set of \emph{%
pseudo} Bloch sums,%
\begin{eqnarray}
&&\left\vert m;\QATOP{\mathbf{k}}{\mathbf{k+c}^{\ast }}\right\rangle = 
\TCItag*{I (52)} \\
&&\frac{1}{\sqrt{2}}\sum_{\mathbf{T}}e^{2\pi i\mathbf{k\cdot T}}\left[
w_{m}\left( \mathbf{r-T}\right) \pm e^{\pi i\left( k_{c}+k_{b}\right)
}W_{m}\left( \mathbf{r-T}\right) \right] ,  \notag
\end{eqnarray}%
which are periodic functions of $\mathbf{k}$ in the \emph{double zone}
(FIG.~I~\ref{FIGDoubleZone}, repeated as FIG.$\ $II$~$\ref{FIGPhysicalZones2}%
) and evaluated in two different single zones, i.e. at $\mathbf{k}$ and $%
\mathbf{k+c}^{\ast }$. Each pseudo-Bloch sum is over \emph{both }ribbons
with the phase factor multiplying $W\left( \mathbf{r-T}\right) $ chosen as
if there were \emph{no} displacement dimerization, \textbf{d} [see Eq.s~I$~($%
\ref{undim}) and (\ref{inv})], i.e. as $e^{2\pi i\mathbf{k\cdot }\frac{%
\mathbf{c+b}}{2}}.$ In this representation, the six-band Hamiltonian is I~(%
\ref{HRecip}). It is now conceivable that \emph{the} \emph{ARPES intensity}
from an occupied band follows its $\left\vert \mathbf{k}\right\rangle $
character, i.e. what we called its fatness in Sect.~I$~$\ref{SectPureBands}
and showed in FIG.~I~\ref{FIGUxy^2} for the two $xy$-bands at $k_{b}$=$0.225$
and as functions of $k_{c}$. In the absence of dimerization, the two pseudo
Bloch sums are linearly dependent, and the $\left\vert \mathbf{k}%
\right\rangle $- and $\left\vert \mathbf{k+c}^{\ast }\right\rangle $%
-projected bands are double-zone periodic and translated by $\mathbf{c}%
^{\ast }$ with respect to each other. \emph{With} dimerization, the pseudo
Bloch sums become linearly \emph{in}dependent and will mix near the
crossings of the undimerized $\mathbf{k}$- and $\left( \mathbf{k+c}^{\ast
}\right) $-bands, where the dimerized bands will gap and thus restore the
single-zone periodicity. Correspondingly, the $\left\vert \mathbf{k}%
\right\rangle $-projection follows the undimerized $\mathbf{k}$-band, except
near its crossing with the undimerized $\left( \mathbf{k+c}^{\ast }\right) $%
-band, where it looses half its intensity to the $\left\vert \mathbf{k+c}%
^{\ast }\right\rangle $-projected band (see FIG.s~I~\ref{ThreePureBands}-\ref%
{FIGDoubleZone}).

The first task of the present Paper II is to derive an expression for the
variation of ARPES intensity with $\mathbf{k}$ which \emph{includes} its 
\emph{distortion} caused by $c$-axis displacement- and inversion
dimerizations. In order to do so, we neglect the coupling between $t_{2g}$
WOs with different $m,$ a good approximation near the FS $(\left\vert
k_{b}\right\vert \approx \frac{1}{4})$, and in fact everywhere, except near
the $\Gamma \mathrm{Z}$-line ($k_{b}$=0)$.$ In this \emph{pure-}$m$-band
approximation introduced in Sect.~I$~$\ref{SectPureBands}, the six-band
Hamiltonian factorizes in three 2$\times 2$ Hamiltonians, $H_{m},$ with
eigenvalues,%
\begin{equation}
\pm \sqrt{A^{2}+G^{2}},\,\pm \sqrt{\bar{A}^{2}+\bar{G}^{2}},\,\text{\textrm{%
and}}\,\tau \pm \sqrt{t^{2}+u^{2}},  \tag*{I (57)}
\end{equation}%
for $m$=$xz,~yz$, or $xy,$ respectively, and with $\mathbf{k}$ dependencies
given by Eq.s~I$~$(\ref{AG}), (\ref{tautu}), and (\ref{tau}). The crossings,%
\emph{\ }I$~$(\ref{xzzone}), (\ref{yzzone}) and (\ref{xyzone}), of these
pure-$m$\ bands define the boundaries of the so-called \emph{physical} zones
shown in FIG.~\ref{FIGPhysicalZones2}. The eigenfunctions of the pure-$m$
bands were given by:%
\begin{eqnarray}
&&w_{\QATOP{2}{1}}\left( \mathbf{k,r}\right) =  \notag \\
&&\frac{1}{\sqrt{2}}\left[ w\left( \mathbf{k,r}\right) e^{-i\phi \left( 
\mathbf{k}\right) }\mp W\left( \mathbf{k,r}\right) e^{\pi i\left(
k_{c}+k_{b}\right) }\right]   \TCItag*{I (58)} \\
&=&\frac{1}{2}\left[ 
\begin{array}{l}
\left\vert w;\mathbf{k}\right\rangle \left( e^{-i\phi \left( \mathbf{k}%
\right) }\mp 1\right) + \\ 
\left\vert w;\mathbf{k+c}^{\ast }\right\rangle \left( e^{-i\phi \left( 
\mathbf{k}\right) }\mp 1\right) 
\end{array}%
\right] ,  \TCItag*{I (59)}
\end{eqnarray}%
where the \emph{band-structure phase,} $\phi \left( \mathbf{k}\right) ,$ is
the phase of the complex Bloch-summed inter-ribbon hopping integral whose
imaginary part gives the asymmetry between the hopping in- and outside the
bi-ribbon, i.e. the \emph{initial-state dimerization}:%
\begin{equation}
e^{i\phi }\equiv \frac{-A-iG}{\sqrt{A^{2}+G^{2}}},\frac{-\bar{A}-i\bar{G}}{%
\sqrt{\bar{A}^{2}+\bar{G}^{2}}},\mathrm{and}\text{ }\frac{-t-iu}{\sqrt{%
t^{2}+u^{2}}}.  \tag*{I (60)}
\end{equation}%
For quasi-1D structures, the band-structure phase $\phi \left( \mathbf{k}%
\right) $ varies from 0 for $\mathbf{k}$ at the centre of the physical zone
to $\pm \frac{\pi }{2}$ at $\QATOP{\mathrm{right}}{\mathrm{left}}$ zone
boundaries (ZB), and to $\pm \pi $ at the centers of the second zones.

According to Eq.~I~(59), the $\left\vert \mathbf{k}\right\rangle $ character
[fatness, $f_{\QATOP{2}{1}}\left( \mathbf{k}\right) ]$ of the $\QATOP{%
\mathrm{upper}}{\mathrm{lower}}\,m$\ band illustrated in FIG.~I$\,$\ref%
{FIGUxy^2} is $\left[ 1\mp \cos \phi _{m}\left( \mathbf{k}\right) \right] /2.
$ In Sect.~\ref{Sectzoneselect} we shall show that the intensity of
photo-emission from the $\QATOP{\mathrm{upper}}{\mathrm{lower}}\,m$\ band is
simply $\left[ 1\mp \cos \left\{ \phi _{m}\left( \mathbf{k}\right) -\eta
_{m}\left( \mathbf{\kappa }\right) \right\} \right] /2,$ which means that
the band-structure phase is shifted by $\eta _{m}\left( \mathbf{\kappa }%
\right) ,$ the phase shift (\ref{eta}) experienced by an electron emitted
with momentum $\mathbf{\kappa }$ and caused by the $c$-axis inversion- and
displacement dimerizations (see Sect. I \ref{SectDims}). This phase shift \emph{%
distorts} the switching curves in FIG.~I$\,$\ref{FIGUxy^2} to what is shown
in the middle panel of FIG.~\ref{FIGxyZoneSelectul}.

On top of these \emph{fine-grained} intensity variations caused by the
near-translational equivalence Eq.~I$~$(\ref{undimprim}) of the WOs on the
upper and lower ribbons, there are \emph{coarse-grained} variations due to
the approximate translational equivalence Eq.~I$~$(\ref{xyz}) of the tails
on the Mo neighbors in the plane of the WO (see FIG. I~\ref{FIGyzb&xyamc})
and therefore described by the WO form factor (Sect.~\ref{Coarse}).

The understanding of the ARPES intensity variations gained in Sect.~\ref%
{SectIntensity} enabled us to obtain the new, detailed ARPES results for the
occupied part of the $t_{2g}$-bands which we present in Sect.~\ref%
{SectARPESData}, compare with the WO band theory (Sect.s \ref{SectCEC} and %
\ref{SectEk}), and in Sect.~\ref{SectAgreement} use to adjust the parameters
of the six-band TB Hamiltonian presented in Sect.~\ref{SectH} of Paper I.
Important information on the technical experimental details are given in
Sect.s~\ref{SectARPESMethod} and \ref{SectKappa}. The ARPES FS measurements
turned out to have a rather strong photon-energy dependence (Sect.~\ref%
{Secthv}) whose origin seems to be the narrowness along $\kappa _{c}$ of the
form factor for the $\widetilde{xy}$ WO, i.e. the widening along $\mathbf{c}$
of $w_{\widetilde{xy}}\left( \mathbf{r}\right) $ caused by the hybridization
with the valence band (see Sect.$~$\ref{twoband} in Paper III).

\section{Theory of ARPES intensity variations in LiPB \label{SectIntensity}}

Our ARPES data to be presented in Sect.$~$\ref%
{SectARPESData} show (fine-grained) intensity variations between equivalent
zones, similar to the BZ-selection effects observed in graphite and
explained by Shirley et al. \cite{Shirley1995}. The primitive cell of LiPB,
Li$_{2}$Mo$_{12}$O$_{34},$ is\ however much larger than that of graphite, C$%
_{2},$ and its ARPES intensity exhibits not only fine-grained zone
selection, but also coarse-grained structures in reciprocal space. On the
other hand, like in graphite, there are only four occupied bands in LiPB,
and this together with the reduction of the dimensionality from 2D to
quasi-1D enables a simple description for LiPB in terms of WOs. As we shall
explain, it is the tails of the Mo1 and MO1-centered $t_{2g}$ WOs, $%
w_{m}\left( \mathbf{r}\right) $ and $W_{m}\left( \mathbf{r}\right) $ seen in
FIG.$\,$I$~$\ref{Wannier}, which give rise to the coarse-grained structure%
\footnote{%
This effect is not included in the recent \textit{experimentalist's guide to
the matrix element in ARPES} \cite{Moser2017} because the WOs used there did
not extend over several atoms per primitive cell.}, and it is the
approximate equivalence I$~$(\ref{undim}) by half a lattice translation
which gives rise to the fine-grained structure. This zone-selection effect
is simpler than the one found for the $\pi $-band in graphite because it is
due to the existence of a hypothetical, un-dimerized form [LiMo$_{6}$O$_{17}$
in Charts~I~(\ref{undimac}) and (\ref{undimxy})] of LiPB, which cannot exist
for C$_{2}$ where the two carbons are far from being separated by half a
lattice vector.

\subsection{Preliminaries}

\label{Section:PrelimInt}

We shall first follow the treatment of Shirley et al.~\cite{Shirley1995},
but in the next subsection switch from their representation of the
initial-states in terms of atomic orbitals (AOs) to one in terms of WOs.
Hence, we start from the one-step (Fermi's golden rule) expression:%
\begin{eqnarray}
I\left( \mathbf{\kappa },\omega \right) \, &\propto &\,\theta \left( \omega
\right) \sum\nolimits_{j\mathbf{k}}^{\mathrm{zone}}\delta \left[ E_{j}\left( 
\mathbf{k}\right) +\omega \right]  \label{intensity} \\
&&\times \left\vert \left\langle e^{2\pi i\mathbf{{\kappa }\cdot r}%
}\left\vert \mathbf{p}\cdot \mathbf{E}\right\vert \psi _{j}\left( \mathbf{k,r%
}\right) \right\rangle \right\vert ^{2},  \notag
\end{eqnarray}%
for the photoemission intensity as functions of the electrons binding energy 
$\omega $ and momentum $\left( 2\pi \right) \mathbf{\kappa }$ inside the
sample. Surface effects are neglected. We have used the one-electron
approximation with initial-state Bloch functions $\psi _{j}\left( \mathbf{k,r%
}\right) $ and energy bands $E_{j}\left( \mathbf{k}\right) $ with respect to
the Fermi level. The sum in (\ref{intensity}) is over all occupied states $j%
\mathbf{k}$ with $\mathbf{k}$ in a (single) zone and $\sum\nolimits_{\mathbf{%
k}}$ denoting the average over this zone. For simplicity --and lack of
knowledge-- the final state inside the sample is taken as the plane wave $%
e^{2\pi i\mathbf{\kappa }\cdot \mathbf{r}},$ the least specific choice
possible. For our purpose, it suffices to express the matrix element as:%
\begin{equation}
\left\langle e^{2\pi i{\mathbf{\kappa }}\mathbf{\cdot r}}\left\vert \mathbf{p%
}\cdot \mathbf{E}\right\vert \psi _{j}\left( \mathbf{k,r}\right)
\right\rangle \propto \left( \mathbf{\kappa \cdot \hat{e}}\right)
\,\left\langle e^{2\pi i{\mathbf{\kappa }}\mathbf{\cdot r}}|\psi _{j}\left( 
\mathbf{k,r}\right) \right\rangle  \label{me}
\end{equation}%
as obtained by, first of all, operating with the dipole operator $\mathbf{%
p\cdot E}$\textbf{\ }to the left, that is on the plane wave, and then by
pulling the polarization-dependent factor from Eq.$\,$(\ref{polarization})
outside the integral, exploiting the fact that the photon wavelength is long
compared with inter-atomic distances. The proportionality constants in (\ref%
{intensity}) and (\ref{me}) are independent of the initial states.

Hence, photoemission at energy $\omega $ occurs for $\mathbf{\kappa }$ at
the constant-energy contours $\omega =-E_{j}\left( \mathbf{\kappa }\left[ 
\func{mod}\mathrm{zone}\right] \right) $ with an intensity which, contrary
to $E_{j}\left( \mathbf{k}\right) ,$ is aperiodic in the reciprocal lattice
and depends on the polarization of the photons and on the initial states, $%
\psi _{j}\left( \mathbf{k,r}\right) ,$ of the electrons.

The initial states may be expanded in Bloch sums I$\,$(\ref{Bloch}) of
localized orbitals $\chi _{RL}\left( \mathbf{r}\right) \mathbf{,}$ e.g. AOs,
NMTOs, or WOs$:$%
\begin{eqnarray}
&&\psi _{j}\left( \mathbf{k,r}\right) =\sum\nolimits_{RL}^{\mathrm{cell}%
}\chi _{RL}\left( \mathbf{k,r}\right) \,u_{RL,j}\left( \mathbf{k}\right) = 
\notag \\
&&\sum\nolimits_{RL}^{\mathrm{cell}}\sum\nolimits_{\mathbf{T}}\chi
_{RL}\left( \mathbf{r-T}\right) e^{2\pi i\mathbf{k\cdot T}}u_{RL,j}\left( 
\mathbf{k}\right) .  \label{LCLO}
\end{eqnarray}%
Since for a Bloch sum I$\,$(\ref{Bloch}) of orbitals, $\chi _{R}\left( 
\mathbf{r}\right) $ centered$^{\text{I}\,\ref{Rcenter}}$ at $\mathbf{r%
\mathrm{=}R}$,%
\begin{eqnarray}
&&\left\langle e^{2\pi i\mathbf{{\kappa }\cdot r}}|\chi _{R}\left( \mathbf{%
k,r}\right) \right\rangle =  \notag \\
&&e^{-2\pi i\mathbf{{\kappa }\cdot R}}\int \chi _{R}\left( \mathbf{r}\right)
e^{-2\pi i\mathbf{{\kappa }\cdot }\left( \mathbf{r-R}\right) }d^{3}r\sum_{%
\mathbf{T}}e^{-2\pi i\left( \mathbf{{\kappa }-k}\right) \mathbf{\cdot T}} 
\notag \\
&=&e^{-2\pi i\mathbf{{\kappa }\cdot R}}\tilde{\chi}_{R}\left( \mathbf{\kappa 
}\right) \sum\nolimits_{\mathbf{G}}\delta \left( \mathbf{\kappa }-\mathbf{k}-%
\mathbf{G}\right) ,  \label{FTBloch}
\end{eqnarray}%
where the sum is over all points, \textbf{G,} of the reciprocal lattice and%
\begin{equation}
\tilde{\chi}_{R}\left( \mathbf{\kappa }\right) \equiv \int \chi _{R}\left( 
\mathbf{r}\right) e^{-2\pi i\mathbf{{\kappa }\cdot }\left( \mathbf{r-R}%
\right) }d^{3}r  \label{FTO}
\end{equation}%
is the Fourier transform (FT)\footnote{%
The small tilde denoting the FT has nothing to do with the large tilde
denoting the downfolding of the $xz$ and $yz$ characters into the $%
\widetilde{xy}$ states in the gap.\label{tilde}} of $\chi _{R}\left( \mathbf{%
r}\right) .$\ Note that we use a notation according to which real-space
functions such as $\chi _{R}\left( \mathbf{r}\right) $\ are centered at $r$=$%
R$, but their Fourier transforms, $\tilde{\chi}_{R}\left( \mathbf{\kappa }%
\right) $\ defined by (\ref{FTO}), only depend on their shape and not on
where they are centered.

The second factor of the matrix element (\ref{me}) thus factorizes as:%
\begin{eqnarray*}
\left\langle e^{2\pi i\mathbf{{\kappa }\cdot r}}|\psi _{j}\left( \mathbf{k,r}%
\right) \right\rangle &=&\sum\nolimits_{RL}^{\mathrm{cell}}e^{-2\pi i\mathbf{%
{\kappa }\cdot R}}\tilde{\chi}_{RL}\left( \mathbf{\kappa }\right)
u_{RL,j}\left( \mathbf{k}\right) \\
&&\times \sum\nolimits_{\mathbf{G}}\delta \left( \mathbf{\kappa }-\mathbf{k}-%
\mathbf{G}\right) ,
\end{eqnarray*}%
whereby expression (\ref{intensity}) for the photoemission intensity
becomes: 
\begin{eqnarray}
&&I\left( \mathbf{\kappa },\omega \right) \propto \left( \mathbf{{\kappa }%
\cdot \hat{e}}\right) ^{2}\theta \left( \omega \right) \sum_{j\mathbf{k}}^{%
\mathrm{zone}}\delta \left[ E_{j}\left( \mathbf{k}\right) +\omega \right]
\times  \label{IntLO} \\
&&\left\vert \sum_{RL}^{\mathrm{cell}}e^{-2\pi i\mathbf{{\kappa }\cdot R}}%
\tilde{\chi}_{RL}\left( \mathbf{\kappa }\right) u_{RL,j}\left( \mathbf{k}%
\right) \sum_{\mathbf{G}}\delta \left( \mathbf{\kappa }-\mathbf{k}-\mathbf{G}%
\right) \right\vert ^{2},  \notag
\end{eqnarray}%
which vanishes unless the wave-vector, $\mathbf{\kappa ,}$ of the final
electronic state inside the sample (see Sect.~\ref{SectARPESMethod}) equals
the Bloch vector, $\mathbf{k,}$ of the initial electronic state, plus an
arbitrary reciprocal lattice vector, $\mathbf{G}$. In (\ref{IntLO}), $\sum_{%
\mathbf{G}}$ is a periodic function of $\mathbf{\kappa -k}$ in the $\mathbf{G%
}$-lattice, but the other factors are not.

An AO factorizes as: $\chi _{RL}\left( \mathbf{r}\right) \equiv Y_{L}\left( 
\widehat{\mathbf{r}_{R}}\right) \varphi _{Rl}\left( r_{R}\right) ,$ whereby
its FT (\ref{FTO}), which depends on its shape, $L,$ including its
orientation, but \emph{not} on its center, $\mathbf{R,}$ can be taken
outside the sum over translationally equivalent AOs:%
\begin{eqnarray}
&&\sum_{R\in \mathrm{eq}}^{\mathrm{cell}}e^{-2\pi i\mathbf{{\kappa }\cdot R}}%
\tilde{\chi}_{RL}\left( \mathbf{\kappa }\right) u_{RL,j}\left( \mathbf{k}%
\right) =  \label{PSF} \\
&&\tilde{Y}_{L}\left( \mathbf{\hat{\kappa}}\right) \int j_{l}\left( 2\pi
\kappa r\right) \varphi _{Rl}\left( r\right) r^{2}dr\sum_{R\in \mathrm{eq}}^{%
\mathrm{cell}}e^{-2\pi i\mathbf{k}\cdot \mathbf{R}}u_{RL,j}\left( \mathbf{k}%
\right) ,  \notag
\end{eqnarray}%
thus leaving the sums $\tsum_{R\notin \mathrm{eq}\,L}^{\mathrm{cell}}$ over
the translationally \emph{in}equivalent AOs to be performed later. Hence,
the sum (\ref{PSF}) over translationally equivalent AOs factorizes into a $%
\mathbf{\kappa }$-dependent AO form factor, times a so-called (Ref.$\,$%
\onlinecite{Matsui2014,Daimon1995,Nishimoto1996}) photoemission structure
factor (PSF). The latter is similar to the geometrical structure factor in
x-ray diffraction, but depends on the initial-state wave function [via its
LCAO coefficients, $u_{RL,j}\left( \mathbf{k}\right) $]. This factorization
holds for the $\pi $-band in C$_{2}$, because this band is singly degenerate
and contains only one type of orbital, $Y_{10}\mathrm{=}p_{z},$ so that the
entire sum, $\sum_{RL}^{\mathrm{cell}}e^{-2\pi i\mathbf{{\kappa }\cdot R}}%
\tilde{\chi}_{RL}\left( \mathbf{\kappa }\right) u_{RL,j}\left( \mathbf{k}%
\right) ,$ reduces to the factorized form (\ref{PSF}). The PSF for the
graphene $\pi $-band thus depends merely on the $\mathbf{k}$-dependent phase
between the $p_{z}$ orbitals on the two atoms \cite{Shirley1995}. However,
to the three $\sigma $-bands in graphene \emph{three} translationally
inequivalent AOs on each of the two C atoms ($s,$ $p_{x},$ and $p_{y},$ or
the three equivalent $sp^{2}$ orbitals directed towards the three nearest
neighbors) contribute, so that for the $\sigma $-bands, $\sum_{RL}^{\mathrm{%
cell}}e^{-2\pi i\mathbf{{\kappa }\cdot R}}\tilde{\chi}_{RL}\left( \mathbf{%
\kappa }\right) u_{RL,j}\left( \mathbf{k}\right) $ does \emph{not} factorize
into an orbital form factor and a geometrical structure factor.

Whereas 8 AOs are needed to describe the occupied bands in graphene, LiPB
needs more than 300 atomically localized AOs (see Sect.$~$I$~$\ref%
{SectElCalc}), but -by virtue of its quasi-1D structure- merely the six $%
t_{2g}$ WOs shown in FIG.$~$I$~$\ref{Wannier}. Using those in expression (%
\ref{IntLO}) for the ARPES intensity and neglecting inter-$m$ mixing, leads
to great simplification, as we shall now see:

\subsection{Using the six $t_{2g}$ WOs\label{Usingt2gWOs}}

For LiPB, we shall predict, and in Sect.$\,$\ref{SectEk} confirm, that the
near-translational equivalence of the $t_{2g}$ WOs, $w_{m}\left( \mathbf{r}%
\right) $ and $W_{m}\left( \mathbf{r}\right) ,$ by half a lattice
translation causes the ARPES to have double-period fine-grained intensity
variations which approximately follow the $\left\vert \mathbf{k}%
\right\rangle $ character I~(\ref{fat}) shown in FIG.s$~$I$\,$\ref%
{ThreePureBands}, \ref{ThreeBands}, and \ref{FIGUxy^2} of the occupied
bands. This implies that the $m$-band with the lower energy appears in the
1st- and is extinguished near the centre of the 2nd physical zone, and
conversely for the $m$-band with the higher energy. On top of this, comes
that the upper $xz$- and $yz$-bands are unoccupied and therefore cause no
intensity in their respective 2nd physical zones. The metallic $xy$-band
with the lower energy should best be seen in the 1st- and the $xy$-band with
the higher energy in the 2nd BZ. This zone-selection effect will allow us to
resolve the perpendicular dispersion and splitting of the quasi-1D $%
\widetilde{xy}$-bands predicted in the bottom panel of FIG.$~$\ref%
{ThreeBandskonly}. We shall see that the deviations from the translational
equivalence of $w_{m}\left( \mathbf{r}\right) $ and $W_{m}\left( \mathbf{r}%
\right) $, i.e. the inversion- and displacement dimerizations I$~$(\ref{inv}%
), will distort the ARPES intensity variation from that of the initial-state 
$\left\vert \mathbf{k}\right\rangle $ character, $\frac{1}{2}\left[ 1\mp
\cos \phi \left( \mathbf{k}\right) \right] ,$ to $\frac{1}{2}\left[ 1\mp
\cos \left\{ \phi \left( \mathbf{k}\right) -\eta \left( \mathbf{\kappa }%
\right) \right\} \right] $ where $\eta \left( \mathbf{\kappa }\right) $ is
the dimerization phase shift of the emitted electron with momentum $\mathbf{%
\kappa }$.\ By suitable choice of $\kappa _{a},$ via control of the photon
energy [see e.g. Eq.~(\ref{kappaa(0,hv)})], this distortion can be
negligible for the $xy$-bands in a large part of the double zone$,$\ and may
also be so for the occupied $xz$\ and $yz$\ bands.

In addition to the \emph{fine-grained zone-selection} effect (Sect.~\ref%
{Sectzoneselect}), well-known from the geometrical \emph{structure factor}
in x-ray crystallography, we shall predict (Sect.~\ref{Coarse}) and find
(Sect.~\ref{SectARPESData}) that the \emph{internal} structure of the $%
w_{m}\left( \mathbf{r}\right) $ $t_{2g}$ WO, spreading out to about the four
nearest molybdenums in its plane (see FIG. I \ref{FIGyzb&xyamc}), makes the 
\emph{form factor of the WO} approximately factorize into a \emph{%
coarse-grained structure factor,} times the \emph{form factor} of the \emph{%
local} partial-wave projection, $Y_{2m}\left( \mathbf{\hat{r}}\right)
\varphi _{2}\left( r\right) ,$ of the WO tail.

\subsubsection{Zone selection; the fine-grained structure\label%
{Sectzoneselect}}

The DFT calculation resulting in the low-energy six-band $t_{2g}$ TB
Hamiltonian and in the band structures in FIG.s~\ref{CEC}~(b) and \ref%
{ARPES_Bandstructure_LDA} shows that --except near band crossings such as
those below $-0.5$ eV and the ones near the top of the valence bands near
Z-- each band is dominated by \emph{one} $m$ character. For describing the
ARPES matrix elements, but not the bands, we shall neglect by-mixing of WOs
with other $m$-values, i.e. use the pure-$m$ approximation (Sect.~I$~$\ref%
{SectPureBands}). Therefore, in the general expression (\ref{LCLO}) for the
the initial-state wave function, $L$ takes \emph{one} value $\left( m\right)
,$ $R$ takes two values (Mo1 and MO1), and expression (\ref{LCLO}) becomes
Eq.$~$I$~$(\ref{wW}) which is repeated in the Introduction to the present
paper.

\begin{figure}[!tbh]
\includegraphics[width=0.48\textwidth]{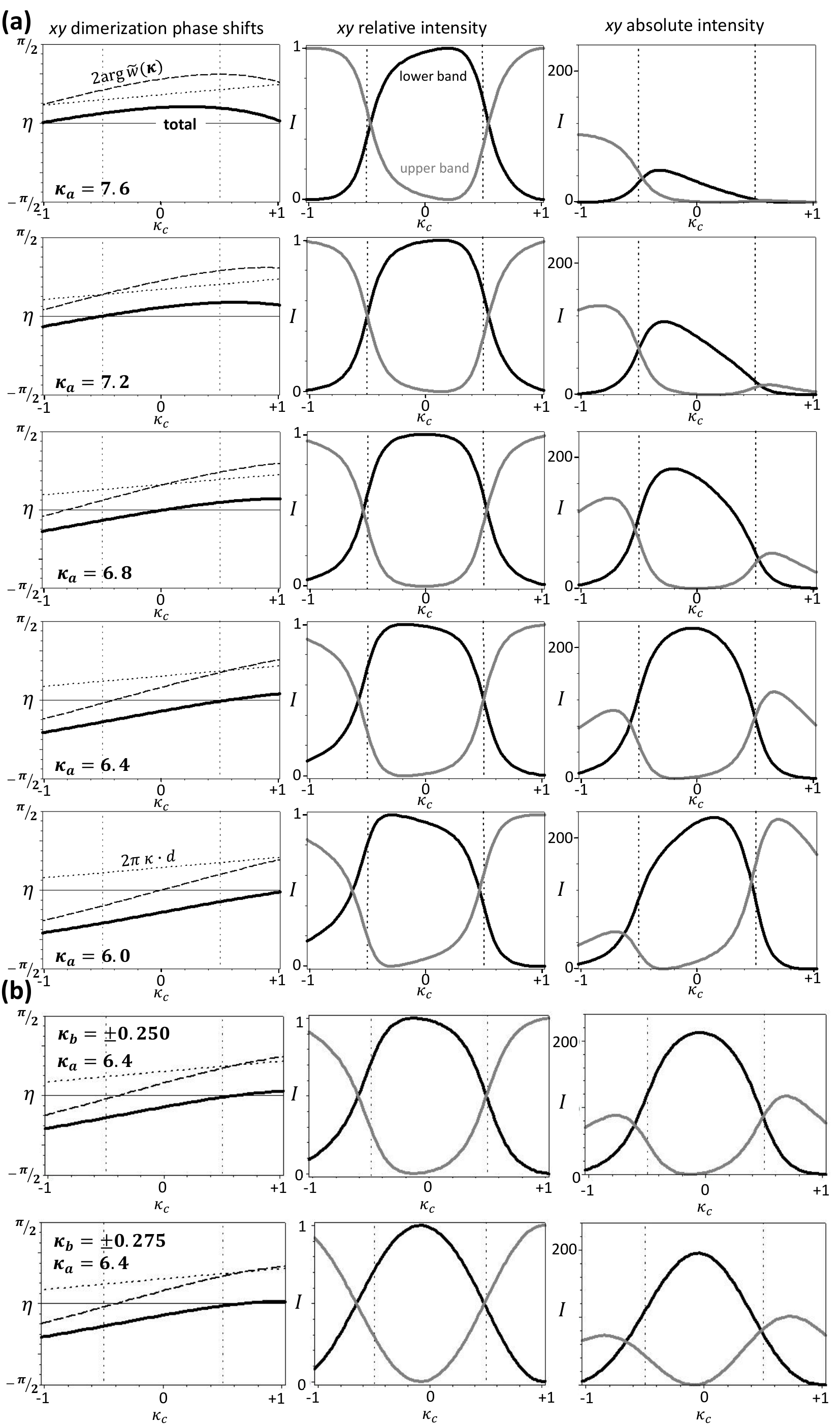}
\caption{\textbf{(a)} Photoemission from the $xy$-bands (red in FIG. \ref{ThreeBandskonly}) for $\left\vert \protect%
\kappa _{b}\right\vert $=$\left\vert k_{b}\right\vert =0.225$ and as
functions of $\protect\kappa _{c}$=$k_{c}$ in the double zone, i.e. along a
brown dot-dashed line in FIG. \ref{FIGPhysicalZones2}. See also FIG.$~$I$\,$\protect\ref{FIGUxy^2}.  
\emph{Column 1:} Inversion- ($2\arg \tilde{w}\left( \mathbf{\protect\kappa }%
\right) ,$ dash), displacement- ($2\protect\pi \mathbf{\protect\kappa \cdot
d,}$ dots), and total ($\protect\eta \left( \mathbf{\protect\kappa }\right) ,
$ full) dimerization phase shift for $\protect\kappa _{a}=6.0-7.6$. \emph{Column 2: }Relative intensities of
photoemission, $\left[ 1\mp \cos \left\{ \protect\phi \left( \mathbf{k}%
\right) -\protect\eta \left( \mathbf{\protect\kappa }\right) \right\} \right]
/2,$ from the upper (light) and lower (dark) bands. These intensities exhibit the fine-grained structure. \emph{Column 3:}
Absolute photoemission intensities, i.e. the relative intensities in column
2, times the polarization- and WO form factors, $\left( \mathbf{{\protect%
\kappa }\cdot \hat{e}}\right) ^{2}\left\vert \tilde{w}\left( \mathbf{\protect%
\kappa }\right) \right\vert ^{2},$ evaluated in Sect.~\protect\ref{Coarse}
and shown in the top panel of FIG.~\protect\ref{ThreeBandskonly}. These
polarization- and WO form factors provide the coarse-grained structure. 
\textbf{(b) }same as \textbf{(a)}, but along $\left\vert \protect\kappa %
_{b}\right\vert $=$\left\vert k_{b}\right\vert =0.250,$ a red dot-dashed
line in FIG. 1 (top), and along $\left\vert \protect\kappa _{b}\right\vert $=%
$\left\vert k_{b}\right\vert =0.275,$ an olive dot-dashed line (bottom), and
only for $\protect\kappa _{a}=6.4.$ The boundaries of the 1st zone are
indicated by vertical, dotted lines.}
\label{FIGxyZoneSelectul}
\end{figure}

Hence, the second factor of the matrix element (\ref{me}) for photoemission
from the upper ($j$=2) or lower ($j$=1) $m$-band state is the FT of the
respective eigenfunction I$~$(\ref{wW}):

\begin{widetext}

\begin{eqnarray}
\left\langle e^{2\pi i\mathbf{{\kappa }\cdot r}}|w_{\QATOP{2}{1}}\left( 
\mathbf{k,r}\right) \right\rangle &=&\frac{1}{\sqrt{2}}\sum_{\mathbf{G}%
}\delta \left( \mathbf{\kappa }-\mathbf{k}-\mathbf{G}\right) \left[ \tilde{w}%
\left( \mathbf{\kappa }\right) e^{-i\phi \left( \mathbf{k}\right) }\mp 
\tilde{W}\left( \mathbf{\kappa }\right) e^{2\pi i\mathbf{\kappa \cdot d}}%
\right]  \notag \\
&=&\frac{1}{\sqrt{2}}\sum_{\mathbf{G}}\delta \left( \mathbf{\kappa -k-G}%
\right) \tilde{w}\left( \mathbf{\kappa }\right) \left[ e^{-i\phi \left( 
\mathbf{k}\right) }\mp e^{-i\eta \left( \mathbf{\kappa }\right) }\right] .
\label{FTw}
\end{eqnarray}
\end{widetext}
For the FTs of $w(\mathbf{k,r})$ and $W(\mathbf{k,r})$ we have
used Eq.~(\ref{FTBloch}$)$ with the sum being over all points $\mathbf{G}$
of the reciprocal lattice\footnote{The reciprocal-lattice points, $\mathbf{G,}$ and the Bloch sum of hopping
integrals, $G,$ should not be confused.\label{G}}. Inside this sum, the
product of the phase factors, $e^{\pi i\left( k_{c}+k_{b}\right) }$ and $%
e^{-2\pi i\mathbf{\kappa \cdot }\left( \frac{\mathbf{c+b}}{2}-\mathbf{d}%
\right) },$ from respectively Eq.s$~$I$~($\ref{wW}$)$ and (\ref{FTBloch}$),$
is simply $e^{2\pi i\mathbf{\kappa \cdot d}}$ with $\mathbf{d}$ being the
displacement dimerization I$~$(\ref{dim}) and $\mathbf{\kappa }$ the
momentum of the emitted electron. On the second line of Eq.~(\ref{FTw}) we
have used that, due to the inversion dimerization I$~$(\ref{inv}), $\tilde{W}%
\left( \mathbf{\kappa }\right) =$ $\tilde{w}\left( -\mathbf{\kappa }\right) $
and that $\tilde{w}\left( -\mathbf{\kappa }\right) =$ $\tilde{w}\left( 
\mathbf{\kappa }\right) ^{\ast }=$ $\left\vert \tilde{w}\left( \mathbf{%
\kappa }\right) \right\vert e^{-i\arg \tilde{w}\left( \mathbf{\kappa }%
\right) }$ because $w\left( \mathbf{r}\right) $ is a real-valued $t_{2g}$
function. As a consequence, $e^{2\pi i\mathbf{\kappa \cdot d}}\tilde{W}%
\left( \mathbf{\kappa }\right) /\tilde{w}\left( \mathbf{\kappa }\right)
=e^{-i\eta \left( \mathbf{\kappa }\right) },$ where%
\begin{equation}
\eta _{m}\left( \mathbf{\kappa }\right) \equiv 2\arg \tilde{w}_{m}\left( 
\mathbf{\kappa }\right) -2\pi \mathbf{\kappa \cdot d},  \label{eta}
\end{equation}%
is the \emph{dimerization phase shift }whose first term is due to the
inversion- and the second to the displacement dimerization. The latter, $%
2\pi \left( 0.012\kappa _{a}+0.033\kappa _{c}\right) $ Eq.~I~(\ref{dim})$,$
is a weak function of $\kappa _{c},$ a very weak function of $\kappa _{a},$
and independent of $\kappa _{b}$ and $m$. It is the dotted, linear curves
shown in the first columns of FIG.s \ref{FIGxyZoneSelectul} and \ref%
{FIGyzZoneSelect}. The first term in expression (\ref{eta}), the phase shift
due to inversion dimerization, is the dashed curves in the first columns
which we have evaluated as will be explained in the next subsection \ref%
{Coarse} using the rough digitalizations (\ref{Sxy}) and (\ref{Syz}) of
FIG.s I~\ref{FIGyzb&xyamc} and I~\ref{Wannier}.

\begin{figure}[!tbh]
\includegraphics[width=0.48\textwidth]{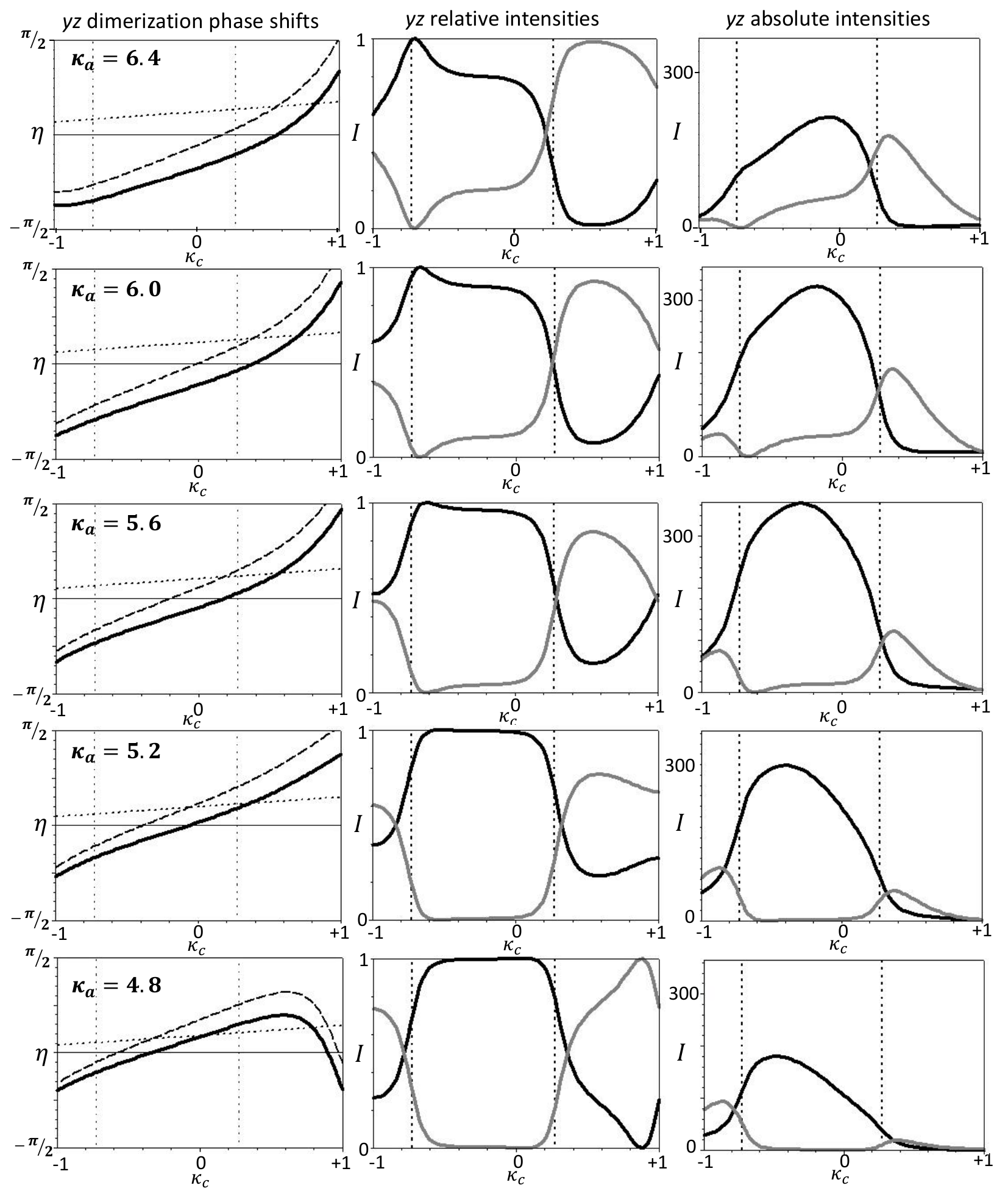}
\caption{Same as in FIG.~\protect\ref{FIGxyZoneSelectul}, but for the $yz$
bands (green in FIG. \ref{ThreeBandskonly}) and for $\protect\kappa _{a}=4.8-6.4.$ The green
zone boundaries in FIG. \ref{FIGPhysicalZones2} intersect the brown dot-dashed near-FS line, $%
\protect\kappa _{b}$=$0.225,$ at $\protect\kappa _{c}$=$0.275$ and at $%
0.275-1=\allowbreak -0.725,$ which is where the swiching curves would have
crossed, had there been no dimerization distortion, i.e. if $\protect\eta %
_{yz}\left( \mathbf{\protect\kappa }\right) $=0. Since the upper\emph{\ }$yz$
band is \emph{empty,} the light switching curves should have been deleted.}
\label{FIGyzZoneSelect}
\end{figure}

The intensity (\ref{IntLO}) of photoemission from two dimerized $m$-bands $%
\left( j\text{=}1,2\right) $ in LiPB is thus: 
\begin{widetext}

\begin{eqnarray}
I\left( \mathbf{\kappa },\omega \right) &\propto &\left( \mathbf{{\kappa }%
\cdot \hat{e}}\right) ^{2}\theta \left( \omega \right)
\dsum\limits_{j=1}^{2}\dsum\limits_{\mathbf{k}}^{\mathrm{zone}}\delta \left[
E_{j}\left( \mathbf{k}\right) -E_{F}+\omega \right] \left\vert \sum_{\mathbf{%
G}}\delta \left( \mathbf{\kappa }-\mathbf{k}-\mathbf{G}\right) \tilde{w}%
\left( \mathbf{\kappa }\right) \left[ e^{-i\phi \left( \mathbf{k}\right)
}\mp e^{-i\eta \left( \mathbf{\kappa }\right) }\right] \right\vert
^{2}\propto  \notag \\
&\sim &\left( \mathbf{{\kappa }\cdot \hat{e}}\right) ^{2}\left\vert \tilde{w}%
\left( \mathbf{\kappa }\right) \right\vert ^{2}\theta \left( \omega \right)
\dsum\limits_{j=1}^{2}\dsum\limits_{\mathbf{k}}^{\mathrm{zone}}\delta \left[
E_{j}\left( \mathbf{k}\right) -E_{F}+\omega \right] \,\delta \left[ \mathbf{%
\kappa }-\left( 6,0,0\right) -\mathbf{k}\right] \,\frac{1\mp \cos \left[
\phi \left( \mathbf{k}\right) -\eta \left( \mathbf{\kappa }\right) \right] }{%
2},  \label{Indep}
\end{eqnarray}%

\end{widetext}where the second line holds when --hinging on the
coarse-grained structure of $\tilde{w}\left( \mathbf{\kappa }\right) $ and
the experimental set up$-$ the $\mathbf{G}$-sum is \emph{dominated by one
term}, the one for $\mathbf{G}=\left( 6,0,0\right) $ in our case, as we
shall see in the following section, FIG.s \ref{SYk} and \ref{Re^2ka6068} in
particular. We can then use that $\frac{1}{4}\left\vert e^{-i\phi }\mp
e^{-i\eta }\right\vert ^{2}=$ $\left[ 1\mp \cos \left( \phi -\eta \right) %
\right] /2,$ which is like in Eq.~I$~($\ref{fat}) and illustrated in FIG.~I~%
\ref{FIGUxy^2}, but shifted by $\eta ,$ and obtain the simple result that
the intensity of photoemission from two dimerized $m$-bands $\left( j\text{=}%
1,2\right) $ in LiPB follows the $\left\vert \mathbf{k}\right\rangle $
character (fatness), but with the band-structure phase, $\phi \left(
k_{b},k_{c}\right) $ [Eq. I$\,$(60) in the Introduction] \emph{shifted }by
the dimerization phase shift, $\eta \left( k_{a}+6,k_{b},k_{c}\right) .$
Remembering that $\phi $ is independent of $k_{a}$ due to the long paths for
hopping between slabs$^{\text{I}\,\text{\ref{slabhop}}}$ and varies from --$%
\frac{\pi }{2}$ to $+\frac{\pi }{2}$ across the 1st zone, we realize that
dimerization phase shift shown by the full curves in the first columns of
FIG.s \ref{FIGxyZoneSelectul} and \ref{FIGyzZoneSelect} can merely \emph{%
distort} the switching curves, and less so for the $xy$-bands than for the
equivalent $yz$- and $xz$-bands. This is seen in the middle panels of FIG.s~%
\ref{FIGxyZoneSelectul} and \ref{FIGyzZoneSelect}. For the $xy$-bands along
the $\kappa _{b}$=$0.225$ line slightly inside the FS the distortions
increase with the deviation of $\kappa _{a}$ from the value $6.8.$ Similarly
for switching curves for the $yz$-bands, also along the $\kappa _{b}$=$0.225$
line, in FIG.~\ref{FIGyzZoneSelect} the distortions increase with the
deviation of $\kappa _{a}$ from the value $5.3.$

As seen in the 1st columns of FIG.s~\ref{FIGxyZoneSelectul} and \ref%
{FIGyzZoneSelect}, the inversion (dash) and displacement (dots) phase shifts 
\emph{tend to cancel out}: For the $xy$-bands along $\kappa _{b}$=$0.225$, a
zero of $\eta \left( \mathbf{\kappa }\right) $ (full) increases through the
1st $xy$-zone ($-\frac{1}{2}|\frac{1}{2})$ for $\kappa _{a}$ decreasing from
7.4 to 6.4, and for the $yz$-bands along the same line, the zero \emph{de}%
creases through the 1st $yz$-zone for $\kappa _{a}$ decreasing from 5.5 to
4.5. In these intervals of $\kappa _{a},$ the $c$-axis dimerization hardly
distorts the zone selection as seen in the 2nd column. That the distortions
due to displacement and inversion dimerizations tend to cancel over a $%
\kappa _{a}$ region of order 1 for both the $xy$- and $yz$-bands is
surprising considering the fact that the $dd\delta $ coupling between the $%
xy\left( \mathbf{r}\right) $ and $XY\left( \mathbf{r-}\frac{\mathbf{c+b}}{2}+%
\mathbf{d}\right) $ WOs, is much smaller $dd\pi $ coupling between $yz\left( 
\mathbf{r}\right) $ and $YZ\left( \mathbf{r-}\frac{\mathbf{c+b}}{2}+\mathbf{d%
}\right) ,$ but for the latter the distortions are of course larger.

As mentioned already in connection with FIG.~I~\ref{FIGUxy^2}, the $xy$
switching curves are even functions of $\kappa _{b}$ whose sharpness
decreases with increasing $\left\vert \kappa _{b}\right\vert $. That this
holds, also when dimerization distortion is included, may be seen from FIG.~%
\ref{FIGxyZoneSelectul} for $\kappa _{a}$=6.4 by comparison of the switching
curves for $\kappa _{b}$=0.225, 0.250 and 0.275.

\begin{figure}[tbh]
\includegraphics[width=0.9\linewidth]{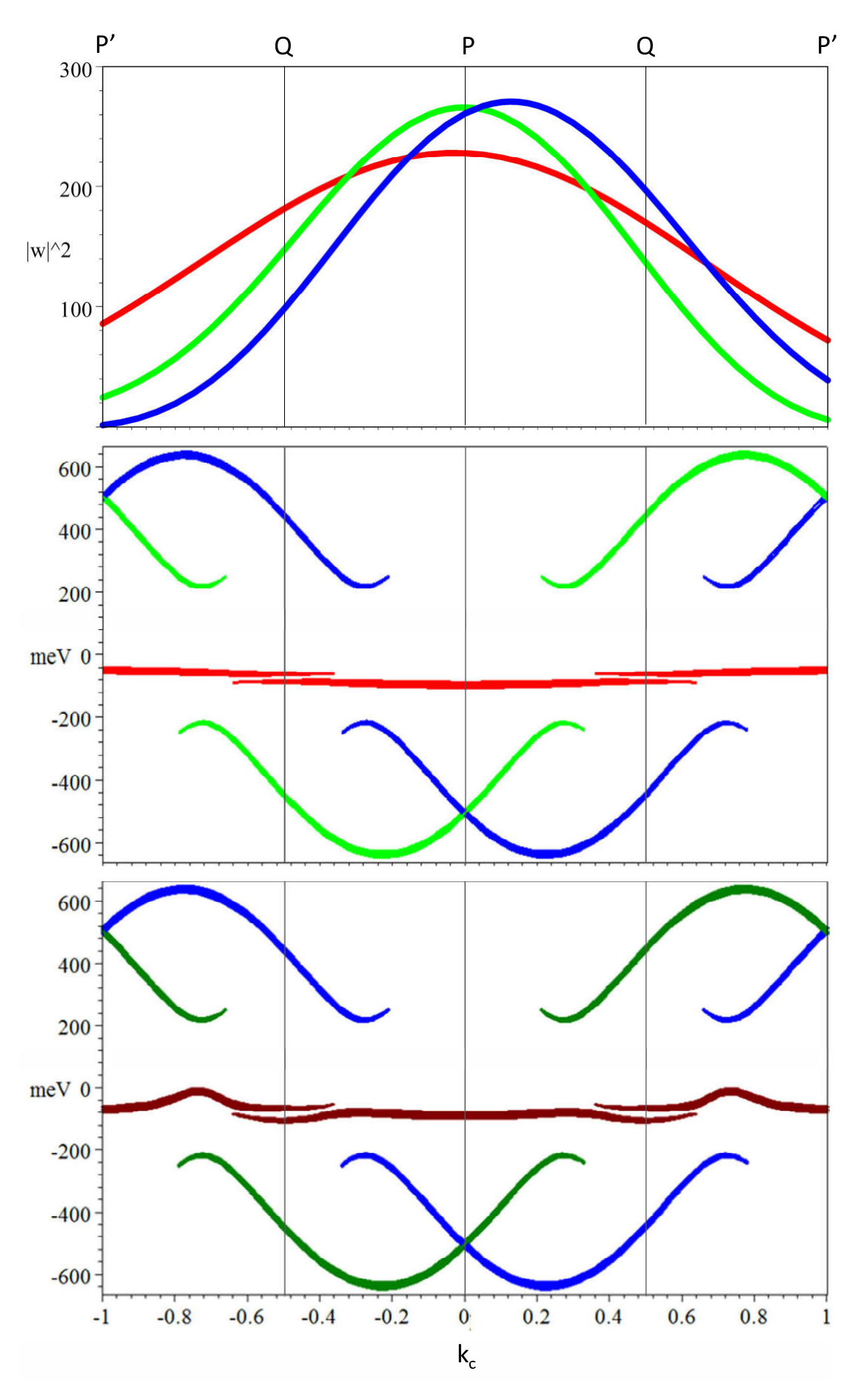}
\caption{\emph{Top:} Polarization- and form factor intensities, $\left( 
\mathbf{{\protect\kappa }\cdot \hat{e}}\right) ^{2}\left\vert \tilde{w}%
_{m}\left( \mathbf{\protect\kappa }\right) \right\vert ^{2}$, for $m$=$xy$
red, $xz$ blue, and $yz$ green along the P'QPQP' line, $\protect\kappa _{b}$%
=0.225, in the double zone $-1<\protect\kappa _{c}\leq 1.$ These aperiodic,
coarse-grained intensities depend on $\protect\kappa _{a}$ --chosen here to
be 6.4-- as explained in Sect.~\protect\ref{SectKappa}. \emph{Middle and
bottom:} Band-factor intensities according to Eq.~(\protect\ref{Indep})
without dimerization, i.e. with $\protect\eta _{m}\left( \mathbf{\protect%
\kappa }\right) =0.$ They are the $\left\vert \mathbf{k}\right\rangle $%
-projected bands [see Eq.\ I~(\protect\ref{fat})] and give rise to the
fine-grained intensity distributions with period 2 in $\protect\kappa _{c}.$
Their dependence on $\protect\kappa _{a}$ is negligible. Like in FIG.s~I\ 
\protect\ref{ThreePureBands} and I~\protect\ref{ThreeBands} the red $xy$
bands in the \textit{middle} are pure, while the dark-red $\widetilde{xy}$
bands at the \textit{bottom} are perturbed by the $xz$ and $yz$ valence and
conduction bands. The $\left( \mathbf{k,k+c}^{\ast }\right) $-hybridization
of the $\widetilde{xy}$-bands is seen to effectively extend over a $k_{c}$%
-region of width 0.3 around the BZ boundaries, $k_{c}=\frac{1}{2}+\func{%
integer}.$ The zero of energy is the center of the gap in the $xz$ and $yz$
bands, whereby $E_{F}$=75~meV (see Sect.~I~\protect\ref{SectH}). The
occupation factor, $\protect\theta \left( \protect\omega \right) $ in Eq. (%
\protect\ref{Indep}), will remove the intensity from the bands with $E>E_{F}$%
. }
\label{ThreeBandskonly}
\end{figure}

In Sects. \ref{SectKappa} and \ref{Secthv} we shall explain how $\kappa _{a}$
is deduced and how it is controlled by the photon energy, $h\nu $. From all
FIG.s \ref{FIGxyZoneSelectul}-\ref{Re^2xykakbkc002550}, we thus conclude that with
proper $m$-dependent choice of the photon energy, the selection rule works
well for each of the three $\left\{ \mathbf{k,k+c}^{\ast }\right\} $-pairs
of quasi-1D $t_{2g}$-bands in LiPB.

In Paper III, we shall study the dispersion of the two quasi-1D bands in the
gap and --as demonstrated by comparison of the red and dark-red bands in
respectively the middle and bottom parts of FIG.$\,$\ref{ThreeBandskonly}--
this requires that we take the weak hybridization of these $xy$-like bands
with the $xz$ and $yz$ valence and conduction bands into account. This we
shall do by downfolding those characters into the tails of the $xy$ WOs,
which thereby attain longer range and become what we call $\widetilde{xy}$
WOs. The concomitant modification of the WO form factor --and, hence,
dimerization phase shift-- we shall neglect. What the weak hybridization
primarily changes is the \emph{dispersion} of the $\widetilde{xy}$
energy-bands.\textbf{\ }As mentioned before, a main goal of the present and
the following Paper III is to detect with ARPES the predicted dispersion in
the perpendicular $\left( k_{c}\right) $ direction, most noticeably the
resonance peaks in the upper $\widetilde{xy}$-band caused by the weak
hybridization with the $yz$- and $xz$-bands\ and expected to be seen near
respectively $\kappa _{c}$=$-0.75$ and 0.75. To see \emph{both} the former
resonance peak \emph{and} its cause, the lower $yz$-band, in the \emph{same}
ARPES experiment like in FIG.$\,$\ref{ThreeBandskonly} is, however,
impossible because the former requires using $\kappa _{a}\sim 7.2$ and the
latter $\kappa _{a}\sim 4.5.$

The Brillouine-zone-dependent photoemission intensity derived and discussed here based on a structure and form factor should be generally applicable to different materials. The main approximation made is in Eq.(\ref{intensity}). We approximate the wave-function of the photo-emitted electron by a plane wave. The free electron state, should in principle be a Bloch wave with an additional modification due to the sample surface. The approximation can become noticeable if the {\bf k} vector of the free electron is close to a zone boundary, where small periodic potentials are expected to gap degenerate free electron states. A full understanding of the importance of describing the state of the photo-electron beyond a plane wave and including surface effects can be investigated with multiple scattering techniques \cite{Strocov2023} but a simple intuitive picture is currently missing and should be further investigated. At the same time, calculations for graphene \cite{Shirley1995} or Bi$_2$Se$_3$ \cite{Zhu2014,Zhu2013} assuming the photo-electron to be a plane wave seem to capture a large part of the Brillouin zone dependent intensity and agree well with experiment. 

\subsubsection{WO form factor, $\tilde{w}_{m}\left( \mathbf{\protect\kappa }%
\right) ,$ inversion dimerization, and the coarse-grained structure \label%
{Coarse}}

A $t_{2g}$ WO, $w_{m}\left( \mathbf{r}\right) ,$ has a halo with
contributions (tails) on the near Mo neighbors in the plane of this flat%
\emph{\ }WO with \emph{the same} $m$ character as that of its head (see
FIG.s~\ref{FIGyzb&xyamc} and$~$\ref{Wannier} in Paper I). For the present
purpose, we shall neglect the hybridization between WOs with different
values of $m,$ as well as the details of the oxygens which $pd\pi $
anti-bond with the WO head and bond with the tail. What is important, is
that the head has a partial-wave shape, $Y_{m}\left( \mathbf{\hat{r}}\right)
\varphi _{2}\left( r\right) ,$ which is \emph{translated} to the Mo$\,n$
neighbors and multiplied by a factor, $c_{n}$\TEXTsymbol{<}1. This makes the
FT, $\tilde{w}_{m}\left( \mathbf{\kappa }\right) ,$ of the WO \emph{%
factorize approximately} into an orbital-dependent structure factor, $%
\mathcal{S}_{m}\left( \mathbf{\kappa }\right) ,$ times the FT of $%
Y_{m}\left( \mathbf{\hat{r}}\right) \varphi _{2}\left( r\right) :$%
\begin{equation}
\tilde{w}_{m}\left( \mathbf{\kappa }\right) \propto \mathcal{S}_{m}\left( 
\mathbf{\kappa }\right) \tilde{Y}_{m}\left( \mathbf{{\hat{\kappa}}}\right)
\int_{0}^{0.55}j_{2}\left( 2\pi \kappa r\right) \varphi _{\mathrm{Mo}%
\,d}\left( r\right) r^{2}dr.  \label{FTWO}
\end{equation}%
We have not computed the form factor by accurate, numerical FT of the WO,%
\footnote{%
It may be kept in mind [see text after Eq.~I~(\ref{WOxy})] that our $t_{2g}$%
WOs are given by the exact crystal structure and are insensitive to the
orientation of the $xyz$ system I~(\ref{xyz}), which is only approximately
cartesian.} but shall use the factorization (\ref{FTWO}) together with the
real-space figures to provide a qualitative \emph{understanding} of the
ARPES data to be presented in the following section.

Whereas the radial and angular factors, $\int_{0}^{0.55}j_{2}\left( 2\pi
\kappa r\right) \varphi _{\mathrm{Mo}\,d}\left( r\right) r^{2}dr$ and $%
\tilde{Y}_{m}\left( \mathbf{{\hat{\kappa}}}\right) ,$ are even, real
functions of $\mathbf{\kappa ,}$ the structure factor, $\mathcal{S}%
_{m}\left( \mathbf{\kappa }\right) ,$ is a complex function whose real and
imaginary parts, like those of $\tilde{w}_{m}\left( \mathbf{\kappa }\right)
, $ are respectively even and odd. From the WO figures I$\,$\ref%
{FIGyzb&xyamc} and \ref{Wannier}, we estimate these structure factors (the $%
c_{n}$ coefficients$)$ to be:%
\begin{eqnarray}
\mathcal{S}_{xy}\left( \mathbf{\kappa }\right) &\sim &\left[ 
\begin{array}{ccc}
& +\frac{1}{2}e^{2\pi i\kappa _{y}} & +\frac{1}{6}e^{2\pi i\left( \kappa
_{x}+\kappa _{y}\right) } \\ 
+\frac{1}{4}e^{-2\pi i\kappa _{x}} & 1 & +\frac{1}{2}e^{2\pi i\kappa _{x}}
\\ 
& +\frac{1}{4}e^{-2\pi i\kappa _{y}} & 
\end{array}%
\right]  \notag \\
&&\left/ \left( 1+\frac{1}{2}+\frac{1}{2}+\frac{1}{4}+\frac{1}{4}+\frac{1}{6}%
\right) \right. ,  \label{Sxy}
\end{eqnarray}%
\begin{eqnarray}
\mathcal{S}_{yz}\left( \mathbf{\kappa }\right) &\sim &\left[ 
\begin{array}{ccc}
+\frac{1}{8}e^{2\pi i\left( \kappa _{z}-\kappa _{y}\right) } & +1e^{2\pi
i\kappa _{z}} & 0 \\ 
+\frac{1}{3}e^{-2\pi i\kappa _{y}} & 1 & +\frac{1}{2}e^{2\pi i\kappa _{y}}
\\ 
0 & +\frac{1}{4}e^{-2\pi i\kappa _{z}} & 0%
\end{array}%
\right]  \notag \\
&&\left/ \left( 1+1+\frac{1}{2}+\frac{1}{3}+\frac{1}{4}+\frac{1}{8}\right)
\right. ,  \label{Syz}
\end{eqnarray}%
and $\mathcal{S}_{xz}\left[ \kappa _{x},\kappa _{z}\right] $=$\mathcal{S}%
_{yz}\left[ \kappa _{y},\kappa _{z}\right] .$

In FIG.~I$~$\ref{FIGyzb&xyamc} we clearly see that the inversion symmetry of
the $xy$\ WO around its center, Mo1, is far better preserved than that of
the $yz$\ WO. The inversion dimerization I$\,$(\ref{inv}) is $w_{m}\left( -%
\mathbf{r}\right) -w_{m}\left( \mathbf{r}\right) ,$\ and the phase shift (%
\ref{eta}) due to inversion dimerization is $2\arg \tilde{w}\left( \mathbf{%
\kappa }\right) =2\arg S\left( \mathbf{\kappa }\right) .$\ The latter
function is easily found from Eq.~(\ref{Sxy}) for $xy$\ and from (\ref{Syz})
for $yz,$\ and was used to produce FIG.s~\ref{FIGxyZoneSelectul} and \ref%
{FIGyzZoneSelect}. Since $\mathcal{S}_{xz}\left[ \kappa _{x},\kappa _{z}%
\right] $=$\mathcal{S}_{yz}\left[ \kappa _{y},\kappa _{z}\right] ,$ we have: 
$\eta _{xz}\left[ \kappa _{x},\kappa _{z}\right] $=$\eta _{yz}\left[ \kappa
_{y},\kappa _{z}\right] .$

\begin{figure*}[tbh]
\includegraphics[width=0.9\linewidth]{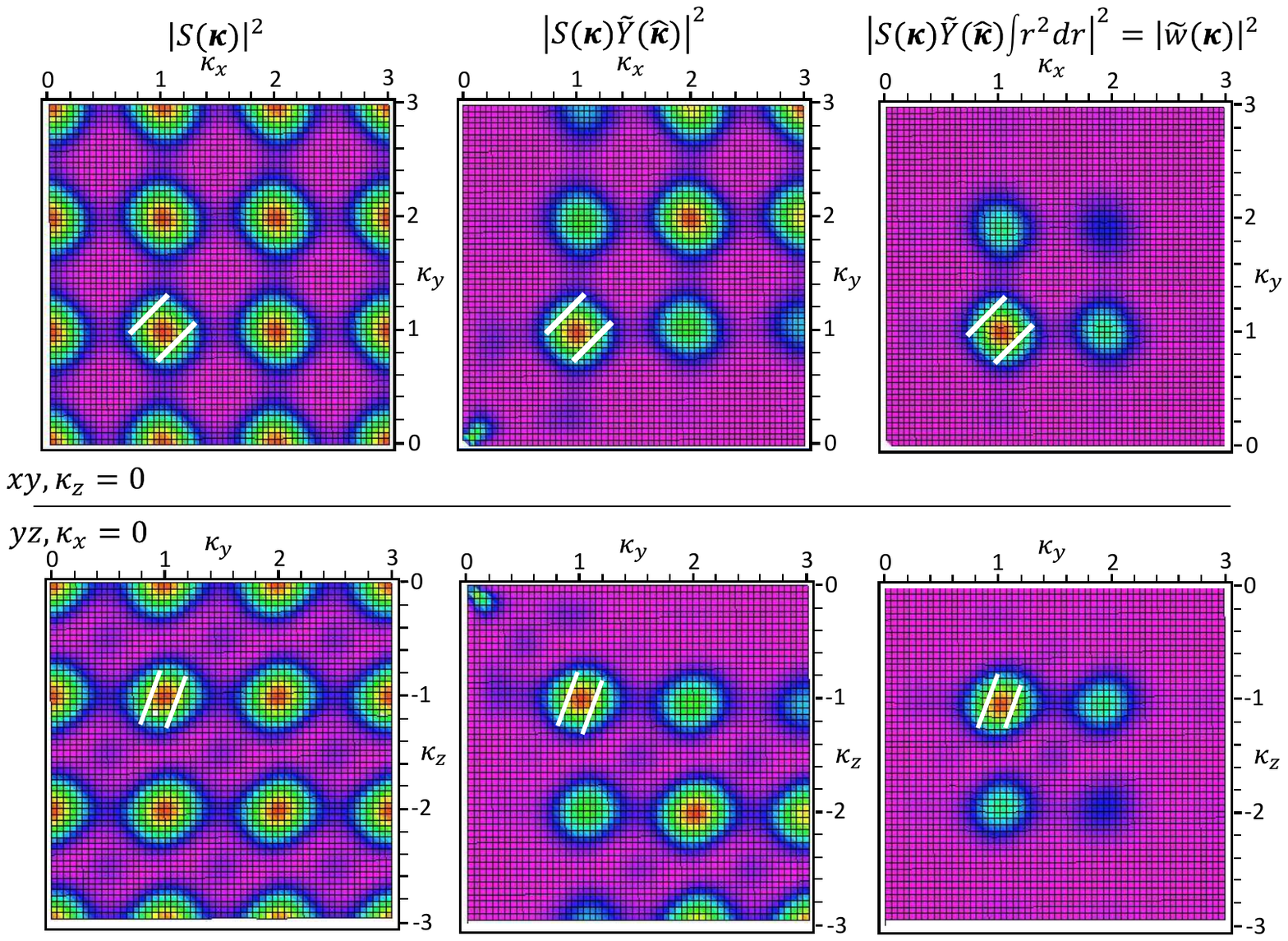}
\caption{Form-factor intensities, $\left\vert \tilde{w}_{m}\left( \mathbf{%
\protect\kappa }\right) \right\vert ^{2},$ i.e. the coarse-grained
structure. \emph{Top row:} $xy$ form-factor intensities in the $\protect%
\kappa _{z}$=0 plane. \emph{Bottom row:} $yz$ form-factor intensities in the 
$\protect\kappa _{x}$=0 plane. $\tilde{w}_{xz}\left( \protect\kappa _{x},%
\protect\kappa _{y},\protect\kappa _{z}\right) $\thinspace $\,$=$\,$%
\thinspace $\tilde{w}_{yz}\left( \protect\kappa _{y},\protect\kappa _{x},%
\protect\kappa _{z}\right) .$ \emph{Columns 1-3:}\textit{\ }Turning on
factors following Eq.~(\protect\ref{FTWO}) with $\mathcal{S}_{m}\left( 
\mathbf{\protect\kappa }\right) $ given by Eq.s (\protect\ref{Sxy}) and~ (%
\protect\ref{Syz}). \emph{\ }The little white lines are the projections onto
the $\protect\kappa _{z}$=0 plane or the $\protect\kappa _{x}$=0 plane of
the near-FS lines $\left\vert \protect\kappa _{b}\right\vert \sim \frac{1}{4}%
,$ limited to the double zone $-1\leq \protect\kappa _{c}\leq 1,$ along
which the bands and fine-grained intensities for $\protect\kappa _{a}$=6.4
were plotted in FIG.~\protect\ref{ThreeBandskonly}. The $\protect\kappa %
_{a}, $ $\protect\kappa _{b},$ and $\protect\kappa _{c}$ directions are
those of respectively $\protect\kappa _{x}+\protect\kappa _{y}-\protect%
\kappa _{z},$ $\protect\kappa _{y}-\protect\kappa _{x},$ and $\protect\kappa %
_{x}+\protect\kappa _{y}+2\protect\kappa _{z};$ see Eq.~(\protect\ref{kabc}%
). }
\label{SYk}
\end{figure*}

The structure factors, $\mathcal{S}_{xy},$ $\mathcal{S}_{yz},$ and $\mathcal{%
S}_{xz},$ peak --with value 1 if the normalizations are as above-- along the
respective lines: $\left[ \kappa _{x},\kappa _{y},\kappa _{z}\right] $=$%
\left[ \text{\textsc{l,m}},\kappa _{z}\right] $, =$\left[ \kappa _{x},\text{%
\textsc{m,n}}\right] $, and =$\left[ \text{\textsc{l,}}\kappa _{y},\text{%
\textsc{n}}\right] ,$ passing through the respective points, \textsc{l}$%
\mathbf{x}^{\ast }+$\textsc{m}$\mathbf{y}^{\ast }+$\textsc{n}$\mathbf{z}%
^{\ast },$ of the lattice reciprocal to the lattice I$\,$(\ref{xyz}) with 1
Mo per primitive cell. In other words: $\left\vert \mathcal{S}%
_{xy}\right\vert ^{2},$ $\left\vert \mathcal{S}_{yz}\right\vert ^{2},$ and $%
\left\vert \mathcal{S}_{xz}\right\vert ^{2}$ form 2D square lattices of
"beams" running in respectively the $\kappa _{z},$ $\kappa _{x},$ and $%
\kappa _{y}$ directions (see FIG.~\ref{SYk}). This gives rise to intensity
patterns that are \emph{coarser} than the zone-selection patterns ($M+N$
even or odd) whose origin is the smallness of the dimerization that
increases the size of the primitive cell from 6 to 12 molybdenums.

The relation between the $\left[ \kappa _{x},\kappa _{y},\kappa _{z}\right] $%
- and the $\left( \kappa _{a},\kappa _{b},\kappa _{c}\right) $-components
--used to describe respectively the WO and the band structure and, hence,
respectively the coarse and the fine-grained structure-- is:

\begin{equation}
\QATOP{\kappa _{x}}{\kappa _{y}}=\frac{\kappa _{a}+\kappa _{c}}{6}\mp \frac{%
\kappa _{b}}{2}\;\mathrm{and}\;-\kappa _{z}=\frac{\kappa _{a}+\kappa _{c}}{6}%
-\frac{\kappa _{c}}{2},  \label{kxyz}
\end{equation}%
which is the same as the transformation I$~$(\ref{xyz}) between the
primitive translations of the approximately cubic Mo$_{1}$- and the exact Mo$%
_{12}$-lattice [see also Charts I$~$(\ref{ac}) and I$~$(\ref{xy})]. The
inverse transformation --the same as I$~$(\ref{abc})-- is: 
\begin{eqnarray}
\kappa _{a} &=&2\left( \kappa _{x}+\kappa _{y}-\kappa _{z}\right) ,
\label{kabc} \\
\kappa _{b} &=&\kappa _{y}-\kappa _{x},\;\mathrm{and}\;\kappa _{c}=\kappa
_{x}+\kappa _{y}+2\kappa _{z}.  \notag
\end{eqnarray}%
Since exchange of $\kappa _{x}$ and $\kappa _{y}$ merely causes $\kappa _{b}$
to change sign, $\mathcal{S}_{xz}\left( \kappa _{a},\kappa _{b},\kappa
_{c}\right) $ $=$ $\mathcal{S}_{yz}\left( \kappa _{a},-\kappa _{b},\kappa
_{c}\right) .$

The angular factors $\tilde{Y}_{m}\left( {\mathbf{\kappa }}\right) $ in
expression (\ref{FTWO}) have the same orientation with respect to the $%
a^{\ast }b^{\ast }c^{\ast }$ system as $Y_{m}\left( \mathbf{\hat{r}}\right) $
has with respect to the $abc$ system seen in FIG.$\,$I$~$\ref{FIGyzb&xyamc}
and Charts I$~$(\ref{ac})-(\ref{xy}). This is so because the $abc$ system is
(almost) orthogonal (Sect.$\,$I$~$\ref{crystal_structure}), whereby $\mathbf{%
a}$ is parallel with $\mathbf{a}^{\ast },$ $\mathbf{b}$ with $\mathbf{b}%
^{\ast },$ and $\mathbf{c}$ with $\mathbf{c}^{\ast }.$ As a result:%
\begin{equation}
\tilde{Y}_{xy}\propto -\frac{\kappa _{x}\kappa _{y}}{\kappa ^{2}}=\left( 
\frac{\kappa _{b}}{2\kappa }\right) ^{2}-\left( \frac{\kappa _{c}+\kappa _{a}%
}{6\kappa }\right) ^{2},  \label{Yxy}
\end{equation}%
where: 
\begin{equation}
\kappa ^{2}\equiv \kappa _{x}^{2}+\kappa _{y}^{2}+\kappa _{z}^{2}=\frac{1}{12%
}\kappa _{a}^{2}+\frac{1}{2}\kappa _{b}^{2}+\frac{1}{6}\kappa _{c}^{2},
\label{kappa2}
\end{equation}%
and similarly for $\tilde{Y}_{yz}$ and $\tilde{Y}_{xz}.$ The angular factor
extinguishes the intensity around the $\left[ 0,0,\kappa _{z}\right] $-, $%
\left[ \kappa _{x},0,0\right] $-, or $\left[ 0,\kappa _{y},0\right] $-lines,
as is clearly seen when proceeding from the 1st to the 2nd column in FIG.$~$%
\ref{SYk}.

The last, radial factor in Eq.$~$(\ref{FTWO}) is independent of $m$ and
merely gives the overall shape of the ARPES intensity. We found it
sufficient to mimic the main part of the Mo $4d$ radial function, continued
for $r>0.5$ (in units of $3.82\mathrm{\mathring{A})}$ as the $t_{2g,m}$%
-average of the neighboring $pd\pi $ anti-bonds, by $\varphi _{2}\left(
r\right) \propto \left( e^{-14r}-e^{-7}\right) r^{2}.$ This function peaks
at $r\approx $0.14, has at node at 0.5, and is truncated at 0.55. The
negative part mimics the contribution from the $pd\pi $ antibonds. The last
factor in (\ref{FTWO}) thus raises quadratically from $\kappa $=0 and peaks
at $\kappa \approx \sqrt{3},$ which is at the sphere passing through $\left[
1,1,-1\right] .$ For larger values of $\kappa $, the radial factor decreases
monotonically and for $\kappa \gtrsim 3,$ i.e. outside the sphere passing
through $\left[ 1,2,-2\right] $ and $\left[ 2,1,-2\right] ,$ it has fallen
to below one third its value at the peak. The radial factor has been
included in the two last columns of FIG.$~$\ref{SYk}.

We thus realize that the angular and radial factors leave intensity in only
small parts of reciprocal space.

\begin{figure}[tbh]
\includegraphics[width=\linewidth]{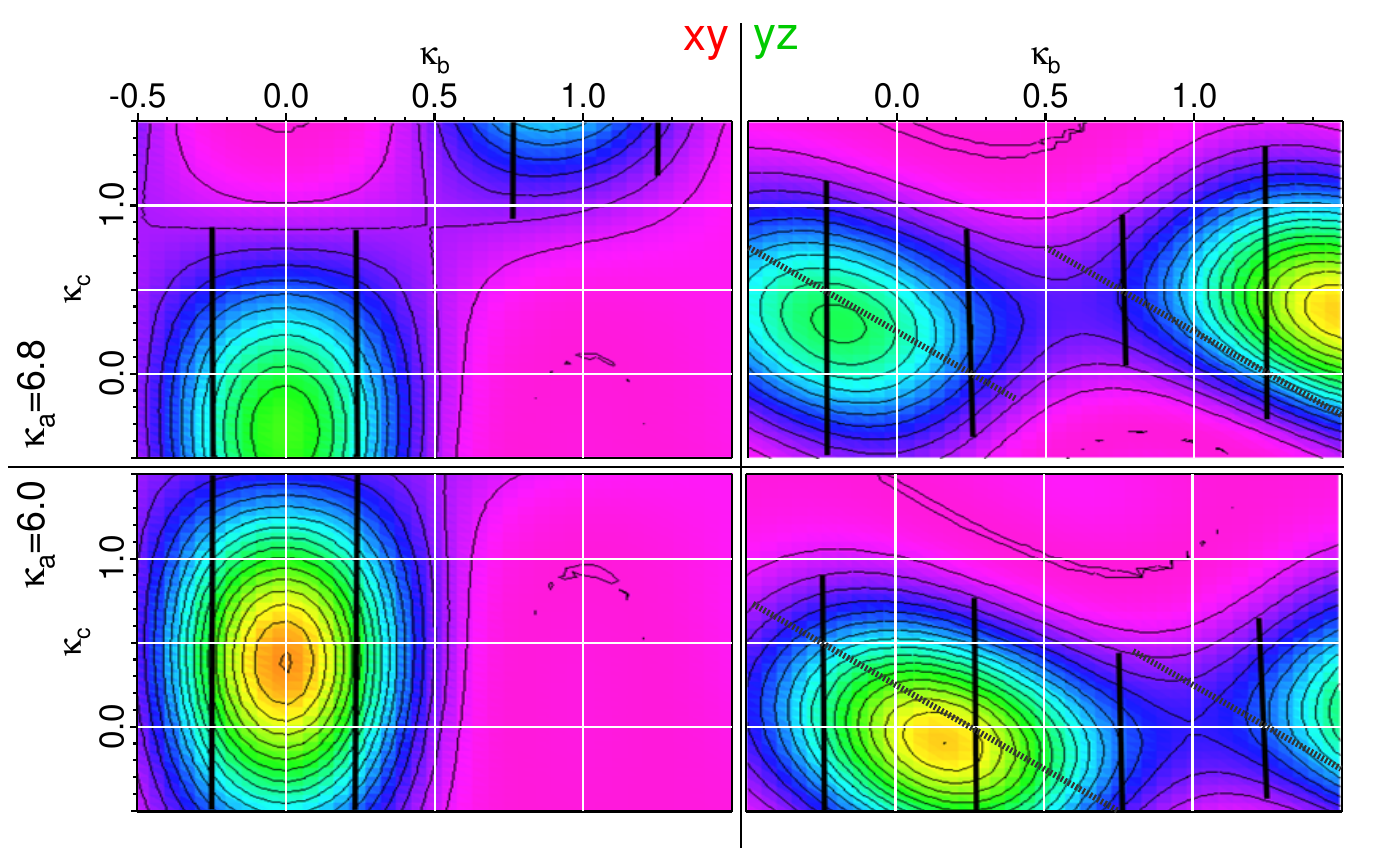}
\caption{Theoretical $xy$ (\textit{left}) and $yz$ (\textit{right})
coarse-grained ARPES intensity distributions, $\protect\kappa %
_{a}^{2}\left\vert \tilde{w}_{m}\left( \mathbf{\protect\kappa }\right)
\right\vert ^{2},$ in the $\protect\kappa _{b}\protect\kappa _{c}$ planes
with $\protect\kappa _{a}$=6.0 (bottom) and 6.8 (top) planes. The $xz$
intensities equal the $yz$ ones mirrored around $\protect\kappa _{b}$=0. See
top part of FIG.~\protect\ref{ThreeBandskonly} and Eq.s~(\protect\ref{Indep}%
) and (\protect\ref{max1xy})-(\protect\ref{max2yz}). The contours go from 0
to 400 in steps of 20. The black lines extending where the intensity exceeds
40 indicate $xy$ and $yz$ CECs with energy respectively 0.1 and 0.5~eV below
the Fermi level (see FIG. \protect\ref{CEC}) and in the periodic zone
scheme. }
\label{Re^2ka6068}
\end{figure}

The $xy$ form-factor intensity, $\left\vert \tilde{w}_{xy}\left( \mathbf{%
\kappa }\right) \right\vert ^{2},$ shown in the 1st row and 3rd column in
FIG.$\,$\ref{SYk}, has \emph{one dominant peak}; its position is at $\left[
\kappa _{x},\kappa _{y}\right] =\left( 1+\frac{\epsilon }{6}\right) \left[
1,1\right] ,$ where the small shift away from [1,1], proportional to $%
\epsilon \sim $0.4, is mainly due to the angular factor. In 3D, and
according to Eq.$~$(\ref{kabc}), this peak becomes a beam centered on the
line given by:%
\begin{equation}
\kappa _{a}+\kappa _{c}=6+\epsilon \;\mathrm{and}\;\kappa _{b}=0.
\label{max1xy}
\end{equation}%
This holds as long as the $\kappa _{z}$ dependence from the radial factor
can be neglected. There are two, less intensive peaks near $\left[ \kappa
_{x},\kappa _{y}\right] $=$\left[ 1,2\right] $ and $\left[ 2,1\right] ,$
which in 3D become beams around the lines given by: 
\begin{equation}
\kappa _{a}+\kappa _{c}=9\;\mathrm{and\;}\kappa _{b}=\pm 1.  \label{max2xy}
\end{equation}%
In FIG.$~$\ref{Re^2ka6068} we show the intensity distributions\footnote{%
We take the polarization factor $\mathbf{{\kappa }\cdot \hat{e}}$ as $\kappa
_{a}$ because this is about six times larger than $\kappa _{b}$ and $\kappa
_{c},$ and because the $e_{a}$ component of the polarization (\ref%
{polarization}) is much larger than $e_{b}$ and $e_{c}.$\label{polfac}}, $%
\kappa _{a}^{2}\left\vert \tilde{w}_{m}\left( \mathbf{\kappa }\right)
\right\vert ^{2},$ in the planes with $\kappa _{a}$=6.0 and 6.8 in the
region $-0.5\leq \left( \kappa _{b},\kappa _{c}\right) \leq 1.5$. In the
left-hand panel, we see the $xy$ beam (\ref{max1xy}) form spots around the
points $\left( \kappa _{b},\kappa _{c}\right) =(0,\epsilon )$ and $%
(0,\epsilon -0.8)$ when hitting the $\kappa _{a}$=6.0 and 6.8 planes$.$ In
the latter plane, the less intensive spot centered near $\left( \kappa
_{b},\kappa _{c}\right) $=$\left( 1,2.2\right) $ and formed by the positive-$%
\kappa _{b}$ beam (\ref{max2xy}) can be barely seen.

The $yz$ form-factor intensity of zone selection, shown in the 2nd row and
3rd column in FIG.$\,$\ref{SYk}, has its \emph{dominant peak} near $\left[
\kappa _{y},\kappa _{z}\right] $=$\left[ 1,-1\right] .$ In 3D, this peak
becomes a beam centered on the line given by:%
\begin{equation}
\kappa _{a}-2\kappa _{c}=6\;\mathrm{and}\;\kappa _{b}+\kappa _{c}=0.
\label{max1yz}
\end{equation}%
There are two weaker peaks at $\left[ \kappa _{y},\kappa _{z}\right] $=$%
\left[ 2,-1\right] $ and $\left[ 1,-2\right] ,$ which in 3D become beams
centered along the lines given by respectively: 
\begin{equation}
\kappa _{a}-2\kappa _{c}=6\;\mathrm{and\;}\kappa _{b}+\kappa _{c}=2,
\label{max3yz}
\end{equation}%
and 
\begin{equation}
\kappa _{a}-2\kappa _{c}=12\;\mathrm{and\;}\kappa _{b}+\kappa _{c}=-2
\label{max2yz}
\end{equation}%
In the right-hand panel of FIG.$\,$\ref{Re^2ka6068} we see the $yz$ beam (%
\ref{max1yz}) form spots around $\left( \kappa _{b},\kappa _{c}\right) $=$%
(0,0)$ and $\left( -0.4,0.4\right) $ upon hitting the $\kappa _{a}$=6.0 and
6.8 planes, as well as the beam (\ref{max3yz}) hitting these two planes at
respectively $\left( \kappa _{b},\kappa _{c}\right) $=$\left( 2,0\right) $
and $\left( 1.6,0.4\right) .$

The vertical and diagonal black lines, periodically repeated, but limited to
the regions where the intensity exceeds 40 in the units used in FIG.$\,$\ref%
{ThreeBandskonly}, indicate the CECs with binding energy 0.1 eV for $xy$ and
0.5~eV for $yz.$ The CECs will be discussed in the experimental section \ref%
{SectCEC}.

Since the $xz$ form-factor intensity is related to the one shown for $yz$ in
FIG.$~$\ref{Re^2ka6068} by a sign change of $\kappa _{b},$ we can see that
only for $\kappa _{a}\sim 6.4$ do the $zx$ and $yz$ intensities along e.g.
the \textrm{P'QPQP'} $\left( 0.225,\kappa _{c}\right) $-line reach roughly
the same maximum value. This is the maximum at $\sim $ 250 seen in the top
part of FIG.$\,$\ref{ThreeBandskonly} and in the 3rd column of FIG.~\ref%
{FIGyzZoneSelect}.

Our ARPES measurements, to be presented and discussed in the following
section, were mainly performed in the neighborhood of the point $\left[
\kappa _{x},\kappa _{y},\kappa _{z}\right] =\left[ 1,1,-1\right] $ which is
close to the peaks of all three form factors and which is the $\Gamma $
point $\left( \kappa _{a},\kappa _{b},\kappa _{c}\right) =\left(
6,0,0\right) $. It may be noted that upon going from the 1st to the 2nd
column in the 1st row of FIG.$\,$\ref{SYk}, the angular factor sharpens up
the [1,1] peak of the $xy$ intensity. The traces of the two FS sheets, $%
\left\vert \kappa _{b}\right\vert $=$\frac{1}{4},$ on the $\kappa _{a}$=6.4
plane and bound between the $\kappa _{c}$=$\pm $1 planes are the little
white lines in FIG.$\,$\ref{SYk}, when projected onto the $\kappa _{z}$=0
plane in the 1st row and onto the $\kappa _{x}$=0 plane in 2nd row. These
are the \textrm{P'QPQP'} lines along which the intensity distributions were
shown in FIG.$\,$\ref{ThreeBandskonly}. Since the $xy$-band disperses in the
direction of the $\kappa _{b}$-lobe of its WO, this direction is
perpendicular to the little white lines.

At the end of this long section \ref{SectIntensity}, we emphasize the
following two points:

Whereas the $k_{a}$-dependence of the $\left\vert \mathbf{k}\right\rangle $%
-projected bands, i.e. of the fine-grained structure, is negligible compared
with the $k_{b}$- and $k_{c}$-dependencies, the $\kappa _{a}$-dependence of
the aperiodic form factors, $\left\vert \tilde{w}_{m}\left( \mathbf{\kappa }%
\right) \right\vert ^{2},$ i.e. of the coarse-grained structure, is strong,
as strong as the dependence on $\kappa _{c}$ for $xy,$ and half as strong
for $yz/xz$; see Eq.s (\ref{max1xy})-(\ref{max1yz}). This is due to the form
factors being 2D$,$ $\propto \kappa _{x}\kappa _{y},$ $\kappa _{y}\kappa
_{z},$ or $\kappa _{x}\kappa _{z},$ and to the orientation of the $t_{2g}$
orbitals with respect to the crystallographic axes I$~$(\ref{abc}).

Since the low-energy WOs in LiPB are relatively extended (covering several
atoms), the radial part of the WO form factors make the beams narrow and
thus cause the ARPES intensity to depend rather strongly on the photon
energy $h\nu ,$ as we shall see in Sect. \ref{Secthv}.

\section{Presentation and discussion of the ARPES data\label{SectARPESData}}

From the basics of the electronic structure in Sect. I$\,$\ref{SectElStruc}
we expect the two $\widetilde{xy}$-bands crossing the Fermi level to have by
far the largest dispersion with $k_{b}\mathbf{,}$ i.e. along the ribbon,
weak dispersion with $k_{c},$ i.e. up and down the staircase, weak $k_{c}$%
-dependent splitting caused by direct inter-ribbon hopping and by
hybridization with the gapped $xz$- and $yz$-bands, and essentially no
dispersion with $k_{a}$ due to the lack of hopping between staircases. In
the following, we want to demonstrate that this strong one-dimensionality\
is indeed confirmed by our ARPES experiment. We also pay attention to the
aperiodic variations of the ARPES intensity between equivalent zones and
compare them with the coarse-grained intensity variations predicted in the
preceding Sect.~\ref{Coarse}. The fine-grained intensity variations (zone
selection) predicted in Sect.~\ref{Sectzoneselect} will be observed in Sects.%
$~$\ref{SectCEC} and \ref{SectEk} where, most importantly, we also compare
the band dispersions in detail with those predicted by the LDA as
parametrized by the $t_{2g}$ TB Hamiltonian (Sect.~I~\ref{SectH}).

Intensity variations and one-dimensionality are also features of the ARPES $%
yz$- and $xz$-bands, but of course not near the Fermi level where they are
gapped. Moreover, their one-dimensionality is with $k_{c}+k_{b}$ or $%
k_{c}-k_{b},$ rather than with $k_{b}.$

\subsection{ARPES method\label{SectARPESMethod}}

We measured several samples for the conclusions presented in this work. The
samples came from two different crystal growers and were all prepared by a
temperature gradient flux growth technique \cite{McCarroll1984}. In the
text, we refer to two samples G (M. Greenblatt) and H (J. He) representing
the variation detectable in our experiments.

Photoemission measurements were performed at the MERLIN endstation (BL
4.0.3) of the Advanced Light Source with a Scienta R4000 electron detector.
The polarization was set to linear vertical, i.e. in FIG.$\,$\ref{CEC} the
vector of the electric field horizontal. The temperature was set to $T$=26$~$%
K for sample G and $T$=6~Kfor sample H and the samples were cleaved while
attached to the cold manipulator at $p\approx 4\times 10^{-11}$torr . The
overall energy resolution was set to around 16~meV\ at photon energy $h\nu $%
=30~eV\ going up to around 40~meV\ at $h\nu $=100~eV. At $h\nu $=30~eV, the
momentum resolution in the \textbf{b*}-direction is 2\% of $k_{Fb}$ at
half-filling (i.e. at nominal Li$_{1}$ stoichiometry), i.e. 0.005. With the
solid-state definition of reciprocal space [see Sect.~I~\ref%
{crystal_structure} below Eq.~(\ref{k})], this is $0.006$ \AA $^{-1}$. The
polarization vector in sample coordinates is: 
\begin{equation}
\frac{\mathbf{E}}{\left\vert \mathbf{E}\right\vert }=\left( 
\begin{array}{c}
e_{a} \\ 
e_{b} \\ 
e_{c}%
\end{array}%
\right) =\left( 
\begin{array}{c}
\sin (65^{\circ }-\phi )\cos \theta \\ 
\sin (65^{\circ }-\phi )\sin \theta \\ 
\cos (65^{\circ }-\phi )%
\end{array}%
\right) .  \label{polarization}
\end{equation}%
Here, $\phi $ is the polar rotation, and the $\theta $ the tilt angle. In
our measurements, $\left\vert \phi \right\vert \lesssim 5%
{{}^\circ}%
$ and $\left\vert \theta \right\vert \lesssim 5%
{{}^\circ}%
$ resulting overall in a strong component along the $\mathbf{a}$-axis,
normal to the cleavage plane, a weak component along the $\mathbf{c}$-axis,
and a very weak component along the $\mathbf{b}$-axis.

LiMo$_{6}$O$_{17}$\ is susceptible to oxygen loss caused by intense
ultraviolet light where the desorption is due to the Knotek-Feibelman
mechanism including a core level excited resonant Auger decay \cite%
{Knotek1978}. It shares this behavior with, i.e. NaMo$_{6}$O$_{17}$\cite%
{Denlinger1998}, K$_{0.3}$MoO$_{3}$\cite{Breuer1995}, or oxides like TiO$%
_{2} $\cite{Moser2013} and SrTiO$_{3}$\cite{Walker2015}. We were not able to
prevent this damage by oxygen dosing, as i.e. possible for SrTiO$_{3}$\cite%
{Dudy2016}. The reason might be the existence of both, tetrahedral and
octahedral coordinated molybdenums. A small oxygen loss causes a slight
electron doping, but, as time progresses, the ARPES signal eventually blurs.
In order to prevent the loss, one concept might be to keep the photon energy
below that of the lowest energy core level resonance. However, often there
is higher-order light that still causes a slow degradation (with the
timescale in the hours instead of minutes) and therefore our concept is to
use a large homogenous area and slightly alter the position of the beam spot
when the sample degradation begins.

\begin{figure}[tbh]
\includegraphics[width=1\linewidth]{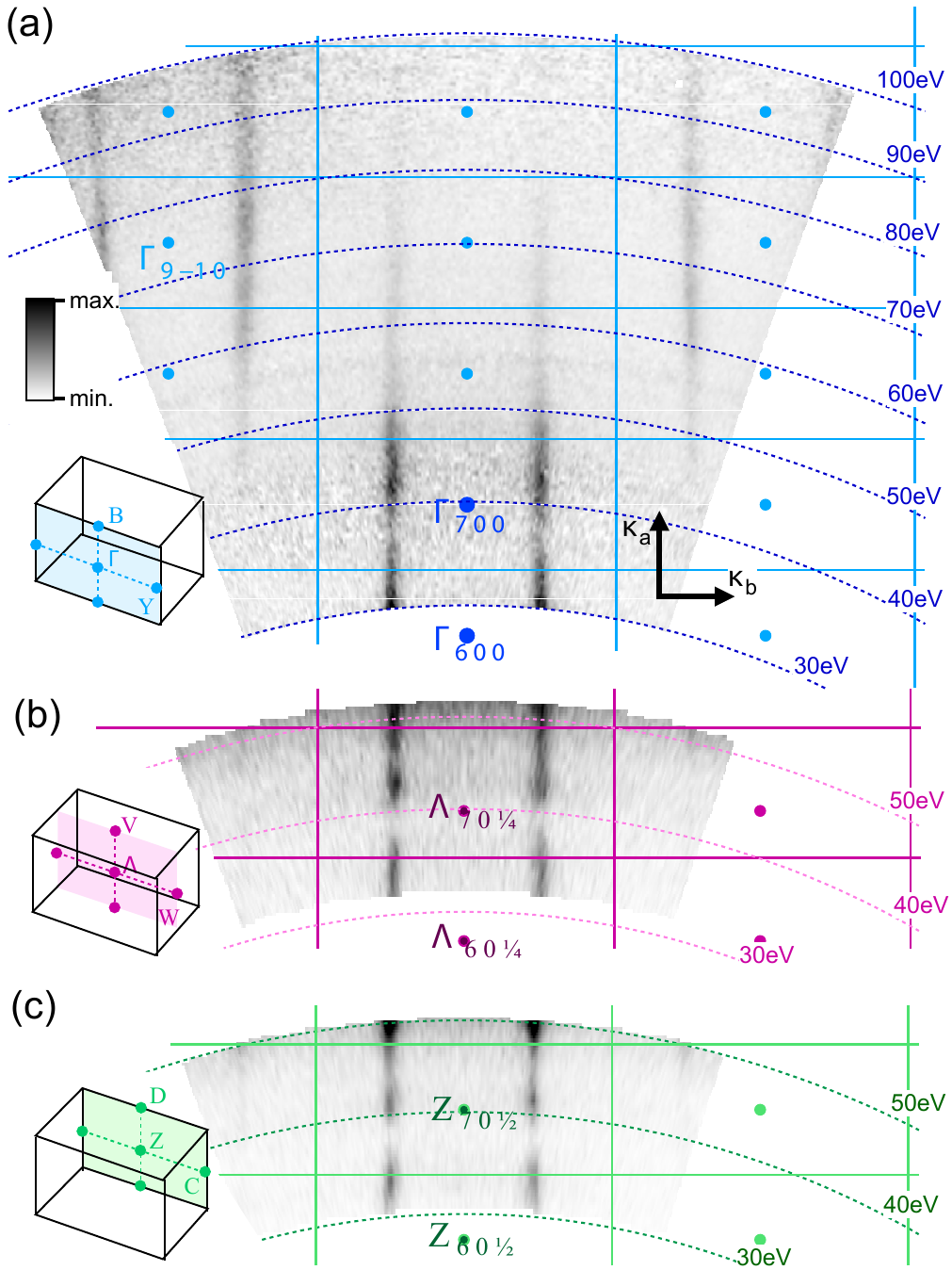}
\caption{Photon energy scan of sample H at $T$=6$~$K for $\mathbf{\protect%
\kappa }$ in three\textbf{\ }$\mathbf{a}^{\ast }\mathbf{b}^{\ast }$-planes:
(a) FS map through $\protect\kappa _{c}$=0 $\left( \mathrm{\Gamma BY}\right) 
$ (b) $\protect\kappa _{c}$=$\frac{1}{4}$ $\left( \mathrm{\Lambda WV}\right) 
\mathrm{,}$ and (c) $\protect\kappa _{c}$=$\frac{1}{2}$ $\left( \mathrm{ZDC}%
\right) \mathrm{.}$ The colored full lines along $\mathbf{a}^{\ast }$ and $%
\mathbf{b}^{\ast }$ are the intersections with the BZs. The blue dotted
lines intersect at the reciprocal lattice points $\Gamma _{LMN}.$ The
high-intensity traces represent the 1D FS showing no dispersion in the $%
\mathbf{a}^{\ast }$-direction. There are coarse-grained intensity variations
whose origin is the structure factor of the form factor for the $xy$ WOs;
see Sect. \protect\ref{Coarse} and FIG.~\protect\ref{Re^2xykakbkc002550}.
This figure and the following are approximately to scale, i.e. consistent
with $a/b=2.31.$}
\label{Photon_Energy_Scan}
\end{figure}

\begin{figure}[tbh]
\includegraphics[width=\linewidth]{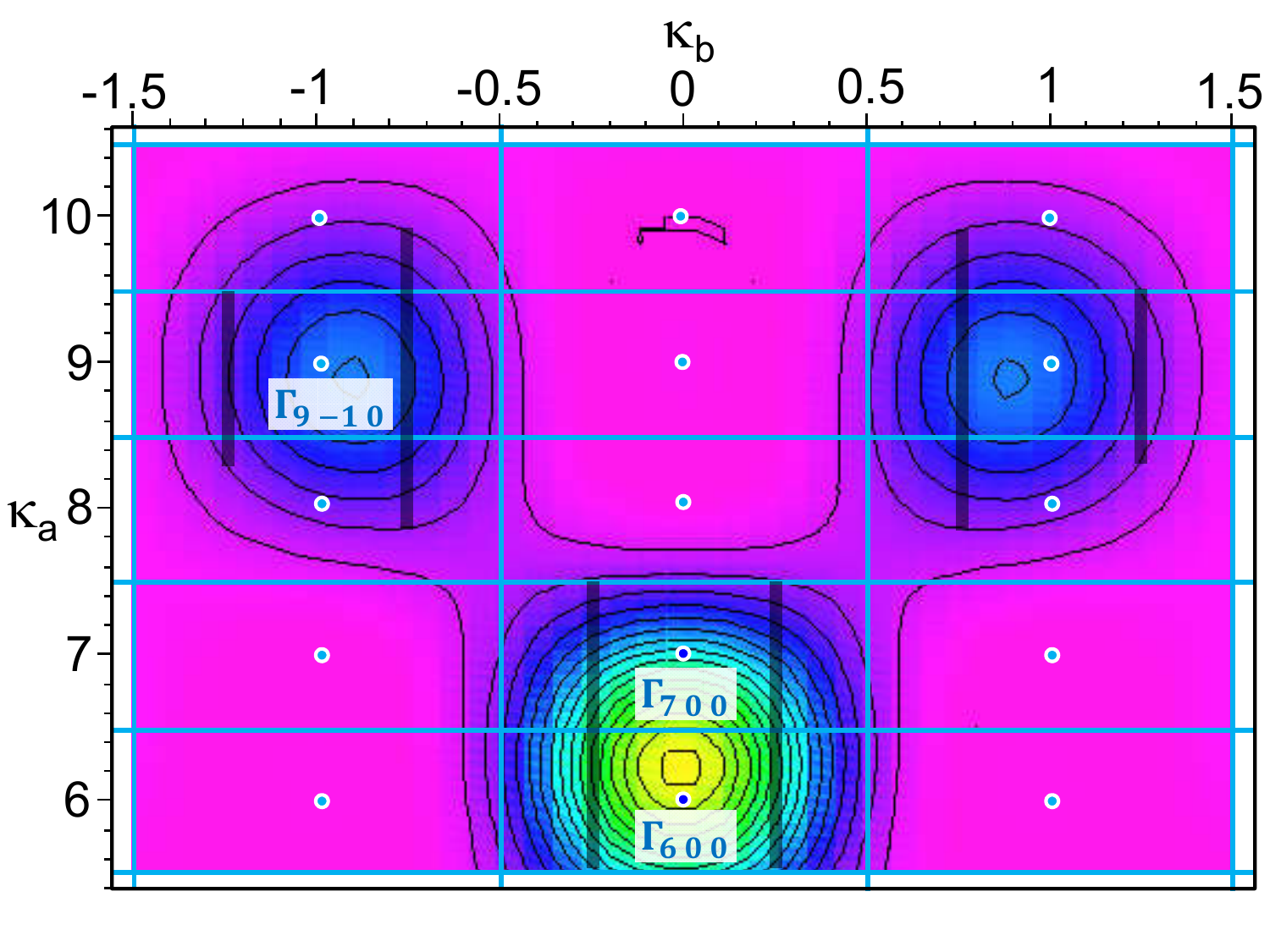}
\caption{Theoretical ARPES intensity distribution $\protect\kappa %
_{a}^{2}\left\vert \tilde{w}_{xy}\left( \protect\kappa _{a},\protect\kappa %
_{b},\protect\kappa _{c}\text{=}0\right) \right\vert ^{2}$ to be compared
with the coarse-grained part of the experimental intensity distribution in
FIG.~\protect\ref{Photon_Energy_Scan}. The contours go from 0 to 400 in
steps of 20. The black lines indicate the traces of the near-FS sheets, $%
\left\vert \protect\kappa _{b}\right\vert \sim \frac{1}{4}$, in the periodic
zone scheme and extend where the intensity exceeds 40. }
\label{Re^2xykakbkc002550}
\end{figure}

Even though the ARPES lineshapes have the general 1D holon-peak and
spinon-edge features \cite{Dudy2013}, they were analyzed by a model-free
method described in detail in Paper III Sect.$~$\ref{SectFSExperiment}. This
procedure was necessitated because the low-$T$ ARPES lineshape is not sharp
enough to agree in detail with the low-$T$ TL lineshape \cite{Wang2009}. If
the TL lineshape is broadened ad hoc it can be made to fit \cite{Dudy2013},
but we did not want to use that ad hoc procedure in the current work.

\subsection{Ansatz for deducing $\protect\kappa _{a}$\label{SectKappa}}

We repeat here some basics of ARPES, cf. \cite{Huefner2013, Damascelli2004}
As in Sect.~\ref{Section:PrelimInt} we use a notation according to which $%
\mathbf{k}$ denotes the Bloch vector in a (periodically repeated) single
zone of an initial-state, and $\mathbf{\kappa }$ is the momentum of the
final, plane-wave state \emph{inside} the crystal. According to Eq.~(\ref%
{Indep}), $\mathbf{\kappa }\,\left[ \func{mod}\mathrm{zone}\right] =\mathbf{%
k.}$ Upon leaving the crystal, the photoelectron is diffracted in the
direction away from the surface, whereby the normal momentum component, $%
\mathbf{\kappa }_{\perp }=\kappa _{a}\mathbf{a}^{\ast },$ jumps
discontinuously to a smaller value, $\mathbf{\kappa }_{\perp o}.$ Parallel
to the surface, the momentum is conserved: 
\begin{equation*}
\mathbf{\kappa }_{\mathbf{\shortparallel }}\equiv \kappa _{b}\mathbf{b}%
^{\ast }+\kappa _{c}\mathbf{c}^{\ast }=\mathbf{\kappa }_{\mathbf{%
\shortparallel }o}.
\end{equation*}

What is measured in ARPES is, for a given photon energy $h\nu ,$ the yield, $%
I,$ the angle of exit, $\theta ,$ and the kinetic energy, $T=\frac{h^{2}}{2m}%
\left\vert \mathbf{\kappa }_{o}\right\vert ^{2},$ of the photoelectrons in
the analyzer \emph{outside} the crystal. The value of $\kappa _{a}$ inside
the sample is deduced from the two latter quantities by assuming that inside
the crystal, the energy of the final state is $V+\frac{h^{2}}{2m}\left\vert 
\mathbf{\kappa }\right\vert ^{2},$ i.e. the energy of an electron, free with
respect to a potential floor, $V,$ and that outside the crystal, the
momentum is:%
\begin{equation}
\mathbf{\kappa }_{o}=\left( \mathbf{\kappa }_{\perp o},\mathbf{\kappa }%
_{\shortparallel }\right) =\frac{\sqrt{2mT}}{h}\left( \cos \theta ,\sin
\theta \right) ,  \label{kappao}
\end{equation}%
and the energy is $\Phi +T.$ Here, $\Phi $ is the work function of the
sample. It should not be confused with $\Delta \Phi $ which is a constant
given by the apparatus and is essentially the difference between work
function of sample and analyzer. $\Delta \Phi $ allows to relate the
measured kinetic energy to the binding energy $\omega $ within the sample 
\begin{equation}
\omega \equiv -E_{j}\left( \mathbf{k}\right) =h\nu -\left( \Delta \Phi
+T\right) .  \label{omega}
\end{equation}

Equating the inside and outside energies, yields the desired relation for $%
\kappa _{a}:$%
\begin{equation}
\frac{\kappa _{a}}{a}=\left\vert \mathbf{\kappa }_{\perp }\right\vert =\frac{%
\sqrt{2m}}{h}\sqrt{T\cos ^{2}\theta +V_{0}},  \label{kappaa}
\end{equation}%
where $V_{0}\equiv \Phi -V$ is the so-called \emph{inner potential}. Taking $%
V$ and $\Phi $ with respect to the Fermi level, which is common for the
crystal and the analyzer, $V$ is negative and $\Phi $ is positive, whereby $%
V_{0}$ is positive. Its value is determined empirically.

\subsection{Photon energy dependence\textrm{\label{Secthv}}}

\subsubsection{FS intersection with $\mathbf{a}^{\ast }\mathbf{b}^{\ast }$
planes}

We begin by showing ARPES for photoelectrons coming from slightly below the
Fermi level, i.e.$\,$from the $\widetilde{xy}$ electrons. FIG.$\,$\ref%
{Photon_Energy_Scan} shows for sample H at $T$=6\thinspace K the
photoelectron yield as a grey-scale intensity, a so-called FS map, in the
three $\kappa _{a}\kappa _{b}$ planes with $\kappa _{c}$=0, $\frac{1}{4}$,
and $\frac{1}{2},$ colored respectively blue, red, and green. In the $\kappa
_{c}$=0 plane, the scan covers many BZs. We see the traces of the two FS
sheets separating the occupied states, $\left\vert k_{b}\right\vert <k_{F},$
between the sheets from the empty states, $\left\vert k_{b}\right\vert
>k_{F},$ outside the sheets. These traces appear as straight lines and are
thus consistent with being the intersections with a $\kappa _{a}\kappa _{b}$
plane of a 1D, nearly half-filled FS, $\left\vert k_{b}\right\vert
=k_{F}\sim \frac{1}{4}$. This FS is seen to be periodic in the $\mathbf{a}%
^{\ast }\mathbf{b}^{\ast }\mathbf{c}^{\ast }$ lattice and to have aperiodic,
coarse-grained intensity variations.

Since the $\kappa _{a}$ direction is perpendicular to the plane of the
sample, it must be accessed by variation of the photon energy. For
presenting these measurements, we have converted our raw yield $\left(
I\right) $ and kinetic-energy $\left( T\right) $ data as functions of angle $%
\left( \theta \right) $ and photon energy $\left( h\nu \right) $ as
explained above to binding-energy $\left( \omega \right) $ and momentum
vector $\left( \mathbf{\kappa }\right) $, using the value $V_{0}$=11~eV for
the inner potential. With the work function of the analyzer being $\Phi $%
=4eV, the potential floor is thus 7 eV below the Fermi level, i.e. $V$=$-$%
7~eV. The dotted constant-$h\nu $ circles in FIG.$\,$\ref{Photon_Energy_Scan}
are the cross-sections of the sphere $V+\frac{h^{2}}{2m}\left\vert \mathbf{%
\kappa }\right\vert ^{2}=h\nu $ with the constant-$\kappa _{c}$ planes. For
a given $h\nu $, normal emission $\left( \theta \text{=0=}\kappa _{b}\text{=}%
\kappa _{c}\right) $ from the Fermi level\footnote{\label{bind}For finite
binding energy, $h\nu $ should be substituted by $h\nu -\omega $ in Eq.$\,$(%
\ref{kappaa(0,hv)}). But even going to the bottom of the band, where $\omega 
$=0.7 eV, this lowers $\kappa _{a}$ by merely $\approx $0.06.} thus has the $%
\kappa _{a}$ value:%
\begin{equation}
\kappa _{a}\left( 0,0,h\nu \right) =\frac{a\sqrt{2m}}{h}\sqrt{h\nu -V}%
\approx 7.0\sqrt{\frac{h\nu +7~\mathrm{eV}}{40+7~\mathrm{eV}}},
\label{kappaa(0,hv)}
\end{equation}%
which, as expressed on the right-hand side, is 7.0 for $h\nu $=40$~$eV.

FIG.$~$\ref{Photon_Energy_Scan}$~$(a) shows intensity variations in central (%
$\kappa _{c}$=0) plane: Strong intensity in the region $\left( \kappa
_{a},\kappa _{b}\right) =$ $\left( 5.5\mathrm{-}7.5,\,\pm k_{F}\right) $ and
weak intensity in the regions $(8.0\mathrm{-}10.5,\,\pm \left( 1\pm
k_{F}\right) ).$ In the former, we see an intensity variation with minimum
at the zone boundary ($\kappa _{a}$=6.5). In the last-mentioned regions, the
intensity along the FS is weak and fairly constant. In FIG.$\,$(b), the $%
\kappa _{c}$=$\frac{1}{4}$ plane, which contains the resonance peak (see
bottom part of FIG.$\,$\ref{ThreeBandskonly}), the intensity variation seems
to be shifted a bit, and again in (c) to be in register with the variation
in (a).

The qualitative theory of the coarse-grained intensity variations presented
in Sect.$\,$\ref{Coarse} associate them with the form factor of the $xy$ WO
in the present case. The theory yields the result shown in FIG.$\,$\ref%
{Re^2xykakbkc002550} which compares quite well with the experimental FIG.$\,$%
\ref{Photon_Energy_Scan}~(a). We thus realize that the region of strong
intensity is due to the beam (\ref{max1xy}) whose center hits the $\kappa
_{c}$=0 plane at $\left( \kappa _{a},\kappa _{b}\right) $=$\left(
6.4,0\right) $, and that the two regions of weak intensity are due to the
beams (\ref{max2xy}) causing the spots at $\left( \kappa _{a},\kappa
_{b}\right) =\left( 9,\pm 1\right) .$

Apart from the observed coarse-grained intensity variations, we confirm that
slightly below the Fermi level there is no dispersion along $\kappa _{a}$,
i.e. there is strongly reduced dimensionality in the $\mathbf{a}^{\ast }$%
-direction$^{\text{I}\,\ref{slabhop}}$. The intensity variation seen as a
function of $\kappa _{a}$ in FIG.$\,$\ref{Photon_Energy_Scan} has period 1
in $\kappa _{a}$ and is, therefore, \emph{not} the fine-grained intensity
variation (zone selection) described in Sect. \ref{Sectzoneselect}, which
has period 2 in $\kappa _{c}$ and is due to emission from \emph{one} of the
two $xy$-bands. The intensity seen could, in principle, be due to $k_{a}$
dispersion$^{\text{I}\,\ref{slabhop}}$, but this, we judge, is far too weak;
so we currently have no explanation.

In the $\mathbf{c}^{\ast }$-direction, i.e. from figure (a) to (b) to (c),
the dispersion is very weak. That there are two bands at the Fermi level\
and the FS therefore has two close-lying Fermi vectors, $k_{Fu}$=$k_{F2}$
and $k_{Fl}$=$k_{F1}$ along $\kappa _{b},$ cannot be seen in these figures,
but might, in principle, be resolved because the resolution is 16~meV when $%
h\nu $=30 eV (and decreases to 40 meV when $h\nu $=100~eV).

\subsubsection{FS intersection with $\mathbf{b}^{\ast }\mathbf{c}^{\ast }$
planes; $h\protect\nu $=30,\thinspace 33, and 37 eV\label{SectDrop}}

\begin{figure}[tbh]
\includegraphics[width=1\linewidth]{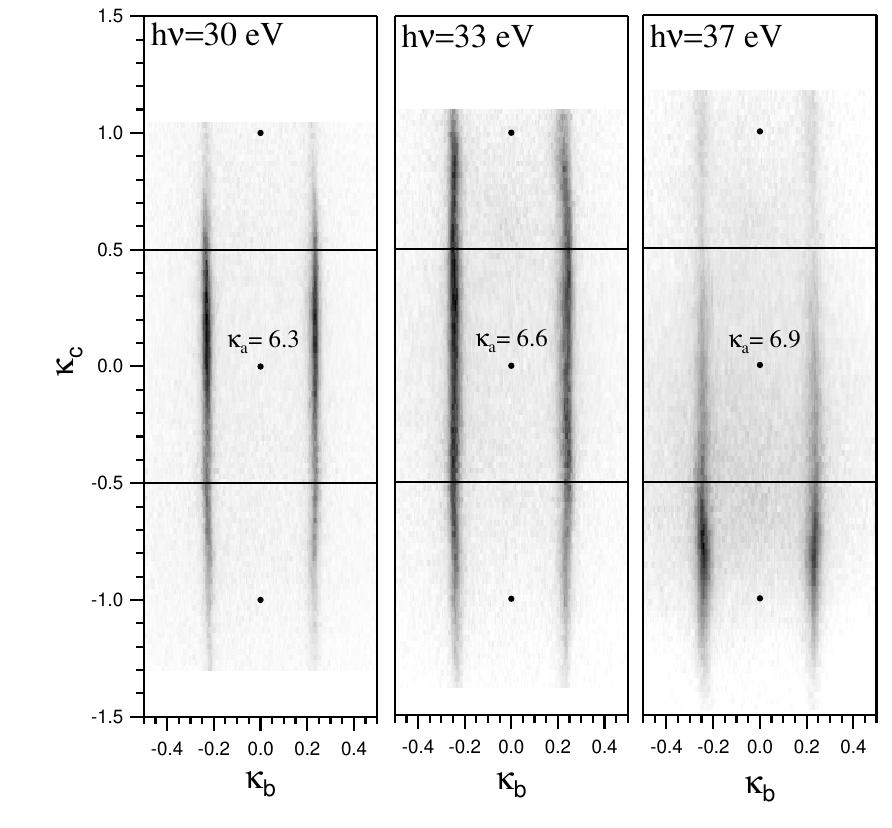}
\caption{FS map at different photon energies $h\protect\nu =$30, 33, 37 eV.
For $\protect\kappa _{b}$=$\protect\kappa _{c}$=$0$, this is at $\protect%
\kappa _{a}$=6.3, 6.6, and 6.9 as indicated. These figures have been
stretched along $\protect\kappa _{c}$ by a factor 1.7 (compare with FIG.s~%
\protect\ref{FIGPhysicalZones2} and \protect\ref{CEC} (b), which are to
scale). }
\label{hv_30_33_37_kbkc}
\end{figure}

Like in FIG.$\,$\ref{Photon_Energy_Scan}, the intensity for photoelectrons
coming from slightly below the Fermi level is shown in FIG.$\,$\ref%
{hv_30_33_37_kbkc}, but now in the $\kappa _{b}\kappa _{c}$ plane, over two
BZs $\left( \left\vert \kappa _{b}\right\vert \leq \frac{1}{2},\text{ }%
\left\vert \kappa _{c}\right\vert \lesssim 1\right) ,$ and for $h\nu =30,$
33, and 37$~$eV. The coarse-grained intensity is seen to change
significantly over this range of photon energies, and only for $33~$eV does
it extend over both BZs. This we can partly understand from the $h\nu $%
-dependence of $\kappa _{a}$:

For the three photon energies, Eq.$~$(\ref{kappaa(0,hv)}) yields
respectively $\kappa _{a}\left( 0,0,h\nu \right) =5.9$, 6.2, and 6.5, which
according to the simplest prediction (\ref{max1xy}) of the $\kappa _{c}$%
-position of the intensity maximum (seen on the left-hand side of FIG.~\ref%
{Re^2ka6068} for $\kappa _{a}$=6.0 and 6.8) for the $xy$ beam gives: $\kappa
_{c}=0.5$, 0.2, and $-$0.1, as compared with 0.2, $\sim $0, and $-$0.6
estimated from the experimental FIG.$\,$\ref{hv_30_33_37_kbkc}.\ The latter $%
\kappa _{c}$-values correspond via Eq.$\,$(\ref{max1xy}) to the somewhat
larger $\kappa _{a}$-values: 6.2, 6.4, and 7.0, whose intensity
distributions can easily be imagined from those for $\kappa _{a}$=0 and 6.8
shown on the left-hand side of FIG. \ref{Re^2ka6068}. These do exhibit the
remarkable contraction along $\kappa _{c}$ of the experimental intensity
distribution seen in FIG.$\,$\ref{hv_30_33_37_kbkc} for the highest energy.
On top of this comes the narrowing of the beam due the delocalization of the 
$\widetilde{xy}$\ WO caused by the downfolding of the valence- and
conduction-band orbitals [see Eq.~III$\,$(\ref{xytilde})].

\subsection{CECs in the $\mathbf{b}^{\ast }\mathbf{c}^{\ast }$ plane\label%
{SectCEC}}

\begin{figure}[tbh]
\includegraphics[width=0.85\linewidth]{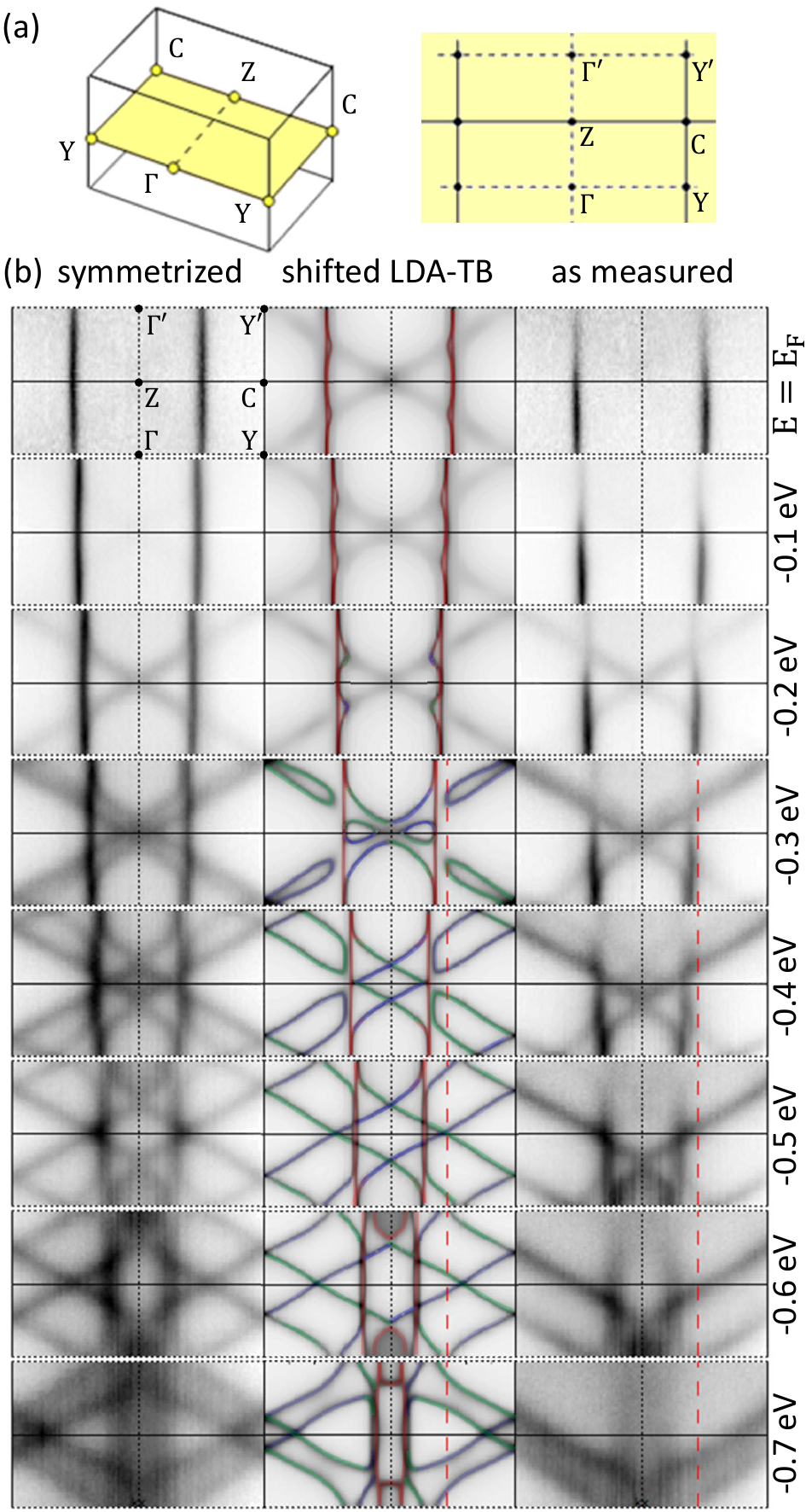}
\caption{\textbf{(a)} Orientation in the zone of the $\protect\kappa _{b}%
\protect\kappa _{c}$-plane with $\protect\kappa _{a}$=6. The CECs are shown
in the upper part of the 1st and the lower part of the 2nd BZ: $\left\vert 
\protect\kappa _{b}\right\vert \leq \frac{1}{2},$ $0\leq \protect\kappa %
_{c}<1.$ See also FIG.~\protect\ref{FIGPhysicalZones2}. \textbf{(b)}
Comparison of the CECs computed using the LDA-TB parameters, listed in
Eq.s~I~(\protect\ref{taup})-(\protect\ref{lp}) including the 100 meV shift,
with those measured by ARPES for sample H ($T$=6$\,$K, $h\protect\nu $%
=30\thinspace eV). Red, blue, and green indicate dominating $xy,$ $xz,$ and $%
yz$ character. The red dashed line $\left( k_{b}\text{=0.225}\right) $ is
the one along which we often show bands, such as the $\left\vert \mathbf{k}%
\right\rangle $-projected ones in FIG.~\protect\ref{ThreeBandskonly}. The
match with the symmetrized ARPES is nearly perfect, and so is the prediction
that in ARPES as measured, the $xz$ and $yz$ CECs are extinguished in the
respective 2nd zone, i.e. above the respective $\mathrm{ZY}^{\prime }$ line.
Indicated on the right are the energies with respect to the Fermi level.}
\label{CEC}
\end{figure}

In FIG.$\,$\ref{CEC} we present ARPES results for binding energies, $\omega
, $ which takes us from the Fermi level to the bottom of the Mo $4d$-bands
so that we also get to see the $xz$ and $yz$ valence bands. The results are
shown in the 2D region $\left( 0\leq \kappa _{b}\leq \frac{1}{2},\text{ 0}%
\leq \kappa _{c}\lesssim 1\right) $ which with reference to FIG.s$\,$\ref%
{CEC} (a) and \ref{FIGPhysicalZones2} includes the upper half of the BZ
centered at $\Gamma _{00}\equiv \Gamma ,$ plus the lower half of the one
centered at $\Gamma _{01}\equiv \Gamma ^{\prime },$ We shall refer to these
BZs as respectively the 1st and the 2nd. Together, they form the upper part
of the double zone centered at $\Gamma .$ Here, "upper" and "lower" refer to
the orientation which has $\kappa _{c}$ pointing upwards.

\bigskip

\bigskip

The right-most panel in FIG.$\,$\ref{CEC}~(b) shows ARPES as measured and
the left-most panel shows ARPES with the fine-grained (plus some of the
coarse-grained) intensity variations symmetrized away by adding the
intensities from both sides of the BZ boundary (CZC,$\,\kappa _{c}$=$\frac{1%
}{2}),$ exploiting the equivalence\footnote{%
Note that whereas the band structure is invariant to the individual mirror
operations $k_{b}\rightarrow -k_{b}$ and $k_{c}\rightarrow -k_{c}$, the
eigenfunctions are merely invariant to the inversion $\left(
k_{b},k_{c}\right) \rightarrow -\left( k_{b},k_{c}\right) $. As a
consequence, the above-mentioned even/odd symmetries around $k_{c}$=--1,0,1
and $k_{c}$=$\frac{1}{2}$ do not hold for the blue and green curves
individually, but only for their sums.\label{inversion}} of $xz$ and $yz$.
This corresponds to adding the intensities at $\mathbf{k}$ and $\mathbf{k+c}%
^{\ast }.$

The middle panel shows the CECs from the LDA-TB bands with the common energy
of the four $xz$ and $yz$ WOs shifted downwards by 100 meV with respect to
the energy of the two $xy$ WOs [see FIG.s$~$\ref{ARPES_Bandstructure_LDA}
(a) and (b)] in order to improve the agreement with the symmetrized ARPES in
the right-most panel. This agreement --down to every detail-- is
astonishing, and so is the straightness of the three sets of lines, even
close to where they cross.\textbf{\ }The theoretical CECs have been colored
red, blue, or green according to their respective $xy,$ $xz,$ or $yz$
character, and in order to mimic the spectral-function broadening of the
dispersion, the shifted LDA-TB bands were broadened by a Lorentzian with
energy-independent width.

The CECs from the \emph{symmetrized} ARPES and the shifted LDA-TB bands
behave as described in Paper I, Sect.$~$\ref{subsubsubCEC} together with
FIG.s$~$\ref{3Dt2gBands}, \ref{ThreePureBands}, \ref{ThreeBands}, and \ref%
{FIGDoubleZone}. The upper part of the double zone is shown also in FIG.~\ref%
{FIGPhysicalZones2} of the present paper.

In FIG.$\,$\ref{CEC} we recognize the CECs of the quasi-1D, degenerate $%
xy\left( \mathbf{k}\right) $ and $xy\left( \mathbf{k+c}^{\ast }\right) $
bands (red) dispersing in the $k_{b}$ direction with the distance $%
2k_{b}\left( E\right) $ between the two sheets increasing like $\sim \left(
2/\pi \right) \sqrt{\left( E-B\right) /\left\vert t\right\vert }$ with $%
t\sim -1~$eV [see Sect.~\ref{Sectt2g} Eq.~I$\,$(\ref{exy})] and heading
towards $\sim \frac{1}{2}$ at the Fermi energy.

The quasi-1D $xz\left( \mathbf{k}\right) $ (blue) and $yz\left( \mathbf{k}%
\right) $ (green) bands dispersing with respectively $k_{c}-k_{b}$ and $%
k_{c}+k_{b}$ [see Eq.$~$I$\,$(\ref{eyz})], are degenerate at their common
bottom at $\Gamma $ where they are also nearly degenerate- and hybridize
with the two $xy$-bands thus giving rise to CECs which are complicated near $%
\Gamma .$ For energies a bit above the common bottom of all three $t_{2g}$
bands$,$ $B\approx E_{F}-0.75$ eV, the blue and green pair of CECs are
parallel with and lie on either side of respectively the $k_{c}-k_{b}=0$ and 
$k_{c}+k_{b}=0$ lines. As the energy increases, so does the distance $\sim
\left( 2/\pi \right) \sqrt{\left( E-B\right) /\left\vert A_{1}\right\vert }$
between each pair of parallel blue or green CECs. This distance is
approximately $\sqrt{t/A_{1}}\sim 1.8$ times the one between the red $xy$
CECs.

The blue and the green $\mathbf{k+c}^{\ast }$-bands are shifted by $\Delta
k_{c}$=$1$ and thus behave in the same way as the respective blue and green $%
\mathbf{k}$-bands. For energies above $\sim -0.5$ eV, the closest CEC pairs
are those on either side of the respective zone boundary, $\left\vert
k_{c}+k_{b}\right\vert =\frac{1}{2}$ or $\left\vert k_{c}-k_{b}\right\vert =%
\frac{1}{2},$ onto which they coalesce when $E\sim E_{F}-2\left\vert
G_{1}\right\vert =E_{F}-0.2\,$eV. Here, 2$G_{1}$ is the electronic
dimerization causing a gap of $\pm 2G_{1}$[see Eq.~I$~$(\ref{gap}) and FIG.~I%
$~$\ref{ThreePureBands} for $k_{b}$=0.225].

We are particularly interested in the hybridization of the $xy$-bands inside
the gap around $E_{F}.$ It can be seen in theory by comparison of the light-
and the dark red bands in FIG.~I~\ref{ThreePureBands} and I~\ref{ThreeBands}%
, respectively, or in the middle and bottom parts of FIG.~\ref%
{ThreeBandskonly} of the present Paper II, and we shall study it in detail
in Paper III. Although bands --and not CECs-- hybridize, we can see the
effect of this hybridization in the LDA-TB part of FIG.$\,$\ref{CEC} at 0.1
and 0.2 eV below $E_{F}$ as four \emph{"notches"} pointing inwards, towards
Z, and we can follow them as the energy is lowered into the valence bands.
The origin of the notches is clearly the energy repulsion between the
hybridizing valence-band edge and one of the two degenerate $xy$-bands; the
other band is unaffected. Since the notch
is sharp, it can only come from an edge of a nearby $yz$- or $xz$-band (but
not from a far-away $yz$- or $xz$-band) with a weak matrix element, and as
long as the notch points towards Z rather than Y, it comes mainly from the
edge of the valence (V) rather than from the edge of the conduction (C)
band. The corresponding peak is in the upper $\widetilde{xy}$\ band (FIG.~I$%
\,$\ref{ThreeBands}) and we shall call it a \emph{resonance peak. }(If we
take the nearly dispersionless $k_{a}$-dimension into account, the peak is%
\textbf{\ }a "mountain ridge" extending along $\mathbf{a}^{\ast })$.

According to the \emph{selection rules} derived in Sect.~\ref{Sectzoneselect}
and illustrated in FIG.~\ref{ThreeBandskonly}, ARPES\emph{\ as measured} in
FIG. \ref{CEC} (b) should see the occupied $\left\vert \mathbf{k}%
\right\rangle $-projected bands in the 1st physical zone and the occupied $%
\left\vert \mathbf{k+c}^{\ast }\right\rangle $-projected bands in the 2nd
physical zone. With the blue $xz$- and green $yz$-bands gapped around the
Fermi level, these bands should be seen in the 1st physical zone only, and
comparison with FIG. \ref{FIGPhysicalZones2} shows this to be the case.
Specifically, the blue and green bands with $E<E_{F}-0.2~$eV are
extinguished in their respective 2nd zone. The dimerization distortion of
the $yz$ (and $xz$) intensities for negative $\kappa _{c}$ predicted in FIG.~%
\ref{FIGyzZoneSelect} for $\kappa _{a}$=6.4, specifically the intensity
enhancement near $\kappa _{c}$=--0.4 for the upper band was seen in the
data; but this is outside the range of positive $\kappa _{c}$ shown in FIG.~%
\ref{CEC} (b).

For the red, metallic $xy$-bands, the Fermi sea inside the outer sheet
should be seen in the 1st BZ and the sea inside the inner sheet should be
seen in the 2nd BZ, with a switch near the BZ boundary, $k_{c}$=$\frac{1}{2}%
. $ This means that the notches should be seen only in the 2nd BZ, but the
drop of intensity for $\kappa _{c}>0.5$ in ARPES as measured with $h\nu $=$%
30~$eV (FIG.$\,$\ref{hv_30_33_37_kbkc}) makes this observation difficult. We
shall return to it in Sect. III$\,$\ref{SectFSExperiment}.

Herewith, we have arrived at the influence of the $yz$ and $xz$ form
factors. In FIG.$\,$\ref{CEC} (b) we see that the intensity as measured for
given binding energy (deep inside the valence bands) is slightly stronger
for $yz\left( \mathbf{k}\right) $ than for $xz\left( \mathbf{k}\right) $ and
increases with $\kappa _{c}$ in the range $\left( 0|1\right) $. As may be
seen from the theoretical FIG.$\,$\ref{Re^2ka6068}, the former property is
consistent with the behavior of $\kappa _{a}^{2}\left\vert w_{yz}\left( 
\mathbf{\kappa }\right) \right\vert ^{2}$ for $\kappa _{a}\sim 6.4$, but the
latter requires $\kappa _{a}\sim 7$ .

\subsection{Energy bands $E_{j}\left( k_{b},k_{c}\right) $\label{SectEk}}

In the preceding section, the CECs \emph{as measured} were shown in the
right-hand panel of FIG.$\,$\ref{CEC}~(b) and were explained as the
fine-grained, double-periodic $\left\vert \mathbf{k}\right\rangle $%
-projection of the occupied part of the lower $m$-band in the 1st- and of
the upper $m$-band in the 2nd physical zone (see Sect.~\ref{Sectzoneselect}%
), on the background of the coarse-grained, aperiodic polarization- and WO
form factor intensity $\kappa _{a}^{2}\left\vert \tilde{w}_{m}\left( \mathbf{%
\kappa }\right) \right\vert ^{2}$~(see Sect.~\ref{Coarse})$.$ Hence, the $xz$
and $yz$-bands were seen only in the respective 1st physical zone, because
their upper bands are empty, and so were the metallic $xy$-bands due to the
drop of their polarization- and form factor intensity for $h\nu $=30~eV ($%
\kappa _{a}$=6.3$)$ in the 2nd BZ.

\bigskip

\begin{figure*}[tbh]
\includegraphics[width=1\linewidth]{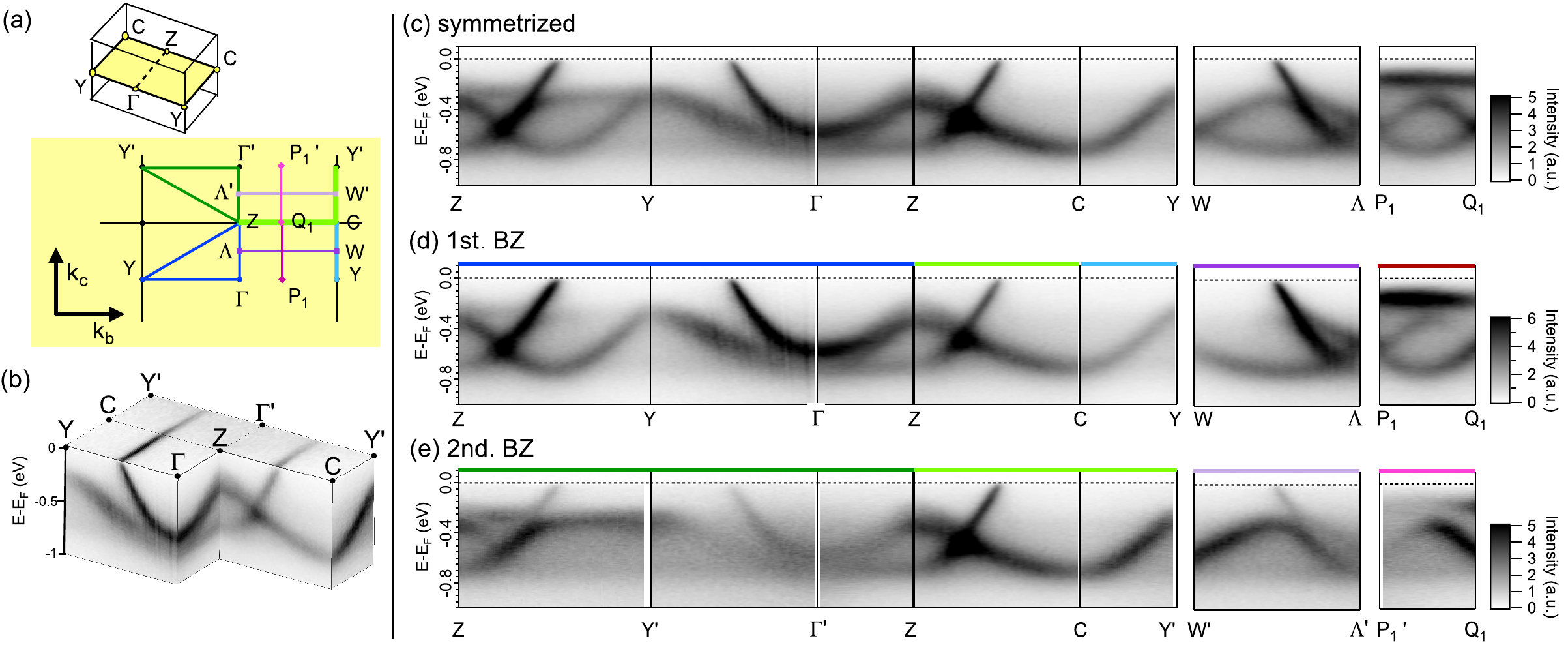}
\caption{ARPES band structure. \textbf{(a)} Orientation of half the 1st BZ
in $\protect\kappa _{a}\protect\kappa _{b}\protect\kappa _{c}$-space and, in
the $\protect\kappa _{b}\protect\kappa _{c}$-plane, half the 1st BZ $(0\leq 
\protect\kappa _{c}\leq \frac{1}{2})$ centered at $\Gamma $, and half the
2nd BZ $\left( \frac{1}{2}\leq \protect\kappa _{c}\leq 1\right) $ centered
at $\Gamma ^{\prime }$. The BZ is the physical zone for the $xy$-bands and
the physical zones for all three $t_{2g}$-bands were shown in FIG.  \protect
\ref{FIGPhysicalZones2}. \textbf{(b)} Data \emph{as measured} from sample H ($%
T$=6\thinspace K, $h\protect\nu $=30\thinspace eV) presented in an $E\left( 
\protect\kappa _{b},\protect\kappa _{c}\right) $-box where from cuts along
the \textbf{$\protect\kappa $}-paths colored in \textbf{(a)} produce the
band structures shown in \textbf{(c)}-\textbf{(e)}. \textbf{(d)\ }Some
features can be better seen in the 1st BZ, \textbf{(e)} while others are
more pronounced in the 2nd BZ, and vice versa. This is so because band
selection follows the physical- rather than the Brillouin zone. \textbf{(c)\ 
}The symmetrized band structure shows all spectral features.}
\label{ARPES_Bandstructure_Cuts}
\end{figure*}

\bigskip

FIG.$~$\ref{ARPES_Bandstructure_Cuts}~(b) now displays the ARPES band
structure as measured on the faces of a box with the basal $\left(
k_{b},k_{c}\right) $ plane extending over the upper half of the double zone,
like in FIG.s$~$\ref{CEC} and \ref{FIGPhysicalZones2}. On the top face, i.e.
for $E\sim E_{F},$\ we recognize the 1D $xy$ FS, $\left( k_{b},k_{c}\right)
\sim \left( \pm \frac{1}{4},k_{c}\right) ,$ with its intensity drop in the
2nd BZ. Not only the $xy$-bands are quasi 1D, but so are the $xz$ and the $%
yz $-bands: Had the box been cut at the top of the valence bands, we would
--like in the right-most panel of FIG.$\,$\ref{CEC}~(b)-- have seen their
ridges follow the boundaries of their physical zones (FIG.~\ref%
{FIGPhysicalZones2}).

The band structures shown in (d)-(e) are obtained by cutting the data along
the lines colored in (a). In order to avoid the fine-grained intensity
variations, we first show (c) --like in the left-most panel of FIG.$~$\ref%
{CEC}~(b)-- the band structure symmetrized over the 1st and 2nd zones; the
benefit of this symmetrization is evident! Had there been no coarse-grained
intensity variations, this band structure would have been periodic in the BZ
and have the rectangle $\mathrm{\Gamma ZCY}$ $\left( 0\leq k_{b}\leq \frac{1%
}{2},0\leq k_{c}\leq \frac{1}{2}\right) $ as its irreducible part. The
symmetrized bands (c) will be compared with the LDA band structure (FIG.~\ref%
{ARPES_Bandstructure_LDA}) in the following Sect.~\ref{SectAgreement} and,
subsequently, its TB parameters will be fine-tuned to achieve almost perfect
agreement with the occupied ARPES bands.

The fine-grained intensity variations, on the other hand, hold the key to
resolving the sofar elusive splitting and warping of the nearly degenerate
quasi-1D $xy$-like FS. But let us first test our understanding of the
intensity variations, by using it to explain the band structure \emph{as
measured} along the lines colored in FIG.~\ref{ARPES_Bandstructure_Cuts}~(a)
and shown in FIG.{s}~(b), (d), and (e).

Along $\mathrm{\Gamma Z\Gamma }^{\prime }$ $(k_{b}$\textrm{=0}$)$\textbf{\ }%
we see the rise of the degenerate $xz\left( \mathbf{k}\right) $ and $%
yz\left( \mathbf{k}\right) $ valence bands from their bottom at $E_{F}-0.75~$%
eV at $\Gamma $ to their highest point, $E_{F}-0.25~$\textrm{eV,} where the
blue and green valence-band ridges cross at $\mathrm{Z.}$ Here, half their $%
\left\vert \mathbf{k}\right\rangle $ characters --and hence ARPES
intensities-- have been lost. On the downturn in the 2nd zones, to $\Gamma
^{\prime }$, the $\left\vert \mathbf{k+c}^{\ast }\right\rangle $ characters
take over and the intensities drop accordingly. At $E_{F}-0.75~$eV we also
see the dispersionless bottom of the two degenerate $xy$-bands and expect
the intensity to shift from the lower to the upper band as we pass from the
1st to the 2nd BZ. That the measured total $xy$ intensity nevertheless
drops, we ascribe to the above-mentioned drop of the form factor in the 2nd
BZ.

When going inside the 1st zone from $\mathrm{\Gamma }$ to $\mathrm{Y}$, the
other crossing point of the blue $xz$ and green $yz$ valence-band ridges
(see FIG.~I~\ref{FIGDoubleZone} or II~\ref{FIGPhysicalZones2}), the $xz\left( 
\mathbf{k}\right) $ and $yz\left( \mathbf{k}\right) $ valence bands are seen
to rise and lose intensity in a similar same way as they did towards $%
\mathrm{Z}$, except that there --due to spin-orbit splitting (see FIG.~\ref%
{LAPWwoSOC})-- the maximum was higher. The $\mathrm{\Gamma Y}$-cut in the
band structure is also shown on the front face of the box in (b). Intensity
is prominently seen from the (lower) $xy$-band rising parabolically from its
bottom along $\Gamma \mathrm{Z}$ to the FS along $k_{b}\approx \frac{1}{4}$.
The aforementioned drop of $xy$ intensity in the 2nd BZ is clearly seen on
the top face of the box. If we go inside the 2nd zones from $\Gamma ^{\prime
}$ to $\mathrm{Y}^{\prime }$ at the zone boundary, this is the only place
where intensity from the valence bands is seen.

Along the green diagonal from $\mathrm{Y}^{\prime }$ to $\mathrm{Z,}$ where
the gapped $yz\left( \mathbf{k}\right) $- and $yz\left( \mathbf{k+c}^{\ast
}\right) $-bands form the valence-band ridge, this ridge is clearly seen; in
fact far better than the blue ridge from $\mathrm{Y}$ to $\mathrm{Z}$ formed
by the gapped $xz\left( \mathbf{k}\right) $- and $xz\left( \mathbf{k+c}^{\ast
}\right) $-bands. A similar intensity difference between $yz$ and $xz$ was
also observed in FIG.$\,$\ref{CEC} and was explained at the end of Sect.$~$%
\ref{SectCEC} as due to different form factors. The $xz$-band gives no
intensity along the green $\mathrm{Y}^{\prime }\mathrm{Z}$ line because it
is in the 2nd $yz$ zone, which is empty. Returning now to the blue
valence-band ridge along $\mathrm{YZ,}$ we see the $yz$-band fall, reach its
bottom halfway towards $\mathrm{Z,}$ and then rise again to the highest
point on the ridge at $\mathrm{Z.}$ Here, the $yz$-band not only attains $%
\left\vert \mathbf{k+c}^{\ast }\right\rangle $ character, but also
hybridizes with the $xz$-band via spin-orbit coupling. We also see the
parabolic rise of the lower $xy$-band from its bottom at $\mathrm{Z}$
towards $\mathrm{Y}$, reaching $E_{F}$\ at half the way$.$ Towards $\mathrm{Y%
}^{\prime },$ the $xy$ intensity is reduced by the form factor.

\bigskip

\begin{figure*}[tbh]
\includegraphics[width=\linewidth]{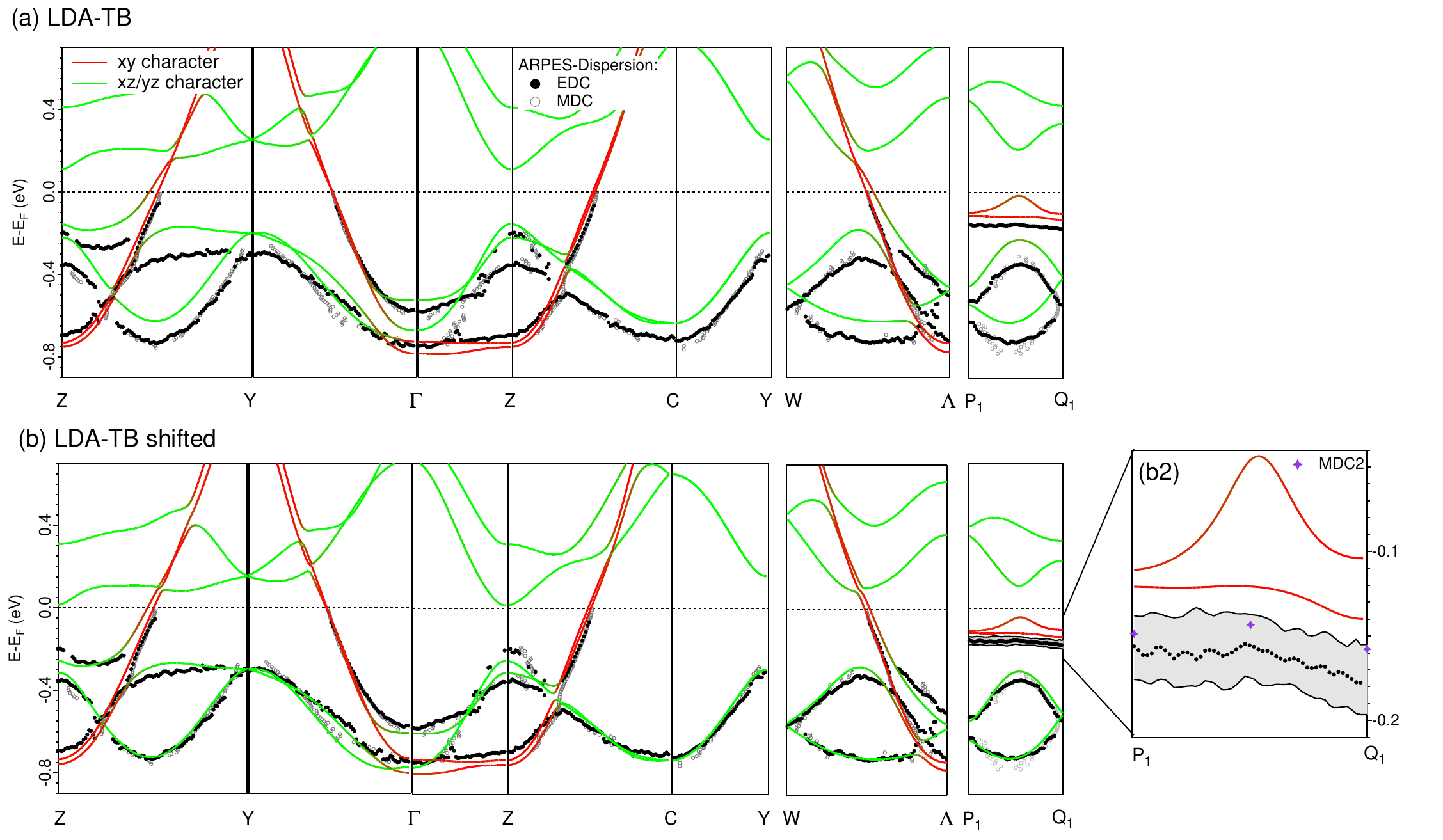}
\caption{ ARPES (black dots and grey circles) and LDA ($xy$\ red and $xz/yz$
green) band structures. The line \textrm{P}$_{1}$\textrm{Q}$_{1}$ has $k_{b}$%
=0.225 and $k_{c}$=0 to 0.5, see FIG.s~\protect\ref{FIGPhysicalZones2}, 
\protect\ref{CEC}, and \protect\ref{ARPES_Bandstructure_Cuts}\thinspace (a).
The experimental bands come from the \emph{symmetrized} ARPES measurements ($%
T$=6K, $h\protect\nu $=30~eV) on sample H whose metallic bands are 51\%
filled (see Sect.~\protect\ref{SectFS} in Paper III) and displayed in FIG{.}%
~ \protect\ref{ARPES_Bandstructure_Cuts} (c). They were determined by
searching the intensity maximum along the direction of either energy (EDC,
black dots) or momentum (MDC, grey circles). (\textbf{a}) The LDA bands are
the eigenvalues of the six-band TB Hamiltonian defined in Sect.~I$~$  
\protect\ref{SectH} with the parameters listed in Tables I$\,$(\protect\ref%
{taup})-(\protect\ref{lp}) and (\protect\ref{EFARPES}). \textbf{(b)} As
above, but with the energy of the $xy$ WOs shifted 100 meV upwards. The
blow-up along \textrm{P}$_{1}$\textrm{Q}$_{1}$ in (b2) compares the shifted
LDA-TB theory with the experimental band (black dots with uncertainty in
grey). The diamonds, labeled 'MDC2', indicate the band determined from the
maximum along the momentum direction $\mathbf{b}^{\ast }$ perpendicular to $%
\mathrm{P}_{1}\mathrm{Q}_{1},$ where the band disperses strongly.}
\label{ARPES_Bandstructure_LDA}
\end{figure*}

\bigskip

Going along the BZ boundary from $\mathrm{Z}$ to $\mathrm{C}$ for $k_{b}$
positive, the nearly degenerate $xy$-bands rise and cross the Fermi level
near $\left( \frac{1}{4},\frac{1}{2}\right) $ and the $xz\mathbf{\ }$valence
band falls from the highest point on the ridge to its bottom at $\mathrm{C.}$
At $\mathrm{Z,}$ the green $yz$-band is degenerate with the blue $xz$
valence band and has intensity there, which however vanishes after leaving
the 1st green zone.

Upon going from\textrm{\ }$\mathrm{C}$ to $\mathrm{Y}^{\prime }$ ($k_{b}$=$%
\frac{1}{2})$ inside the blue $xz$ zone, we see the $xz$-band increase to
the blue ridge at $\mathrm{Y}^{\prime }$. Going instead to $\mathrm{Y,}$ we
see the other branch of the $xz$-band increase to the blue ridge at $\mathrm{%
Y.}$

Along $\mathrm{\Lambda W}\,$($k_{c}$=$\frac{1}{4}$), the lowest band --with
minimum at ($\frac{1}{4}$,$\frac{1}{4}$)-- is the blue $xz\left( \mathbf{k}%
\right) $-band. The $yz\left( \mathbf{k}\right) $-band is degenerate with $%
xz\left( \mathbf{k}\right) $ at $\Lambda ,$ but then increases, until at ($%
\frac{1}{4}$,$\frac{1}{4}$) it reaches the top of the green $\mathrm{ZY}$
ridge where it mixes with the $yz\left( \mathbf{k+c}^{\ast }\right) $-band
coming from $\mathrm{W,}$ and thereby looses its intensity. Going instead
from $\mathrm{\Lambda }^{\prime }$ towards $\mathrm{W}^{\prime }\mathrm{\ }$(%
$k_{c}$=$\frac{3}{4}$), it is the blue $xz\left( \mathbf{k+c}^{\ast }\right) 
$-band which --with weak intensity-- increases until at ($\frac{1}{4}$,$%
\frac{3}{4}$) it reaches the top of the blue $\mathrm{ZY}^{\prime }$ ridge
where the $\left\vert \mathbf{k}\right\rangle $ character takes over and the 
$xz\left( \mathbf{k}\right) $-band continues with full intensity downhill.
The lower $xy$-band disperses parabolically upwards from $\mathrm{\Lambda }$
and reaches the Fermi level half the way to $\mathrm{W,}$ and the same is
seen --with reduced intensity-- along $\mathrm{\Lambda }^{\prime }\mathrm{W}%
^{\prime }\,$($k_{c}$=$\frac{3}{4}$) for the higer $xy$-band. The
non-vanishing hopping integrals I~(\ref{ap}) between an $xy$ WO and an $xz$
or $yz$ WO causes the corresponding bands to repel where they run close.
This is the case for the two $xy$-bands and the $yz\left( \mathbf{k}\right) $
band between $\Lambda $ and $\left( \frac{1}{4},\frac{1}{4}\right) ,$ and
for the two $xy$-bands and the $xz\left( \mathbf{k}\right) $-band between $%
\Lambda ^{\prime }$ and $\left( \frac{1}{4},\frac{3}{4}\right) .$ It is
remarkable that of the three close bands, \emph{two} repel around the third
band, which remains unaffected and thereby ends up as the band of
intermediate energy, the lowest of the two $xy$-bands in the present case.
This can be seen in FIG.$\,$\ref{ARPES_Bandstructure_Cuts} (c)-(e) and will
be referred to often in the following.

The $\mathrm{P}_{1}\mathrm{Q}_{1}\mathrm{P}_{1}^{\prime }$-line ($k_{b}$%
=0.225) is perpendicular to the $\mathrm{\Lambda W}$ and $\mathrm{\Lambda }%
^{\prime }\mathrm{W}^{\prime }$ lines and is parallel to, but slightly
inside the FS so that the trace of the $xy$-bands is $\mathrm{\sim }170$ meV
below the Fermi level and thereby clearly visible in ARPES --albeit with the
usual $xy$ form-factor reduction of the intensity towards $\mathrm{P}%
_{1}^{\prime }$. Starting from $\mathrm{P}_{1},$ we see the blue $xz\left( 
\mathbf{k}\right) $-band reaching its bottom midways to $\mathrm{Q}_{1}$ and
--above it, with slightly less intensity,-- the green $yz\left( \mathbf{k}%
\right) $-band rising towards the top of the green ridge where 50\% of its
character and intensity are lost and losing even more on the downhill side
towards $\mathrm{Q}_{1}$. Proceeding from here towards $\mathrm{P}%
_{1}^{\prime }$ in the 2nd green zone, we vaguely see the green $yz$-band
--now $\left\vert \mathbf{k+c}^{\ast }\right\rangle $-like-- reach its
bottom midways to $\mathrm{P}_{1}^{\prime }.$ From $\mathrm{Q}_{1}$ in the
blue 1st zone, we see the blue $xz$-band rise to the $\mathrm{ZY}^{\prime }$
ridge, where it --like the green $yz$-band at the green ridge-- has lost
half its intensity and thereafter vanish on the downhill side towards $%
\mathrm{P}_{1}^{\prime }.$ All of this agrees with the $\left\vert \mathbf{k}%
\right\rangle $ characters of the four lowest LDA-TB bands shown at the
bottom of FIG.$\,$\ref{ThreeBandskonly} and, at the top, with the $xy$
form-factor dropping in the 2nd BZ thus hiding the resonance peak predicted
to exist in the upper $\widetilde{xy}$-band.

Before attempting to extract ARPES data beyond the 100 meV scale, we need to
assess the degree of agreement between the LDA and the ARPES
dispersions.\bigskip

\begin{figure}[tbh]
\includegraphics[width=1\linewidth]{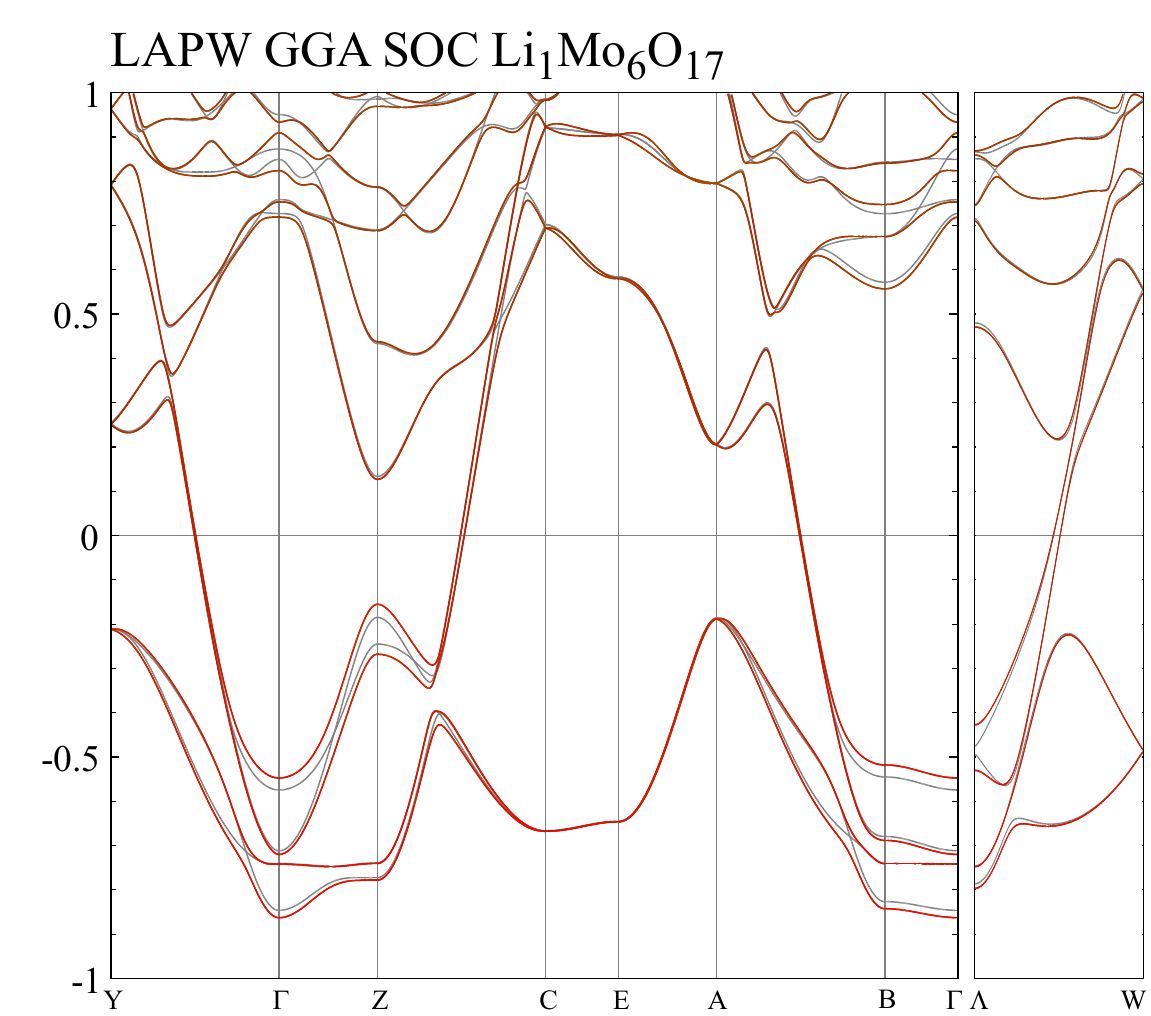}
\caption{Relativistic LAPW GGA control calculation of the LiPB band
structure using the Perdew–Burke–Ernzerhof (PBE) functional~\cite{PBE1996}.
We show the bands along the same lines in the BZ as shown at the top of FIG.~I$\,$\protect\ref%
{FIG2} whose bottom shows the 336 NMTO LDA band structure. The 6 NMTO TB
band structure in the Y$\Gamma $ZC plane is shown in FIG.~I~\protect\ref%
{3Dt2gBands}. LAPW band energies are in eV with respect to $E_{F}.$ Red
bands are with- and grey bands are without spin-orbit coupling. We see that
the crossings between the 3rd and 4th bands along $\mathrm{\Gamma Z}$ and $%
\mathrm{ZC}$ and between the 1st and 2nd bands between $\mathrm{Y\Gamma }$
and $\mathrm{AB}$ are spin-orbit split by about $80$ meV.}
\label{LAPWwoSOC}
\end{figure}

\section{Agreement between ARPES and the LDA\label{SectAgreement}}

The band structure derived from the symmetrized ARPES data is shown in FIG.$%
~ $\ref{ARPES_Bandstructure_LDA}~(a) and (b). The points indicated by grey
circles were extracted from the peak locations in the momentum distribution
curves (MDCs), $I(|\mathbf{k}|$\ in a specified direction, fixed energy $E)$%
, and the points indicated by black dots were extracted from peak locations
in energy distribution curves (EDCs), $I$(fixed~$\mathbf{k},E)$. The peak
maxima in these curves were found from the zeros in the smoothed first
derivative with respect to $|\mathbf{k}|$ for an MDC, or $E$ for an EDC,
under the condition that the smoothed second derivative is smaller than zero.

The bands resulting from the first-principles LDA calculation (in the TB
representation of the six $t_{2g}$ WOs; see Sect.~\ref{SectLowE} in paper I) are
shown in (a), with the sum of the $xz$ and $yz$ characters in green and the $%
xy$ character in red. Note the strong hybridization of the nearly degenerate
bottoms of the two $xy$ and the $xz$ and $yz$ valence bands near $\Gamma .$
Note also the hybridizations between $\mathrm{Z}$ and $\mathrm{Y}$ and
between $\mathrm{W}$ and $\mathrm{\Lambda .}$ Near the middle of the latter
line where the two degenerate $xy$-bands in the gap come close to the\
valence and conduction (V\&C) bands, and their repulsion is therefore strong
--but in opposite directions-- one of the $xy$ states stays unaffected and
the other is pushed up or down in energy, depending on whether the repulsion
from the valence or the conduction band is stronger. Since this balance tips
as we move up through the gap, the hybridization shifts from the upper to
the lower $xy$-band and causes the resonance peak to shift from upwards
pointing in the upper band, to downwards pointing in the lower band. The
fact that the matrix element, $\alpha \left( \mathbf{k}\right) +a\left( 
\mathbf{k}\right) ,$ for hybridization of the $xy$- and $xz/yz$-bands
decreases with increasing $k_{b}$ causes the rather strange-looking
dispersion of the two $xy$-bands along $\Lambda \mathrm{W}$. This asymmetry
will be explained in Sect.$~$III$\,$\ref{SectOrigins}; specifically, in
connection with Eq.$~$III$\,$(\ref{Goff}) and with the left-hand sides in
the 6th row of FIG.$~$III$\,$\ref{Analysis}. Along $\mathrm{P}_{1}\mathrm{Q}%
_{1}$, we clearly see the resonance peak in the upper $\widetilde{xy}$-band.

\subsection{Shifting the LDA $xz$- and $yz$-bands downwards with respect to
the $xy$-bands \label{SectShifting}}

Overall in FIG.$~$\ref{ARPES_Bandstructure_LDA}$\,$(a), there is good
agreement with the occupied ARPES bands, the main discrepancy being that the
LDA valence bands lie 100 meV too high with respect to the $xy$-bands and
thereby with respect to the Fermi level. This may be partly a surface
effect: The $xz$ and $yz$ WOs reach farther into the vacuum and therefore
feel a higher LDA potential than the $xy$ WOs which are well inside the
staircase. In addition, there is undoubtedly an LDA error; for instance, 
LDA bandgaps in semiconductors are too small, and FS measurements for bulk $4d$
metals indicate that the accuracy with which the LDA describes the energy
separation between inequivalent $t_{2g}$ levels is $\sim $50~meV.\footnote{{%
For elemental transition metals, 100 meV is the typical size of the $s$ to $%
d $ energy shift needed to bring the LDA and experimental (dHvA) Fermi
surfaces into agreement, see \cite{Mackintosh1980}.}} For LiPB, we therefore
correct the bulk LDA bands by shifting the energy of the green V\&C bands
downwards by 100 meV [more precisely, we shift the on-site energy, $\tau
_{0},$ of the $xy$ and $XY$ WOs 100 meV upwards, i.e. from 47 to 147 meV in
Table I$\,$(\ref{taup}), with respect to the common on-site energy of the $%
xz,XZ,yz,$ and $YZ$ WOs, and subsequently recalculate the Fermi level]. The
result shown in (b) agrees very well with ARPES, as was seen already in FIG.$%
~$\ref{CEC}$\,($b$)$ for the CECs. An exception is near $\mathrm{Z,}$ where
the splitting of the valence band is too small and the lowest conduction
band nearly touches the Fermi level, thus asking for a fine-adjustment of
the TB parameters.

The LDA and ARPES band structures in FIG.s$\,$\ref{ARPES_Bandstructure_LDA}
(a) and (b), are lined up with respect to the Fermi level, which for the LDA
calculation for stoichiometric LiMo$_{6}$O$_{17}$ corresponds to 50\%
filling of the metallic bands. In the experiment, Li and O vacancies make
the filling uncertain and is estimated from the measured $k_{Fb}$ value
(Sect.~III$\,$\ref{SectFSExperiment}) to be $51\pm 1$~$\%,$ i.e. to have the
effective stoichiometry Li$_{1.02\pm 0.02}$. Using the measured
Fermi-velocity, this then gives a Fermi level which with respect to the band
structure is between 50 and 0 meV above the level for the stoichiometric
crystal assumed in the calculation, which means that the metallic ARPES
bands could lie 50 to 0 meV below the LDA bands in FIG.$\,$\ref%
{ARPES_Bandstructure_LDA}\thinspace (b). But this can only account for the $%
\sim $40~meV distance to the lower LDA $\widetilde{xy}$-band seen in a
direction perpendicular to $\mathbf{b}^{\ast },$ such as along $\mathrm{P}%
_{1}\mathrm{Q}_{1},$ in particular in the blow up (b2).

The black dots in (b2) were obtained as the position of the EDC maxima (%
\emph{one} for each $\mathbf{k}$\textbf{) }and the grey area indicates the
uncertainty of the experiment, as well as the uncertainty in determining the
position of the one EDC maximum. The three purple diamonds labeled 'MDC2'
are from MDCs along respectively $\mathrm{\Gamma Y,}$ $\Lambda \mathrm{W,}$
and $\mathrm{ZC}$, perpendicular to $\mathrm{P}_{1}\mathrm{Q}_{1}.$

The upper $\widetilde{xy}$-band predicted by the LDA, seems to be missing in
the symmetrized ARPES data (b2). This, we can understand by going back to
the ARPES bands "as-measured" along $\mathrm{P}_{1}\mathrm{Q}_{1}\mathrm{P}%
_{1}^{\prime }$ in FIG.$~$\ref{ARPES_Bandstructure_Cuts} (d) and (e) where
intensity just below the Fermi level was seen in the 1st- but hardly in the
2nd BZ. As illustrated in FIG.$~$\ref{ThreeBandskonly}, the lower $xy$-band
is selected in the 1st- and the upper band in the 2nd BZ. Since for $k_{b}$%
=0.225, the $xy$-bands are closer to the valence than to the conduction
bands, the repulsion from the former dominates and pushes a resonance peak 
\emph{up} in the \emph{upper} band. This peak thus has ARPES intensity in
the 2nd zone where it gets strongly reduced by the $xy$ form factor. The
symmetrization of the $\mathrm{P}_{1}\mathrm{Q}_{1}$ and $\mathrm{Q}_{1}%
\mathrm{P}_{1}^{\prime }$ ARPES data finally adds to hiding the band in the
2nd BZ behind the one seen in the 1st BZ, and that is the reason why in FIG.$%
\,$\ref{ARPES_Bandstructure_LDA}$\,$(b2) only the lower ARPES band is seen.
From the ARPES data, we can therefore only uncover the upper $\widetilde{xy}$
band if we \emph{avoid} the symmetrization.

\bigskip

\begin{figure*}[tbh]
\includegraphics[width=\linewidth]{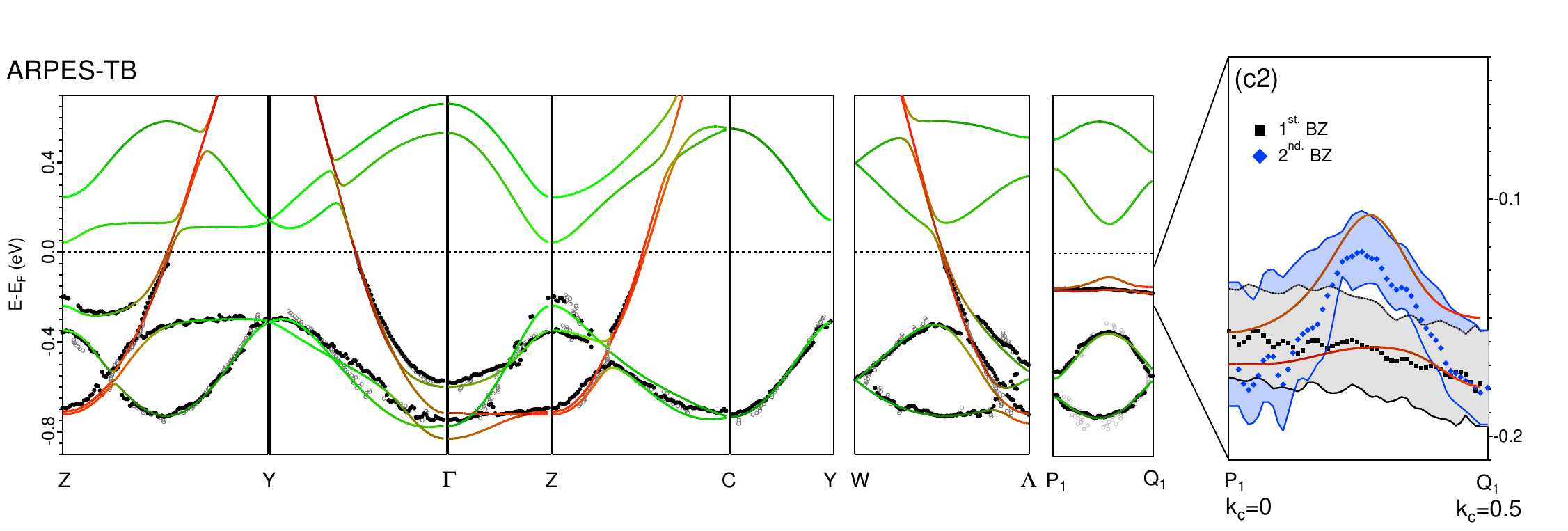}
\caption{Band structure obtained as the eigenvalues of the six-band TB
Hamiltonian, Eq.s I$~$(\protect\ref{Hsub}) or I$\,\,$(\protect\ref{HRecip}),
with parameter values fine-tuned to improve the fit to the ARPES bands and
given in the square parentheses in Tables I$\,$(\protect\ref{taup})-( 
\protect\ref{lp}). The Fermi level is $\sim $0.1~eV above the center of the
gap [see text following Eq.~I~(\protect\ref{EFARPES})]. The ARPES data here
are those already shown in FIG. \protect\ref{ARPES_Bandstructure_LDA},
except for those in the blow-up (c2) along $\mathrm{P}_{1}\mathrm{Q}_{1}$ ($%
k_{b}$=0.225), which are \emph{non}-symmetrized EDC data from respectively
the 1st (black) and 2nd BZ. The latter data is subsequently mirrored around
the $k_{c}$=0.5 zone boundary into the 1st BZ (blue).}
\label{ARPES_Bandstructure_TB}
\end{figure*}

\subsection{Fine-tuning the TB Hamiltonian and extracting\ both quasi-1D
bands from ARPES as measured \label{SectRefining}}

Our first-principles LDA TB description of the six lowest energy bands given
in Sect.~I~\ref{SectH} has about 40 TB parameters (WO energies and hopping
integrals). Their LDA values, derived as described in Sect.~I~\ref%
{SectElCalc}, are given in Tables I$\,$(\ref{taup})-(\ref{lp}) and yield the
band structure shown in FIG.$\,$\ref{ARPES_Bandstructure_LDA} (a). As
explained above, the improved agreement with ARPES seen in (b) was achieved
by merely shifting the value of $\tau _{0},$ the energy of the degenerate $%
xy $ and $XY$ WOs, up with respect to that of the degenerate $xz,$ $XZ,$ $%
yz, $ and $YZ$ WOs. The latter is the energy at the center of the gap and is
taken as the zero of energy in the TB Hamiltonian. Still, the Fermi velocity
is too small and the resonance peak along P$_{1}$Q$_{1}$ is too high.
Moreover, the levels near $Z,$ where the V\&C bands come closest, remain
inaccurate: the splitting of the valence band levels is too small while that
of the conduction band levels is too large. This is due to bad convergence
and truncation of the $xz$-$yz$ hybridizations I$\,$(\ref{lambdalmum}), to
our neglect of the spin-orbit coupling in the NMTO --but not in the LAPW
(FIG.~\ref{LAPWwoSOC})-- calculation, and to the LDA yielding too small a
gap.

We therefore refined the parameter values$,$ with the results given in
square parentheses in Tables I$\,$(\ref{taup})-(\ref{lp}). Specifically, we
found it necessary to modify the values of the intra-ribbon hopping
integrals, first of all, between respectively the on-site, 1st-, and
2nd-nearest $xy$ WOs, $\tau _{0},$ $\tau _{1},$ and $\tau _{2}.$ This
increases the Fermi velocity by about 15\%, increases the upward curvature
of the bands near half filling, and lowers the half-filling Fermi level to
53 meV above the center of the gap; see Eq.s I$\,$(\ref{taup}) and I$\,$(\ref%
{lin})-(\ref{vF}) in Sect.$~$I$~$\ref{SectH}. Secondly, we refined the
values of the $xz$-$yz$ hopping integrals, $m_{1},$ $\lambda _{2},$ and $%
\lambda _{3}$ in Eq.~I$\,$(\ref{lp}). In addition, the value of the gap
parameter, $G_{1}$, originating in the asymmetry (dimerization) between the
forwards and backwards hoppings, $xz\leftrightarrow XZ$ $\left(
yx\leftrightarrow YZ\right) ,$ as explained after Eq.$~$I$\,$(\ref{eyz}) and
given in Eq.~I$\,$(\ref{A&Gp}), was increased by 10\%. The resulting band
structure is displayed in FIG.$\,$\ref{ARPES_Bandstructure_TB} and is seen
to agree almost perfectly with the experiment.

Most importantly, we have succeeded in extracting \emph{both} metallic bands
by using the ARPES data \emph{as measured} along $\mathrm{P}_{1}\mathrm{Q}%
_{1}\mathrm{P}_{1}^{\prime }$ ($k_{b}$=0.225) for the EDCs. These ARPES
bands are displayed in the blow-up (c2) where the band obtained in the 1st
zone from $\mathrm{P}_{1}$ to $\mathrm{Q}_{1}$ has been plotted in black
squares with the uncertainty in grey, and the band obtained in the 2nd zone
from $\mathrm{Q}_{1}$ to $\mathrm{P}_{1}^{\prime }$ has been plotted in
reverse order, from $\mathrm{Q}_{1}$ to $\mathrm{P}_{1}$ in the 1st zone,
and in blue. Even though the 2nd BZ emission is relatively weak in this h$\nu$=30 eV data (recall from Sec.~\ref{SectDrop}),  nonetheless, we now see at mid-zone the resonance peak in the blue upper band, well
separated from the grey lower band, while near $k_{c}$$=0$, $0.5$, and $1$, the
splitting of the two bands is less clear. The reasons for the latter are
technical:

1) The extraction was done by assuming only one (possibly broad) maximum per
EDC whereby in the zone-boundary (ZB) region $\left( 0.4<k_{c}<0.6\right) $\
where both bands have a significant $\left\vert \mathbf{k}\right\rangle $
projection, only one band is found (see red bands in the middle and bottom
parts of FIG.$\,$\ref{ThreeBandskonly}). In FIG.~III$~$\ref{Fig:NewFSExtract}
we shall see an example of how the one-peak extraction method switches
between the two bands when the ZB is crossed.

2) With the ARPES intensity fading away upon approaching $k_{c}$$=$$1$ ($\mathrm{%
P}_{1}^{\prime }$) smoothening the EDCs before taking the derivative (to
locate the maximum) pushes the intensity to increasingly higher binding
energies, $\omega ,$ and thereby causes the band near $\mathrm{P}%
_{1}^{\prime },$ which is the (blue) upper band, to fall below the one near $%
\mathrm{P}_{1},$ which is the (grey) lower band\footnote{%
The uncertainty of this is not included in the blue uncertainty interval.}.

FIG.$\,$\ref{ARPES_Bandstructure_TB} thus demonstrates that the refinement
of merely 7 out of the more than 40 TB parameters to fit the ARPES bands,
lying more than 0.15~eV below the Fermi level, achieves nearly perfect
agreement \emph{also} for the $k_{c}$-dispersion of the quasi-1D bands
closer to the Fermi level. This includes agreement with the size and shape
of the resonance peak in the upper band, \emph{without} having modified any
of the 17 $\left( a,g,\alpha ,\gamma \right) $-parameters I~(\ref{ap})
describing hybridization between the $xy$ and the $xz$- and $yz$-bands. This
clearly shows the decisive role that resonant coupling to the gapped $xz$
and $yz$-bands plays in determining the splitting and dispersion of the
metallic $xy$-bands.

Besides the above-mentioned peculiarities and experimental uncertainties,
the agreement between ARPES and the refined TB bands is astonishingly good.
In both cases, the peak caused by the resonance with the $xz$ valence band
at $\left( k_{b},k_{c}\right) $=(0.225, 0.725) --mirrored around the $k_{c}$%
=0.5 line to the resonance with the $yz$ valence band at (0.225, 0.275)--
comes out clearly in the upper $\widetilde{xy}$-band, and so does the \emph{%
lack} of a visible resonance peak in the \emph{lower} band at (0.225,0.275).

\subsection{True Value of the Fine Tuning Fit}

While it is intrinsically very satisfying to obtain a good fit to the ARPES data, the true value of the fit is the very fact that it was possible to do.  Thus we know that our analytic TB representation of the bands, as inferred from the LDA results, is fundamentally correct at the qualitative level, and is not missing any essential underlying physics.  Otherwise, merely varying the magnitudes of parameters could not achieve a good fit.  In particular, as emphasized already, obtaining a good description of the low-energy FS features by fitting only the high-energy features serves to validate our essential insight that the details of these FS features do indeed result from their couplings to the higher energy valence and conduction bands.

Therefore we proceed to use the refined TB model in the following Paper III to focus on the metallic $\widetilde{xy}$-bands in the gap, the origins of their observed $k_c$-dispersion on the 10 meV-scale, and their $k_c$-dispersion as a function of their position in the gap, i.e. of the $k_b$-value. This will enable us to study the details of the FS. We shall see that the theoretical splitting between the two metallic bands near the $\vert k_c\vert$=0.5 zone boundary increases as the energy moves away from the valence - and towards the conduction band. With the h$\nu$=33 eV data (see FIG. \ref{hv_30_33_37_kbkc}) the zone-selection rule, and a detailed analysis of our high-resolution data, we shall be able also with ARPES to separate the two FS sheets.

\section{Summary and Outlook}

In Sect. \ref{SectIntensity} of this Paper II we derived a one-electron
theory of the ARPES intensity variations in LiPB, which was then used in
Sect. \ref{SectARPESData} to understand and analyze the extensive data
presented there.

For the intensity of photoemission with momentum $\mathbf{\kappa }$ and
binding energy $\omega $ $\left( \geq 0\right) ,$ we used the one-step
expression (\ref{intensity}) with the least specific choice, plane waves $%
e^{2\pi i\mathbf{\kappa \cdot r}},$ for the final states, and approximation (%
\ref{me}) for the matrix element. As a basis for the initial states, we used
the six $t_{2g}$ WOs, $w_{m}\left( \mathbf{r}\right) $ and $W_{m}\left( 
\mathbf{r}\right) ,$ centered on respectively Mo1 and MO1, the most central
molybdenums of the lower and the upper strings, $^{\text{Mo2}}\diagdown _{%
\text{Mo1}}\diagup ^{\text{Mo4}}\diagdown _{\text{Mo5}}$ and $^{\text{MO5}%
}\diagdown _{\text{MO4}}\diagup ^{\text{MO1}}\diagdown _{\text{MO2}},$ shown
in FIG.$\,$I~\ref{FIG1} (a). Due to the approximate translational
equivalence [see Eq.s I$\,($\ref{undim}$)$ and (\ref{inv})] of $w_{m}\left( 
\mathbf{r}\right) $ and $W_{m}\left( \mathbf{r}\right) ,$ the photoemission
intensity is essentially the projection of the initial-state band with
energy $E\left( \mathbf{k}\right) =E_{F}-\omega ,$ onto the \emph{pseudo}
Bloch sum, $\left\vert w;\mathbf{k}\right\rangle $ in Eq.$\,$I$\,$(\ref{kket}%
) with $\mathbf{k=\kappa -G}$ in the \emph{double} zone (see FIG.s~\ref%
{FIGPhysicalZones2} and FIG.~\ref{ThreeBandskonly} \emph{bottom}). This would
have been the initial-state band had the vector distance between Mo1 and MO1
been $\frac{\mathbf{c+b}}{2},$ rather than $\frac{\mathbf{c+b}}{2}-\mathbf{d,%
}$ and had $W\left( \mathbf{r}\right) $ been equal to $w\left( \mathbf{r}%
\right) ,$ rather than inverted around $\frac{\mathbf{c+b}}{2}-\mathbf{d.}$
For a band with dominant $m$ character, we thus expected that ARPES will see
the lower band if $\mathbf{\kappa }$ is in the 1st physical zone, and the
upper band if $\mathbf{\kappa }$ is in the 2nd physical zone and the band is
occupied. Specifically, that the relative intensity of emission from the $%
\QATOP{\mathrm{upper}}{\mathrm{lower}}$ $m$\ band is $\frac{1}{2}\left[ 1\mp
\cos \phi \right] ,$ where $\phi _{m}\left( \mathbf{k}\right) $ is given by
the $m$-band structure in Eq. I$\,($\ref{Phi}$).$ Taking now the $c$-axis
dimerization into account, we found that this expression is modified to: $%
\frac{1}{2}\left[ 1\mp \cos \left( \phi -\eta \right) \right] ,$ where $\eta
_{m}(\mathbf{\kappa )\equiv }$ $2\arg \tilde{w}_{m}\left( \mathbf{\kappa }%
\right) -2\pi \mathbf{\kappa \cdot d}$ is the difference between the phase
shifts due to the inversion- and displacement dimerizations. These phase
shifts were shown in the 1st column of FIG.s \ref{FIGxyZoneSelectul} and \ref%
{FIGyzZoneSelect} for respectively the $xy$ and $yz$-bands. Whereas the
inversion phase shift depends stronger on $\mathbf{\kappa }$ for the $yz$ WO
than for the more structurally protected $xy$ WO, the displacement phase
shift is independent of $m$ and rather constant. We had therefore expected
the selection rule to hold better for the $xy$-bands. However, in both cases
and with $\kappa _{a}$ (which does not influence the band structure)
suitably chosen via the photon energy, the inversion- and displacement phase
shifts tend to cancel with the result that the selection rule holds
surprisinly well, provided that $\kappa _{a}$ is properly chosen. This was
seen in the 2nd column of FIG.s \ref{FIGxyZoneSelectul} and \ref%
{FIGyzZoneSelect}.

On top of this fine-grained structure of the photoemission intensity, there
is a coarse-grained aperiodic structure given by the WO \emph{form factor} $%
\left\vert \tilde{w}_{m}\left( \mathbf{\kappa }\right) \right\vert ^{2}$.
Since in real space the $t_{2g}$ WO (FIG.~I$\,$\ref{Wannier}) spreads with
the same $m$ character onto the 4-5 nearest Mo sites in its plane and on the
simple cubic $xyz$ lattice I$~$(\ref{xyz}) with 1 Mo per cell, its FT (\ref%
{FTWO}) factorizes approximately into a \emph{structure factor, }$\mathcal{S}%
_{m}\left( \mathbf{\kappa }\right) ,$ times the FT of the "atomic" part of
the $t_{2g}$ orbital, factorizing into an angular and a radial part. As
explained in Sect.~\ref{Coarse}, the structure factors $\left\vert \mathcal{S%
}_{xy}\right\vert ^{2},$ $\left\vert \mathcal{S}_{yz}\right\vert ^{2},$ and $%
\left\vert \mathcal{S}_{xz}\right\vert ^{2}$ form 2D square lattices of
beams running in respectively the $\kappa _{z},$ $\kappa _{x},$ and $\kappa
_{y}$ directions in reciprocal space thus giving rise to intensity patterns
which are roughly 6 times coarser than the zone-selection patterns. The
square lattices formed by the beams from $\left\vert \tilde{w}_{xy}\left( 
\mathbf{\kappa }\right) \right\vert ^{2}$ and $\left\vert \tilde{w}%
_{yz}\left( \mathbf{\kappa }\right) \right\vert ^{2}$ were shown in column 1
of FIG.~\ref{SYk} and the following columns showed how the angular and
radial factors limit the intensities to the extent that only the beams
passing through $\left[ \kappa _{x},\kappa _{y},\kappa _{z}\right] =\left[
1,1,1\right] $ should be useful for ARPES investigation of all three WOs. A
closer look --and in the crystallografic $\left( \kappa _{a},\kappa
_{b},\kappa _{c}\right) $ space-- was given in FIG.$~$\ref{Re^2ka6068}. The
central parts of the three form factors were shown in the \emph{top} part of
FIG.~\ref{ThreeBandskonly} along the line $\left( \kappa _{a},\kappa
_{b}\right) =\left( 6.4,0.225\right) .$ As seen in the last columns of FIG.s %
\ref{FIGxyZoneSelectul} and \ref{FIGyzZoneSelect}, the narrowness of the
form factors ($\Delta \kappa _{c}\sim 1)$ washes out details of the
dimerization distortions of the zone selection, except near the suitably
chosen values of $\kappa _{a}$.

Our extensive ARPES data confirmed the LDA-based WO theory of the energy bands and the ARPES intensity variations. The agreement between the band structures obtained by ARPES and by the LDA is already good (FIG. \ref{ARPES_Bandstructure_LDA}) and refinement of the LDA-TB parameters can make the fit almost perfect for the large energy features (Fig. \ref{ARPES_Bandstructure_TB}, main panel). So doing automatically improves the results for the small-energy features such as the resonance peaks in the upper metallic $\widetilde{xy}$-band, which are caused by repulsion from the top of the $xz$- or $yz$- valence bands (FIG. \ref{ThreeBandskonly} bottom). Taking advantage of the BZ selection rule enabled observation of the upper metallic $\widetilde{xy}$-band resonance peak (second-BZ data, FIG. \ref{ARPES_Bandstructure_TB} (c2)).  Although the resonance peak is quite weak in this ARPES data measured at h$\nu$=30 eV, we understood this difficulty (Sect. \ref{SectDrop}) to stem from the rapid fall-off of the form factor of the $\widetilde{xy}$-WO (see FIG.s \ref{ThreeBandskonly} top, and \ref{Re^2ka6068} left) when moving away from the center of the beam given by Eq. (\ref{max1xy}). The latter depends on $\kappa_a$ which is controlled by h$\nu$ [FIG. \ref{Photon_Energy_Scan} and Eq. (\ref{kappaa(0,hv)})], causing the intensity distribution along the Fermi surface (FS) to depend sensitively on the photon energy (FIG. \ref{hv_30_33_37_kbkc}). To observe both metallic $\widetilde{xy}$-bands equally well, we need strong emission in both the first and second BZs and this occurs only for our h$\nu$=33 eV data.

Our trust in the $t_{2g}$ Hamiltonian with the refined parameter values has
thus been strengthened to the extent that in the following Paper III we go
on using it together with ARPES data taken at $h\nu $=33~eV to study the
splitting and warping of the FS. Such deviations from one-dimensionality,
crucial for the physical properties, are tiny and can in FIG.$\,$\ref{CEC}
only be seen in the theoretical bands. Moreover, since these deviations are
largely induced by the $xz$- and $yz$-bands, as may be realized by comparing
the red with the dark-red bands in FIG.~\ref{ThreeBandskonly}, they depend
sensitively on the Fermi level's distances from these V\&C bands and, hence,
on the doping.

The resonance peaks are pushed up/down in the upper/lower $\widetilde{xy}$-band by the V/C-band edge, whose character is 50\% \emph{mixed} $\left\vert 
\mathbf{k}\right\rangle $ and $\left\vert \mathbf{k+c}^{\ast }\right\rangle $%
. It was therefore not obvious to what extent the character of the original $xy$-bands (the red ones in the middle panel of FIG.~\ref{ThreeBandskonly})
near $k_{c}$=$\pm $0.75 or $\pm 0.25$ are retained in the $\widetilde{xy}$-bands, and, hence, how strong the ARPES intensity, proportional to the $%
\left\vert \mathbf{k}\right\rangle $ character, should be. We therefore
needed to compute the $\left\vert \mathbf{k}\right\rangle $ characters of
the $\widetilde{xy}$-bands, the dark-red ones in the bottom panel of FIG.~%
\ref{ThreeBandskonly}. This was done using a two-band Hamiltonian obtained
by L\"{o}wdin downfolding of the V\&C blocks of the six-band Hamiltonian I$%
\, $(\ref{HRecip}) in the $\left\{ \mathbf{k,k+c}^{\ast }\right\} $%
-representation. The derivation of this two-band Hamiltonian will be our
first task in the following Paper III.

\mbox{} \clearpage

\title{Wannier-Orbital theory and ARPES for the quasi-1D conductor LiMo$_{6}$%
O$_{17}$. \\
Part III: The two metallic bands in the gap}

\begin{abstract}
This is the third paper of a series of three papers presenting a combined
study by band theory and angle-resolved photoemission spectroscopy (ARPES)
of lithium purple bronze. The first paper laid the foundation for the
theory, and the second paper discussed a general comparison between theory
and experiment, including deriving an ARPES selection rule. The present
paper III focuses in detail on the two metallic, quasi-1D $xy$-like bands
left in the 0.4~eV dimerization gap between the $xz$ and $yz$ valence and
conduction (V\&C) bands. The hybridizations with the latter change the
perpendicular dispersions of --and splitting between-- the resulting $%
\widetilde{xy}$ bands. The edges of the $\QATOP{\mathrm{V}}{\mathrm{C}}$
bands, in particular, push resonance peaks $\QATOP{\mathrm{up}}{\mathrm{down}%
}$ in the $\widetilde{xy}$ bands which are now described by a two-band
Hamiltonian (\ref{H2}) whose two first terms consist of the pure $xy$-block
of the six-band TB Hamiltonian, I Eq. (\ref{HRecip}), and whose 4 following terms
describe the resonant coupling to (i.e., indirect hopping via) the V\&C
bands. The two-band Hamiltonian extends the selection rule derived in the
previous paper to the hybridized $\widetilde{xy}$ bands, which enables, for
the first time, extracting the split quasi-1D Fermi surface (FS) from the
raw ARPES data. The complex shape of the FS, verified in detail by our
ARPES, depends strongly on the Fermi energy position in the gap, implying a
great sensitivity to Li stoichiometry of properties dependent on the FS,
such as FS nesting or superconductivity. The strong resonances prevent
either a two-band TB model or a related real-space ladder picture from
giving a valid description of the low-energy electronic structure. Down to a
temperature of 6$\,$K, we find no evidence for a theoretically expected
downward renormalization of perpendicular single-particle hopping due to LL
fluctuations in the quasi-1D chains.
\end{abstract}

\date{\today }
\pacs{Valid PACS appear here}
\maketitle


\section{Introduction}

This last paper in a series of three about ARPES and Wannier orbital (WO)
theory of the quasi-1D conductor LiMo$_{6}$O$_{17}$ deals with the Fermi
surface (FS) formed by the two metallic $xy$-like bands (the $\widetilde{xy}$
bands) in the gap caused by the $c$-axis dimerization (Sect.~I~\ref{SectDims}%
)\footnote{%
I or II refers to sections, figures, and equations in Paper I or II.}
between the $xz$- and $yz$-like valence and conduction (V\&C) bands. All six 
$t_{2g}$ bands, half of them filled in the stoichiometric compound, were
studied by band theory in Paper I. In Paper II, we compared the theory with
the experiment (by ARPES) for the filled bands. There, also a WO theory of
the experimentally found ARPES intensity variations was derived and applied.

As basis for the initial states we used the six $t_{2g}$ WOs, $w_{m}\left( 
\mathbf{r}\right) $ and $W_{m}\left( \mathbf{r}\right) $ with $m=xy,xz,yz,$
centered on respectively Mo1 and MO1, the most central molybdenums of the
lower and the upper strings, $^{\text{Mo2}}\diagdown _{\text{Mo1}}\diagup ^{%
\text{Mo4}}\diagdown _{\text{Mo5}}$ and $^{\text{MO5}}\diagdown _{\text{MO4}%
}\diagup ^{\text{MO1}}\diagdown _{\text{MO2}},$ wiggling around the $\mathbf{%
c+a}$ direction perpendicular to $\mathbf{b,}$ the direction of the quasi-1D
conductivity [see Chart~I$\,$(\ref{xy}), and FIG.s~\ref{FIG1} and \ref%
{Wannier}]. Due to the approximate translational equivalence of $w_{m}\left( 
\mathbf{r}\right) $ and $W_{m}\left( \mathbf{r}\right) $ [see Eq.s~I$~($\ref%
{undim}$)$ and (\ref{inv})], the photoemission intensity is essentially the
projection of the initial-state band onto the \emph{pseudo} Bloch sum $%
\left\vert \mathbf{k}\right\rangle ,$ defined by Eq.~I$~$(\ref{kket}). Such
a pseudo Bloch sum is a periodic function of $\mathbf{k}$ in the \emph{double%
} zone (see FIG.~I~\ref{FIGDoubleZone}), i.e. the zone of the un-dimerized
lattice (the one with primitive translations\textbf{\ }$\mathbf{a,}$ $\frac{%
\mathbf{c+b}}{2},$ and $\frac{\mathbf{c-b}}{2},$ and reciprocal translations 
$\mathbf{a}^{\ast }\mathbf{,}$ $\mathbf{c}^{\ast }\mathbf{+b}^{\ast }\mathbf{%
,}$ and $\mathbf{c}^{\ast }\mathbf{-b}^{\ast }).$ The other basis function
is simply $\left\vert \mathbf{k+c}^{\ast }\right\rangle .$ In the absence of 
$c$-axis dimerization, $w_{m}\left( \mathbf{r}\right) $ and $W_{m}\left( 
\mathbf{r}\right) $ are identical, apart from a phase factor, and the $%
\left\vert \mathbf{k}\right\rangle $ and the $\left\vert \mathbf{k+c}^{\ast
}\right\rangle $ band structures are each periodic in the double zone and
translated by $\mathbf{c}^{\ast },$ i.e. by $\Delta k_{c}$=1, with respect
to each other. In the presence of dimerization, the basis functions become
linearly independent and will mix near the crossings of the $\left\vert 
\mathbf{k}\right\rangle $ and $\left\vert \mathbf{k+c}^{\ast }\right\rangle $
band structures, which will now gap and thereby restore the single-zone
periodicity (see FIG.s~I$~$\ref{ThreePureBands}, \ref{ThreeBands}, and \ref%
{3Dt2gBands}).

This description of the band structure in Paper I, followed by the theory of
the ARPES intensity variations in Paper II, explained our comprehensive set
of ARPES data for the occupied bands (FIG.s~II~\ref{CEC}, \ref%
{ARPES_Bandstructure_Cuts}, \ref{ARPES_Bandstructure_LDA}, and \ref%
{ARPES_Bandstructure_TB}), including the observations that the intensity
follows the $\left\vert \mathbf{k}\right\rangle $ character (FIG.$\,$II~\ref%
{ThreeBandskonly}) and of a surprisingly strong photon-energy dependence of
the photoelectron intensity from the FS (FIG.$~$II$~$\ref{hv_30_33_37_kbkc}%
). The former means that the ARPES intensity is enhanced for the
lower-energy band in the 1st- and is extinguished in the 2nd physical zone;
conversely for the higher-energy band, if occupied [Sect.$~$II~\ref%
{Sectzoneselect} and FIG.~\ref{ThreeBandskonly}]. The latter leads us to use 
$h\nu $=33$\,$eV for the FS studies to be described in the present Paper III
Sect.~\ref{SectFSExperiment}.

\begin{figure}[tbh]
\includegraphics[width=\linewidth]{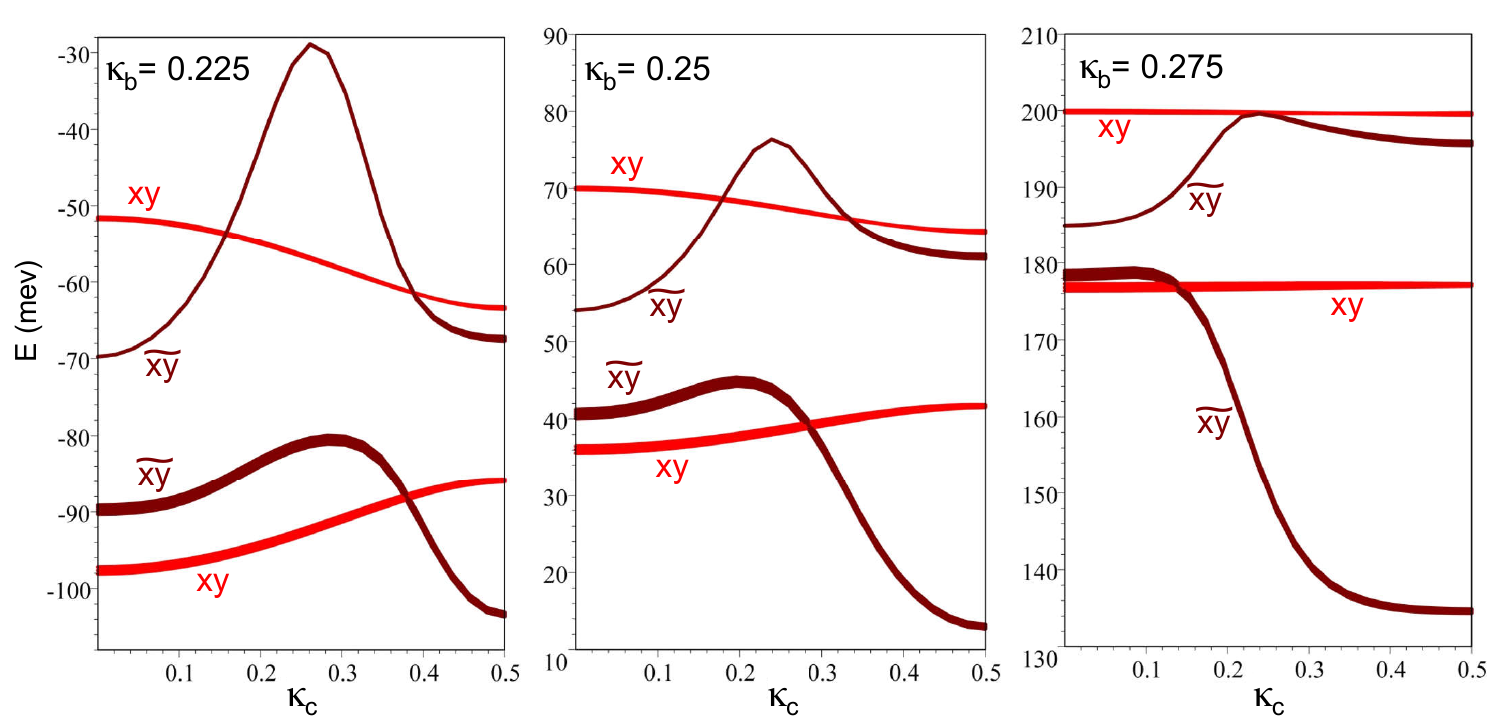}
\caption{The two quasi-1D, metallic bands in the gap between the valence (V)
and conduction (C) bands as functions of the perpendicular $k_{c}$ in the
irreducible BZ, $0\leq k_{c}<0.5,$ and for $k_{b}$ fixed at 0.225, 0.250,
and 0.275. The red bands are the pure $xy$ bands and the dark-red $%
\widetilde{xy}$ bands include the hybridization with the four V\&C bands.
The extra fatness (band decoration) is proportional to the $\left\vert 
\mathbf{k}\right\rangle $ character, which for the dark-red bands was
obtained from the two-band Hamiltonian (\protect\ref{H2}). The red and the
dark-red fat bands for $k_{b}$=0.225 were already shown in respectively
FIG.s I \protect\ref{ThreePureBands} and \protect\ref{ThreeBands}, and their 
$\left\vert \mathbf{k}\right\rangle $ characters in the middle and bottom
parts of FIG. II~\protect\ref{ThreeBandskonly}. The upper and lower $%
\widetilde{xy}$ bands as functions of $\left( k_{b},k_{c}\right) $ are shown
on the left-hand side of FIG.~\protect\ref{Theo_CEC}. The ARPES-refined TB
parameter values I~(\protect\ref{taup})-(\protect\ref{ap}) were used and the 
$xz$-$yz$ hybridization [parameters I~(\protect\ref{lambdalmum})] was
neglected.$^{\protect\ref{NeglectLambda}}$ The value of $E$ in the
denominators were iteratively adjusted to equal the eigenvalue in question
of the two-band Hamitonian whereby it equals the appropriate eigenvalue of
the six-band Hamiltonian (with the values I~(\protect\ref{lp}) of the $xz$-$%
yz$ parameters set to zero). The energy is in meV with respect to the center
of the gap. The top of the valence band and the bottom of the conduction
band are $\pm 2G_{1}=\mp 218$ meV and the ARPES samples had $E_{F}$=75~meV.}
\label{9to11over40irrBZ}
\end{figure}

FIG.~II$~$\ref{ARPES_Bandstructure_TB} demonstrated that refinement of
merely 7 of the over 40 LDA-TB parameters in our detailed six-band $t_{2g}$
Hamiltonian (Sect.~I$~$\ref{SectH}$)$ to fit the ARPES bands lying more than
0.15~eV below the Fermi level, achieves nearly perfect agreement \emph{also}
for the $k_{c}$-dispersion of the quasi-1D bands closer to the Fermi level.
This includes agreement with the size and shape of the resonance peak in the
upper band --blue in (c2)-- \emph{without} having modified any of the 17 $%
\left( a,g,\alpha ,\gamma \right) $-parameters I~(\ref{ap}) describing the
hybridization between the $xy$ and the $xz\,$and $yz$ V\&C bands, parameters
which we shall see decisively influence the structure of the metallic $%
\widetilde{xy}$ bands via the second order terms (\ref{Gdiag}) and (\ref%
{Goff}) in the two-band Hamiltonian (\ref{H2}).

The resonance peaks are pushed $\QATOP{\mathrm{up}}{\mathrm{down}}$ in the $%
\QATOP{\mathrm{upper}}{\mathrm{lower}}$ $\widetilde{xy}$ band by the edges
of $\QATOP{\mathrm{V}}{\mathrm{C}}$-bands, whose characters before
hybridization with the $xy$\ band are 50\% \emph{mixed} $\left\vert xz;%
\mathbf{k}\right\rangle $ and $\left\vert xz;\mathbf{k+c}^{\ast
}\right\rangle ,$ or $\left\vert yz;\mathbf{k}\right\rangle $ and $%
\left\vert yz;\mathbf{k+c}^{\ast }\right\rangle ;$ it is therefore not
obvious to what extent the character (fatness) of the original $xy$ bands
(the red ones in the middle panel of FIG.~II~\ref{ThreeBandskonly}) near $%
k_{c}$=$\pm $0.75 or $\pm 0.25$ are retained in the hybridized $\widetilde{xy%
}$ bands, and, hence, how strong the ARPES intensity should be. The
selection rule that the ARPES intensity follows the $\left\vert \mathbf{k}%
\right\rangle $ character holds also for the two weakly hybridized $%
\widetilde{xy}$ bands, because --to lowest order-- the effect of weak
hybridization is to distort the bands. The $\left\vert \mathbf{k}%
\right\rangle $ character we then compute by expressing the two bands in the 
$\{\mathbf{k},\mathbf{k+c}^{\ast }\}$ representation and add to the diagonal
elements, $\left\langle \mathbf{k}\left\vert H\right\vert \mathbf{k}%
\right\rangle $, of the two-band Hamiltonian a few small, equidistant
energies. This produces fat bands which include the effects of weak
hybridization with the V\&C bands [but neglects the relatively small effects
of dimerization distortion, $\eta \left( \kappa \right) $ Eq. II (\ref{eta})]. 

We therefore need to compute the $\left\vert \mathbf{k}\right\rangle $
character of the $\widetilde{xy}$ bands. This we do by using the Hamiltonian
(\ref{H2}) obtained by downfolding \cite{Lowdin1951} of the V\&C blocks of
the six-band Hamiltonian I$~$(\ref{HRecip}) in the $\left\{ \mathbf{k,k+c}%
^{\ast }\right\} $-representation. For simplicity of the formalism, we
neglect the mixing between the $xz$ and $yz$ WOs, which is a good
approximation near the FS, $\left\vert k_{b}\right\vert \sim 0.25,$ far away
from the Y and Z points (see FIG.~I~\ref{FIGDoubleZone}). The derivation of
this two-band Hamiltonian will be our first task in the theoretical Sect.~%
\ref{SectDispInGap} of the present paper.

The two-band Hamiltonian is also the one from which we can best understand
the origin of the splitting and perpendicular dispersion of the two metallic
bands shown in FIG.$~$\ref{9to11over40irrBZ}, including their development
with increasing $k_{b},$ which is surprisingly unsymmetric around
half-filling, $k_{b}$=$0.25$. Having discussed the two bands in great detail
in Sect.~\ref{SectOrigins} and understood that merely the light-red $xy$
--but \emph{not} the dark-red $\widetilde{xy}$-- bands can be described by a 
$2\times 2$ TB Hamiltonian, we shall in the final theoretical subsection \ref%
{SectBandsandFS} present and discuss their constant-energy contours (CECs),
the FS in particular.

The second part of this paper, starting from Sect. \ref{SectFSExperiment},
is devoted to the demanding task of using ARPES to determine the FS and its
velocities. First, the challenges, issues, and methods are discussed in
Sect.~\ref{Challenges}. The results of the FS and velocity extraction are
then shown in Sect.~\ref{SectFS}. They are compared with the TB theory using
the ARPES refined parameters and are finally presented in FIG.~\ref%
{Fig:NewFSExtract}.

In theory, the (fine-grained) ARPES intensity from each metallic band is
periodic in the double zone ($|\kappa _{c}|\leq 1$) with the intensity from
the lower band (outer sheet) dominating in the 1st zone ($|\kappa _{c}|\leq
0.5$) and the intensity from the upper band (inner sheet) dominating in the
2nd zone ($|\kappa _{c}|\geq 0.5$). Between zone centers ($\kappa _{c}$
integer), the origin of the dominating intensity thus switches from one band
(sheet) to the other, with the intensities being equal at the zone boundary (see Fig. \ref{Fig:RepeatPureK}).
Since the measured spectra of the sheets possess a finite width in momentum,
which is much larger than the splitting of the sheets, we can only detect
the center of gravity of the sum of intensities from the occupied bands. By
taking this into account, the experimentally obtained FS fits the
theoretical one very well. At the zone boundary, we determined an upper
bound for the splitting, which is in that sense, in agreement with the
theory, that it is larger than the theoretical value. Comparing the
experimentally extracted velocities with the theoretical ones, we reach a
similar perfect fit as for the FS. Interestingly, the velocities are
generally greater than those for pure LDA by about 15\%, as we discuss in
detail in Sect.~\ref{Section:ExpFermiVelo}.\\

\section{Theoretical splitting and perpendicular dispersion of the two
metallic bands in the gap\label{SectDispInGap}} 

\begin{widetext}

\subsection{Hamiltonian for the two metallic bands in the $\left\{ \mathbf{%
k,k+c}^{\ast }\right\} $-representation and of resonance form\label{SectH2}}

We start from the six-band Hamiltonian I~(\ref{HRecip}) in the $\left\{ 
\mathbf{k,k+c}^{\ast }\right\} $-representation:%
\begin{equation}
\fbox{$%
\begin{array}{cccccccc}
H &  & \left\vert xy;\mathbf{k}\right\rangle & \left\vert xy;\mathbf{k+c}%
^{\ast }\right\rangle & \left\vert xz;\mathbf{k}\right\rangle & \left\vert
xz;\mathbf{k+c}^{\ast }\right\rangle & \left\vert yz;\mathbf{k}\right\rangle
& \left\vert yz;\mathbf{k+c}^{\ast }\right\rangle \\ 
&  &  &  &  &  &  &  \\ 
\left\langle xy;\mathbf{k}\right\vert &  & \tau +t & iu & \alpha +a & 
i\left( \gamma +g\right) & \bar{\alpha}+\bar{a} & i\left( \bar{\gamma}+\bar{g%
}\right) \\ 
\left\langle xy;\mathbf{k+c}^{\ast }\right\vert &  & -iu & \tau -t & i\left(
\gamma -g\right) & \alpha -a & i\left( \bar{\gamma}-\bar{g}\right) & \bar{%
\alpha}-\bar{a} \\ 
\left\langle xz;\mathbf{k}\right\vert &  & \alpha +a & -i\left( \gamma
-g\right) & A & iG & \lambda +l & -i\left( \mu -m\right) \\ 
\left\langle xz;\mathbf{k+c}^{\ast }\right\vert &  & -i\left( \gamma
+g\right) & \alpha -a & -iG & -A & -i\left( \mu +m\right) & \lambda -l \\ 
\left\langle yz;\mathbf{k}\right\vert &  & \bar{\alpha}+\bar{a} & -i\left( 
\bar{\gamma}-\overline{g}\right) & \lambda +l & i\left( \mu +m\right) & \bar{%
A} & i\bar{G} \\ 
\left\langle yz;\mathbf{k+c}^{\ast }\right\vert &  & -i\left( \bar{\gamma}+%
\bar{g}\right) & \bar{\alpha}-\bar{a} & i\left( \mu -m\right) & \lambda -l & 
-i\bar{G} & -\bar{A}%
\end{array}%
$}.  \tag*{I (56)}
\end{equation}%
Like in Eq.~I$~$(\ref{Hsub}), the argument, $\mathbf{k,}$ of the Bloch sums
of hopping integrals I~(\ref{tau})-(\ref{lambdalmum}) is omitted for brevity
and an overbar, used when switching from an $xz$ to a $yz$ orbital,
indicates the mirror operation $k_{b}\leftrightarrow -k_{b}$.

\subsubsection{Six-band Hamiltonian in the V\&C subband representation\label%
{MixRepr}}

Since the largest off-diagonal elements I~(\ref{taup})-(\ref{lp}) are $\bar{G%
}$ and $G,$ which gap the $yz$ bands as in Eq.$\,$I~(\ref{gap}) and the $xz$
bands in the same way, but with $k_{b}$ substituted by $-k_{b}$, we shall
now turn to a representation in which the $xz$-$xz$ and $yz$-$yz$ blocks are
diagonal. This mixed representation is the natural one to use for
downfolding these blocks of the six-band Hamiltonian to a two-band
Hamiltonian which then describes merely the two $\widetilde{xy}$ bands in
the gap and --provided that we neglect the hybridization I~(\ref{lambdalmum}%
) between the $xz$ and $yz$ bands-- has the simple resonance form (\ref{H2}%
). As said above, this approximation of using the $\left( m,m^{\prime
}\right) $-unhybridized, socalled \emph{pure} $xz$ and $yz$ bands (Sect.~I$~$%
\ref{SectPureBands}) is a good one far away from Y and Z, especially near
the FS, $\left\vert k_{b}\right\vert \sim 0.25$ (FIG.~I$~$\ref{FIGDoubleZone}). The eigenvalues, $\pm \sqrt{\bar{A}^{2}+\bar{G}^{2}},$ of the $yz$-$yz$ 2$%
\times $2 diagonal block of the six-band Hamiltonian, Eq. I (\ref{HRecip}), are the $%
yz$ conduction (C)- and valence (V)-band energies I~(\ref{gap}), and so are $%
\pm \sqrt{A^{2}+G^{2}}$ for the $xz$ $\QATOP{C}{V}$-band energies. The structure of
the six bands for $k_{b}$=0.225, 0.250, and 0.275 as a function of $k_{c}$ may
be seen on the 3rd line to the left in FIG.~\ref{Analysis} and we shall
describe it in Sect.~\ref{SectOrigins}.  \\ 

The orthonormal $xz$ C\&V-band
orbitals are:%
\begin{equation}
\left( 
\begin{array}{cc}
\left\vert xz_{C}\left( \mathbf{k}\right) \right\rangle & \left\vert
xz_{V}\left( \mathbf{k}\right) \right\rangle%
\end{array}%
\right) =\left( 
\begin{array}{cc}
\left\vert xz;\mathbf{k}\right\rangle & \left\vert xz;\mathbf{k+c}^{\ast
}\right\rangle%
\end{array}%
\right) \times \frac{1}{2}\left( 
\begin{array}{cc}
e^{-i\phi }-1 & e^{-i\phi }+1 \\ 
e^{-i\phi }+1 & e^{-i\phi }-1%
\end{array}%
\right) ,  \tag*{I (59)}
\end{equation}%
where the $xz$ band-structure phase, $\phi \left( \mathbf{k}\right) ,$ is
that of $-A\left( \mathbf{k}\right) -iG\left( \mathbf{k}\right) .$ The $%
\left\vert \mathbf{k}\right\rangle $ $\left( =1-\left\vert \mathbf{k+c}%
^{\ast }\right\rangle \right) $ characters --or fatnesses-- of the $\QATOP{C%
}{V}$ bands are:%
\begin{equation}
f_{\QATOP{C}{V}}\left( \mathbf{k}\right) =\left\vert \frac{e^{-i\phi }\mp 1}{%
2}\right\vert ^{2}=\frac{1\mp \cos \phi }{2},  \tag*{I (61)}
\end{equation}%
and they are shown on the 3rd line to the right in FIG.~\ref{Analysis} in $%
\frac{\mathrm{light}}{\mathrm{dark}}$ blue. While the (C and V) 
bands are periodic, their fatnesses are periodic in the double zone (see FIG.~I~\ref%
{ThreePureBands}).

Using the unitary matrix I~(59) to transform the six-band Hamiltonian I~(56)
to the pure-$xz$ representation in which the V\&C block is diagonal yields:%
\begin{equation}
\fbox{$%
\begin{array}{ccccc}
H & \left\vert xy;\mathbf{k}\right\rangle & \left\vert xy;\mathbf{k+c}^{\ast
}\right\rangle & \left\vert xz_{C}\left( \mathbf{k}\right) \right\rangle & 
\left\vert xz_{V}\left( \mathbf{k}\right) \right\rangle \\ 
\left\langle xy;\mathbf{k}\right\vert & \tau +t & iu & cc & cc \\ 
\left\langle xy;\mathbf{k+c}^{\ast }\right\vert & -iu & \tau -t & cc & cc \\ 
\left\langle xz_{C}\left( \mathbf{k}\right) \right\vert & \left( \alpha
+a\right) \frac{e^{i\phi }-1}{2}-i\left( \gamma +g\right) \frac{e^{i\phi }+1%
}{2} & \left( \alpha -a\right) \frac{e^{i\phi }+1}{2}-i\left( \gamma
-g\right) \frac{e^{i\phi }-1}{2} & \sqrt{A^{2}+G^{2}} & 0 \\ 
\left\langle xz_{V}\left( \mathbf{k}\right) \right\vert & \left( \alpha
+a\right) \frac{e^{i\phi }+1}{2}-i\left( \gamma +g\right) \frac{e^{i\phi }-1%
}{2} & \left( \alpha -a\right) \frac{e^{i\phi }-1}{2}-i\left( \gamma
-g\right) \frac{e^{i\phi }+1}{2} & 0 & -\sqrt{A^{2}+G^{2}}%
\end{array}%
$}  \label{mixed}
\end{equation}%
for the first four rows and columns. The last two ($yz_{C}$ and $yz_{V})$
rows and columns obtained by transformation to the pure-$yz$ representation,
equal those given above for $xz_{C}$ and $xz_{V},$ but with $k_{b}$
substituted by $-k_{b},$ or $A$ and $G$ substituted by $\bar{A}$ and $\bar{G}
$.

\subsubsection{L\"{o}wdin-downfolded two-band Hamiltonian\label{twoband}}

The WOs for the two $\widetilde{xy}$ bands in the gap have much longer range
than the $xy$-WOs shown in FIG.s~I~\ref{FIGyzb&xyamc} and I~\ref{Wannier},
and similarly for the elements in the effective two-band Hamiltonian
compared with the range of the Bloch sums in the six-band Hamiltonian. For
this reason, we do not perform the downfolding of the $xz,$ $XZ,$ $yz,$ and $%
YZ$ WOs into the "tails" of the $\widetilde{xy}$ and $\widetilde{XY}$ WOs in
real space, but in reciprocal space. Real-space pictures of these downfolded
orbitals would be unwieldy and would crucially depend on the position of the
energies, $E,$ of the bands in the gap. For the same reason, tables of
hopping integrals would be unwieldy and energy dependent. For the
downfolding from six to two narrow bands near the centre of the gap, we can
use simple, analytical L\"{o}wdin\cite{Lowdin1951}- rather than numerical (NMTO)
downfolding because the splitting of the two $\widetilde{xy}$ bands is far
less than their distance to the C\&V-band edges (see the left-hand panels on
the 3rd line in FIG.~\ref{Analysis}). Hence, the  explicit dependence of the
two-band Hamiltonian (4) on the TB parameters I (\ref{taup})-(\ref{lp}) is preserved.

With the hybridization between the $xz$ and $yz$ orbitals neglected, the $xz$
and $yz$ downfoldings are additive:%
\begin{equation}
\left( 
\begin{array}{ll}
\left\vert \widetilde{xy};E,\mathbf{k}\right\rangle & \left\vert \widetilde{%
xy};E,\mathbf{k+c}^{\ast }\right\rangle%
\end{array}%
\right) =\left( 
\begin{array}{ll}
\left\vert xy;\mathbf{k}\right\rangle & \left\vert xy;\mathbf{k+c}^{\ast
}\right\rangle%
\end{array}%
\right) +\delta _{xz;E}\left( 
\begin{array}{ll}
\left\vert xy;\mathbf{k}\right\rangle & \left\vert xy;\mathbf{k+c}^{\ast
}\right\rangle%
\end{array}%
\right) +\delta _{yz;E}\left( 
\begin{array}{ll}
\left\vert xy;\mathbf{k}\right\rangle & \left\vert xy;\mathbf{k+c}^{\ast
}\right\rangle%
\end{array}%
\right) .  \label{xytilde}
\end{equation}%
Here, the $\delta _{xz;E}$ and $\delta _{yz;E}$ perturbations involve the
Green function for the $xz$-$xz$ or $yz$-$yz$ block of the six-band
Hamiltonian I~(56) times the corresponding $xz$-$xy$ or $yz$-$xy$
hybridization matrix. The representation chosen for the $xz$ and $yz$ states
to be downfolded (integrated out) matters for the formalism, but \emph{not}
for the resulting two-band Hamiltonian. Choosing the pure-$m$ representation
(\ref{mixed}) in which the $xz$-$xz$ and the $yz$-$yz$ blocks of the
Hamiltonian are diagonal we get: 
\begin{eqnarray}
&&\delta _{xz;E}\left( 
\begin{array}{cc}
\left\vert xy;\mathbf{k}\right\rangle & \left\vert xy;\mathbf{k+c}^{\ast
}\right\rangle%
\end{array}%
\right) =  \label{dwnfxy} \\
&&\frac{\left\vert xz_{C}\right\rangle }{E-\sqrt{A^{2}+G^{2}}}\left( 
\begin{array}{cc}
\left\langle xz_{C}\left\vert H\right\vert xy;\mathbf{k}\right\rangle & 
\left\langle xz_{C}\left\vert H\right\vert xy;\mathbf{k+c}^{\ast
}\right\rangle%
\end{array}%
\right) +\frac{\left\vert xz_{V}\right\rangle }{E+\sqrt{A^{2}+G^{2}}}\left( 
\begin{array}{cc}
\left\langle xz_{V}\left\vert H\right\vert xy;\mathbf{k}\right\rangle & 
\left\langle xz_{V}\left\vert H\right\vert xy;\mathbf{k+c}^{\ast
}\right\rangle%
\end{array}%
\right) ,  \notag
\end{eqnarray}%
and similarly for $\delta _{yz;E}.$ It is by virtue of this choice that the $%
E$-dependence of the downfolded orbitals enters solely through the
denominators in (\ref{dwnfxy}).

In the $\left\{ \widetilde{xy};E,\mathbf{k,k+c}^{\ast }\right\} $%
-representation (\ref{xytilde}), the two-band Hamiltonian is finally seen to
be:%
\begin{eqnarray}
&&\left( 
\begin{array}{cc}
\left\langle \widetilde{xy};E,\mathbf{k}\left\vert H\right\vert \widetilde{xy%
};E,\mathbf{k}\right\rangle  & \left\langle \widetilde{xy};E,\mathbf{k}%
\left\vert H\right\vert \widetilde{xy};E,\mathbf{k+c}^{\ast }\right\rangle 
\\ 
c.c. & \left\langle \widetilde{xy};E,\mathbf{k+c}^{\ast }\left\vert
H\right\vert \widetilde{xy};E,\mathbf{k+c}^{\ast }\right\rangle 
\end{array}%
\right) =\tau \left( k_{b}\right) \left( 
\begin{array}{cc}
1 & 0 \\ 
0 & 1%
\end{array}%
\right) +\left( 
\begin{array}{cc}
t\left( \mathbf{k}\right)  & iu\left( \mathbf{k}\right)  \\ 
-iu\left( \mathbf{k}\right)  & -t\left( \mathbf{k}\right) 
\end{array}%
\right)   \label{H2} \\
&&+\frac{\Gamma _{C}\left( \mathbf{k}\right) }{E-\sqrt{A^{2}\left( \mathbf{k}%
\right) +G^{2}\left( \mathbf{k}\right) }}+\frac{\Gamma _{V}\left( \mathbf{k}%
\right) }{E+\sqrt{A^{2}\left( \mathbf{k}\right) +G^{2}\left( \mathbf{k}%
\right) }}+\frac{\overline{\Gamma }_{C}\left( \mathbf{k}\right) }{E-\sqrt{%
\bar{A}^{2}\left( \mathbf{k}\right) +\bar{G}^{2}\left( \mathbf{k}\right) }}+%
\frac{\overline{\Gamma }_{V}\left( \mathbf{k}\right) }{E+\sqrt{\bar{A}%
^{2}\left( \mathbf{k}\right) +\bar{G}^{2}\left( \mathbf{k}\right) }},  \notag
\end{eqnarray}%
with the directly coupled terms on the 1st line and the resonance terms for
the coupling via the C\&V bands on the 2nd line. The poles at the C\&V $xz$
and $yz$ bands, $\pm \sqrt{A^{2}\left( \mathbf{k}\right) +G^{2}\left( 
\mathbf{k}\right) }$ and $\pm \sqrt{\bar{A}^{2}\left( \mathbf{k}\right) +%
\bar{G}^{2}\left( \mathbf{k}\right) },$ are numbers while the residues, $%
\Gamma _{C}\left( \mathbf{k}\right) ,$ $\Gamma _{V}\left( \mathbf{k}\right) $%
, $\overline{\Gamma }_{C}\left( \mathbf{k}\right) ,$ and $\overline{\Gamma }%
_{V}\left( \mathbf{k}\right) $, are 2$\times $2 matrices of the following
form for the perturbation by the $xz$ C band:%
\begin{equation}
\Gamma _{C}\left( \mathbf{k}\right) =\left( 
\begin{array}{cc}
\left\vert \left\langle xz_{C}\left( \mathbf{k}\right) \left\vert
H\right\vert xy;\mathbf{k}\right\rangle \right\vert ^{2} & \left\langle xy;%
\mathbf{k}\left\vert H\right\vert xz_{C}\left( \mathbf{k}\right)
\right\rangle \left\langle xz_{C}\left( \mathbf{k}\right) \left\vert
H\right\vert xy;\mathbf{k+c}^{\ast }\right\rangle  \\ 
cc & \left\vert \left\langle xz_{C}\left( \mathbf{k}\right) \left\vert
H\right\vert xy;\mathbf{k+c}^{\ast }\right\rangle \right\vert ^{2}%
\end{array}%
\right) ,  \label{GC}
\end{equation}%
and similarly for the perturbation by the $xz$ V band. Since translation of $%
\mathbf{k}$ by $\mathbf{c}^{\ast }$ yields:%
\begin{equation}
\widehat{\mathbf{c}^{\ast }}\left\langle \mathbf{k}\left\vert \Gamma
\right\vert \mathbf{k}\right\rangle =\left\langle \mathbf{k+c}^{\ast
}\left\vert \Gamma \right\vert \mathbf{k+c}^{\ast }\right\rangle ,
\label{TcResOn}
\end{equation}%
the two diagonal elements with $\mathbf{k}$ in the single zone reduce to 
\emph{one} real-valued function of $\mathbf{k,}$ periodic in the double
zone. For the purely imaginary off-diagonal element:%
\begin{equation}
\widehat{\mathbf{c}^{\ast }}\left\langle \mathbf{k}\left\vert \Gamma
\right\vert \mathbf{k+c}^{\ast }\right\rangle =\left\langle \mathbf{k+c}%
^{\ast }\left\vert \Gamma \right\vert \mathbf{k}\right\rangle =\left\langle 
\mathbf{k}\left\vert \Gamma \right\vert \mathbf{k+c}^{\ast }\right\rangle
^{\ast }=-\left\langle \mathbf{k}\left\vert \Gamma \right\vert \mathbf{k+c}%
^{\ast }\right\rangle ,  \label{TcResOff}
\end{equation}%
i.e. it is an anti-periodic function of $\mathbf{k}$ in the single zone.

The elements of the matrix (\ref{GC}) are given by those of Eq.~(\ref{mixed}%
) in terms of the $xz$-band phase $\phi \left( \mathbf{k}\right) $ [I~(60)]
and the Bloch sums of the $xy$-$xz$ intra- and inter-ribbon hopping
integrals $\alpha \pm a$ and $\gamma \pm g$ $\left[ \text{I}~\text{(39)}%
\right] ,$ and may be expressed as:%
\begin{equation}
\left\langle \mathbf{k}\left\vert \Gamma _{\QATOP{C}{V}}\left( \mathbf{k}%
\right) \right\vert \mathbf{k}\right\rangle =\left\vert \left\langle xz_{%
\QATOP{C}{V}}\left( \mathbf{k}\right) \left\vert H\right\vert xy;\mathbf{k}%
\right\rangle \right\vert ^{2}=\left[ \left( \alpha +a\right) f_{\QATOP{C}{V}%
}\mp \left( \gamma +g\right) \frac{\sin \phi }{2}\right] ^{2}+\left[ \left(
\gamma +g\right) f_{\QATOP{V}{C}}\mp \left( \alpha +a\right) \frac{\sin \phi 
}{2}\right] ^{2}  \label{Gdiag}
\end{equation}%
for the diagonal elements, and as:%
\begin{eqnarray}
\left\langle \mathbf{k}\left\vert \Gamma _{\QATOP{C}{V}}\left( \mathbf{k}%
\right) \right\vert \mathbf{k+c}^{\ast }\right\rangle  &=&\left\langle xy;%
\mathbf{k}\left\vert H\right\vert xz_{\QATOP{C}{V}}\left( \mathbf{k}\right)
\right\rangle \left\langle xz_{\QATOP{C}{V}}\left( \mathbf{k}\right)
\left\vert H\right\vert xy;\mathbf{k+c}^{\ast }\right\rangle   \label{Goff}
\\
&=&i\left[ -\left( \alpha +a\right) \left( \gamma -g\right) f_{\QATOP{C}{V}%
}+\left( \gamma +g\right) \left( \alpha -a\right) f_{\QATOP{V}{C}}\pm \left(
\left( \alpha +a\right) \left( \alpha -a\right) +\left( \gamma +g\right)
\left( \gamma -g\right) \frac{\sin \phi }{2}\right) \right]   \notag
\end{eqnarray}%
for the off-diagonal elements, which are purely imaginary. They are shown on
the 4th line of FIG.$\,$\ref{Analysis} as functions of $k_{c}$ for $%
k_{b}=0.225,$ 0.250, and 0.275 and will be discussed in Sect \ref{indictRes}%
. On the 5th line, we show the $xy$-$xz$ intra and inter-ribbon hopping
integrals: to the left $\alpha +a$ in purple and $\alpha -a$ in grey, and to
the right $\gamma +g$ in turquise and $\gamma -g$ in grey. These Bloch sums
are periodic in the double zone and each grey Bloch sum equals the colored
one inside the same frame, but translated by $\mathbf{c}^{\ast }.$ This
follows from Eq.s~I~(\ref{Greek}) and (\ref{Latin}). Together with the $%
\left\vert \mathbf{k}\right\rangle $ characters of the C and V bands, $f_{C}$
and $f_{V},$ in respectively light and dark blue on the 3rd line to the
right, $\frac{1}{2}\sin \phi $ is shown in grey.

The residues for the perturbations by the $yz$ C\&V bands are respectively $%
\overline{\Gamma }_{C}\left( \mathbf{k}\right) $ and $\overline{\Gamma }%
_{V}\left( \mathbf{k}\right) $ with the overbar indicating the mirror
operation $k_{b}\leftrightarrow -k_{b}$.

\end{widetext}

$E$ is the energy of the $\widetilde{xy}$ state that we are seeking, i.e.$\,$%
the upper or lower eigenvalue of the two-band Hamiltonian (\ref{H2}), and
should therefore be found self-consistently. For states deep inside the gap,
we may substitute $E$ by $\tau \left( k_{b}\right) $ from Eq. I$~$(\ref{tau}%
) because the splitting of the two $\widetilde{xy}$ bands is far less than
half the gap. Note that $E$ is with respect to the center of the gap and
that it enters the two-band Hamiltonian (\ref{H2}) only through the
denominators of the four resonance terms. Keeping $E$ as a free parameter
therefore provides insight to study how the perpendicular dispersion of the $%
\widetilde{xy}$-bands depend on their placement in the gap and on the $%
\mathbf{k}$-dependence of the four residues.

The simplest way to understand the energy dependence of a L\"{o}%
wdin-downfolded Hamiltonian, $\widetilde{H}(E),$ is to consider downfolding
of a $2\times 2$ Hamiltonian, $H:$ Its exact $E$-eigenvalues are the roots
of the secular determinant $\left\vert H-E1\right\vert =\left(
H_{11}-E\right) \left( H_{22}-E\right) -\left\vert H_{12}\right\vert ^{2},$
which satisfy: $E=H_{11}+\left\vert H_{12}\right\vert ^{2}\left/
(E-H_{22})\right. \equiv \widetilde{H}(E),$ showing that $\widetilde{H}%
(H_{11})$ is the well-known 2nd order estimate of the eigenvalue closest to $%
H_{11}$ and that $\widetilde{H}(E)$ has a pole at the energy, $H_{22}.$ It
should now be obvious that the six-band Hamiltonian (\ref{mixed}) can be
downfolded to a two-band Hamiltonian with the form (\ref{H2}).

\subsection{Origin of the splitting and perpendicular dispersion\label%
{SectOrigins}}

Having derived a Hamiltonian (\ref{H2}) for the two metallic $\widetilde{xy}$
bands in the gap (FIG.\thinspace \ref{9to11over40irrBZ}) consisting of TB 
\emph{plus resonance} terms, we now take up the thread and trace the
non-trivial features of the bands back to the Bloch-sums I$\,$(\ref{tau})-(%
\ref{tautu}), $\tau \left( k_{b}\right) ,$ $t\left( \mathbf{k}\right) ,$ and 
$u\left( \mathbf{k}\right) ,$ of the $xy$-$xy$ hopping integrals, to the
Bloch sums I$\,$(\ref{AG}), $A\left( k_{c}\mp k_{b}\right) $ and $G\left(
k_{c}\mp k_{b}\right) ,$ of the $xz$-$xz$ or $yz$-$yz$ hopping integrals,
and to the Bloch sums I$\,$(\ref{ag}), $\alpha \left( \mathbf{k}\right) \pm
a\left( \mathbf{k}\right) $ and $\gamma \left( \mathbf{k}\right) \pm g\left( 
\mathbf{k}\right) ,$ of the $xy$-$xz$ hopping integrals. This is a long
route and the essence may be extracted from the synthesis in Sect.\thinspace %
\ref{SectSynth}.

We start by extending FIG.$\,$\ref{9to11over40irrBZ} from the irreducible $%
\left( 0\leq k_{c}\leq 0.5\right) $ to the double $\left( -1<k_{c}\leq
1\right) $ zone in which all Bloch sums are periodic (the Greek- and
Latin-lettered Bloch sums are periodic in respectively the single and the
double zone). This is done in FIG.$\,$\ref{Analysis} on the top line to the
right (the figure to the left will be described in the last paragraph of
this subsection), in each of three panels, for $k_{b}$=0.225, 0.250 and
0.275, i.e. along the brown, red, and olive dot-dashed lines in FIG.~I$~$\ref%
{FIGDoubleZone}. For clarity, the color of the $\widetilde{xy}$ bands has
been changed from dark-red to black in FIG.$~$\ref{Analysis}.

Upon increasing $k_{b}$ from 0.225 to 0.275, we see the $\widetilde{xy}$
bands develop from having strong upwards-pointing peaks in the upper band
near $k_{c}$=$\pm $0.75 and $\pm $0.25, plus small downwards bulges in the
lower band around $k_{c}$=$\pm $0.5, over having reduced peaks and increased
bulges --and thus minimal total width-- near midgap, to having large
downwards-pointing peaks connected pairwise by large bulges in the lower
band, plus reminiscences of the upwards-pointing peaks in the upper band.
This development is far from symmetric around the mid-gap energy $(\equiv 0)$%
, despite the fact that the V and C bands on either side of the direct gap
have the same character, apart from being respectively $xz$-$XZ$ (or $yz$-$%
YZ $) bonding and anti-bonding. In the present section we shall show in
didactic detail that the origin lies in the complicated bi-products (\ref{GC}%
) forming the residues, $\Gamma \left( \mathbf{k}\right) ,$ of the matrix
elements for the resonant couplings.

On lines 2-5 in Fig.~\ref{Analysis}, we identify and analyze the individual
contributions from the direct $xy$-$xy$ hops (red) and the indirect hops via
the $xz$ (blue) and $yz $ (green) V\&C bands to the diagonal and
off-diagonal elements, $\left\langle \widetilde{xy};\mathbf{k}\left\vert
H\right\vert \widetilde{xy};\mathbf{k}\right\rangle $ and $\left\langle 
\widetilde{xy};\mathbf{k}\left\vert H\right\vert \widetilde{xy};\mathbf{k+c}%
^{\ast }\right\rangle ,$ of the two-band Hamiltonian (\ref{H2}). We end on
line 5 with $\alpha \left( \mathbf{k}\right) \pm a\left( \mathbf{k}\right) $
in $\QATOP{\mathrm{purple}}{\mathrm{grey}}$ and $\gamma \left( \mathbf{k}%
\right) \pm g\left( \mathbf{k}\right) $ in $\QATOP{\mathrm{turquoise}}{%
\mathrm{grey}}$.

For simplicity in FIG.$\,$\ref{Analysis}, we have substituted $E$ in the
denominators of the resonance terms (\ref{H2}) by $\tau \left( k_{b}\right) $
and shall use a notation in which we drop this argument from e.g. $%
\left\langle \widetilde{xy};\tau \left( k_{b}\right) ,\mathbf{k}\left\vert
\,H\,\right\vert \widetilde{xy};\tau \left( k_{b}\right) ,\mathbf{k+c}^{\ast
}\right\rangle .$ This approximation slightly enhances the peak features, as
may be seen by comparison of FIG.s$~$\ref{9to11over40irrBZ} and \ref%
{Analysis}.

The bands --but not their $\left\vert \mathbf{k}\right\rangle $ decoration
(extra fatness)-- have the proper single-zone period $1$ in $k_{c}.$ Where
one band is fat and the other not, those bands have respectively pure $%
\left\vert \mathbf{k}\right\rangle $ and pure $\left\vert \mathbf{k+c}^{\ast
}\right\rangle $ character. This is the case for integer values of $k_{c}$,
whereas for half-integer values, the two bands are of 50\% mixed character.

To the left on the 1st line, we show in respectively fat and thin lines the
unhybridized $\widetilde{xy}\left( \mathbf{k}\right) $ and $\widetilde{xy}%
\left( \mathbf{k+c}^{\ast }\right) $ bands. These are the diagonal elements, 
$\left\langle \widetilde{xy};\mathbf{k}\left\vert H\right\vert \widetilde{xy}%
;\mathbf{k}\right\rangle $ and $\left\langle \widetilde{xy};\mathbf{k+c}%
^{\ast }\left\vert H\right\vert \widetilde{xy};\mathbf{k+c}^{\ast
}\right\rangle $ of the two-band Hamiltonian (\ref{H2}) and have the
double-zone period $2$ in $k_{c}$.

\subsubsection{Peak-, bulge-, and contact features\label{PB&C}}

The \emph{primary feature} of the $\widetilde{xy}$ bands in the gap, the 
\emph{resonance peaks}, originate from either an $xz$-band edge, which runs
along a blue \textrm{YZY'} line in FIG. I$~$\ref{FIGDoubleZone}, or from a $%
yz$-band edge, which runs along a green \textrm{YZY' }line in the same
figure, and are therefore located at the crossings between such a line and
the two red $\widetilde{xy}$-band CECs seen in the uppermost panels of FIG.
II~\ref{CEC}$\,$(b). In the CECs, the resonance peaks appear as \emph{notches%
}. With reference to the bands for fixed values of $k_{b}$ in FIG.~\ref%
{9to11over40irrBZ}, a resonance peak is located where a band edge crosses
the appropriate constant $k_{b}$-line (dot-dashed in FIG.~I$\,$\ref%
{FIGDoubleZone}). The resonance features are therefore well separated in $%
k_{c}$ as seen in FIG.$\,$\ref{Analysis} on the 1st line to the right.

From the fatnesses of the bands we see that the resonance peaks have almost
pure $\left\vert \widetilde{xy};\mathbf{k}\right\rangle $ or $\left\vert 
\widetilde{xy};\mathbf{k+c}^{\ast }\right\rangle $ character, although the
V\&C-band edges have $\sim $50\% mixed $\left\vert xz;\mathbf{k}%
\right\rangle $ and $\left\vert xz;\mathbf{k+c}^{\ast }\right\rangle $ (or $%
\left\vert yz;\mathbf{k}\right\rangle $ and $\left\vert yz;\mathbf{k+c}%
^{\ast }\right\rangle )$ characters, as we saw along $\Lambda \mathrm{W}$ ($%
k_{c}$=$\frac{1}{4})$ and $\Lambda ^{\prime }\mathrm{W}^{\prime }$ ($k_{c}$=$%
\frac{3}{4})$ in FIG.\thinspace II~\ref{ARPES_Bandstructure_Cuts}~(d) and
(e). The strong $\left\vert \mathbf{k}\right\rangle $ character is what
enabled us, in Sect.~\ref{SectRefining} of Paper II, to detect with ARPES
the large peak in the upper $\widetilde{xy}$ band from the resonance with
the blue $xz$ valence band at $\mathbf{k}$=$\left( 0.225,0.725\right) ,$
mirrored ("symmetrized") around $k_{c}$=0.5 to $\left( 0.225,0.275\right) $.
That the lower $\widetilde{xy}$ band merely exhibits a shoulder at $\left(
0.225,0.725\right) $ will be explained later, at the beginning of Sect. \ref%
{indictRes}.

The understanding is quite different for the \emph{secondary feature, }seen
around the BZ boundaries $\left( k_{c}\text{=}\pm 0.5\right) $ on the 1st
line in FIG.$\,$\ref{Analysis} to the right (but absent to the left). This
feature consists of a \emph{bulge} in the lower band and the concomitant
filling-in of the valleys between the neighboring resonance peaks repelled
by the $xz$ and $yz$ V or C bands, whichever is closer in energy. These
neighboring resonance peaks are therefore in the upper $\widetilde{xy}$ band
when $k_{b}$=0.225 and 0.250, and in the lower $\widetilde{xy}$ band when $%
k_{b}$=0.275. The bulge is caused by the hybridization between the $%
\widetilde{xy}\left( \mathbf{k}\right) $ and $\widetilde{xy}\left( \mathbf{%
k+c}^{\ast }\right) $ bands --displayed to the left-- which cross at $k_{c}$=%
$\pm 0.5$ and split by $\pm \left\vert \left\langle \widetilde{xy};\mathbf{k}%
\left\vert H\right\vert \widetilde{xy};\mathbf{k}+\mathbf{c}^{\ast
}\right\rangle \right\vert .$ The latter, off-diagonal matrix element of the
two-band Hamiltonian (\ref{H2}) is shown in black on the 2nd line to the
right. This element is seen to attain its largest absolute value near $k_{c}$%
=$\pm 0.5$ and, here, to have equal contributions from the indirect hops via
the $xz$ (blue) and $yz$ (green) bands, and to be amplified by the direct $%
xy $-$xy$ (red) contribution. To the left, and in the same colors, are shown
the diagonal element, $\left\langle \widetilde{xy};\mathbf{k}\left\vert
H\right\vert \widetilde{xy};\mathbf{k}\right\rangle ,$ and its three
contributions.

Also the \emph{third} characteristic \emph{feature} of the $\widetilde{xy}$
bands, the \emph{near contact} between the two bands --and in particular
between their CECs [FIG.~\ref{Theo_CEC}~(d)]-- on the $\Gamma \mathrm{Y}$
and $\Gamma ^{\prime }\mathrm{Y}^{\prime }$ lines ($k_{c}$=$\func{integer}$%
), is connected with the hybridization between the $\widetilde{xy}\left( 
\mathbf{k}\right) $ and $\widetilde{xy}\left( \mathbf{k+c}^{\ast }\right) $
bands, albeit with its zero rather than its maximum [FIG.$~$\ref{Analysis},
black curve on the 2nd line to the right]. In ARPES [FIG.\thinspace II$~\,$%
\ref{ARPES_Bandstructure_TB}~(c2)], as well as in previous calculations \cite%
{Satpathy2006}, an apparent crossing on the $\Gamma \mathrm{Y}$ line $\left(
k_{c}\text{=}0\right) $ was noted and a TB description attempted \cite%
{Chudzinski2012}. With our improved resolution, this peculiarity is now seen
[FIG.~\ref{Theo_CEC}~(d)] as a splitting between the two CECs, which along $%
\Gamma \mathrm{Y}$ is anomalously small\footnote{\label{NeglectLambda}
Including the $xz$-$yz$ hybridization, i.e. the $\left( \lambda ,l,\mu
,m\right) $ hopping integrals, as was done in FIG.s I~\ref%
{ARPES_Bandstructure_TB}~(c2) and \ref{Theo_CEC}~(d), but was neglected in
the two-band Hamiltonian and, hence, in FIG.s \ref{Analysis} and \ref%
{9to11over40irrBZ}, \emph{de}creases the splitting between the $\widetilde{xy%
}$ bands for integer $k_{c}$ by a factor $\sim 2.$} and even \emph{de}%
creases with energy. This is in contrast to the relatively large splitting
along ZC ($k_{c}=\frac{1}{2}+\func{integer})$ which is caused by $\left( 
\mathbf{k,k+c}^{\ast }\right) $-hybridization and increases with energy. The
splitting at integer $k_{c}$ is even \emph{smaller} than that of the
directly-coupled, red bands. This is simple to understand: First of all, the
splitting, 2$t\left( \mathbf{k}\right) =$ $8\left( t_{1}\cos \pi
k_{b}+t_{2}\cos 3\pi k_{b}\right) ,$ of the red bands decreases from 46~meV
for $k_{b}$=0.225 to 24~meV for $k_{b}$=0.275. Secondly, along $\Gamma 
\mathrm{Y}$ ($k_{c}$=0) the pure $xy\left( \mathbf{k}\right) ,$ $xz\left( 
\mathbf{k}\right) ,$ and $yz\left( \mathbf{k}\right) $ bands are all bonding
between ribbons while the $xy\left( \mathbf{k+c}^{\ast }\right) ,$ $xz\left( 
\mathbf{k+c}^{\ast }\right) ,$ and $yz\left( \mathbf{k+c}^{\ast }\right) $
bands are all anti-bonding [see FIG.\thinspace I~\ref{ThreePureBands}].
Since both $xy$ bands in the gap lie above the $xz\left( \mathbf{k}\right) $
and $yz\left( \mathbf{k}\right) $ bands, but below the $xz\left( \mathbf{k+c}%
^{\ast }\right) $ and $yz\left( \mathbf{k+c}^{\ast }\right) $ bands, the
valence bands will push the bonding $xy\left( \mathbf{k}\right) $ band up,
and the conduction bands will push the anti-bonding $xy\left( \mathbf{k+c}%
^{\ast }\right) $ band down in energy. Hence, the hybridization with the $xz$
and $yz$ V\&C bands will \emph{diminish} the separation between the $xy$
bands\footnote{{The result along $\Gamma ^{\prime }\mathrm{Y}^{\prime }$ ($%
k_{c}$=$\pm $1) is of course the same although, there, the $\left\vert 
\mathbf{k}\right\rangle $ bands are anti-bonding and the $\left\vert \mathbf{%
k+c}^{\ast }\right\rangle $ bands bonding.}}.

It is remarkable that in a region around $k_{c}$=integer, the lower $%
\widetilde{xy}$ band runs parallel with the red $xy$ band, and that the $%
k_{c}$-region over which this happens, as well as the distance above the $xy$
band, decreases with increasing $k_{b}$.

We emphasize that neither of the three characteristic features of the two
metallic bands in the gap can be described by merely a 2$\times $2 TB
Hamiltonian, but need the resonance terms. The three characteristic features
are seen as functions of $\left( k_{b},k_{c}\right) $ on the left-hand side
of FIG.~\ref{Theo_CEC} in Sect.~\ref{SectBandsandFS} to which we shall
return.

\subsubsection{Directly coupled terms\label{Direct}}

We now systematically identify the different terms of the two-band
Hamiltonian (\ref{H2}).

\begin{figure*}[tph]
\includegraphics[width=\linewidth]{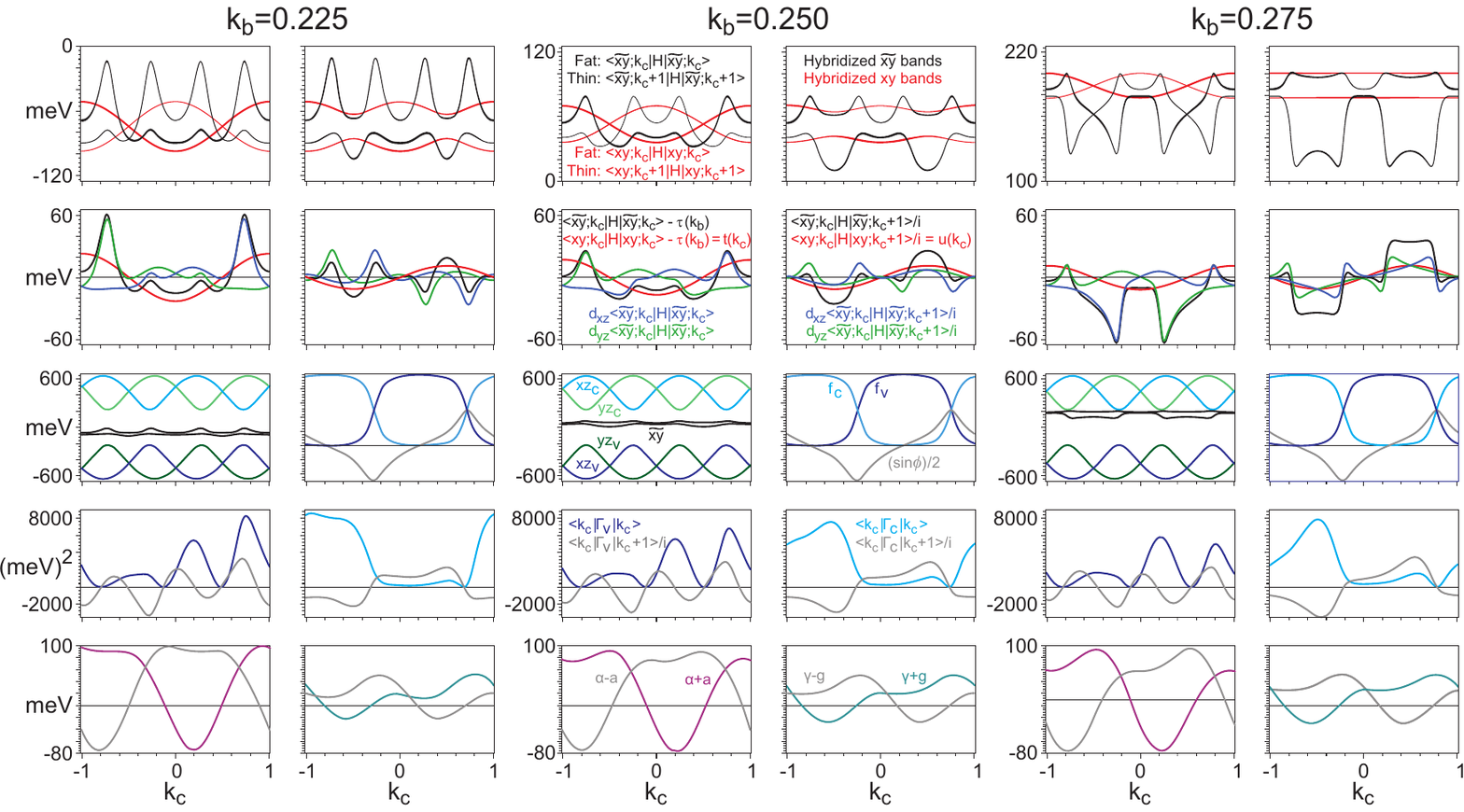}
\caption{The following caption is for each of the three $k_{b}$-panels: 
\textbf{1st line:} The two metallic $xy$ (red) and $\widetilde{xy}$ (black)
bands, decorated with their $\left\vert \mathbf{k}\right\rangle $ character,
as fct.s of $k_{c}$ in the double zone. $\left( \mathbf{k,k+c}^{\ast
}\right) $-unhybridized (\textbf{left}) and -hybridized (\textbf{right}).
The latter are as in FIG.~\protect\ref{9to11over40irrBZ}, but extended to
the double zone. As functions of $\left( k_{b},k_{c}\right) ,$ the six bands
were shown on a 2 eV scale in FIG. I~\protect\ref{3Dt2gBands} and the two
metallic bands are shown on a 0.3 eV scale in FIG. \protect\ref{Theo_CEC}
(a)-(c). The \textbf{2nd line:} Diagonal, $\left\langle \widetilde{xy};%
\mathbf{k}\left\vert H\right\vert \widetilde{xy};\mathbf{k}\right\rangle -%
\protect\tau \left( k_{b}\right) $ (\textbf{left}), and off-diagonal, $%
\left\langle \widetilde{xy};\mathbf{k}\left\vert H\right\vert \widetilde{xy};%
\mathbf{k+c}^{\ast }\right\rangle /i$ (\textbf{right}), elements of the
two-band Hamiltonian (\protect\ref{H2}). In black, the sum of the
contributions from the direct, inter-ribbon $xy$-$XY$ hops (red) and from
the indirect hops via the $xz$ (blue) and $yz$ (green) valence (V) and
conduction (C) bands. [Not shown are the $xy\left( \mathbf{k+c}^{\ast
}\right) $ and $\widetilde{xy}\left( \mathbf{k+c}^{\ast }\right) $ bands
minus $\protect\tau \left( k_{b}\right) $ in thin black]. \textbf{3rd line
left:} The black $\widetilde{xy}$ bands in the gap between the blue $xz$ and
green $yz$ V (dark) and C (light) bands; the hybridizations of the two
latter with the $xy$ bands were neglected, i.e. $\protect\varepsilon _{xz}$=$%
\mp \protect\sqrt{A^{2}+G^{2}},$ but not vice versa. \textbf{3rd line right: 
}$\QTATOP{\mathrm{Light}}{\mathrm{Dark}}$blue\textrm{:} the probability that
at $\mathbf{k,}$ the $\left\vert xz;\mathbf{k}\right\rangle $ character is
in the $\QATOP{\mathrm{C}}{\mathrm{V}}$ band is $f_{\QATOP{\mathrm{C}}{%
\mathrm{V}}}\equiv \left( 1\mp \cos \protect\phi \right) /2$ Eq.~I~(61).
Grey: $\left( \sin \protect\phi \right) /2.$ \textbf{4th line:} Diagonal
(blue) and off-diagonal (grey) matrix elements, (\protect\ref{Gdiag}) and (%
\protect\ref{Goff}), in meV$^{2}$ of the residue for the perturbation of the
two-band Hamiltonian via the $xz$ V band (\textbf{left}) or C band (\textbf{%
right}). \textbf{5th line left: }Bloch sums of $xy$-$xz$ hopping integrals
I~(\protect\ref{ag}): $\protect\alpha +a\mathbf{\ }$(purple), $\protect%
\alpha -a$ (grey); \textbf{right:} $\protect\gamma +g$ (turquoise), $\protect%
\gamma -g$ (grey). These four Bloch sums are periodic in the double zone and
each grey Bloch sum equals the colored one inside the same frame, but
translated by $\mathbf{c}^{\ast }.$ Energies are in meV with respect to the
centre of the gap. The ARPES-refined parameter values were used. $E$ in the
resonance terms was approximated by $\protect\tau \left( k_{b}\right) .$ See
Sects.\thinspace I~\protect\ref{SectH} and III \protect\ref{SectH2}. }
\label{Analysis}
\end{figure*}

Its first term, the energy $\tau \left( k_{b}\right) $ of the two degenerate
1D intra-ribbon $xy$ bands, is included only on the 1st line of
FIG.\thinspace \ref{Analysis}, where it is the average of the two red,
directly-coupled $xy$ bands, $xy\left( \mathbf{k}\right) $ and $xy\left( 
\mathbf{k+c}^{\ast }\right) $ to the left, or of the $\left( \mathbf{k,k+c}%
^{\ast }\right) $-hybridized bands to the right. This average is independent
of $k_{c}.$ On the 2nd line to the left, $\tau \left( k_{b}\right) $ is
neither included in the red, directly-coupled $xy\left( \mathbf{k}\right) $
band, nor in the black, directly plus indirectly-coupled $\widetilde{xy}%
\left( \mathbf{k}\right) $ band.

The second term in Eq.$\,$(\ref{H2}) is the $xy$-block of the six-band
Hamiltonian I~(56) and it gives the perpendicular dispersions and splitting
of the red $xy$ bands shown on the 1st line in FIG.$\,$\ref{Analysis}. The
corresponding diagonal and off-diagonal matrix elements, $t\left( \mathbf{k}%
\right) $ and $iu\left( \mathbf{k}\right) ,$ are the Bloch sums of
respectively the average $xy$-$XY$ hoppings and their dimerizations. They
are given in Eq.$~$I$\,$(\ref{tautu}) and are shown in red on the 2nd line
to respectively the left and the right. These Bloch sums of direct hoppings
are seen to depend little on $k_{b}$ in the $\pm 10\%$ interval around $%
k_{F}.$

Also the TB model \cite{Chudzinski2012} upon which current TLL theories \cite%
{Merino2015}\cite{Chudzinski2017} are based, includes 1st- and
2nd-nearest-neighbor terms. But in the attempt to fit the peak, bulk, and
band-crossing features of the LDA FS \cite{Satpathy2006} without recognizing
their resonant nature, the resulting TB model had an unphysical form
(containing e.g. $\sin \pi k_{c}$ and $\sin 2\pi k_{c}$ terms) and, as a
consequence, its parameter values are incompatible with ours. That the
magnitude of its FS warping is several times ours is partly because the
stoichiometry was taken to be Li$_{0.90}$ rather than Li$_{1.02}.$

\subsubsection{Symmetries \label{Sym}}

The red, directly-coupled and the black, directly plus indirectly-coupled $%
\left\vert \mathbf{k}\right\rangle $ bands shown to the left on the 2nd line
are even around $k_{c}$=$\func{integer}$. The red and black matrix elements
shown to the right couple each of these bands to itself after a translation
of $k_{c}$ by $1.$ These off-diagonal elements, divided by $i,$ are odd
around $k_{c}$=$\func{integer}$ and even around $k_{c}$=$\pm $0.5. The
indirect couplings alone, i.e. the perturbations of the $\widetilde{xy}$%
-band Hamiltonian by the $xz$ or $yz$ V\&C bands, are shown in respectively
blue and green. They are related to each other by a sign change$^{\text{%
II\thinspace \ref{inversion}}}$ of $k_{b}$, and those blue and red curves to
the left/right are related to each other by a mirror/anti-mirror operation
around $k_{c}$=$\func{integer}.$ Moreover, each of the blue and green curves
to the right change sign upon $k_{c}$-translation by 1, i.e. they are
anti-periodic, Eq.$\,$(\ref{TcResOff}).

\subsubsection{Indirectly coupled terms; role of the denominators\label%
{indirectlyDen}}

The indirect couplings via the $xz$ and $yz$ bands are additive and given by
respectively the 3rd\&4th and the 5th\&6th terms in expression (\ref{H2}),
provided that the hybridization between the $xz$ and $yz$ bands is neglected.%
$^{\ref{NeglectLambda}}$ The perturbations via the $xz$ and $yz$ bands have
been subdivided into V and C bands whereby each of them takes the form of a
single resonance (pole) with the denominator being the distance between the
narrow $\widetilde{xy}$ band and one of the four $xz$ or $yz$ V or C bands.
The energy of the former is $E$ $\left[ \sim \tau \left( k_{b}\right) \right]
,$ the energy of an $yz$ band is given by expression$~$I$\,$(\ref{gap}), and
that of an $xz$ band by the same expression with $k_{b}$ substituted by $%
-k_{b}.$ While each denominator is a single-periodic, scalar function of 
\textbf{k}$,$ each nominator (residue) is a double-periodic 2$\times $2
matrix function given by Eq.s (\ref{GC})-(\ref{Goff}).

From now on, we shall take advantage of the mirror/anti-mirror symmetries
mentioned in the previous Sect.~\ref{Sym} between the blue $xz$- and the
green $yz$ perturbations of the two-band Hamiltonian to consider merely the
blue $xz$ perturbation, which we shall trace back from the 2nd line in FIG.~%
\ref{Analysis} to the Bloch sums $A$ and $G$ (of respectively the symmetric
and asymmetric $xz$-$xz$ integrals for hopping in- and out-side a bi-ribbon)
giving the $xz$ C\&V-band energies to the left and their characters, $f_{C}$
to the right on the 3rd line, and to the Bloch sums of the integrals for
hopping between the $xy$ and $xz$ WOs on the 5th line.

The most important factor influencing the shapes of the diagonal and
off-diagonal matrix elements shown in blue on the 2nd line to the left and
the right, and given by the 3rd\&4th terms in Eq.$\,$(\ref{H2}), is their
common energy denominator. This is the distance seen on the 3rd line to the
left between the $\widetilde{xy}$ bands in black and the $xz$ V or C bands
in respectively dark and light blue. Also shown are the $yz$ V and C bands
in respectively dark and light green. The edges, $\mp 2\left\vert
G_{1}\right\vert ,$ of the $xz$ bands (blue in FIG.~I$\,$\ref{FIGDoubleZone}%
) are along $k_{c}=$ $k_{b}\mp \frac{1}{2}+2n,$ which for the three chosen
values of $k_{b},$ and for $k_{c}$ in the $\left( -1|1\right) $ double-zone
are at $k_{c}\sim -$0.25 and 0.75$.$ This is where the $xz$-band edges may
cause resonance peaks in the $\widetilde{xy}$ bands. The $\left\vert \mathbf{%
k}\right\rangle $ characters of the $xz$ Vand C bands, $f_{V}$ and $f_{C},$
are given by respectively the dark- and light-blue curves to the right on
the 3rd line in FIG.~\ref{Analysis}.

When $k_{b}$=0.225, there is a large peak in the $\left\vert \mathbf{k+c}%
^{\ast }\right\rangle $-unhybridized $\widetilde{xy}\left( \mathbf{k}\right) 
$ band on the 1st line to the left near 0.75 and a small one near $-$0.25.
On the 2nd line, both peaks are blue and point upwards, i.e. are caused by
repulsion from the $xz$ valence band. Their sizes decrease strongly as
--with $k_{b}$ increasing$-$ the $\widetilde{xy}$ bands move upwards, away
from the valence band. For $k_{b}$=0.250, the peaks can still be seen in the
unhybridized $\widetilde{xy}\left( \mathbf{k}\right) $ bands, as well as in
the fully hybridized $\widetilde{xy}$ band on the 1st line to the right. But
for $k_{b}$=0.275, when the upper $\widetilde{xy}$ band is touching the
bottom of the C-band edge, only the peak from the $xz$ V-band resonance near
0.75 has survived. The small peak near $-$0.25 has been overpowered by a
large, downwards pointing C-band resonance, shaped like a \emph{canine tooth}%
. Going back to $k_{b}$=0.250, this tooth is reduced to a "hole" on the low-$%
k_{c}$ side of the small, blue V-band peak near $k_{c}$=$-$0.25.

The contribution from the $xz$ V\&C bands to the hybridization between the $%
\widetilde{xy}\left( \mathbf{k}\right) $ and $\widetilde{xy}\left( \mathbf{%
k+c}^{\ast }\right) $ bands is shown in blue (and divided by $i)$ on the 2nd
line to the right. Whereas the diagonal element (\ref{Gdiag}) of the residue
matrix (\ref{GC}) is never negative, its $\left( \mathbf{k,k+c}^{\ast
}\right) $-mixing off-diagonal elements (\ref{Goff}) are purely imaginary
and anti-periodic, i.e. they change sign upon translation of $k_{c} $ by 1.
For $k_{b}$=0.225 the blue peak pointing downwards near 0.75 is similar to
that of the diagonal element pointing upwards, but its magnitude is reduced
by roughly a factor 2. For $k_{b}$ increasing, this peak decreases further
and it gets superposed by the growing, anti-periodic canine-tooth structure.
At the zone boundaries (ZB), $k_{c}$=$\pm $0.5, the $\widetilde{xy}\left( 
\mathbf{k}\right) $ and $\widetilde{xy}\left( \mathbf{k+c}^{\ast }\right) $
bands are degenerate, but get split by $\pm $ the numerical value of the
off-diagonal element of the two-band Hamiltonian (\ref{H2}), which is seen
to increase strongly with $k_{b}$. The reason is --as we shall see below--
that the contribution from the V band nearly vanishes at the zone
boundaries. The uncompensated repulsion from the C band is then what causes
the development of the bulge in the lower $\widetilde{xy}$ band.

The V band thus causes peaks in the upper $\widetilde{xy}$ band, and the C
band causes merging canine teeth plus ZB-centered bulges in the lower $%
\widetilde{xy}$ band. The peaks and the teeth are resonance features
occurring where the FS, $\left\vert k_{b}\right\vert \sim \frac{1}{4},$
crosses between the edges, $\left\vert k_{c}\pm k_{b}\right\vert $=$\frac{1}{%
2},$ of $xz$ and $yz$-like V\&C bands.

\subsubsection{Indirectly coupled terms; role of the residues\label%
{indictRes}}

The sign of a resonance term is that of its denominator, i.e. it is \emph{%
repulsive}. As a consequence, \emph{if} the Hamiltonian minus $\tau \left(
k_{b}\right) $ is dominated by \emph{one} of the resonance terms$,$ e.g. due
to a small denominator, that term will repel \emph{one} of the two $xy$
bands, and leave the other band unperturbed. Examples are seen on the 1st
line to the right in FIG.\thinspace \ref{Analysis}: Where a resonance peak
exists in one of the $\widetilde{xy}$ bands, there is merely a tiny peak or
shoulder in the other band. Taking, first, the resonance peak as the one
caused by the blue $xz$ valence band near $k_{c}$=0.75 for $k_{b}$=0.225 or
0.250, the resonance terms parts relevant for the blue $\delta _{xz}$%
-perturbation on the 2nd line to the left, $\frac{\left\langle \mathbf{k}%
\left\vert \Gamma _{V}\right\vert \mathbf{k}\right\rangle }{\tau \left(
k_{b}\right) +\sqrt{A^{2}\left( \mathbf{k}\right) +G^{2}\left( \mathbf{k}%
\right) }}$ and $\frac{\left\langle \mathbf{k}+c^{\ast }\left\vert \Gamma
_{V}\right\vert \mathbf{k}+c^{\ast }\right\rangle }{\tau \left( k_{b}\right)
+\sqrt{A^{2}\left( \mathbf{k}\right) +G^{2}\left( \mathbf{k}\right) }},$ are
those near $k_{c}$=0.75 and --0.25. The closeness of the black and blue
curves confirms that the Hamiltonian is,\ in fact, dominated by this \emph{%
one} resonance term. Next, we go to $k_{b}$=0.275 where the $\widetilde{xy}$
bands are located just below the bottom of the C bands. The peak in the
upper band caused by the repulsion from the V-band edge can still be seen on
the top line to the right near $k_{c}$=0.75, but the nearby C-band edge
repels the lower band much further. In fact, it is now the upper band which
is the flatter and has an energy near the upper red, pure $xy$ band. The
closeness of the black and blue curves on the 2nd line to the left confirms
that the Hamiltonian is dominated by the $xz$-band resonances, with a minor,
peak-shaped contribution from the V band near $k_{c}$=0.75 and a major,
contribution with the shape of a canine tooth from the C band near $k_{c}$%
=--0.25. This behavior is also clearly seen in the band structures, FIG.{s}%
~II$\,$\ref{ARPES_Bandstructure_LDA} and~\ref{ARPES_Bandstructure_TB}, along
the $\mathrm{ZY}$ and $\mathrm{W\Lambda }$ lines as was described in
Sect.\thinspace II$\,$\ref{SectAgreement}. This could be another reason for
the "non-linearity" seen in FIG.$\,$\ref{Fig:Experiment_Anal2}.

The residues $\Gamma \left( \mathbf{k}\right) $ of the 4 resonance terms in
expression (\ref{H2}) are 2$\times $2 matrices (\ref{GC}) with diagonal
elements forming a real-valued, non-negative function of $\mathbf{k}$ which
is periodic in the double zone (\ref{TcResOn}) and imaginary off-diagonal
elements forming an anti-periodic function of $\mathbf{k}$ in the single
zone (\ref{TcResOff}). These properties are clearly exhibited by the plots
of expressions (\ref{Gdiag}) and (\ref{Goff}) on the 4th line of FIG.~\ref%
{Analysis} showing $\Gamma _{C}\left( \mathbf{k}\right) $ on the right-hand
sides with the diagonal element $\left\langle \mathbf{k}\left\vert \Gamma
_{C}\right\vert \mathbf{k}\right\rangle $ in light blue and the off-diagonal
element $\left\langle \mathbf{k}\left\vert \Gamma _{C}\right\vert \mathbf{k+c%
}^{\ast }\right\rangle /i$ in grey. Similarly on the left-hand sides: $%
\Gamma _{V}\left( \mathbf{k}\right) $ with the diagonal element $%
\left\langle \mathbf{k}\left\vert \Gamma _{V}\right\vert \mathbf{k}%
\right\rangle $ in dark blue and the off-diagonal element $\left\langle 
\mathbf{k}\left\vert \Gamma _{V}\right\vert \mathbf{k+c}^{\ast
}\right\rangle /i$ in grey. Whereas the light blue conduction band residue
follows $\left\vert \mathbf{k}\right\rangle $ character of the conduction
band, $f_{C}$ (3rd line to the right), as expected, the dark blue valence
band residue drops to zero near $k_{c}$=0.5, which is in the middle of the
region where the $\left\vert \mathbf{k}\right\rangle $ character of the
valence band dominates, and the same happens near $-0.1$. Expression (\ref%
{Gdiag}) together with the plots of $\alpha \pm a$ and $\gamma \pm g$ on the
5th line show that the reason for the unexpected behavior of $\Gamma _{V}$
is the u-shape with two zeroes of $\alpha \left( \mathbf{k}\right) +a\left( 
\mathbf{k}\right) .$ The zero of $\Gamma _{V}$ near $-0.85$ is caused by the
zero of $\gamma \left( \mathbf{k}\right) +g\left( \mathbf{k}\right) $ shown
in turquoise on the bottom line to the right.

The peaks due to resonances with the $xz$ bands occur near the $xz$-band
edges. Exactly at the edges, $f_{C}$=$f_{V}$=$\frac{1}{2}$ and $\sin \phi
=\pm 1.$ The $\mathbf{k}$-conserving part of the residues therefore takes
the values:%
\begin{equation}
\left\langle \mathbf{k}\left\vert \Gamma _{\QATOP{\mathrm{C}}{\mathrm{V}}%
}\right\vert \mathbf{k}\right\rangle =\frac{1}{2}\left[ \left( \alpha
+a\right) \pm \left( \gamma +g\right) \sin \phi \right] ^{2}.  \label{4peaks}
\end{equation}%
The magnitudes and signs of $\alpha +\alpha $ and $\gamma +g$ shown in
respectively purple and turquoise on the 5th line, cause the $\Gamma _{V}$
coupling at the $k_{c}$=0.75 edge (where $\sin \phi $=$-1)$ and the $\Gamma
_{C}$ coupling at the $k_{c}$=--0.25 edge (where $\sin \phi $=1) to be much
stronger than the two others, i.e. than the $\Gamma _{V}$ coupling at --0.25
and the $\Gamma _{C}$ coupling at 0.75, which even has a "hole" here. This
is exactly the behavior of the blue peaks seen on the 2nd line to the left.

We finally come to the grey, off-diagonal elements, $\left\langle \mathbf{k}%
\left\vert \Gamma \right\vert \mathbf{k+c}^{\ast }\right\rangle ,$ on the
4th line. They are --roughly speaking-- anti-periodic (\ref{TcResOff})
versions of the double-periodic blue, diagonal elements. The reason why the
grey $\left\langle \mathbf{k}\left\vert \Gamma _{V}\right\vert \mathbf{k+c}%
^{\ast }\right\rangle $ to the left is far more wiggly than the grey $%
\left\langle \mathbf{k}\left\vert \Gamma _{C}\right\vert \mathbf{k+c}^{\ast
}\right\rangle $ to the right, is that not only does the former possess the
two "extra" zeroes from $\alpha +a$ near $-0.1$ and $0.5,$ as well as the
one from $\gamma +g$ at $-0.85,$ but also those translated by $1$, i.e.
those near $-0.5,$ $0.9,$ and $0.15.$

These very different $k_{c}$-dependencies seen on the 4th line of the blue $%
\left\langle \mathbf{k}\left\vert \Gamma _{C}\right\vert \mathbf{k}%
\right\rangle $ and $\left\langle \mathbf{k}\left\vert \Gamma
_{V}\right\vert \mathbf{k}\right\rangle $ curves, and of the grey $%
\left\langle \mathbf{k}\left\vert \Gamma _{C}\right\vert \mathbf{k+c}^{\ast
}\right\rangle $ and $\left\langle \mathbf{k}\left\vert \Gamma
_{V}\right\vert \mathbf{k+c}^{\ast }\right\rangle $ curves, i.e. of the
conduction- and valence-band residues, are the causes of the strong
asymmetry of the perpendicular dispersion and splitting of the metallic
bands around the center of the gap.

\subsubsection{$xy$-$xz$ and $xy$-$yz$ hopping integrals\label{Sectxy-xz}}

The purple $\alpha \pm a$ and turquoise $\gamma \pm g$ Bloch sums I~(\ref{ag}%
) shown on the bottom line are determined by the hopping integrals, $a_{n},$ 
$g_{n},$ $\alpha _{n},$ and $\gamma _{n},$ computed as matrix elements I (%
\ref{FT}) of the LDA Hamiltonian between $n$th-nearest-neighbor $xy$ and $xz$
(or $XZ)$ WOs (see FIG.\thinspace I$~$\ref{Wannier}) with the results given
in Table I~(\ref{ap}). Specifically, the integrals for hopping between $xy$
and $XZ$ WOs on \emph{different} sublattices are $a_{n}\pm g_{n}.$ Here, $%
a_{1}$ is the average of- and $g_{1}$ half the difference between the
integrals for hopping from $xy$ at the origin to $XZ$ on the neigboring
ribbon, inside or outside the same bi-ribbon, i.e. to $XZ$ at respectively $%
-0.012\mathbf{a}-0.5\mathbf{b+}0.467\mathbf{c}$ and $-\left( 0.012\mathbf{a-}%
0.5\mathbf{\mathbf{b+}}0.533\mathbf{c}\right) $; similarly for $%
a_{1}^{\prime }$ and $g_{1}^{\prime },$ except that the $XZ$ orbital is
translated by $\mathbf{b}$. For $a_{2}$ and $g_{2},$ the $XZ$ WO is
translated by $-2\mathbf{b,}$ and for $a_{2}^{\prime }$ and $g_{2}^{\prime
}, $ by $2\mathbf{b.}$ For the Greek-lettered hopping integrals, the two
orbitals are on the \emph{same} sublattice. Specifically, the integrals for
hopping from $xy$ at the origin to $xz$ at $\pm \mathbf{b}$ are $\alpha
_{1}\pm \gamma _{1},$ to $xz$ at $\pm \mathbf{c}$ are $\alpha _{2}\pm \gamma
_{2},$ to $xz$ at\textbf{\ }$\pm \left( \mathbf{c+b}\right) $ are $\alpha
_{3}\pm \gamma _{3},$ and to $xz$ at $\mathbf{\pm }\left( \mathbf{c-b}%
\right) $ are $\alpha _{3}^{\prime }\pm \gamma _{3}^{\prime }.$ Calling $%
\gamma $ an electronic dimerization is really a misnomer, because the reason
for its existence is simply the difference of relative orientation of the
two orbitals. Finally, $\alpha _{0}$ is the $xy$-$xz$ on-site
(crystal-field) term$.$

The parameters dominating the behavior of the $\alpha +a$ Bloch sum are: the
integral for hopping between the $xy$ and $XZ$ nearest-neighbor WOs, $a_{1}$=%
$-$49~meV, and the crystal-field term, $\alpha _{0}$=31~meV. Had the former
been the only non-vanishing parameter in the $\alpha +a$ Bloch sum, the
corresponding term, $2a_{1}\cos \pi \left( k_{c}-k_{b}\right) ,$ would have
killed the peak from the valence-band resonance at $\left\vert
k_{c}-k_{b}\right\vert \mathrm{=}\frac{1}{2}.$ So, clearly, this peak
--convincingly observed with ARPES-- is sensitive to the value of the
crystal-field term caused by the ribbon-inversion (see Sect.\thinspace I$~$%
\ref{SectDims}) and to the details of the $xy$-$XZ$ and $xy$-$xz$ hoppings.
Note that none of these parameter values were adjusted to fit the ARPES.

\subsubsection{Synthesis\label{SectSynth}}

From the bottom three lines in FIG.$\,$\ref{Analysis} we have seen that the $%
k_{c}$-dependencies of the $A,$ $G,$ $\alpha +a,$ and $\gamma +g$ Bloch sums
of the $xz$-$xz$ and $xy$-$xz$ hopping integrals change relatively little
for $k_{b}$ in the $\pm 10\%$ range around $k_{F}$=$\frac{1}{4}.$

By far the strongest $k_{b}$-variation of the black $\widetilde{xy}$ bands
displayed on the top two lines is the one coming from the denominators of
the four resonance terms via $E\approx \tau \left( k_{b}\right) ,$ to be
seen on the 3rd line to the left, in combination with the very different
shapes of the V- and C-band residues seen on the 4th line to respectively
the left and the right.

What makes the blue resonance peak caused by the edge of the $xz$ V band
--seen on the 2nd line to the left near $k_{c}$=0.75-- differ in shape from
the (unhybridized) blue canine-tooth resonance near $k_{c}$=--0.25 caused by
the edge of the $xz$ C band$,$ is the zero of the purple $\alpha +a$ near $%
k_{c}$=0.5 seen on the bottom line. This zero is a bit inside the frame of
the dark blue $f_{V}$ window $\left( -0.25|0.75\right) $ and therefore "cuts
a hole" in $\Gamma _{V}$ on the low-$k_{c}$ side of the resonance, which is
thereby sharpened up. Nothing like this happens for $\Gamma _{C}$ near $%
k_{c} $=--0.25$,$ because the zero of $\alpha +a$ near --0.1 is outside the $%
f_{C}$ window $\left( -1.25|-0.25\right) $. Hence, it is the shape of the
canine tooth which is the simpler!

On the other hand, as seen for $k_{b}$=0.275 on the 1st line to the left,
the backside of the tooth at $k_{c}$=--0.25 reaches across the ZB at --0.5,
where it is crossed symmetrically by the backside of the $\widetilde{xy}%
\left( \mathbf{k+c}^{\ast }\right) $-band tooth caused by the resonance with
the $yz$ C band at $-$0.75. To the right and in black, we now see that
strong $\left( \mathbf{k,k+c}^{\ast }\right) $-hybridization around $-$0.5
merges the canine teeth in the lower band at $-0.75$ and $-$0.25, thus
resulting in a 60-meV splitting of the two $\widetilde{xy}$ bands.

We can go back and compare with what happens for $k_{b}$=0.225$.$ Here, we
see on the 2nd line to the left that the blue resonance peak at 0.75 is so
sharp, that it hardly reaches the ZB at 0.5 and therefore hardly overlaps
the peak at 0.25 in the $\widetilde{xy}\left( \mathbf{k+c}^{\ast }\right) $
band (seen above on the 1st line) from the $yz$ valence band. Moreover, the
hybridization at the zone boundary ZB is much weaker than for $k_{b}$=0.275
(black curves to the right on the 2nd line) so that it merely leads to the
formation of a bulge in the lower band, 35 meV below the minimum in the
upper band between its resonance peaks (1st line to the right).

The zero of the purple $\alpha +a$ near $k_{c}$=0.5 which sharpens the peaks
from the V bands, also makes the V bands (dark blue and dark green on line
3) contribute nothing to the $\left( \mathbf{k,k+c}^{\ast }\right) $%
-hybridization at the ZB. The hybridization, therefore, comes exclusively
from the C bands and from the dimerization, $u,$\ of the direct,
perpendicular hops (red curves to the right on the 2nd line). The blue and
the green --equally large-- contributions each have a residue given by the
value at $k_{c}$=0.5 of the grey curve to the right on line 4. For $k_{b}$
increasing from 0.225 to 0.275, this value increases from 2500 to 3500~meV$%
^{2}$ and thereby enhances the dominating effect of the decreasing
denominator.

\begin{figure*}[tbh]
\includegraphics[width=\linewidth]{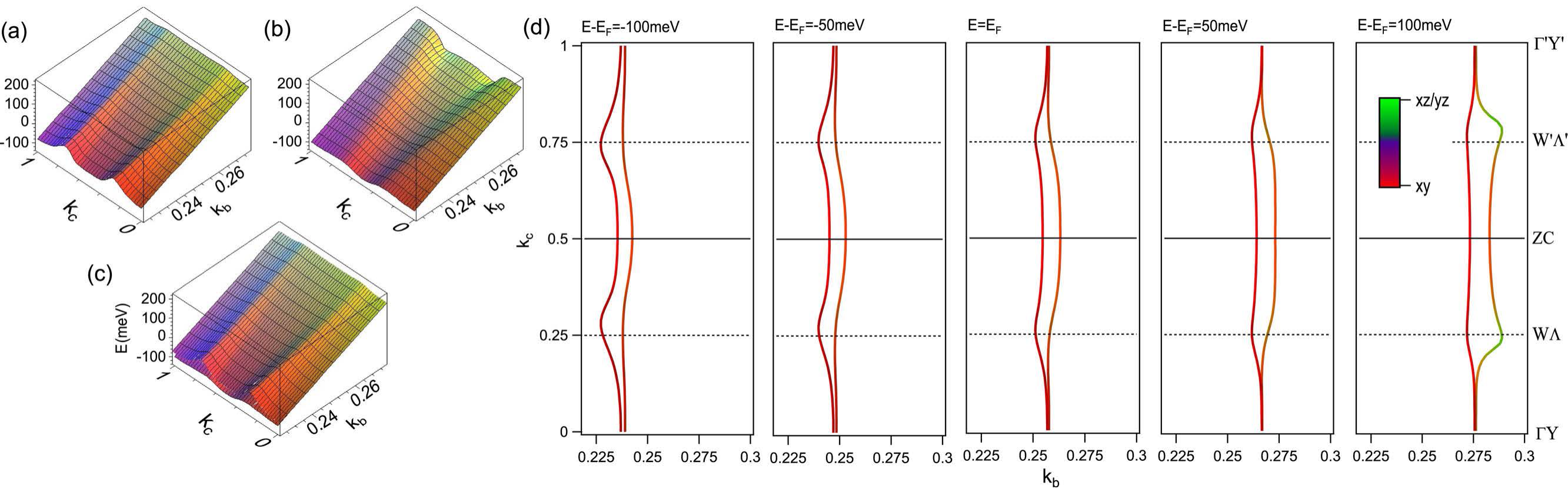}
\caption{\textbf{(a)-(c)}: The two $\widetilde{xy}$\ quasi-1D bands in the
gap for the six-band Hamiltonian I (\protect\ref{HRecip}) with the
ARPES-refined parameter values I~(\protect\ref{taup})-(\protect\ref{lp}).
The $\mathbf{k}$-space region considered is the stripe $\pm 10\%$ around $%
k_{Fb}$ in the upper half of the 1st and the lower half of the 2nd BZ, $%
0\leq k_{c}\leq 1.$ The energies of the upper \textbf{(a)}, the lower 
\textbf{(b)}, and\textbf{\ }both metallic bands \textbf{(c)}, relative to
the center of the gap, defined as $0.$ $E_{F}$=75~meV above the center. 
\textbf{(d): }CECs calculated by tracing the roots of the secular
determinant $\left\vert H\left( k_{a},k_{b}\right) -E\right\vert .$ The
colors indicate the orbital characters of the bands. The notches in the
inner sheet point to $\mathrm{Z}$ and those in the outer sheet to $\mathrm{Y}
$ and $\mathrm{Y}^{\prime }$ (see FIG.s~\protect\ref{FIGPhysicalZones2} and 
\protect\ref{CEC} in Paper II). The CECs have been \emph{compressed} by a
factor 3.3 along $k_{c}$ in order to make their warping visible. The ratio
between the warping $\protect\delta k_{\QATOP{i}{o}Fb}$ of each (inner or
outer) FS sheet and $k_{Fb}\sim 0.25$ is $\sim $0.02. This is $\sim $5 times
more than without the resonant coupling to the V\&C bands (see red bands for 
$k_{b}$=0.25 in FIG.s~\protect\ref{9to11over40irrBZ} and \protect\ref%
{Analysis}).}
\label{Theo_CEC}
\end{figure*}

Finally, we shall explain why around $k_{c}$=0 the \emph{lower} band is so
flat, more than the upper band, and why with increasing $k_{b}$ this
flatness increases and its range decreases$.$

But first, we will explain why the repulsion of the \emph{upper} $\widetilde{%
xy}$ band by the C band increases far less with $k_{b}$ than expected from
the decrease of the denominators (3rd line, left). The reason is found on
lines 4 and 5: For the upper band near $k_{c}$=0 --which is the $\widetilde{%
xy}\left( \mathbf{k+c}^{\ast }\right) $ band and, hence, the $\widetilde{xy}%
\left( \mathbf{k}\right) $ band near $\pm $1-- the light-blue C-band
residue, $\Gamma _{C}\left( k_{b},~k_{c}\sim \pm 1\right) ,$ decreases by a
factor 4 for $k_{b}$ increasing from 0.225 to 0.275, mainly because $\alpha
+a$ decreases by almost a factor 2. This trend is furthermore enhanced by a
non-vanishing repulsion from the V band whose dark-blue residue near $k_{c}$=%
$\pm $1 hardly changes with $k_{b}$ and thus becomes more important when $%
\Gamma _{C}\left( k_{b},\pm \text{1}\right) $ is small.

The reason why around $k_{c}$=0 the black lower $\widetilde{xy}$ band runs
parallel to the red $xy$ band is (see line 2 to the left) that the
repulsions from the blue $xz$ and the green $yz$ V bands disperse in
opposite directions away from $k_{c}$=0, whereby their effects on the
dispersion cancel. The reason why the distance of the black band above the
red band as well as the $k_{c}$-extent of its flat part decreases with
increasing $k_{b},$ is the same as the reason why the blue resonance peak
near $k_{c}$=$-$0.25 is much smaller than the one near 0.75, namely: that
for the dark-blue V-band residues on the 4th line $\Gamma _{V}\left(
k_{b},-0.25\right) $ is much less than $\Gamma _{V}\left( k_{b},0.75\right)
, $ and this --in itself-- is because the zero of $\alpha +a$ at $k_{c}$=$-$%
0.1 is closer to $-0.25$ than the zero at 0.5 is to 0.75 (purple curves on
line 5). As we now --with $k_{b}$ increasing-- move up in the metallic
bands, the V-band perturbation decreases due to the increasing energy
denominator and --as the C band is approached-- canine teeth growing near $%
k_{c}$=$-$0.25 and 0.25 limit the region over which the lower band is flat.

We conclude that the remarkable asymmetry between the contributions from the
V\&C bands to the $k_{c}$-dispersion and splitting of the $\widetilde{xy}$
bands in the gap, is mainly due to the difference between the positions of
the V\&C bands with respect to the structure in the $\mathbf{k}$-conserving $%
\alpha +a$ Bloch sum of the $xy$-$xz$ hopping integrals. Specifically, the
zero of $\alpha +a$ near $k_{c}$=0.5 is inside the region where the V band
is formed by the $xz\left( \mathbf{k}\right) $ band --and the C band by the $%
xz\left( \mathbf{k+c}^{\ast }\right) $ band (see see dark- and light-blue
curves in the figures to the right on line 3)-- and not the other way around.

\subsection{Constant energy contours\label{SectBandsandFS}}

On the left-hand side of FIG.$\,$\ref{Theo_CEC}, we show the upper (a), the
lower (b), and both (c) metallic $\widetilde{xy}$ bands in the gap, which
extends from $-$218~meV to $+$218~meV, as functions of $\left(
k_{b},k_{c}\right) $ in the stripe $0.225\leq k_{b}\leq 0.275$ and $0\leq
k_{c}\leq 1.$ From the description at the beginning of Sect.\thinspace \ref%
{SectOrigins}, we recognize the development of the bands --for $k_{b}$
increasing-- from having strong, upwards-pointing resonance peaks near $%
k_{c} $=0.75 and 0.25 in the upper band (a), plus a small downwards bulge
around $k_{c}$=0.5 in the lower band (b), over having reduced peaks plus a
wider and deeper bulge --and minimal width-- near midgap, to having strong,
downwards-pointing resonance peaks (canine teeth) connected by a large bulge
in the lower band, plus reminiscences of the upwards-pointing resonance
peaks in the upper band. The splitting between the two bands (c) is smallest
at $k_{c}$=integer where the $\left\vert \mathbf{k}\right\rangle $ and $%
\left\vert \mathbf{k+c}^{\ast }\right\rangle $ characters cannot mix and
where the direct and indirect hoppings work in opposite directions.

In (d), we show the constant energy contours (CECs) for $k_{b}$ positive and
energies ranging from 100 meV below-- to 100 meV above the Fermi level which
is, itself, 75~meV above the center of the gap. For $E-E_{F}$=$-$100 meV, we
recognize from the LDA-TB part of FIG.\thinspace II~\ref{CEC} , two notches
pointing towards\textrm{\ }$\mathrm{Z}$ in the inner sheet and, in the outer
sheet, a bulge centered at the ZB, $k_{c}$=0.5. As the energy increases, so
does the distance between the bulge and the inner sheet, the notches shrink,
and new notches develop in the outer sheet, on either side of the bulge, and
pointing towards $\mathrm{Y}$ and $\mathrm{Y}^{\prime }\mathrm{.}$ It is
obviously the resonance peaks pointing $\QATOP{\mathrm{upwards}}{\mathrm{%
downwards}}$ in the $\QATOP{\mathrm{upper}}{\mathrm{lower}}$ band which give
rise to the notches pointing towards $\QATOP{\mathrm{Z}}{\mathrm{Y}\&\mathrm{%
Y}^{\prime }}$ in the $\QATOP{\mathrm{inner}}{\mathrm{outer}}$ sheets. Along 
$\Gamma \mathrm{Y}$ ($k_{c}$=integer), the two sheets are in near contact.

Since the quasi-1D bands disperse far more along $k_{b}$ than along $k_{c},$
the shape of two CECs in (d) resembles that of the two energy bands in FIG.$%
\,$\ref{9to11over40irrBZ} or in FIG.$\,$\ref{Analysis} on the 1st line, to
the right. The $E\leftrightarrow k_{b}$ scaling is approximately: $dE=-\tau
^{\prime }\left( k_{b}\right) dk_{b},$ with $\tau ^{\prime }\left(
k_{b}\right) $ being the dominating part of the Fermi velocity (I~\ref{lin}%
). This resemblance is less good close to the edge of the C- or the V-band
where the hybridization with the edge makes the two bands and the two CECs
behave differently: Whereas one band remains undistorted, the other gets
repelled and, eventually, fuses with the CECs of the edge [see
FIG.\thinspace II~\ref{CEC}].

If we interpret a CEC as a doped FS, an energy increase of 50~meV
corresponds to a 4\% increase of the electron doping, and the undoped FS is
the CEC whose $k_{b}$ averaged over $k_{c}$ equals $\frac{1}{4}$.

We shall now analyze the ARPES data for energies closer to the Fermi level
than the 0.15 eV studied in Paper II and identify further features of the
theory discussed above.

\section{Experimental FS and velocities and comparison with theory\label%
{SectFSExperiment}}

\begin{figure}[tbh]
\includegraphics[width=1.1\linewidth]{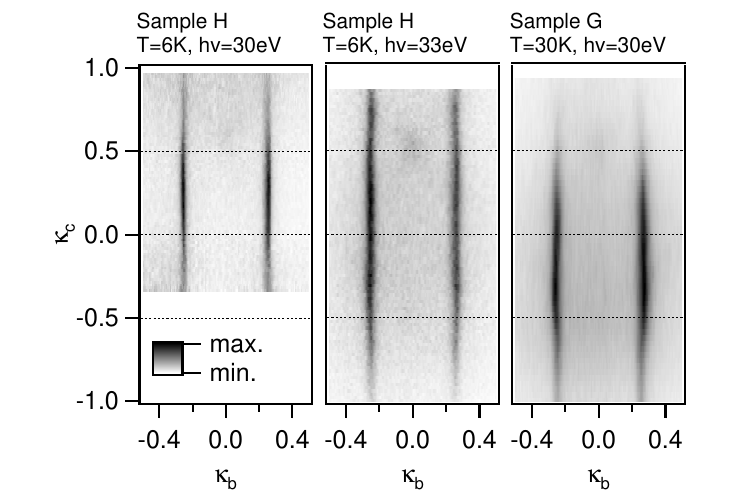}
\caption{Three Fermi-surfaces out of the data which we use in the following
analysis. The two first (similar to FIG.~II~\protect\ref{hv_30_33_37_kbkc})
are from a sample H at T=6K measured at two photon energies $h\protect\nu $%
=30 eV and 33eV, corresponding to $\protect\kappa _{a}$=6.3 and 6.6,
respectively. The last is from a sample G with $h\protect\nu $=30 eV, $%
\protect\kappa _{a}$=6.3. The figures have been \emph{stretched} along $%
\protect\kappa _{c}$ by a factor 1.7; compare with FIG.s II \protect\ref%
{FIGPhysicalZones2} and \protect\ref{CEC} (b), which are to scale. }
\label{Fig:AllFS}
\end{figure}

In this section we describe our analysis of the ARPES data taken from the intensity cube $I(E,\kappa _{b},\kappa _{c})$. Fig.~\ref{Fig:AllFS}  shows the FS obtained from sample H at T=6K measured at the two photon energies $%
h\nu $=30 eV and 33eV, which correspond to $\kappa _{a}$=6.3 and 6.6,
respectively, and from sample G with $h\nu $=30 eV, $\kappa _{a}$=6.3. We know from theory that the FS consists of the four values (left and right, inner and outer sheets) of the Fermi-momentum $\kappa _{Fb}$ as a function of $\kappa _{c}$ when the dispersion with $\kappa _{a}$ is neglected. 

At first glance the experimental FS does not appear to have two sheets.  However, one can clearly see by eye in Fig.~\ref{Fig:AllFS} that the black vertical lines (representing the FS) are not exactly straight, but are slightly “breathing” inwards and outwards in a regular pattern. In the following, we will examine this breathing more meticulously.  We will find that it is the result of our selection rule causing the observed FS to switch from one sheet to the other.  Further, the shape of each sheet is found to be consistent with our theory. 
Nonetheless, as we have shown in past work \cite{Gweon2003,Wang2006, Dudy2013}, and as discussed further below, we emphasize that the FS is defined by dispersing lineshapes with non-Fermi liquid features that are generic to the TL model.  The selection rule explains a puzzle from this past work, that we were able to analyze the lineshapes with a one-band TL spectral theory, even though there are actually two bands.  We now understand that for most of k-space, ARPES sees only one band at a time.
Before presenting and discussing the data in relation to our theory, we discuss the challenges and issues we are facing and explain the method we are using.  This method also enables the $\kappa _{b}$-projected Fermi-velocity to be extracted as a function of $\kappa _{c}$.

\subsection{Challenges, issues, and methods used for the Fermi-surface
determination\label{Challenges}}

\subsubsection{LiPB-specific issues}

As explained in detail in Sect.~II~\ref{SectIntensity}, the ARPES
intensities from the six $t_{2g}$ $\left( m=yz,xz,xy\right) $ bands in LiPB
display fine- and coarse-grained variations.

\begin{figure}[tbh]
\includegraphics[width=1.2\linewidth]{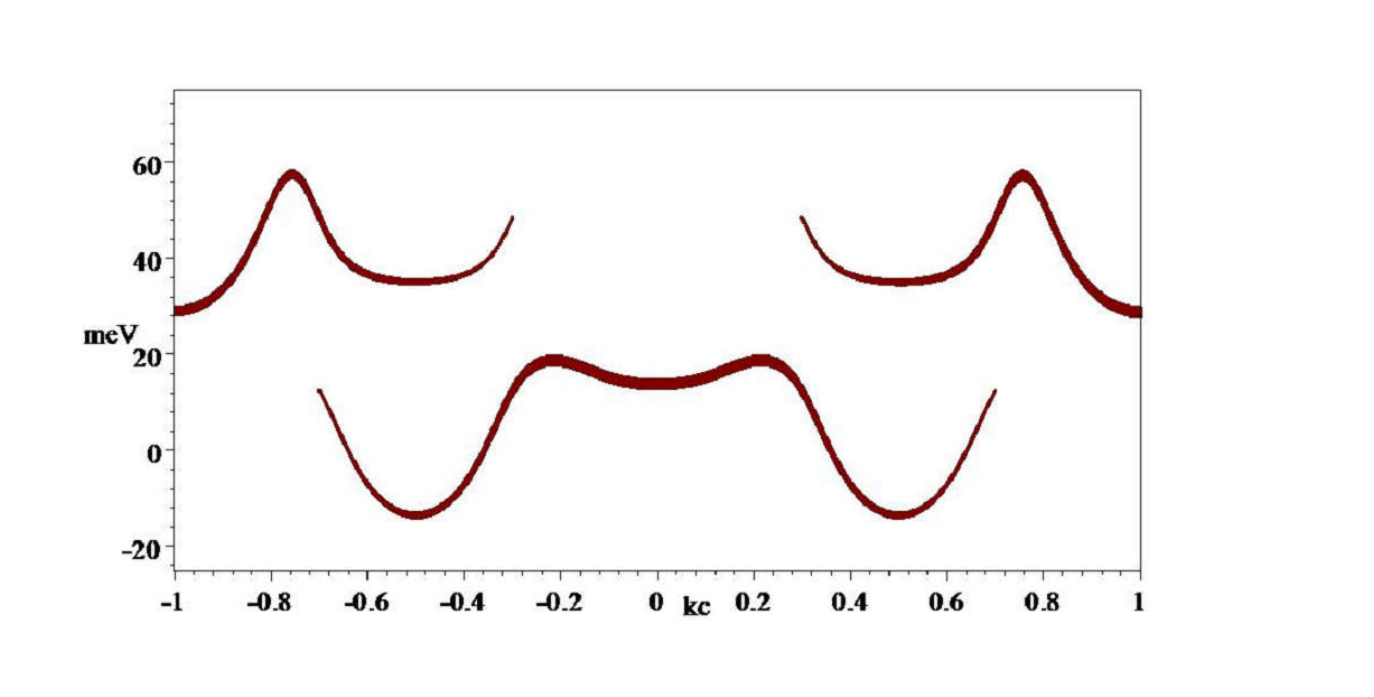}
\caption{The $\widetilde{xy}$ bands in the double zone for the two-band
Hamiltonian (\protect\ref{H2}) with ARPES-refined TB parameters, like in the
bottom panel of Fig.~II~\protect\ref{ThreeBandskonly}, but only for the
dark-red $\widetilde{xy}$ bands, for $k_{b}$=0.245, and on an extended
energy scale with the upper frame at the Fermi level (75 meV). The fatness
is proportional to the fine-grained band-factor ARPES intensity because for
our particular case of $\protect\kappa _{a}$=6.6, the dimerization phase
shift, $\protect\eta $ II$\,$(\protect\ref{eta}), is negligible in the $%
k_{c} $-region of interest (see Fig.$~$II$\,$\protect\ref{FIGxyZoneSelectul}%
).}
\label{Fig:RepeatPureK}
\end{figure}

While the two subbands of a given $m$ are periodic in the single zone, the
fine-grained intensity modulation considered in Sect.~II$~$\ref%
{Sectzoneselect} follows the $\left\vert \mathbf{k}\right\rangle $%
-character, apart from the nearly canceling phase shifts from the inversion
and displacement dimerizations. It is therefore almost periodic in the
double zone (Fig.s$~$II~\ref{FIGPhysicalZones2}, \ref{FIGxyZoneSelectul}, and %
\ref{FIGyzZoneSelect}), as if the ribbons had been translationally
equivalent (Sect.~I~\ref{SectDims}) with the intensity from the lower $m$%
-band in the 1st zone and the intensity from the upper $m$-band in the 2nd
zone. Due to the dimerization of the ribbons, the $m$-band is gapped at the
physical zone boundary and the shift from one subband to the other of the
dominating $\left\vert \mathbf{k}\right\rangle $-character takes place over%
\emph{\ a region around} this boundary. This was illustrated in the bottom
part of Fig.~II~\ref{ThreeBandskonly} for $k_{b}$=0.225 and we now repeat
this in Fig.~\ref{Fig:RepeatPureK}, but only for $k_{b}$=0.245 and the two $%
\widetilde{xy}$-bands, i.e. those with energies less than $\sim $100 meV
below $E_{F}.$ The tilde and the dark-red color indicate that these $xy$%
-like bands are hybridized with the valence(V)- and conduction(C)- $xz$-and $%
yz$-bands which give rise to the peak-, bulge-, and near-contact features
first mentioned in Sect. \ref{PB&C}. The 2nd zone extends from $k_{c}$=$-1$
to $-0.5$ and from $0.5$ to $1,$ with the 1st zone inserted between them,
from $-0.5$ to $0.5$. Only near the zone-boundaries, $k_{c}$=$\pm 0.5,$ does
ARPES see both bands of which the lower has a large bulge whose minimum is
split from the upper minimum by as much as 50 meV. The near contact between
the two bands [sheets in Fig. \ref{Theo_CEC} (d)] at $k_{c}$ integer is not
directly seen in ARPES, because here, one of the bands (sheets) is
extinguished, the upper band (inner sheet) at even $k_{c}$ and the lower
band (outer sheet) at odd $k_{c}$. The resonance peaks in the upper band are
clearly seen at $|k_{c}|$=0.76.

Since without the fine-grained ARPES intensity modulation, each band is
periodic in the single zone (see Fig.s$~$\ref{9to11over40irrBZ} and \ref%
{Analysis} top line to the right), symmetrization of the ARPES bands around
the zone boundaries enabled us in Paper II to reconstruct the dispersions
continuously (Fig.s~II~\ref{CEC} and II~\ref{ARPES_Bandstructure_Cuts}).
However, with $h\nu $=30 eV, the coarse-grained intensity of the $\widetilde{%
xy}$-bands falls off rapidly for $\kappa _{c}>0.5$, and this made it
difficult, but --thanks to the fine-grained modulations-- not impossible to
detect the resonance peaks in the upper band [see Fig.$~$II$~$\ref%
{ARPES_Bandstructure_TB}$($c2)]. Later, we realized that the spread of the
Wannier orbital onto several molybdenums makes the coarse-grained intensity
sensitive to the photon energy, which can therefore be choosen to yield good
visibility over a sufficiently wide range of $\kappa _{c}$ and to produce
cancellation between the inversion- and displacement phase shifts.

The ARPES intensity from the inner and outer sheets of the FS behaves like
the intensity for respectively the upper and lower $\widetilde{xy}$-bands,
as we shall see in Fig.~\ref{Fig:NewFSExtract}. For extracting the FS, our
present method to be described below, however fails near the zone boundaries
because we have no good symmetrization scheme unlike for the bands. We can
merely \emph{estimate} the splitting \emph{at} the boundary from the
so-called Sparrow criterion used in astronomy \cite{Jones1995}

To set the scale, we first recall from the theoretical FS in Fig.~\ref%
{Theo_CEC}~(d) that even at $k_{c}$=0.5, the $k_{b}$-splitting between the
inner and outer sheets amounts to merely 0.01$b^{\ast },$ i.e. to 2\% of the
distance between the left and right-hand FS sheets. The experimental Fig.~%
\ref{Fig:AllFS} shows both left- and right-hand sheets and, here, the inner
and outer sheets cannot be distinguished because the momentum-distribution
widths of the two spectral functions of the bands are larger than their
splitting (see Sect.~II~\ref{SectDrop}). However, already in the data, we
see a slight wrinkle around $\kappa _{c}$=$\pm 0.5$ caused by the shift of
intensity from one band to the other.

There is a second issue specific to LiPB; it is a quasi-1D material and, at
high enough temperatures, manifests a TL-like spectral function with a broad
spinon edge feature and a somewhat sharper holon peak feature \cite%
{Wang2006,Wang2009,Dudy2013}. For the data here, the momentum integrated
EDCs around $k_{Fb}$ gave a power-law like line-shape. Although the LiPB\
ARPES lineshape is well described by the TL spectral function at high
temperature, the spectra do not sharpen as much as expected in the theory%
\cite{Wang2009} at low temperature \footnote{%
It might be important here to remind again that, in the TL model, the Fermi
momentum $k_{F}$ remains well defined.}. Therefore our specific procedure,
described below, follows a route in which no theoretical spectral function
is forced onto the experimental data.

\subsubsection{General issues}

Note that the change of momentum in $\kappa_b$, visible in Fig. \ref%
{Fig:AllFS}, is small - we will see that it is in the range of 0.01 which
translates to the experimentally very demanding range of 0.012\AA $^{-1}$
using the solid-state definition (ssd) of reciprocal space (see Sect.$\,$\ref%
{crystal_structure} of Paper~I). For our measurements, that is about the
size of two detector pixels. We will see that it is very well possible to
extract such a \textit{relative} change from the ARPES data by our method.
The determination of the absolute value of the Fermi-momentum and the
filling is limited to a systematic error of about 1 \% .

\begin{figure}[tbh]
\includegraphics[width=\linewidth]{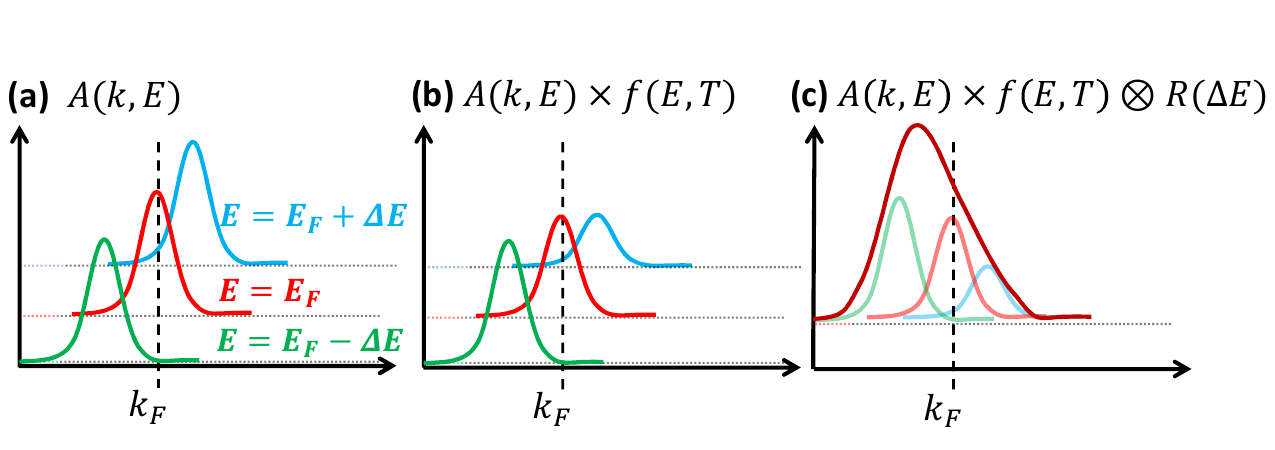}
\caption{Sketch to explain a masking effect hindering the Fermi-momentum
determination when only considering the maximum of the MDC at $E_{F}$. (a)
Shown are three MDC's for different energies as indicated around $E_{F}$. In
(a), the spectral function give perfectly symmetric and Lorentzian-like
MDC's. The combined effect of (b) the distribution function $f(E)$, and (c)
convolution with the energy resolution $R(\Delta E$) results in a an
asymmetric line-shape for the MDC at $E_{F}$ (dark red in (c)).}
\label{Fig:PaedagogicalMethodA}
\end{figure}

In order to obtain the value of the Fermi-momentum, one may think that it
suffices to determine the peak maximum of the MDC at $E_{F}$. In reality,
the \textit{exact} determination of the Fermi-momentum is an important issue
warranting discussions to be found in e.g. Refs.~%
\onlinecite{Lindroos1996, Strocov1998,
Kipp1999, Kaminski2005,Brouet2012}. There are many intrinsic and extrinsic
reasons not to determine the correct $k_{Fb}$. Particularly important for us
is a masking effect in extracting the Fermi momentum which occurs rather
generally and does not depend much on whether the sample is a Fermi-liquid,
marginal Fermi-liquid, or a TL-liquid. This effect is sketched in Fig.~\ref%
{Fig:PaedagogicalMethodA}: Even if (a) the ARPES spectral function is
perfectly symmetric and Lorentzian-like in the momentum direction, the
combined effects of the distribution function (e.g. the Fermi function) (b)
and the convolution with the experimental energy resolution (c), results in
an asymmetric line-shape for MDC's for energies near $E_{F}\pm \Delta E$.
Here, $\Delta E$ represents the experimental energy resolution, which for $%
h\nu $=30~eV is 16 meV (Sect.~\ref{SectDrop}). Hence, the asymmetry causes
the peak momentum to be \emph{inside} the FS.

\subsubsection{Method}

With these considerations in mind, we now explain our methods. It is
important to mention again that theory dictates that we cannot distinguish
the two bands but obtain an average dispersion weighted by the ARPES
band-factor intensity, the $\left\vert \mathbf{k}\right\rangle $-character
shown in Fig.~\ref{Fig:RepeatPureK}. This means that for $|\kappa _{c}|<0.35$%
, we are mainly extracting the outer FS sheet (lower band) while for $%
0.65<\left\vert \kappa _{c}\right\vert $, we mainly extract the inner sheet
(upper band). Thus, like for the bands, the selection rule allows us to
distinguish the two FS sheets, except near the zone-boundary, $\left\vert
\kappa _{c}\right\vert $=0.5, where the intensity shifts from one band to
the other. We first discuss, in Sect.~\ref{Sect:MDCanalysis}, what can be
deduced for the zone-boundary situation, using the Sparrow criterion~\cite%
{Jones1995} to analyze the MDC widths. This exercise will bring out the fact
that the experimental MDC widths greatly exceed the splitting that we wish
to determine and thereby make clear the crucial role of the selection rule,
which we exploit in Sect.~\ref{Sect:DispAna}, using a so-called dispersion
analysis, to determine the Fermi momentum as well as the Fermi velocity of
separate branches.

\paragraph{Sparrow MDC peak-width analysis:\label{Sect:MDCanalysis}}

At the zone boundaries, $\left\vert \kappa _{c}\right\vert $=0.5, although
we have no way to separate the inner and outer FS sheets, we can nonetheless
estimate their possible splitting, as we now describe. As follows from the
fine-grained intensity modulation of the two $\widetilde{xy}$-bands
discussed in Sect.~II~\ref{Sectzoneselect} and displayed in Fig. \ref%
{Fig:RepeatPureK}, the MDC at $E_{F}$ for $\kappa _{c}$=0.5 (or $-$0.5) is
an equally weighted combination of the Lorentzian-like\footnote{%
Here, we use the standard definition of the Lorentzian with $I=\frac{A}{\pi }%
\frac{\Gamma /2}{\left( \kappa -\kappa _{F}\right) ^{2}+\left( \Gamma
/2\right) ^{2}}$.} MDC's of both sheets and, as the combination, has the
measured total width $\Gamma _{MDC}^{\kappa _{c}=0.5}.$ With this, we can
use the so-called Sparrow criterion \cite{Jones1995}, which states that two
identical, separated Lorentzians of the same width $\Gamma
_{MDC}^{j=1}=\Gamma _{MDC}^{j=2}$ are indistinguishable if they add up to
give a flat top with zero slope and curvature at their center of mass. These
conditions allow computing a splitting ($\Delta \kappa _{b}$) as well as the
width of the two Lorentzians ($\Gamma _{MDC}^{1/2}$) from the measured total
width ($\Gamma _{MDC}^{\kappa _{c}=0.5}$). In fact, it will be an upper
limit for the splitting and reads: $\Delta \kappa _{b}^{\kappa
_{c}=0.5}=\Gamma _{MDC}^{\kappa _{c}=0.5}/\sqrt{3}$. The two Lorentzian have
a width of: $\Gamma _{MDC}^{j}=\Gamma _{MDC}^{\kappa _{c}=0.5}\,\sqrt{3}/(1+%
\sqrt{3})$. The results of this Sparrow analysis are given in Tab.~\ref%
{tab:Sparrow} and also indicated by the diamonds in Fig.~\ref%
{Fig:NewFSExtract}. From the table, by seeing that the measured MDC-widths
at zone-boundary ($\Gamma _{MDC}^{\kappa _{c}=\pm 0.5}$) and zone-center ($%
\Gamma _{MDC}^{\kappa _{c}=0}$) are almost the same, we can already see that
there is a limit to the determination of the splitting. The MDC's are
obviously broad. This broadening is much more than the momentum resolution
of the apparatus ($\Delta \kappa _{b}\approx $ 0.005, see Sect. II \ref%
{SectARPESMethod}). The broadening can be for different reasons. There is an 
\textit{intrinsic} component by the spectral function of the corresponding
electron liquid. In the case of a Luttinger-liquid, it is well known (see,
e.g., Ref. \onlinecite{Voit1993}) that the MDC's are typically sharper, and
the EDC's are broader (when compared to a Fermi-liquid). However, there can
also be an \textit{extrinsic} component --by an experimental momentum
resolution which is not produced by the apparatus alone but, for example,
can be caused by the quality of the samples-surface and the so-called $k_{z}$%
-broadening \cite{Grandke1978, Strocov2003} that arises from the limited
probing depth of the photoelectron. In general, it is hard and requires
multiple experiments to distinguish the intrinsic and extrinsic components
with full certainty. Looking at the widths for different samples, displayed
in Tab.~\ref{tab:Sparrow}, we can believe that there is a larger intrinsic
component but also some extrinsic component.

We reiterate that the resulting separation estimate by the Sparrow criterion
is an upper limit and the actual splitting has to be below the smallest
splitting listed in Tab.~\ref{tab:Sparrow}, resulting in $\Delta \kappa
_{b}^{\kappa _{c}=0.5}<0.017$. By taking advantage of the selection rule, we
will see that this is indeed the case, and we note now that even this upper
limit is considerably less than the magnitude of the experimental MDC width
at $\kappa_c=0$, where the selection rule applies and only one branch
contributes.

\begin{table}[th]
\begin{center}
\begin{tabular}{|c||c|c|c|c||c|}
\hline
Sample &  & $\Gamma_{MDC}^{|\kappa_{c}|=0.5}$ & $\Gamma_{MDC}^{j}$ & $%
\Gamma_{MDC}^{\kappa_{c}=0}$ & $\Delta\kappa_{b}^{|\kappa_{c}|=0.5}$ \\ 
\hline\hline
H 6K,30 eV & + & 0.030 & 0.0191 & 0.031 & 0.017 \\ \hline
H 6K,33 eV & + & 0.047 & 0.030 & 0.051 & 0.028 \\ 
& - & 0.044 & 0.028 & 0.051 & 0.025 \\ \hline
G 30K,30 eV & + & 0.060 & 0.038 & 0.052 & 0.035 \\ 
& - & 0.051 & 0.033 & 0.052 & 0.030 \\ \hline
\end{tabular}%
\end{center}
\caption{Result of the MDC peak-width analysis (compare also with Fig.~%
\protect\ref{Fig:AllFS}). The 1st column gives the individual dataset, the
2nd indicates whether $\protect\kappa _{c}$=$+$0.5 or $\protect\kappa _{c}$=$%
-$0.5 was measured. Column 3 gives the Lorentzian-width $\Gamma $ obtained
by a line-fit and is averaged over both branches, at positive and negative $%
\protect\kappa _{b}.$ Column 4 gives the width of the two Lorentzian ($%
\Gamma _{MDC}^{j}$) at $\protect\kappa _{c}$=$\pm $0.5 according to the
Sparrow criterion. The 5th is the Lorentzian-width at $\protect\kappa _{c}$%
=0. The last column, finally, shows the upper limit for the separation in $%
\protect\kappa _{b}$ momentum ($\Delta \protect\kappa _{b}^{\protect\kappa %
_{c}=0.5}$) according to the Sparrow-criterion.}
\label{tab:Sparrow}
\end{table}

\paragraph{Dispersion Analysis:\label{Sect:DispAna}}

\begin{figure}[tbh]
\includegraphics[width=0.8\linewidth]{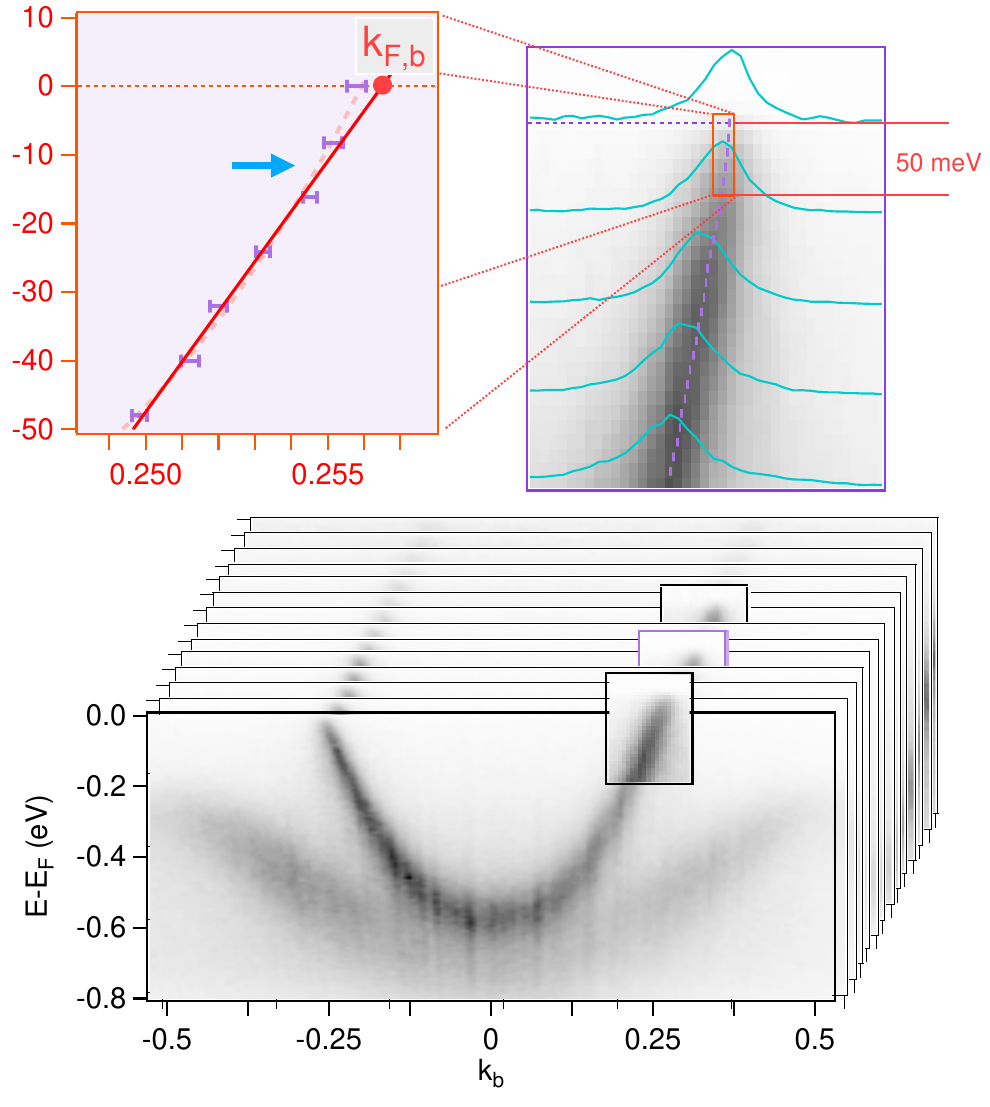}
\caption{Method used for the extraction of the Fermi-momentum. The
experimental data set is sliced along the $\protect\kappa _{c}$-direction
(bottom). Each $\protect\kappa _{c}$-slice can then be viewed as containing
multiple so-called MDC's. The MDC's are intensity curves at fixed energy,
dependent only on $\protect\kappa _{b}$ (see the magenta curves in the upper
right as examples). Then, a line fit with a Lorentzian for all the MDC's
between $E_{F}-50~$meV and $E_{F}$ was performed. The maximum of the
Lorentzian defines the dispersion $E(\protect\kappa _{b},\protect\kappa %
_{c}=const.)$ (upper left). The Fermi-momentum $\protect\kappa _{Fb}$ is
found by linear extrapolation in this interval between $E_{F}-50~$meV and $%
E_{F}$ (see red extrapolation line and red circle at $k_{Fb}$ in upper
left). }
\label{Fig:Experiment_Anal2}
\end{figure}

\begin{table}[th]
\begin{center}
\begin{tabular}{|c||c|c|c|}
\hline
& used &  & Li- \\ 
Sample & $\kappa_{c}$-range & $k_{Fb}^{avg}$ & stoichiometry \\ \hline\hline
H 6K,30 eV & 0 ; 0.995 & $0.255\pm0.006$ & $1.02 \pm 0.02$ \\ \hline
H 6K,33 eV & -0.785; 0.905 & $0.254\pm 0.005 $ & $1.02 \pm 0.02$ \\ \hline
G 30K,30 eV & -0.824; 0.777 & $0.255 \pm 0.008$ & $1.02 \pm0.02$ \\ 
\hline\hline
Average &  & $0.255 \pm 0.006$ & $1.02 \pm0.02$ \\ \hline
\end{tabular}%
\end{center}
\caption{Extracted absolute average Fermi-momentum $k_{Fb}^{avg}$ (3rd
column) and the Li-stoichiometry from the electron filling (last column).
The 1st column identifies the dataset used, the 2nd column the span of $%
\protect\kappa _{c}$ over which the average was taken (cf. FIG.~\protect\ref%
{Fig:NewFSExtract}). The error given here is the error of determining the
average and does not include the variation of $k_{Fb}(\protect\kappa _{c})$
with $\protect\kappa _{c}$. The last column gives the Li-stoichiometry from
the Luttinger count of the FS, including its error in determination. }
\label{tab:vF_and_filling}
\end{table}

\begin{figure*}[bth]
\includegraphics[width=\linewidth]{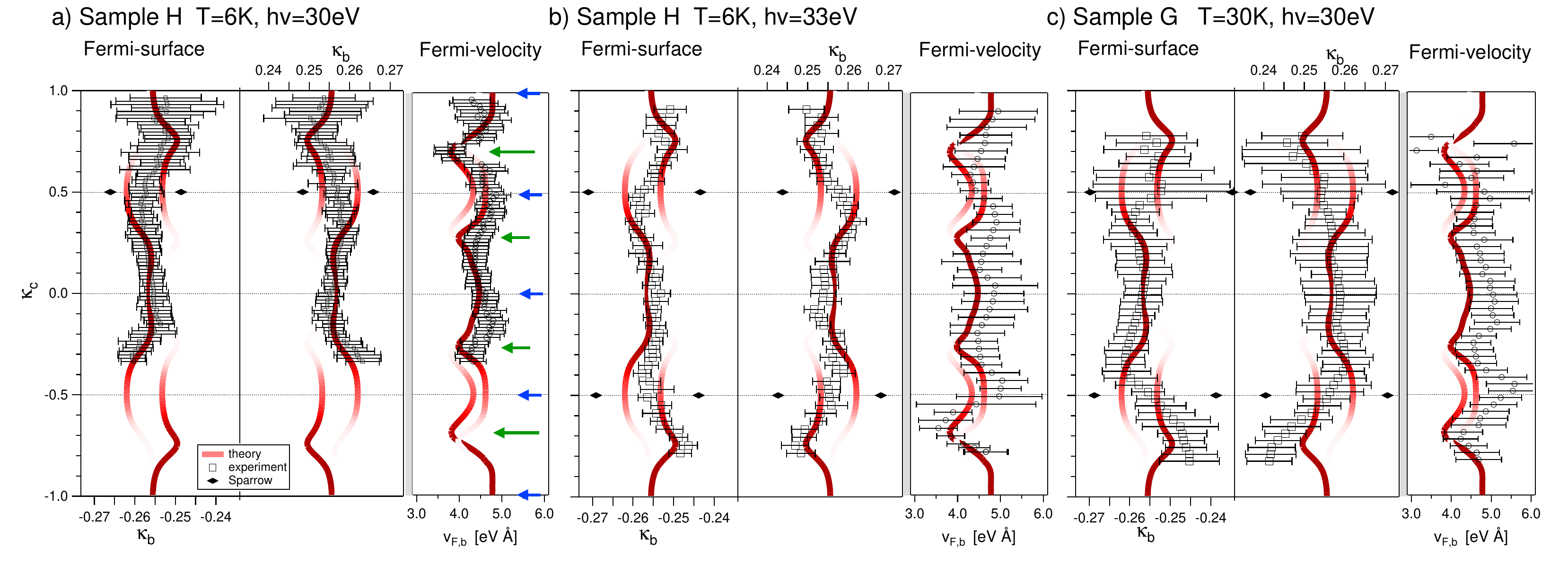}
\caption{Results of the Fermi-momentum extraction (black points with error
bars, see Sect. \protect\ref{Sect:DispAna}) in the double zone and for each
individual dataset, as indicated. For each set, the larger panel shows the
four FS sheets (left and right, outer and inner), while the smaller panel
shows the $k_{b}$-projected velocity (\protect\ref{vFj}) averaged using the
left and right $k_{b}$-projected velocity. Underlayed in dark red with an
intensity proportional to the $\left\vert \mathbf{k}\right\rangle $%
-character [Eq.~II~(\protect\ref{Indep}) with $\protect\eta $=0, i.e. Eq.~I~(%
\protect\ref{fat})] are the theoretical FS [Fig.~\protect\ref{Theo_CEC} (d)]
and its $k_{b}$-projected Fermi-velocity. Also included in the graphs for
the FS are the results (Tab.~\protect\ref{tab:Sparrow}) of the Sparrow
analysis for the maximum splitting at the zone boundary. $\protect\kappa %
_{a} $=6.3, 6.6, 6.3 in respectively \textbf{a)}, \textbf{b)} and \textbf{c)}%
. Each theoretical and experimental FS has been compressed by a factor 11
along $\protect\kappa _{c}.$}
\label{Fig:NewFSExtract}
\end{figure*}

The method used to extract the Fermi-momentum $\kappa _{Fb}$ as a function
of $\kappa _{c}$ is basically to extrapolate the metallic ARPES band along $%
\kappa _{b}$ for fixed $\kappa _{c}$ to the Fermi level (Fig.$~$II$~$\ref%
{ARPES_Bandstructure_Cuts}). The details are sketched in Fig. \ref%
{Fig:Experiment_Anal2}. The experimental data set is sliced along the $%
\kappa _{c}$-direction; see the slices on the lower part of the figure. Each
slice can be seen as a set of so-called momentum distribution curves
(MDC's). An MDC (see also Sect.\ref{SectAgreement} in Paper II) is the
photoelectron-intensity at fixed energy, here only dependent on $\kappa _{b}$
(see the magenta curves in the upper right of Fig. \ref{Fig:Experiment_Anal2}
as examples for MDC's). For the analysis, we choose now to use all MDC's
with energies between $E_{F}-$50~meV and $E_{F}$ (the separation $\Delta E$
was typically 5 meV). We then fit each MDC (on each dispersion branch) with
a Lorentzian. The maximum position of the Lorentzians defines the dispersion
(indicated by the purple lines in the upper left of Fig. \ref%
{Fig:Experiment_Anal2}). We interpolated the Fermi-momentum ($\kappa _{Fb}$)
by a line-fit with the linear function $\kappa _{b}=(E-E_{F})/v_{F}+\kappa
_{Fb}$ (see red line in upper left of Fig. \ref{Fig:Experiment_Anal2}). By
this line-fit we remove the error arising from the masking-effect discussed
above (see Fig.~\ref{Fig:PaedagogicalMethodA}) which appears as a "kink"
near the value of the energy resolution\footnote{%
This "kink" should not be confused with the "kink" resulting from an
interaction of a bosonic mode (for an example see, e.g. Ref. %
\onlinecite{Lanzara2001}). Such a bosonic "kink", for energies lower than
the energy of the bosonic mode, is bending to the momentum on the unoccupied
side- because the Fermi-velocity is re-normalized to smaller values.}. In
accordance with the energy resolution of about 16 meV, the kink sets in at
(see blue arrow) $E-E_{F}\approx 13~$meV in Fig.~\ref{Fig:Experiment_Anal2}.

The dispersion analysis offers a notable advantage in terms of statistical certainty, primarily due to the utilization of a broader data range spanning 50 meV. For the analysis of TL-liquid performed here, this holds particular significance as the density of states exhibits a power-law decay as it approaches E$_F$. Furthermore, the dispersion analysis also gives naturally an estimate of the $\kappa _{c}$-dependence
of the Fermi velocities projected onto the $\kappa _{b}$-direction, 
\begin{equation}
\mathbf{v}_{F\,j}\left( k_{c}\right) \cdot \mathbf{b}^{\ast }/b^{\ast
}\equiv \partial E_{j}\left( k_{b},k_{c}\right) \left/ \left( \partial
k_{b}b^{\ast }\right) \right\vert _{E_{F}}.  \label{vFj}
\end{equation}

The result of this extraction method is displayed in Fig.~\ref%
{Fig:NewFSExtract} and Tab.~\ref{tab:vF_and_filling}.

\subsection{Results of the Fermi-surface determination and Comparison with
Theory \label{SectFS}}

\subsubsection{Fermi-surface}

The result of the Fermi-momentum extraction is displayed in Fig.~\ref%
{Fig:NewFSExtract}. It compares the experimental FS (black points with error
bars) with the theoretical FS (dark-red) calculated using the two-band
Hamiltonian (\ref{H2}) with the ARPES-refined parameter values and drawn
with the fine-grained intensity proportional to the $\left\vert \mathbf{k}%
\right\rangle $-character I~(\ref{fat}) because for $\kappa _{a}$=6.3 and
6.6 the dimerization phase shift $\eta \left( \mathbf{\kappa }\right) $ is
neglible (see Sect.~II~\ref{Sectzoneselect} Eq.~(\ref{Indep}) and Fig.~\ref%
{FIGxyZoneSelectul}).

We see that the Fermi-momentum extracted from ARPES fits beautifully with
the theoretical FS sheet of dominating intensity: In all three measurements,
it aligns with the outer sheet for $|\kappa _{c}|<0.35$ and with the inner
sheet for $|\kappa _{c}|>0.65.$ Near the zone-boundaries, $\left\vert \kappa
_{c}\right\vert $=0.5, where the dominating intensity shifts from one sheet
to the other, so does the experimentally extracted $\kappa _{Fb}$. Also, the
general shape of the experimental FS fits nicely with the prediction by
theory. This is, on the one hand, not astonishing as we use ARPES-refined
parameter values, but on the other hand, the refinement was done for
features well away from the Fermi energy, and yet, all details of the
theoretical FS (seen in Fig.~\ref{Theo_CEC} (d) and described in Sect. \ref%
{SectBandsandFS}) are seen in the experiment, with the exception of the near
contact between the inner and outer sheets, which does require interpolation
of the former to $k_{c}$=0 or extrapolation of the latter to $k_{c}$=1.
Later on, we will discuss this also for the Fermi-velocities.

The experimental upper bound given by the Sparrow criterion on the splitting
between the inner and outer sheets at the ZB, $\left\vert \kappa
_{c}\right\vert $=0.5$,$ is indicated by the black diamonds. It is
consistent with, but considerably larger than, the theoretical splitting.

As can be seen from Tab. \ref{tab:vF_and_filling}, for all three samples, $%
k_{Fb}$ averaged over $k_c$ gives a Luttinger volume which corresponds to an
effective Li$_{1.02\pm 0.02}$ stoichiometry, i.e. an electron filling of $%
0.51\pm 0.01$. This places the Fermi level 75 meV above the center of the
gap in the calculation.

We note that the procedure of refining the values of the TB parameters to
fit the ARPES dispersions for energies more than 0.15 eV below $E_{F}$\ in
one sample does well in describing the dispersions for energies closer to $%
E_{F}$\ than 0.15 eV in all three samples. This is a testimonial, both to
the reproducibility of our findings for samples from different sources and
to the essential role that resonant coupling to the higher-energy gapped $xz$
and $yz$ bands plays in determining the dispersion and splitting of the
metallic $\widetilde{xy}$ bands.

Finally, we take notice of a very recent publication~\cite{Cohn2023}
reporting transport data interpreted as showing a FS reconstruction
resulting in a semi-metal FS below 100K. The modeling for this very
interesting proposal was based on our analysis, reported in Ref.~%
\onlinecite{Dudy2018v1} (Sect. VI E)~\footnote{%
This section was deleted before resubmission to PRB at the insistence of one
of two referees.}, of possible nesting and gapping of the metallic FS, which
however concluded that complete SDW gapping requires an effective exchange
interaction about three times the value given by the local spin-density
approximation (LSDA) and is therefore unlikely.

We point out, as is also acknowledged in Ref.~\onlinecite{Cohn2023}, that
the 6K ARPES data reported here and in Ref.~\onlinecite{Dudy2018v1} do not
show any evidence for such a FS reconstruction and that our measured Fermi
velocity ($\sim$4.6 eV\AA $\approx $7$\times$10$^5$ m/s) which was shown as
the slope of the unreconstructed, weak bands in Fig. 15 of Ref.~%
\onlinecite{Dudy2018v1} and of the dashed bands in Fig. S10 (c) of Ref.~%
\onlinecite{Cohn2023}, is several times larger than those ($\sim$2$\times$10$%
^5$ m/s $\approx$1.3 eV\AA ) for the gapped bands at the Fermi level in
Fig.s 15 and S10 (c).

\subsubsection{Fermi-velocity}

\label{Section:ExpFermiVelo}

To the right of each FS in Fig.~\ref{Fig:NewFSExtract}, we show the $k_{c}$%
-dependence of the $k_{b}$-projected Fermi-velocity, defined by Eq.(\ref{vFj}%
: in black, as extracted from ARPES and in red, as calculated with the
two-band Hamiltonian\footnote{%
The velocity projections of Eq.(\ref{vFj}) were calculated as differences
between the bands for $k_{b}$=0.2505 and 0.2495.}.

Overall, we see a good qualitative correspondence between the theoretical
and the experimentally extracted values. For those $k_{c}$ values where the $%
k_{b}$ direction is normal to the FS, the velocity projection (\ref{vFj})
has extrema. At the zone centers ($k_{c}$=integer) and zone boundaries ($%
\left\vert k_{c}\right\vert $=0.5) --indicated by blue arrows in Fig.~\ref%
{Fig:NewFSExtract}-- these extrema are flat maxima. For the inner-sheet
(upper-band) notches (green long arrows at $\left\vert k_{c}\right\vert
\approx $0.75), the extrema are deep \emph{minima}. Also the outer sheet
(lower band) has velocity minima (green short arrows at $\left\vert
k_{c}\right\vert \approx $0.30). Their origins are the weak resonance peaks
at $\left\vert k_{c}\right\vert \approx $0.25 (see Fig.~\ref{Fig:RepeatPureK}
and Eq.~(\ref{4peaks})), combined with the increase of $\left\vert \mathbf{k}%
\right\rangle $ \textbf{-} $\left\vert \mathbf{k+c}^{\ast }\right\rangle $
hybridization and the concomitant formation of bulges as the zone boundaries
at $\left\vert k_{c}\right\vert $=0.5 are approached.

The velocity of the outer sheet (lower band) decreases from 4.5~eV~\AA\ at
the zone center ($k_{c}$=0) to the 4.0~eV$~$\AA\ deep minima near $%
\left\vert k_{c}\right\vert $=0.30, and rises again to the 4.6~eV~\AA\ %
maxima at the centers of the bulges, $\left\vert k_{c}\right\vert $=0.5. For
the inner sheet (outer band), the velocity decreases from 4.7~eV~\AA\ at the
zone centers ($\left\vert k_{c}\right\vert $=1), to 3.9~ eV~\AA\ deep minima
near the notches, and rises again to 4.3~eV~\AA\ maxima at the zone
boundaries. These sheet- and $k_{c}$-dependent values may be compared with
the dominating value 4.6~eV~\AA\ of $b\tau ^{\prime }\left( k_{b}\right) $
in Eq.s~I~(\ref{vF}) and (\ref{lin}) from the direct hopping along the
ribbon. Due to the indirect hops via the valence and conduction bands giving
rise to the resonance terms in Eq.$~$(\ref{H2}), the band- and $k_{c}$
average of the velocity projections is smaller than $b\tau ^{\prime }\left(
k_{b}\right) $. The velocities extracted from the ARPES data
(black) clearly show both the qualitative behavior and the general magnitude
implied by the theory (dark-red). To a small extent, this is expected since,
as explained in Sect.$~$I$~$\ref{SectAgreement} and specified in Tables~I~%
\ref{taup} and \ref{A&Gp}, a few of the many LDA TB parameters were refined
to make the bands agree with the large- but not the small-energy features of
the ARPES bands.

It is also interesting for the many-body physics of LiPB to compare the
experimental velocities to those for the TB bands based on the LDA
parameters. The LDA dominant velocity value is\footnote{%
Calculated from $b\tau ^{\prime }\left( k_{b}\right) $ and also given in Eq.$%
\,$(\ref{vF}) of Paper I.} 4.0 eV~\AA . The experimental velocities (and
those for the ARPES-refined TB) are generally greater than those for the LDA%
\footnote{%
The band- and $k_{c}$-resolved velocities which result from using the
shifted and the straight LDA parameters (see FIG.~\ref%
{ARPES_Bandstructure_LDA}) have averages more than 15\% below that of the
experimental velocities in FIG.~\ref{Fig:NewFSExtract}, and they have much
larger variations: Near $k_{c}$=0 and 0.5 the LDA velocities do lie around
the 15\% lower LDA value, 4.0 eV\AA , but for intermediate values of $k_{c}$%
, they vary much more, reaching minima at 3.4 and 3.0~eV\AA\ for
respectively the upper and lower bands (inner and outer sheets) in the
shifted LDA and, in the straight LDA, minima at 2.5~eV\AA\ and 3.3~eV\AA\ %
with the deeper minimum now for the upper band.} by about 15\%.

There are two points to be made. First, for a 3D quasi-particle material the
increase of the experimental velocity relative to the LDA value would seem
surprising since the usual effect \cite{Mackintosh1980}, arising from an
energy-dependent single-particle self-energy, e.g., caused by ${e}${-}${e}$
or ${e}${-}${ph}$ interactions, is an \emph{in}crease of the Fermi-\emph{mass%
}, i.e. a decrease of the Fermi velocity \footnote{{For quasi-particles, one
would invoke a very strong k-dependence of the self-energy to understand a
decrease of the mass, as discussed, e.g., in Ref.\onlinecite{Miyake2013} and
Ref.\onlinecite{Wen2018}.}}. Indeed such was found in the single-site DMFT
quasi-particle treatment of LiPB \cite{Nuss2014}. However, LiPB is a
quasi-1D material whose ARPES $k$-averaged lineshapes show TL-model
properties, i.e. quasi-particle suppression and spin-charge separation.
Specifically, the holon-peak and spinon-edge features disperse with
different velocities, $v_{\rho }$ and $v_{\sigma }$, respectively \cite%
{Giamarchi2004, Voit1993, Orgad2001}. Our model-independent ARPES analysis
procedure, if performed on a TL-lineshape, would yield a dispersion
intermediate between $v_{\rho }$ and $v_{\sigma }$, but tending mostly to
that of the holon peak. Within 1D theory, $v_{\rho }$ and $v_{\sigma }$ can
just as well be either larger or smaller than the underlying $v_{F}$ of a
non-interacting system, as can be seen, for example from formulas within the
framework of the \textquotedblleft g-ology\textquotedblright\ formulation 
\cite{Solyom1979,Voit1993}. So if we identify the LDA value of $v_{F}$=4.0$~$%
eV~\AA\ as \textquotedblleft non-interacting,\textquotedblright\ which
ignores the difficulty of disentangling any many-body contribution already
present in LDA, and think of our ARPES lineshape in a TL context, it is well
within general theoretical expectations that our experimental velocity is
larger than the LDA value. In this view, it may well be that our
ARPES-refined TB description is modifying the entire $k_{b}$ dispersion
somewhat in order to reproduce the experimental low-energy scale velocity
near $E_{F}.$

Second, combining our results with a previous ARPES lineshape analysis \cite%
{Gweon2003}, we can be somewhat more precise about the velocity
renormalizations for LiPB. At high temperatures, where the LiPB ARPES
lineshapes are well described by TL lineshape theory \cite{Orgad2001} for
nonzero $T$, the best TL description \cite{Gweon2003} for the $\Gamma $-Y ($%
k_{c}$=0) ARPES lineshapes was achieved for $v_{\rho }/v_{\sigma }$=2. At
that time no definitive LDA value of $v_{F}$ was available. If we now think
of our ARPES velocity as being nearly that of $v_{\rho }$, and take our LDA
velocity as an underlying \textquotedblleft
non-interacting\textquotedblright\ $v_{F}$, then --at least for $\Gamma $Y--
we conclude that $v_{\rho }$ is roughly 1.15$v_{F}$, and that $v_{\sigma }$
is roughly 0.6$v_{F}$ \footnote{{{We take cognizance that the previous high-$%
T$, $\Gamma $-Y, TL lineshape analysis \cite{Gweon2003} found $v_{\rho }$=
4.0~eV\AA , coincidentally, we think, the same as our LDA value. That our
present value of roughly 4.6 eV \AA\ along $\Gamma $-Y ($k_{c}$=0) is
somewhat larger could perhaps be due to the considerable temperature
difference (250K vs. 6K), the considerable difference in the analyzer angle
resolutions along $k_{b}$ (0.016 \AA $^{-1}$ in the early work vs. 0.006 
\AA\ $^{-1}$\ in the present work), or perhaps some small sample dependence.
In any case the ratio $v_{\rho }/v_{\sigma }$ $\approx $ 2 is essentially
the same for the present lineshapes and the text conclusion that $v_{\rho }$
is nontrivially larger than $v_{F}$ is unaltered.}}}. A 1D Hubbard-model
analysis \cite{Chudzinski2012} estimated $v_{\sigma }/v_{F}\approx J/2\tau $
with $J\approx 0.2~$eV being an effective super-exchange interaction, and $%
\tau \approx 0.8~$eV being the primary $k_{b}$ hopping, implying $v_{\sigma
}/$ $v_{F}\approx 1/8.$

\subsubsection{Connecting to the TL critical exponent $\protect\alpha$}

As already written in the introduction of Paper I, LiPB displays Luttinger
liquid properties. Most remarkably, although the FS is well defined and
could be extracted above, the lineshapes are better described by a TL
spectral function. Nevertheless, there are also substantial $T$-dependent
departures from the TL model. The ARPES line shapes at temperatures 250K to
300K are well described by the TL spectral function, showing both spinon and
holon features (broadened by temperature and experimental resolution). With
decreasing $T$ the exponent varies with $T$, from $\alpha =0.9$ at 250K to $%
0.6$ at 50K, and also, the ARPES lineshape no longer agrees with the TL
lineshape, although it does continue to display quantum critical scaling, a
characteristic 1D property \cite{Wang2006}. The spectra for $k$-integration
along the quasi-1D direction, for temperatures $T$=4K and 30 K and
resolution 5 meV, are well described by a power law with $\alpha =0.7$ \cite%
{Dudy2013}.

In the following, we repeat the analysis of Ref.~\onlinecite{Dudy2013} with
our recent data sets used to determine the FS above. However, in the course
of the present study, we found that (i) all previous TB ladder models are
very unrealistic, and (ii) the detectable bands of ARPES vary with
fine-grained intensity proportional to the $\left\vert \mathbf{k}%
\right\rangle $-character. In principle, these findings should inform the
choice of a particular TL model. For the phenomenological extraction of the $%
\alpha $ value, however, we still use the same procedure as in Ref.~%
\onlinecite{Dudy2013} to obtain its value. As before, we use the $\kappa $%
-integrated spectral weight of a one-band, spin-rotational invariant
TL-model \cite{Orgad2001} with $v_{\rho}/v_{\sigma}$=2. The theoretical
spectrum was broadened by the experimental energy resolution of 16 meV.

\begin{figure}[tbh]
\includegraphics[width=0.8\linewidth]{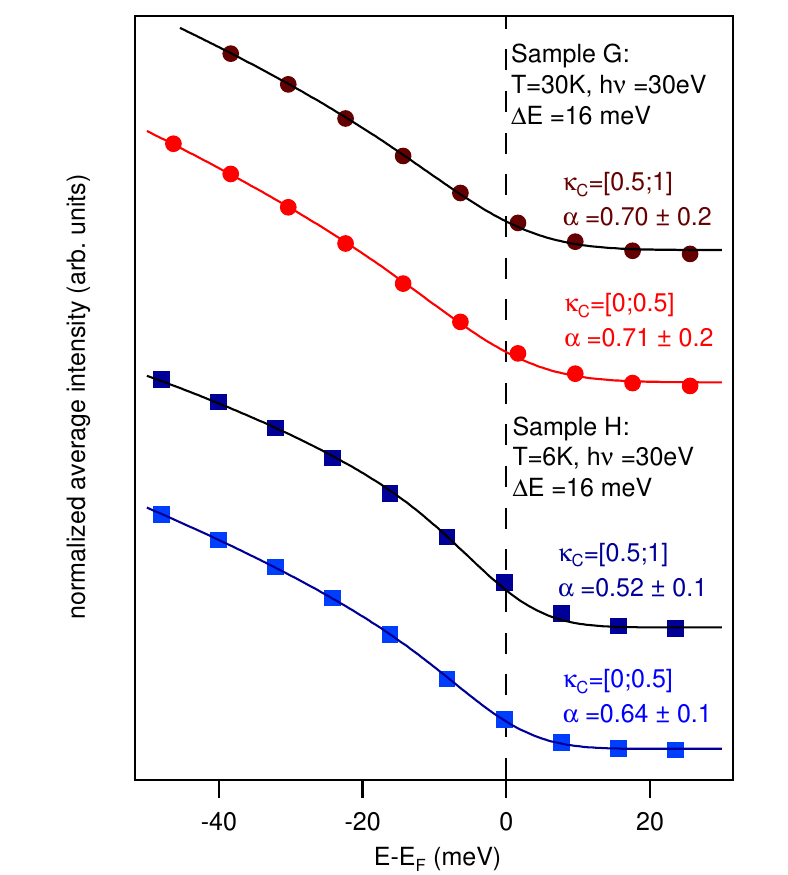}
\caption{Analysis of the TL-exponent of the data for sample G (T=30K) and H
(T=6K). The FS's were presented in Fig.~\protect\ref{Fig:AllFS}. Here the
data is momentum-integrated over two intervals, where the dominating band
character is either $\left\vert \mathbf{k}\right\rangle $ or $\left\vert 
\mathbf{k+c}\right\rangle ^{\ast }.$ }
\label{Fig:NewTL}
\end{figure}

The data displayed in Fig. \ref{Fig:NewTL} are integrated over $\kappa _{b}$
and $\kappa _{c}$ (cf. Fig.\ref{Fig:AllFS}). For the $\kappa _{c}$%
-direction, the data are integrated over two intervals, according to whether
the dominant character is $\left\vert \mathbf{k}\right\rangle $ or $%
\left\vert \mathbf{k+c}^{\ast }\right\rangle $. As visible in Fig. \ref%
{Fig:NewTL}, sample G (T=30K) shows a typical value of $\alpha =0.70$ for
both intervals. The data for sample H (T=6K) vary a bit more on the two
intervals but are still within the error range, yielding in average $\alpha
=0.58$. Connecting to the study of Ref. \cite{Wang2006}, despite a
considerable difference in the determined $\alpha $-values for samples G and
H, these values are within the $\alpha $-ranges of former studies. To shed
light on more details, temperature-dependent measurements in the full range
of relevant momenta would be required as well as the usage of a more
realistic TL model than the one-band, spin-rotational invariant TL-model for
the line-fit.

\section{Interactions and correlations}

\label{Sect:InteractionsAndCorrelations}

As described in the first paragraph of Paper I, for interacting electrons
that can only propagate in one spatial dimension Landaus Fermi-liquid (FL) theory does not hold \cite{Luttinger1963, Tomonaga1950,
Giamarchi2004}. Instead one finds a correlated Luttinger-liquid (LL) system
with a zero density of states at the Fermi energy. Furthermore, the density
of states increases at finite energy away from the Fermi energy following a
Tomonaga-Luttinger (TL) model line-shape characterized by an anomalous
exponent $\alpha$. As shown in Fig. 1 of paper I and Fig. \ref{Fig:NewTL} of
paper III we indeed observe such a line-shape of the density of states below
the Fermi energy in LiMo$_6$O$_{17}$. Although this LL behavior occurs for
even an infinitesimally small Coulomb interaction, nonetheless the actual
magnitude -and also the range- of the interaction is important because these
determine the values of correlation function power laws such as $\alpha$, as
well as other possible properties of the actual interacting state in any
particular material. Thus it is generally important to try to make an
estimate of the interaction strength and range.

Using constrained DFT Popovic and Satpathy\cite{Satpathy2006} showed that
the Coulomb interaction $U$ between two electrons in atomic d-like orbitals
of Mo in LiPB is $U_{Mo-d}$ = 6.4 eV. Constrained RPA calculations for a
dense f.c.c. lattice of atomic Mo yield a value of $U_{Mo-t_{2g}}$ = 3.7 eV
for the atomic-like $t_{2g}$-orbitals of a single Mo atom \cite{Sasioglu2011}%
. However, it is important to realize that the Wannier orbitals that build
the one-dimensional bands are far from being atomic and localized on a
single Mo atom. As we showed the smallest tight binding model one can make
for LiMo$_6$O$_{17}$ considers 6 bands per unit cell, and the resulting
Wannier orbitals spread over several unit cells (see Figs. \ref{FIGyzb&xyamc}
and \ref{Wannier} of paper I). Thus, the parameter determining the Coulomb
interaction strength for the Wannier orbitals will be smaller than the
Coulomb parameter for a single atomic Mo orbital, due both to the dilution
of the charge density to several Mo atoms and to screening. Calculating the
screened Coulomb interaction between such large Wannier orbitals for a
material with many atoms per unit cell is a daunting task. Neither accurate
constrained random phase approximation nor constrained density functional
theory calculations seem feasible with the currently available codes.

Whilst it is computationally difficult to reliably calculate the screened
Coulomb interaction of such large Wannier orbitals, one can very well
calculate the bare Coulomb interaction. We have made this calculation and
the result is $U^{Bare}_{Mo-W_{xy}}= 2$ eV. We can easily rationalize this
result by noting that the bare Coulomb interaction (F0) for the atomic 4d
orbitals in Mo is $U^{Bare}_{Mo-d}= 12.5$ eV and that the Wannier orbitals
in LiPB are spread over several (5+) Mo atoms. Due to the large spread of
the Wannier function the bare value of the nearest neighbor Coulomb
interaction $V$ (and also those for even further distant neighbors) will be
smaller, but still sizeable compared to $U.$ Noting that the bare value of $%
U^{Bare}_{Cu-d} = 25$ eV for strongly correlated cuprates and is $%
U^{Bare}_{C-p} = 14$ eV for graphene gives a useful perspective on the much
smaller value of the bare $U$ in LiPB. We should warn the reader here that
this value $U^{Bare}_{Mo-W_{xy}}= 2$ eV is not the value one should use in a
model calculation. The bare interaction sets a clear upper limit but should
be greatly reduced in the material due to solid state screening. As a guide
to the possible magnitude of the screening effect, we note again that
Popovic and Satpathy calculated a screened value of 6.4 eV for atomic Mo in
LiPB, roughly half the bare atomic value $U^{Bare}_{Mo-d}= 12.5$ eV. A
similar screening reduction down from the bare 2 eV value for our Wannier
orbitals is expected.

The small value of the Coulomb interaction between the extended Wannier
orbitals also explains why we experimentally find that the $xz$ ($yz$) and $%
XZ$ ($YZ$) bonding orbitals are doubly occupied forming an $S = 0$ state
instead of favoring a Hunds rule ground-state triplet
state with $S = 1$ where one electron is in the $xz$ ($XZ$) orbital and one
electron in the $yz$ ($YZ$) orbital with parallel spin. One should not
compare the gap $4G_1\sim$ 400 meV (Eq.s (\ref{gap}), (\ref{Hsub}), and (\ref%
{A&Gp}) in Paper I) to the bare atomic interaction strength, but to the
greatly reduced interaction strength of the Wannier functions due to the
spread of the Wannier function over several Mo atoms within the unit cell.

For LiPB there are a number of previous theoretical studies aimed at
understanding its various interesting strongly correlated quasi-1D
properties \cite{Merino2012,Chudzinski2012,Merino2015,Cho2015,Lera2015,
Platt2016,Chudzinski2017,Sepper2013,Lebed2013}. Some of the various studies
take specific values for $U$- and $V$-type interactions, for $U$ in a range
from the Popovic-Satpathy value of 6.4 eV down to 1 eV, and for $V$-type
interactions in the range 0.5 eV to 1 eV. It is not our purpose here to give
any critique of these various models, which are elegant and creative, but
the magnitude of the Coulomb interaction for our Wannier orbitals is
supportive of the models which take $U$ at the low end of the range. The
experimental $\alpha > 1/2$ can be rationalized with both large and small
values of $U$ \cite{Merino2015, Cho2015}, but in so doing, a role for $V$%
-type neighbor interactions is essential, because for a simple single-chain
Hubbard model the maximum value of $\alpha= 1/8$ is obtained only for
infinite $U$. The $V$-type values used are all generally consistent with our
findings.

We note that Ref.~\cite{Nuss2014}, using the variational cluster
approximation (VCA) and dynamical mean field theory (DMFT), found that for
values of $U_{Mott}$ exceeding 0.7 eV and 2.5 eV, respectively, a
Mott-Hubbard gap opens in their LDA bands, glaringly inconsistent with
experiment, and therefore giving a clear upper limit. They note that the
DMFT (VCA) overestimates (underestimates) $U_{Mott}$, so the two methods
provide a range for an upper limit on $U$ for the 4-orbital set of Ref.~\cite%
{Nuss2014}. Roughly, we expect that a set reduced in size by a factor of $n$
will be $n$-times less localized, with $U$ about $n$-times smaller, implying
for our final 2-orbital set a general reduction in Coulomb energies by a
factor of 2/4=1/2, relative to the 4-orbital set of Ref.~\cite{Nuss2014}.
Thus our conclusion on the smallness of $U$ is generally consistent with the
findings of Ref.~\cite{Nuss2014}.

Ref.~\cite{Nuss2014} also emphasized that the two metallic bands are
half-filled. This question is important because most of the theories cited
above assume no $b$-dimerization, i.e. assume that Mo(1) and Mo(4) are
equivalent, and thereby view the two metallic bands as being quarter-filled.
We agree with this emphasis in Ref.~\cite{Nuss2014}, in the sense that the
dimerization gap is 0.7 eV, nearly 20\% of the entire bandwidth (see Sect
III B in Paper I). But there is some nuance. The motivation in the
theoretical models is to make contact with a property of the quarter-filled
Hubbard model, that of so-called 4k$_F$ charge fluctuations, or even charge
ordering, resulting from the $V$-type interactions. Dimerization is known to
greatly diminish the tendency to such charge fluctuations~\cite{Ejima2006},
i.e. the two compete. The fact that the estimated magnitudes of the $V$-type
interactions are in fact comparable to the dimerization gap implies that the
relevance of the quarter-filling scenarios is not ruled out, presumably
depending on details. So on the theory side, further assessment using more
realistic modeling is essential. On the experimental side, Ref.~\cite%
{Merino2015} proposes measurements to directly search for the 4k$_F$%
-charge-fluctuations.

\section{Conclusion and Implications\label{SectConclusion}}

In conclusion, we have presented in great detail the electronic structure of
LiMo$_{6}$O$_{17}$\ that is experimentally obtainable using ARPES,
emphasizing the degree of one-dimensional behavior of the bands in the
vicinity of $E_{F}$\thinspace\ and the excellent overall agreement with the
LDA band structure. With the aim of fully describing and understanding the
metallic bands found in the ARPES experiment, especially the details of FS
splitting and warping, the LDA electronic structure was downfolded to a
tight-binding description with the three Mo1-centered $t_{2g}$ Wannier
orbitals (WOs) per formula unit (Sect.~I~\ref{SectH}) using the newly
developed full-potential version of the NMTO method (Sect.~I~\ref{SectElCalc}%
). This description is based on analyzing the LiPB\ crystal structure as
built from corner-sharing MoO$_{6}$ octahedra forming a staircase running
along $\mathbf{c}$ of bi-ribbons extending along $\mathbf{b}$ (Sect. ~I~\ref%
{crystal_structure}).

The six $t_{2g}$ WOs per primitive cell accurately describe not only the
four bands seen by ARPES, but all six bands in the 1~eV neighborhood of $%
E_{F}$. This band structure (Sect.\ I~\ref{SectElStruc} and Fig.~I~\ref%
{3Dt2gBands}) is basically 2D and formed by the $xy,$ $xz,$ and $yz$ WOs
(Fig.~I~\ref{Wannier}) giving rise to three 1D bands running along
respectively $\mathbf{b},$ $\mathbf{c}\mathbf{+b},$ and $\mathbf{c}\mathbf{-b%
},$ i.e. at a 120$^{\circ }$ angle to the two other bands (Fig.s~I~\ref%
{FIGDoubleZone} and II~\ref{CEC}). The dimerization from $c/2$ to $c$ of the
ribbons into bi-ribbons gaps the $xz$ and $yz$ bands and leaves the $xy$
band metallic in the gap, but resonantly coupled to its edges and, hence, to
the $\mathbf{c}\mathbf{+b}$ and $\mathbf{c}\mathbf{-b}$ directions.
Inclusion of the $xz$ and $yz$ bands are indispensable in describing the
strong indirect contributions to the $k_{c}$-dispersion and splitting of the
metallic $xy$ bands. These are most prominent (see Fig.s~I~\ref%
{FIGDoubleZone}, ~II~\ref{CEC}, and ~III~\ref{Fig:NewFSExtract}) at the
crossing of the $xy$-band CECs running parallel to the $\mathrm{P}_{1}%
\mathrm{Q}_{1}\mathrm{P}_{1}^{\prime }$-line ($k_{b}$=0.225$)$ in reciprocal
space with those of the $xz$ and $yz$ V\&C-band edges along respectively $%
\mathrm{ZY}^{\prime }$ and $\mathrm{ZY.}$ All the ARPES-measured
dispersions, as well as the FS, indeed confirm the resonant indirect
couplings and thus the essential need for the six-band picture. The TB bands
are very well described by an analytic 6$\times $6 Hamiltonian I~(\ref{Hsub}%
) or (\ref{HRecip}) with parameters optimized to match the ARPES data for
energies more than 0.15 eV below $E_{F}$. Finally, the mix of direct and
resonant indirect couplings along the $\mathbf{c}$-direction can be
explicitly displayed by further analytical downfolding to the effective 2$%
\times $2 Hamiltonian III~(\ref{H2}). This and direct observation by ARPES
is compelling evidence for the existence of pronounced resonance structures
near $E_{F}$\ in LiPB.

In section \ref{Sect:InteractionsAndCorrelations}, we have presented some
implications of our results specifically regarding interactions and
correlations. In addition, there are four important implications for the
general questions posed in the Introduction of Paper I. These implications
follow directly from the central content of the paper, our new knowledge,
and understanding of the size of the splitting and perpendicular dispersions
of the quasi-1D bands in the gap (Fig.s$\,$III~\ref{9to11over40irrBZ} and %
\ref{Analysis}), especially the indirect resonance contributions. They have
been stated already in the flow of the presentation, and we merely summarize
them here.

First, the reality of the resonance contributions casts serious doubts on
theoretical descriptions based on TB bands which are featureless like the
red ones in Fig.s$~$\ref{9to11over40irrBZ} and \ref{Analysis}, i.e. casts
doubt on all previous TB and TL models. Further, in constructing an
appropriate many-body Hamiltonian, it should be taken into serious
consideration that with ARPES, we have now been able to follow the resonance
peak induced by the valence band to energies nearly 150 meV below the Fermi
level (Fig.$\,$II \ref{CEC}) and, there, find the peak to have a magnitude
of about $50$ meV, as predicted by the LDA, cf. Fig.$\,$II~\ref%
{ARPES_Bandstructure_TB}$\,$(c2).

Second, the general magnitude of the $t_{\perp }$ -hoppings would suggest
that 1D to 3D crossover should occur for $T$ as high as at least 150 K,
unless thwarted by the theoretically expected strong downward low-$T$
renormalization due to LL fluctuations on the chains, as pointed out in the
Introduction. However, the good agreement between LDA and ARPES data at $T$%
=6 K implies that this renormalization does not take place. This
circumstance is not only puzzling, given the evidence for LL effects on the
chains at high $T$, but eliminates one very attractive explanation for the
exceptional stability of quasi-1D behavior in this material. Our new
quantitative knowledge of the $t_{\perp }$-hoppings further emphasizes this
puzzle.

Third, the coupling of the quasi-1D bands to the V\&C bands causes the
details of the FS splitting and warping (Fig.s$\,$\ref{Theo_CEC} and \ref%
{Fig:NewFSExtract}) to depend strongly on the position of $E_{F}$\thinspace
which in turn depends on the Li concentration. This implies that any
property sensitive to the details of the FS will be very sensitive to the
stoichiometry. One can then speculate that this FS sensitivity is connected
to the sample dependence of the SC, especially if the SC is the product of
the quasi-1D nature of the FS. We have already noted that the actual
position of $E_{F}$\ in LiPB\ is such as to maximize the quasi-1D nature of
the FS. This could be an important addition to the various previous theories
of the SC \cite{Sepper2013,Lebed2013,Cho2015,Lera2015,Platt2016}.

Fourth, the spatial dependence of the $t_{\perp }$-hoppings argues strongly
against coupled ladder models of the chains. At the simplest level, the
magnitude of the direct terms for hoppings within and between bi-ribbons
(Sect.~I~\ref{SectElStruc}) differ by less than a factor of two,
respectively, $t_{\perp ,1}\equiv $ $-\left( t_{1}+u_{1}\right) =$ $14$ meV
and $t_{\perp ,2}\equiv $ $-\left( t_{1}-u_{1}\right) =$ $8$ meV. Just this
would leave the ladders not very well defined as separable objects. But,
much more importantly, the range of the indirect contributions is at least
an \emph{order of magnitude} longer and even depends crucially on the
position of the $\widetilde{xy}$ bands in the gap (Sect.~III~\ref%
{SectDispInGap}). We conclude that modeling the chains as separable, weakly
coupled ladders is very unrealistic.

To conclude, our results offer both a strong motivation and a concrete
framework for a serious reappraisal of the extent to which the various past
many-body models capture the actual measured one-electron electronic
structure of LiPB sufficiently well to be trusted for rationalizing its
fascinating quasi-1D, many-body, coupled-chain physics. The efforts to
understand the resulting behavior are still ongoing, and we hope that our
new knowledge and highly portable description of the one-electron electronic
structure will contribute to this effort.

\bibliographystyle{apsrev}
\bibliography{LiPB}

\end{document}